\documentclass[pra,aps,twocolumn,superscriptaddress,amsmath,amssymb,showpacs]{revtex4-1}
\pdfoutput=1
\usepackage{graphicx}
\usepackage{dcolumn}
\usepackage{bm}
\usepackage{hyperref}
\usepackage{tikz}
\usepackage{mathtools}
\usepackage{physics}
\usetikzlibrary{matrix,calc}
\usepackage{karnaugh-map}
\usepackage{kvmap}
\usepackage[caption=false]{subfig}
\usepackage{graphicx}
\usepackage{braket}
\usepackage{hyperref}
\usepackage{multirow}
\usepackage{stackengine}
\usepackage{array}
\usepackage{xcolor,color}
\usepackage{lipsum}
\usepackage{multirow}
\usepackage{stackengine}
\usepackage{enumitem}
\usepackage{courier}
\usepackage[normalem]{ulem}
\usepackage{amsthm}

\usetikzlibrary{quantikz}

\newcolumntype{P}[1]{>{\centering\arraybackslash}p{#1}}

\usepackage{array}
\newcolumntype{?}{!{\vrule width 1pt}}
\newcommand\xrowht[2][0]{\addstackgap[.5\dimexpr#2\relax]{\vphantom{#1}}}

\newtheoremstyle{plainindent}
  {\topsep}   
  {\topsep}   
  {\itshape}  
  {\parindent}
  {\bfseries} 
  {.}         
  {5pt plus 1pt minus 1pt} 
  {}          
\theoremstyle{plainindent}

\begin{document}


\title{Clifford Manipulations of Stabilizer States:\\{\small{A graphical rule book for Clifford unitaries and measurements on cluster states, and application to photonic quantum computing}}}
\author{Ashlesha Patil}
\email{ashlesha@arizona.edu}
\affiliation{NSF-ERC Center for Quantum Networks, \\ Wyant College of Optical Sciences, The University of Arizona, 1630 East University Boulevard, Tucson, AZ 85721, USA.}
\author{Saikat Guha}
\affiliation{NSF-ERC Center for Quantum Networks, \\ Wyant College of Optical Sciences, The University of Arizona, 1630 East University Boulevard, Tucson, AZ 85721, USA.}

\begin{abstract}
Stabilizer states along with Clifford manipulations (unitary transformations and measurements) thereof---despite being efficiently simulable on a classical computer---are an important tool in quantum information processing, with applications to quantum computing, error correction and networking. Cluster states, defined on a graph, are a special class of stabilizer states that are central to measurement based quantum computing, all-photonic quantum repeaters, distributed quantum computing, and entanglement distribution in a network. All cluster states are local-Clifford equivalent to a stabilizer state. In this paper, we review the stabilizer framework, and extend it, by: incorporating general stabilizer measurements such as multi-qubit fusions, and providing an explicit procedure---using Karnaugh maps from Boolean algebra---for converting arbitrary stabilizer gates into tableau operations of the CHP formalism for efficient stabilizer manipulations. Using these tools, we develop a graphical rule-book and a MATLAB simulator with a graphical user interface for arbitrary stabilizer manipulations of cluster states, a user of which, e.g., for research in quantum networks, will not require any background in quantum information or the stabilizer framework. We extend our graphical rule-book to include dual-rail photonic-qubit cluster state manipulations with probabilistically-heralded linear-optical circuits for various rotated Bell measurements, i.e., fusions (including new `Type-I' fusions we propose, where only one of the two fused qubits is destructively measured), by incorporating graphical rules for their success and failure modes. Finally, we show how stabilizer descriptions of multi-qubit fusions can be mapped to linear optical circuits.
\end{abstract}

\maketitle

\tableofcontents

\section{Introduction}
\label{sec:intro}
\begin{figure}
    \includegraphics[width=\linewidth]{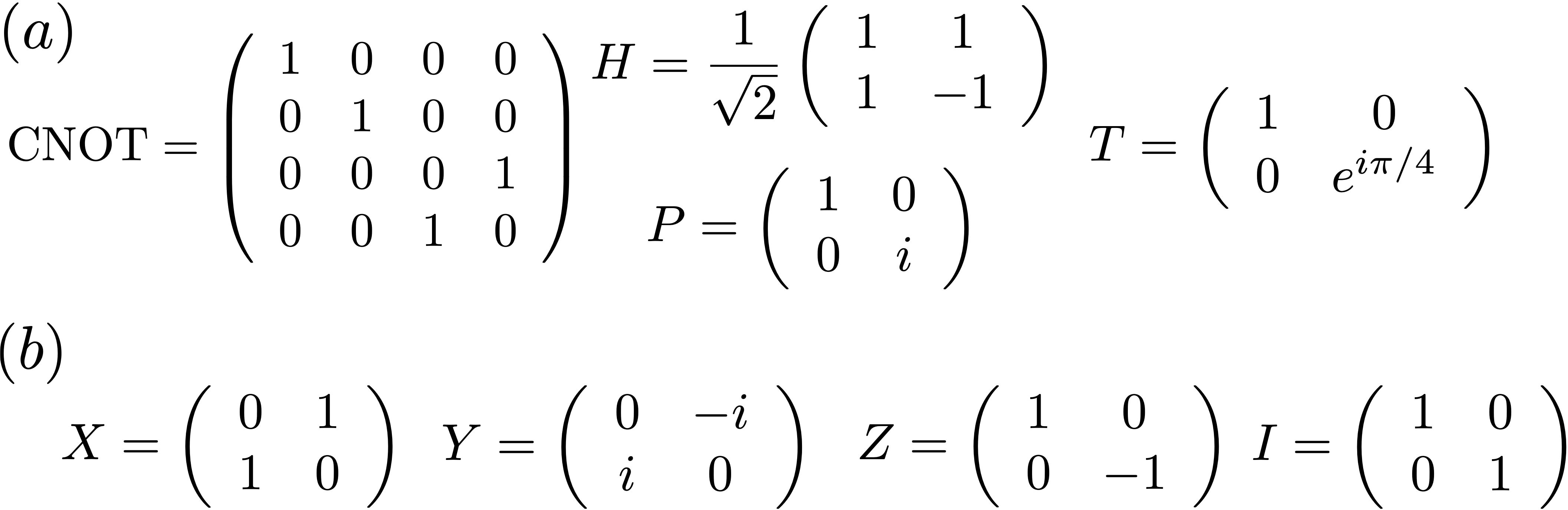}
    \caption{(a) A stabilizer circuit on $n$ qubits is defined to be a quantum circuit composed of the {\em stabilizer gates}---the C (two-qubit controlled-NOT), H (one-qubit Hadamard), and the P ($\pi/4$ phase) gates---also known as {\em Clifford gates}---followed by measurement of each qubit in the computational basis. Universal quantum computing requires in addition at least one non-Clifford gate, e.g., the $T$ ($\pi/8$ phase) gate. (b) Single-qubit Pauli matrices $X$, $Y$ and $Z$, and the identity matrix $I$.}
    \label{fig:stabilizers_Paulis}
\end{figure}

{\em Stabilizer circuits}, introduced by Gottesman and Knill~\cite{gottesman1996class}, are quantum circuits built with {\em stabilizer gates}---unitaries composed of Controlled-NOT (CNOT), Hadamard ($H$) and the $\pi/4$ Phase ($P$) gates (see Fig.~\ref{fig:stabilizers_Paulis})---followed by measurement of each qubit in the ($|0\rangle\langle 0 |, |1\rangle\langle 1 |$) basis, where $|0\rangle = \left(\begin{array}{c}1 \\ 0\end{array}\right)$, and $|1\rangle = \left(\begin{array}{c}0 \\ 1\end{array}\right)$. A {\em stabilizer state} is one that can be produced by a stabilizer gate applied to a computational basis state, e.g., $|0, \ldots, 0\rangle$. Stabilizer circuits acting on a stabilizer state can be simulated efficiently on a classical computer. Such is done by evaluating the $n$-qubit entangled (stabilizer) {\em state} evolving through an $n$-qubit stabilizer circuit, by keeping track of a set of $n$ {\em stabilizers} that uniquely identify the state. Stabilizers identifying a state are independent, commuting $n$-fold tensor-product Pauli operators (see Fig.~\ref{fig:stabilizers_Paulis} for the single-qubit Pauli operators)---that {\em stabilize} the quantum state. An operator $U$ is a {\em stabilizer} of a quantum state $|\psi\rangle$ if and only if $U|\psi\rangle = |\psi\rangle$. Therefore, representing a stabilizer state requires $O(n^2)$ bits. Despite their compact representation, stabilizer states exhibit near-maximal entanglement~\cite{GarciaMC14}, and have many applications, e.g., in the distribution of Bell and Greenberger-Horne-Zeilinger (GHZ) states over quantum networks~\cite{Pant2019,Patil2021,Chakraborty2020,Das2018}, fault-tolerant quantum error-correcting codes~\cite{fowler2009high,raussendorf2005long,raussendorf2006fault,raussendorf2007fault,bolt2016foliated,nickerson2018measurement}, and measurement-based quantum computation (MBQC)~\cite{raussendorf2001one,Raussendorf2006,Gimeno-Segovia2015,Pant2019MBQC}. 

The gates that stabilizer circuits comprise of, i.e., ones that can be pieced together with the CNOT, $H$ and $P$ (or, CHP) gates, are also called {\em Clifford} gates. This is because these gates form the Clifford group, the {\em normalizer}~\footnote{The centralizer of a subset $S$ of group (or semigroup) $G$ is defined as: $\mathrm {C}_{G}(S)=\left\{g\in G \mid gs=sg {\text{ for all }} s \in S\right\} = \left\{g\in G\mid gsg^{-1}=s{\text{ for all }}s\in S\right\}$, where only the first definition applies to semigroups. The normalizer of S in the group (or semigroup) G is defined as $\mathrm {N}_{G}(S)=\left\{g\in G\mid gS=Sg\right\}=\left\{g\in G\mid gSg^{-1}=S\right\}$, where again only the first definition applies to semigroups. The two definitions above are similar but not identical. If $g$ is in the centralizer of $S$ and $s$ is in $S$, then it must be that $gs = sg$, but if $g$ is in the normalizer, then it could be that $gs = tg$ for some $t \in S$, with $t$ possibly different from $s$. That is, elements of the centralizer of $S$ must commute point-wise with $S$, but elements of the normalizer of $S$ need only commute with $S$ as a set. Source: \href{https://en.wikipedia.org/wiki/Centralizer_and_normalizer}{Wikipedia}.} of the Pauli group---the group ${\mathcal G}_n$ generated by $n$-qubit tensor-product Pauli matrices. Since stabilizer circuits can be efficiently simulated on a classical computer, any quantum circuit whose computational power exceeds that of a classical computer, must employ non-Clifford gates, such as the $\pi/8$-phase (or, the $T$) gate. A Clifford (resp., non-Clifford) measurement is a Clifford (resp., non-Clifford) unitary on a set of qubits, followed by computational-basis measurements on each qubit.

This paper is intended to be an all-inclusive guidebook to the stabilizer formalism. It also includes new tools, results and insights relevant to stabilizer manipulations of cluster states, and applications to linear-optical MBQC, using dual-rail photonic qubits. Our main technical contributions are as follows: 
\begin{enumerate}
\item We extend the stabilizer formalism to incorporate {\em fusions}: $k$-qubit projective measurements in a maximally-entangled $k$-qubit GHZ-state basis, the Bell state measurement (BSM) being a special case.
\item We provide a simple and explicit procedure to convert {\em any} multi-qubit stabilizer unitary into a series of {\em tableau} operations---introduced as the `CHP formalism' in Ref.~\cite{aaronson2004improved}---using {\em Karnaugh maps}, which is used to efficiently simplify Boolean expressions.
\item We develop a complete graphical rule-book for manipulating cluster states---a special class of stablizer states that are defined on a graph---via stabilizer unitaries and measurements. Two important new features of our rule-book are:
\begin{itemize} 
\item an explicit prescription of how to bring back the post-stabilizer-manipulation cluster state (original cluster state defined by graph $G$)---which is a stabilizer state---back to a cluster state form (i.e., a post-manipulation graph, $G^\prime$) via a set of single-qubit Hadamard transformations. Our rule-book describes explicitly how $G^\prime$ is obtained from $G$, and the associated stabilizer operations.
\item an explicit description of the stabilizer operation enacted by multiple linear-optical two-qubit fusion operations (rotated BSMs; some of which are new contributions)---both in the cases of when the fusions succeed and when they fail, and both for Type-II and Type-I fusions~\cite{browne2005Fusion}---and associated graphical rules for when the aforesaid fusions act on photonic cluster states of single-photon dual-rail-encoded qubits~\cite{Knill2001}.
\end{itemize}

\item Multiple stabilizer simulators~\cite{Anders2006,Gidney2021-dw,IBM_simulator_website,Krastanov_simulator,NETSQUID_Coopmans2021,Chicago_simulator_Wu2020,Qiskit_stabilizer_sim}, including the one associated with the original CHP paper~\cite{aaronson2004improved} are available online. However, in order to include the additional features described above, we developed a MATLAB implementation of our general cluster-state manipulation algorithm---available as Supplemental Material with this manuscript---which also produces visual graphical representations of the aforesaid stabilizer manipulations of cluster states.

\item We provide a way to reverse engineer any $k$-qubit stabilizer operation (stabilizer unitary followed by computational-basis measurement) on dual-rail photonic qubits, into an explicit linear-optical circuit. We do not claim our method to yield the most resource-efficient circuit, or that which succeeds with the highest probability.

\item We provide new linear optical circuits, leveraging some Type-I fusion operations we propose, which can create photonic cluster states using fewer single photon sources as compared to the existing circuits at the cost of having more unheralded losses in the generated cluster state.
\end{enumerate}

Next, we describe how to read this paper, and the contents of each Section. The reader is expected to be familiar with the basics of quantum information: quantum state of a qubit, entanglement, mixed states, unitary operators (gates) and von Neumann projective measurements. We discuss the basics of Gottesman's stabilizer formalism in Section~\ref{sec:stab_form}. In this Section, we also show how to extend the formalism to incorporate stabilizer {\em measurements}, such as rotated BSMs, or `fusions'. Readers interested in quantum error correction may find this section appealing as stabilizer measurements are an important building block of decoders for quantum error correcting codes. Once the reader is comfortable with the contents of Section~\ref{sec:stab_form}, they can jump to any of the subsequent sections as per their need, whose contents are described below. In Section~\ref{sec:ClusterStates}, we review cluster states, a special class of stabilizer states that can be defined on a graph. We develop graph-theoretic rules for manipulating cluster states via various standard stabilizer operations such as single-qubit Pauli measurements, local complementation, and different two-qubit rotated fusion measurements. This section is aimed at readers interested in measurement based quantum computation (MBQC)~\cite{raussendorf2001one} and entanglement distribution in quantum networks~\cite{pant2017allOptical, Grosshans2019distributing}. In Section~\ref{sec:sta_sim}, we review the {\em tableau} method described in~\cite{aaronson2004improved} to encode and efficiently simulate stabilizer circuits on a classical computer. Stabilizer gates and measurements translate to row and column operations on the tableau. We explicitly relate the tableau operations to stabilizer transformation rules described in Section~\ref{sec:stab_form}, making the correspondence between the two formalisms clear. We then describe our new Karnaugh-map-based method for tableau operations, illustrate how we used it to develop a MATLAB-based stabilizer simulator, and show examples of visualizations of cluster-state manipulations generated by our simulator. Section~\ref{sec:LO} is for readers interested in linear optical (LO) MBQC. Here we describe circuits for heralded LO fusion measurements and gates. We generalize LO fusion circuits by providing a recipe to design a LO circuit given the projection operators of any multi-qubit fusion. Section~\ref{sec:conclusions} concludes the paper with thoughts on applications and further extensions of our results, e.g., to boosted linear-optical stabilizer unitaries and measurements, and to non-stabilizer multi-qubit entangled state manipulations.

\section{The Stabilizer Formalism}
\label{sec:stab_form}

\subsection{Pauli operators}\label{subsec:definitions}
The single qubit Pauli matrices (or operators) $X$, $Y$ and $Z$, and the single-qubit identity matrix, shown in Fig.~\ref{fig:stabilizers_Paulis} form a basis for the real vector space of $2 \times 2$ Hermitian matrices, i.e., any $2 \times 2$ Hermitian matrix has a unique representation as a linear combination of Pauli matrices, with real-valued coefficients. It is easy to verify that $X^2 = Y^2 = Z^2 = I$, $XY = i Z$, $YX = -i Z$, $YZ = i X$, $ZY = -i X$, $ZX = i Y$, and $XZ = -i Y$, where $i = \sqrt{-1}$. This implies the commutators $[X, Y] = XY-YX = 2 i Z$, $[Y, Z] = YZ-ZY = 2 i X$, $[Z, X] = ZX-XZ = 2 i Y$ and anticommutators $\left\{X, Y\right\} = XY+YX = 0 = \left\{Y, Z\right\} = \left\{Z, X\right\}$. 

When referring to Pauli operators on multiple qubits, we will index the Pauli operators by the qubit index. For example, $M = X_1Z_2$ is a $2$-qubit Pauli operator that is a tensor product of $X$ (acting on qubit 1), $Z$ (acting on qubit 2), and identity $I$ (acting on any other qubits involved). Clearly, two $n$-qubit Pauli operators $M_1$ and $M_2$ must either commute, i.e., $[M_1, M_2] = 0$, or anticommute, i.e., $\{M_1, M_2\} = 0$. If $M$ is an $n$-qubit Pauli operator that anticommutes with $M_1$ and with $M_2$, i.e., $M M_1 = - M_1M$ and $M M_2 = - M_2 M$, then $M$ must commute with $M_1M_2$, i.e., $[M, M_1M_2] = 0$. This can be seen as follows: $M (M_1 M_2) = (-M_1 M) M_2 = -M_1 (M M_2) = -M_1 (-M_2 M) = (M_1 M_2) M$.

In addition to the computational basis $\left\{|0\rangle, |1\rangle\right\}$ of the single-qubit Hilbert space, we will also be using the $45^{\text o}$ rotated basis $\left\{|+\rangle, |-\rangle\right\}$, where $|\pm\rangle \equiv \left(|0\rangle \pm |1\rangle\right)/\sqrt{2}$. Since $X$ causes a bit flip, i.e., $X\ket{0} = \ket{1}$ and $X\ket{1} = \ket{0}$, $X|\pm\rangle = (\pm 1) \ket{\pm}$. In other words, $\ket{\pm}$ are eigenstates of $X$, with eigenvalues $+1$ and $-1$ respectively. On the other hand, $Z$ causes a phase flip to the $|1\rangle$ state, i.e., $Z\ket{0} = \ket{0}$ and $Z\ket{1} = -\ket{1}$. Therefore, the computational basis vectors $\left\{|0\rangle, |1\rangle\right\}$ are eigenstates of $Z$, with eigenvalues $+1$ and $-1$ respectively. It is simple to verify that the states $\left(\ket{0} \pm i\ket{1}\right)/\sqrt{2}$ are the eigenstates of $Y$, with eigenvalues $+1$ and $-1$ respectively. The Hadamard operator $H$, shown in Fig.~\ref{fig:stabilizers_Paulis}, satisfies $H\ket{0} = \ket{+}$ and $H\ket{1} = \ket{-}$. Since $H^2 = I$, we have $H\ket{+} = \ket{0}$ and $H\ket{-} = \ket{1}$. Further, $HXH = Z$ and $HZH = X$. 

\subsection{Stabilizer states and stabilizer generators}\label{sec:stabilizeroperators}

A unitary operator $U$ {\em stabilizes} a state $\ket{\psi}$ if $\ket{\psi}$ is a $+1$ eigenvalue eigenstate of $U$, i.e., $U\ket{\psi}=\ket{\psi}$. In this paper, we will consider $n$-qubit states ---viz., {\em stabilizer states} ---that are stabilized by elements of the group ${\mathcal G}_n\backslash [-I^{\otimes n}]$. ${\mathcal G}_n$ is the {\em Pauli group} and consists of the $n$-fold tensor products of single-qubit Pauli matrices, $P = i^m P_1 \otimes \ldots \otimes P_n$, with $P_j \in \{I,X,Y,Z\}$ and $m\in\{0,1,2,3\}$ s.t. $\vert {\mathcal G}_n\vert = 4^{n+1}$. We exclude the negative identity operator $-I^{\otimes n}$ from consideration as it can never have a non-trivial $+1$ eigenvalue, and thus cannot stabilize any quantum state. For brevity, we will omit the `$\otimes$' notation and denote $P$ as a string of single-qubit Pauli literals $I$, $X$, $Y$, and $Z$, along with a separate two-bit integer value $k$ for the overall phase. 
Next, we formally define a stabilizer group. Let $\mathcal{S}$ with $\vert\mathcal{S}\vert=2^k$  be a subgroup of ${\mathcal G}_n\backslash [-I^{\otimes n}]$, such that $\mathcal{S}$ is Abelian, i.e., $\mathcal{S}$ is a set of $2^k$ $n$-qubit Pauli operators which commute with each other (under multiplication). Here, $k\leq n$. The group $\mathcal{S}$ can be uniquely described using $k$ generators, i.e., a set $\mathcal{G}$ of $k$ independent elements of $\mathcal{S}$~\cite{gottesman1997stabilizer}. Every element in $\mathcal{S}$ that is not in $\mathcal{G}$ can be obtained by taking the product of two or more generators. However, any given element of $\mathcal{G}$ cannot be obtained as a product of other elements in $\mathcal{G}$. Further, the set of generators is not unique. Let $\mathcal{V}$ be the set of all $n$-qubit pure states that are stabilized by all elements of the group $\mathcal{S}$, i.e., $\forall \ket{\psi}\in \mathcal{V}, \forall S \in \mathcal{S}, S\ket{\psi}=\ket{\psi}$. We then can completely describe $\mathcal{V}$ using its {\em stabilizer group} $\mathcal{S}$, or more compactly using the $k$ {\em stabilizer generators}. From here onwards, we refer to $\mathcal{S}(\mathcal{V})$ as a stabilizer group, and use the term {\em stabilizer} to refer to its elements.


The single-qubit pure states stabilized by the single-qubit Pauli matrices, expressed as \{stabilizer, state\} pairs, are: $\left\{X, \ket{+}\right\}$, $\left\{-X, \ket{-}\right\}$, $\left\{Y, (\ket{0}+i\ket{1})/\sqrt{2}\right\}$, $\left\{-Y, (\ket{0}-i\ket{1})/\sqrt{2}\right\}$, $\left\{Z, \ket{0}\right\}$, and $\left\{-Z, \ket{1}\right\}$. We observe that the single-qubit identity $I$ stabilizes all single-qubit states. Therefore, specifying one single-qubit non-Identity Pauli operator uniquely identifies one single qubit (stabilizer) state, i.e., the one it stabilizes. 
To generalize, specifying $k=n$ independent $n$-qubit Pauli operators (or stabilizer generators) uniquely identifies one $n$-qubit (stabilizer) quantum state (i.e., the state which they {\em all} stabilize)~\cite{Nielsen2010}. As a result, the stabilizer formalism reduces the exponentially many parameters required to describe a general $n$-qubit quantum state, to only a $n(2n+1)$ bit string needed to identify $n$ independent $n$-qubit Pauli operators, to describe an $n$-qubit stabilizer state. 

Let us consider the two-qubit Pauli operator $M_1 = X_1X_2$. Observe that $X_1X_2 \ket{++} = \ket{++}$, as well as $X_1X_2 \ket{--} = \ket{--}$. Since $X_1X_2$ stabilizes both $\ket{++}$ and $\ket{--}$, it must also stabilize any superposition of these two states, e.g., the Bell state $|\psi^+\rangle = \left(\ket{++} + \ket{--}\right)/\sqrt{2}$, i.e., $X_1X_2 |\psi^+\rangle = |\psi^+\rangle$. So, $X_1X_2$ specifies a {\em stabilizer subspace}, the subspace of the two-qubit Hilbert space spanned by the eigenvectors of $X_1X_2$ with $+1$ eigenvalue. $Z_1Z_2$ stabilizes $\ket{00}$ and $\ket{11}$, and hence also any superposition of the two. Therefore, $Z_1Z_2 \ket{\psi^+} = \ket{\psi^+}$, where $|\psi^+\rangle = \left(\ket{++} + \ket{--}\right)/\sqrt{2} =  \left(\ket{00} + \ket{11}\right)/\sqrt{2}$. The state $|\psi^+\rangle$ is stabilized by both $X_1X_2$ and $Z_1Z_2$. It belongs to two distinct $1$-qubit subspaces of the $2$-qubit Hilbert space, and is, therefore, the unique quantum state in the stabilizer subspaces specified by $M_1 = X_1X_2$ and $M_2 = Z_1Z_2$. Hence, the two $2$-qubit Pauli operators $(M_1, M_2\rangle$ uniquely specifies the $2$-qubit state $\ket{\psi^+}$.

If $M$ is an $n$-qubit Pauli operator, it has $2^n$ eigenstates, each with eigenvalue $+1$ or $-1$. Therefore, $M = (+1)A_+ + (-1)A_- = A_+ - A_-$, where $A_+$ (respectively $A_-$) is the sum of rank-one density operators of eigenstates of $M$ of $+1$ (respectively $-1$) eigenvalues. $A_+$ is a projector onto the {\em stabilizer subspace} of $M$, i.e., a subspace of the $n$-qubit Hilbert space spanned by the $+1$ eigenvectors of $M$. For $M=X$, $A_+ = \ket{+}\bra{+}$ and $A_- = \ket{-}\bra{-}$. For $M = X_1X_2$, $A_+ = \ket{++}\bra{++}+\ket{--}\bra{--}$ and $A_- = \ket{+-}\bra{+-} + \ket{-+}\bra{-+}$. For $M = X_1Z_2$, $A_+ = \ket{+0}\bra{+0} + \ket{-1}\bra{-1}$ and $A_- = \ket{+1}\bra{+1}+\ket{-0}\bra{-0}$. For $M = Z_1Z_2$, $A_+ = \ket{00}\bra{00} + \ket{11}\bra{11}$ and $A_- = \ket{10}\bra{10} + \ket{01}\bra{01}$.

Since $-Y_1Y_2 = (X_1Z_1)(X_2Z_2)$, $-Y_1Y_2$ also stabilizes $\ket{\psi^+}$. The two-qubit identity $I_1I_2$ trivially stabilizes $\ket{\psi^+}$ as well. Therefore, the Bell state $\ket{\psi^+}$ is stabilized by the following two-qubit Pauli operators: $\mathcal{S}(\ket{\psi^+}) = \{I_1I_2, X_1X_2, Z_1Z_2, -Y_1Y_2\}$. $\mathcal{S}(\ket{\psi^+})$ is an Abelian group closed under matrix-multiplication operation, i.e., all stabilizers commute with each other, with generators, $\mathcal{G}(\ket{\psi^+}) = \langle  X_1X_2,Z_1Z_2\rangle$. The element $-Y_1Y_2$ can be obtained by multiplying $X_1X_2$ and $Z_1Z_2$. Note that, $\langle  X_1X_2,  -Y_1Y_2\rangle$ and $\langle  Z_1Z_2, -Y_1Y_2\rangle$ are other possible stabilizer generators of $\ket{\psi^+}$.

Following a similar reasoning, one can show that the Bell state $\ket{\psi^-}=\left(\ket{00}-\ket{11}\right)/{\sqrt{2}}$ is stabilized by $\mathcal{S}(\ket{\psi^-}) = \{I_1I_2,-X_1X_2, Z_1Z_2, Y_1Y_2\}$. The stabilizer generators for this state are $\mathcal{G}(\ket{\psi^-}) = \langle -X_1X_2, Z_1Z_2\rangle$. The element $Y_1Y_2$ is generated by multiplying $-X_1X_2$ and $Z_1Z_2$. If we were to specify both elements of $\mathcal{G}(\ket{\psi^+})$, or both elements of $\mathcal{G}(\ket{\psi^-})$, we would uniquely specify the states $\ket{\psi^+}$ or $\ket{\psi^-}$, respectively. However, if we only specified the single stabilizer generator $Z_1Z_2$, it would specify the one-qubit stabilizer subspace of the two-qubit Hilbert space, identified by the projector $A_+ = \ket{00}\bra{00} + \ket{11}\bra{11}$, as discussed above. 

We discussed above that $n$ independent $n$-qubit stabilizer generators $S_i$, $1\le i \le n$, specifies a unique pure stabilizer state $\ket{\psi}$ s.t. $S_i\ket{\psi}=\ket{\psi}, \forall i$, whereas $k < n$ independent $n$-qubit stabilizer generators $S_i$, $1\le i \le k$, specifies a stabilizer subspace. A set of $k < n$ independent $n$-qubit stabilizer generators $S_i$ also specifies an $n$-qubit mixed state $\rho$ that is stabilized by all the generators, i.e., $S_i \rho S_i = \rho$, $1\le i \le k$. The density matrix of a state with $k$ generators $\langle  S_1,S_2,
\dots, S_k\rangle$ can be written as $\rho = \frac{1}{2^k}\prod_{i=1}^k(I+S_i)$~\cite{aaronson2004improved}.  If $k=n$, $\rho$ describes a pure state. If $k<n$, the state described by $\rho$ is called a {\em stabilizer mixed state}, which can be obtained by tracing out $n-k$ qubits from an $n$-qubit pure stabilizer state~\cite{aaronson2004improved}. Finally, a set of $k<n$ independent $n$-qubit stabilizer generators, which identify an $n-k$-qubit subspace of the $n$-qubit Hilbert space, also specifies a {\em stabilizer code} in quantum error correction, which encodes $k$ logical qubits in $n$ physical qubits. Most of this paper deals with the case $k=n$, i.e., the stabilized state is unique and pure.


One thing we should note at this point is that not all states are unity-eigenvalue eigenstates of elements of ${\mathcal G}_n$. In other words, not every quantum state is a stabilizer state. For example, the $n$-qubit W state, $\{\ket{100\dots 0}+\ket{010\dots 0}+\dots+\ket{00\dots 01}\}/{\sqrt{n}}$ is not a stabilizer state.

\subsection{Stabilizer measurements}\label{subsec:measurements}
Any $K$-outcome measurement on $n$ qubits is defined by a {\em positive operator valued measure} (POVM): a set of operators $\left\{\Pi_k\right\}$, $1 \le k \le K$, where $\sum_{k=1}^K \Pi_k^\dagger \Pi_k = {\mathbb I}_n$ (the $n$-qubit identity operator). If this POVM is used to measure the $n$-qubit state $|\psi\rangle$, the probability of obtaining the $k$-th measurement outcome $p_k = \langle  \psi | \Pi_k^\dagger \Pi_k |\psi\rangle$, resulting in the post-measurement state $\Pi_k|\psi\rangle / \sqrt{p_k}$. A {\em projective measurement} is one for which $\Pi_k^{\dagger} = \Pi_k$ (Hermitian) and $\Pi_k \Pi_j = \delta_{kj}\Pi_k$ (projector), and therefore $\sum_{k=1}^K \Pi_k = {\mathbb I}_n$ are satisfied. Projective measurements with all rank-one operators $\Pi_k = |w_k\rangle \langle  w_k |$, where $\left\{|w_k\rangle\right\}$ is a complete orthonormal basis for the $n$-qubit Hilbert space, are called {\em von Neumann projective measurements}. 

In the above quantum information theorist's treatment of quantum measurement, $k$ is simply an {\em index} for the measurement outcomes. A physicist would often describe a measurement as a Hermitian operator $\Lambda$ (corresponding to an observable), with spectral decomposition $\Lambda = \sum_{k=1}^K \lambda_k |\lambda_k\rangle \langle  \lambda_k |$. The eigenvalues $\left\{\lambda_k\right\}$ are the measurement outcomes. If all the eigenvalues are distinct, this measurement would be a {\em von Neumann projective measurement}, described in the POVM language by operators $\Pi_k = |\lambda_k\rangle \langle  \lambda_k |$, with $k = 1, \ldots, K$ indexing the outcomes---the actual measured value of the observable corresponding to the outcome $k$, being $\lambda_k$. If there is degeneracy, i.e., $J < K$ distinct eigenvalues of $\Lambda$, we can rewrite $\Lambda = \sum_{j=1}^J \mu_j A_j$, where each $A_j$ is either a rank-one projector of the form $|\lambda_k\rangle \langle  \lambda_k |$, or a sum of two or more rank-one projectors of the form $|\lambda_k\rangle \langle  \lambda_k |$, for $k$ values corresponding to an identical eigenvalue. This measurement is a {\em projective measurement}, described in the POVM language by operators $\Pi_j = A_j$, with $j = 1, \ldots, J$ indexing the outcomes---the measured values of the observable corresponding to the outcome $j$ being $\mu_j$, which in turn is one the degenerate eigenvalues $\lambda_k$ of $\Lambda$.

Next, we consider stabilizer measurements, or measurement of Pauli operators, in the physicist's sense discussed above. If $M$ is an $n$-qubit Pauli operator, it has $2^n$ eigenstates, each with eigenvalue $+1$ or $-1$. Therefore, $M = (+1)A_+ + (-1)A_- = A_+ - A_-$, where $A_+$ (respectively $A_-$) is the sum of rank-one density operators of eigenstates of $M$ of $+1$ (respectively $-1$) eigenvalues. So, ``measuring $M$" is a binary-outcome projective measurement described by POVM operators $\left\{A_+, A_-\right\}$. The observed outcome is either $+1$ or $-1$. Here, $A_+$ is the projector onto the $(+1)$-eigenvalue eigenspace of $M$, also known as the {\em stabilizer subspace} of $M$, whereas $A_-$ is the projector onto the $(-1)$-eigenvalue eigenspace of $M$, which is the stabilizer subspace of $-M$. If $n=1$, $A_+$ and $A_-$ are rank-one projectors, making the measurement a von Neumann projective measurement. For example, $X = A_+ - A_-$, with $A_+ = \ket{+}\bra{+}$ and $A_- = \ket{-}\bra{-}$. So, ``measuring $X$" on a single-qubit state $|\psi\rangle$ is a von Neumann projective measurement described by operators: $\left\{\ket{+}\bra{+}, \ket{+}\bra{+}\right\}$. Consider measuring $M = X_1Z_2$ on two qubits. The $+1$ eigenstates of $M$ are $\{\ket{+0},\ket{-1}\}$ and the $-1$ eigenstates are $\{\ket{+1},\ket{-0}\}$. Therefore, $M = A_+ - A_-$, with $A_+ = \ket{+0}\bra{+0} + \ket{-1}\bra{-1}$ and $A_- = \ket{+1}\bra{+1}+\ket{-0}\bra{-0}$. $M$ is thus a projective measurement, but not a von Neumann projective measurement. If $M$ is measured on a two qubit state $\ket{\psi}$, the post-measurement state will be one of the two: $A_\pm \ket{\psi}/\sqrt{p_\pm}$ with probabilities $p_\pm = \bra{\psi}A_\pm\ket{\psi}$.

An alternative interpretation of ``measuring $M$" on an $n$-qubit state, where $M$ is an $n$-qubit Pauli operator, is a projective measurement described by POVM operators $\left\{\Pi_+, \Pi_-\right\}$, where $\Pi_+$ is the projector onto the {\em stabilizer subspace} of $M$, and $\Pi_-$ is the projector onto the {\em stabilizer subspace} of $-M$. When we describe a measurement by two Pauli operators $\langle M_1, M_2\rangle$, it is a $4$-outcome projective measurement described by POVM operators $\left\{\Pi_{++}, \Pi_{+-}, \Pi_{-+}, \Pi_{--}\right\}$, where $\Pi_{++}$ is the projector onto the stabilizer subspace of $\langle M_1, M_2\rangle$ (i.e., intersection of the stabilizer subspaces of $M_1$ and $M_2\rangle$, $\Pi_{+-}$ is the projector onto the stabilizer subspace of $\langle M_1, -M_2\rangle$, $\Pi_{-+}$ is the projector onto the stabilizer subspace of $\langle -M_1, M_2\rangle$, and $\Pi_{--}$ is the projector onto the stabilizer subspace of $\langle -M_1, -M_2\rangle$. Consider a measurement on two qubits specified by $M_1 = X_1Z_2$ and $M_2 = Z_1X_2$. Recall from Section~\ref{sec:stabilizeroperators} that $M_1 = A_+ - A_-$, with $A_+ = \ket{+0}\bra{+0} + \ket{-1}\bra{-1}$ and $A_- = \ket{+1}\bra{+1}+\ket{-0}\bra{-0}$. Similarly, $M_2 = B_+ - B_-$, with $B_+ = \ket{0+}\bra{0+} + \ket{1-}\bra{1-}$ and $B_- = \ket{0-}\bra{0-}+\ket{1+}\bra{1+}$. But also recall from Section~\ref{sec:stabilizeroperators} that the stabilizer subspace of two independent two-qubit Pauli operators $\langle M_1, M_2\rangle$ is a unique two-qubit quantum state, $\ket{\phi_1}=(\ket{0+}+\ket{1-})/\sqrt{2}$. Similarly, $\langle -M_1, M_2\rangle$ uniquely stabilize the state $\ket{\phi_2}=(\ket{0+}-\ket{1-})/\sqrt{2}$, $\langle M_1, -M_2\rangle$ uniquely stabilize the state $\ket{\phi_{3}}=(\ket{1+}+\ket{0-})/{\sqrt{2}}$, and $\langle -M_1, -M_2\rangle$ uniquely stabilize the state $\ket{\phi_{4}}=(\ket{1+}-\ket{0-})/{\sqrt{2}}$. Therefore, ``measuring $\langle M_1$, $M_2\rangle$ on two qubits" amounts to a von Neumann projective measurement described by the POVM operators: $\{\ket{\phi_1}\bra{\phi_1},\ket{\phi_2}\bra{\phi_2},\ket{\phi_3}\bra{\phi_3},\ket{\phi_4}\bra{\phi_4}\}$. As we will see later, $\ket{\phi_1}$ is a two-qubit {\em cluster state} (or graph state), which is a single-qubit Hadamard gate away from the two-qubit Bell state $\ket{\psi^+}$ discussed in Section~\ref{subsec:measurements}. The measurement above is thus a `rotated' Bell-basis measurement, also known as a two-qubit {\em fusion} measurement, which we will discuss in detail in Section~\ref{sub:stab_fusion} below. Using the above notation for stabilizer measurements, measuring $Z$ individually on qubits $1$ and $2$ would be described by: $\langle Z_1I_2, I_1Z_2\rangle$. This is because the simultaneous stabilizer subspaces of $\langle Z_1I_2, I_1Z_2\rangle$, $\langle -Z_1I_2, I_1Z_2\rangle$, $\langle Z_1I_2, -I_1Z_2\rangle$ and $\langle -Z_1I_2, -I_1Z_2\rangle$ are the unique quantum states $\ket{00}$, $\ket{10}$, $\ket{01}$ and $\ket{11}$, respectively.

The reader may have noticed the parallel between stabilizer states and measurements in Sections~\ref{sec:stabilizeroperators} and~\ref{subsec:measurements}. Both stabilizer states and measurements can be compactly described by (collections of) Pauli operators. For specifying a two-qubit stabilizer pure state $\ket{\psi}$ uniquely, we need to specify two independent two-qubit Pauli operators $M_1$ and $M_2$ (both of which stabilize $|\psi\rangle$). If we specify one two-qubit Pauli operator $M$, we identify the stabilizer-subspace (the $(+1)$-eigenvalue eigenspace) of $M$. This sub-space, spanned by two orthogonal two-qubit states, can also be thought of as a code that encodes (embeds) a single qubit into a two-qubit Hilbert space. If we specify a stabilizer measurement by two independent two-qubit Pauli operators $M_1$ and $M_2$, we specify a four-element von-Neumann projective measurement described by $\Pi_1 = \ket{\psi}\bra{\psi}$, and $\Pi_2$, $\Pi_3$, $\Pi_4$ being projectors on the unique two-qubit pure states stabilized by $(M_1, -M_2\rangle$, $(-M_1, M_2\rangle$ and $(-M_1, -M_2\rangle$, respectively. If a measurement is specified by a single two-qubit Pauli $M$, it is a two-element projective measurement with POVM operators $\Pi_+$ and $\Pi_-$, where $\Pi_\pm$ is the projector onto the stabilizer subspace of $\pm M$. Generalizations of the above to states and measurements of more than $2$ qubits is straightforward. Since stabilizer states and measurements can be both described succinctly by Pauli operators, we will rarely write down stabilizer states as kets, or measurements as a collection of POVM operators.

Finally, let us discuss the difference between a (projective) {\em measurement} and a {\em projection}. As discussed above, a $K$-outcome projective measurement on $n$ qubits is described by POVM operators $\Pi_k$, $k = 1, \ldots, K$, each of which is a projector, such as $\sum_{k=1}^K \Pi_k = {\mathbb I}_n$. When this measurement acts on state $\ket{\psi}$, it produces post-measurement state $\Pi_k\ket{\psi}/\sqrt{p_k}$ with probability $p_k = \bra{\psi} \Pi_k \ket{\psi}$. The measurement always produces an outcome, albeit a random one. The state is projected onto one of the $K$ projectors. In this paper, we will encounter gadgets that implement a {\em projection} $\Pi_j$, for a given $j$. Such a projection acting on state $\ket{\psi}$ {\em succeeds} with probability $p_j = \bra{\psi} \Pi_j \ket{\psi}$, and upon success produces the state $\Pi_j\ket{\psi}/\sqrt{p_j}$. A projection can also be described by more than one projector, e.g., a binary-outcome projection described by projectors $\Pi_j$ and $\Pi_l$. This projection, acting on state $\ket{\psi}$ {\em succeeds} with probability $p_j + p_l = \bra{\psi} \Pi_j \ket{\psi} + \bra{\psi} \Pi_l \ket{\psi}$, and conditioned on success, it produces the state $\Pi_j\ket{\psi}/\sqrt{p_j}$ with probability $p_j/(p_j+p_l)$ and $\Pi_l\ket{\psi}/\sqrt{p_l}$ with probability $p_l/(p_j+p_l)$. We will also encounter examples of gadgets which---upon {\em success}---implements a projection on an $n$-qubit state described by projection operators $\{\Pi_l\}$, $l = 1, \ldots, L$, with $\sum_{l=1}^{L} \Pi_l < {\mathbb I}_n$; while---upon {\em failure}---implements {\em another} projection described by projection operators $\{\Sigma_j\}$, $j = 1, \ldots, J$, with $\sum_{j=1}^{J} \Sigma_j < {\mathbb I}_n$. Just like stabilizer measurements can be described by Pauli operators (instead of POVM operators), stabilizer projections can also be described by Pauli operators, but with an added qualification on the subspace(s) onto which the projection projects the state measured. For example, a one-outcome projection (on an $n$-qubit state) can be described by the projector onto the $+1$ eigenspace of an $n$-qubit Pauli $M$. A two-outcome projection can be described by two Paulis $(M_1, M_2\rangle$, with an added qualification of the two projectors onto their joint eigenspaces with $(+1,-1)$ and the $(-1,+1)$ eigenvalues, and so on.

\subsection{Stabilizer or Clifford unitaries}
\label{subsec:UnitaryOps}

Let us consider a state $\ket{\psi}$ stabilized by an operator $S$, i.e., $S\ket{\psi}=\ket{\psi}$. Say, $\ket{\psi}$ undergoes a unitary evolution $U$ resulting in a final state $\ket{\psi'} = U\ket{\psi}$. We can write, 
$$\ket{\psi} = U\ket{\psi}=US\ket{\psi}=USU^{\dagger}U\ket{\psi} \equiv S'\ket{\psi'},$$
with $S' = USU^{\dagger}$. As a result, the transformed state $\ket{\psi'}$ is stabilized by $S'$. Therefore, when a stabilizer state evolves through a unitary quantum circuit, it suffices to evolve its stabilizer(s). This notion of evolving operators rather than states is often associated with the {\em Heisenberg interpretation} of quantum mechanics.

Next, we want to consider what properties $U$ must have such that if $\ket{\psi}$ is a stabilizer state (i.e., $S \in {\mathcal G}_n$), that $\ket{\psi'}$ is also a stabilizer state. We thus define a {\em Clifford} unitary operator $U$ to be one that transforms a Pauli tensor-product operator into another Pauli tensor-product operator under conjugation, i.e., $$UO_1U^{\dagger}=O_2,\;\;\;\;\;\; O_1, O_2\in\mathcal{P}^{\otimes n}.$$
In other words, the {\em Clifford group} is the normalizer of the Pauli group ${\mathcal G}_n$. Clearly, if $U$ is a Clifford unitary, and ${\ket{\psi}}$ is a stabilizer state with stabilizers $S \in {\mathcal G}_n$, the transformed state $\ket{\psi'}$, stabilized by $S' = USU^{\dagger} \in {\mathcal G}_n$ is also a stabilizer state. This is why the term {\em Clifford unitary} is often used synonymously with the term {\em stabilizer unitary}. Transformations of stabilizer states under Clifford operations can be studied within the stabilizer formalism, i.e., by keeping track of the stabilizer generators of the state as it passes through Clifford gates.

As an example, if we apply the Hadamard gate $H$ to the second qubit of the Bell state $\ket{\psi^+}=(\ket{00}+\ket{11})/{\sqrt{2}}$, then the resulting state $\ket{{\psi^-}'}= H\ket{\psi^-} = (\ket{0+}+\ket{1-})/{\sqrt{2}}$. We discussed above that the stabilizer generators of $\ket{\psi^+}$, $\mathcal{G}(\ket{\psi^+}) = \langle X_1X_2, Z_1Z_2\rangle$. Therefore, the stabilizer generators of $\ket{{\psi^-}'}$ are given by: $\mathcal{G'} = \langle H_2(X_1X_2)H_2, H_2(Z_1Z_2)H_2\rangle =\langle X_1Z_2, Z_1X_2\rangle$. This is the state $\ket{\phi_1}$, the two-qubit graph state (or cluster state) we discussed in Section~\ref{subsec:measurements}. 

Consider a few examples of two-qubit Clifford unitaries involving the CNOT (controlled-not) and the CZ (controlled-phase) gates. The CNOT gate flips the second ({\em target}) qubit if the first ({\em control}) qubit's state is $\ket{1}$. Its action on two qubits can be described as: ${\rm CNOT}_{1,2}\ket{00} = \ket{00}$, ${\rm CNOT}_{1,2}\ket{01} = \ket{01}$, ${\rm CNOT}_{1,2}\ket{10} = \ket{11}$, ${\rm CNOT}_{1,2}\ket{11} = \ket{10}$. CNOT is denoted by a vertical line connecting the two qubits it connects in a quantum circuit, with the control qubit shown with a dot, and the target qubit shown with a $\oplus$ sign. The CZ gate flips the sign of the target qubit if the control qubit's state is $\ket{1}$. Its action on two qubits can be described as: ${\rm CZ}_{1,2}\ket{00} = \ket{00}$, ${\rm CZ}_{1,2}\ket{01} = \ket{01}$, ${\rm CZ}_{1,2}\ket{10} = \ket{10}$, ${\rm CZ}_{1,2}\ket{11} = -\ket{11}$. Its action is symmetric with respect to the two qubits it acts on. So, either qubit can be interpreted as the control or the target. CZ is denoted by a vertical line connecting the two qubits it connects in a quantum circuit, with both the control and target qubits shown as dots. CNOT and CZ are both unitary and self adjoint, i.e., ${\rm CZ}_{1,2}^\dagger = {\rm CZ}_{1,2}^{-1} = {\rm CZ}_{1,2}$, and ${\rm CNOT}_{1,2}^\dagger = {\rm CNOT}_{1,2}^{-1} = {\rm CNOT}_{1,2}$.
\begin{figure}
    \centering
\includegraphics[width=\columnwidth]{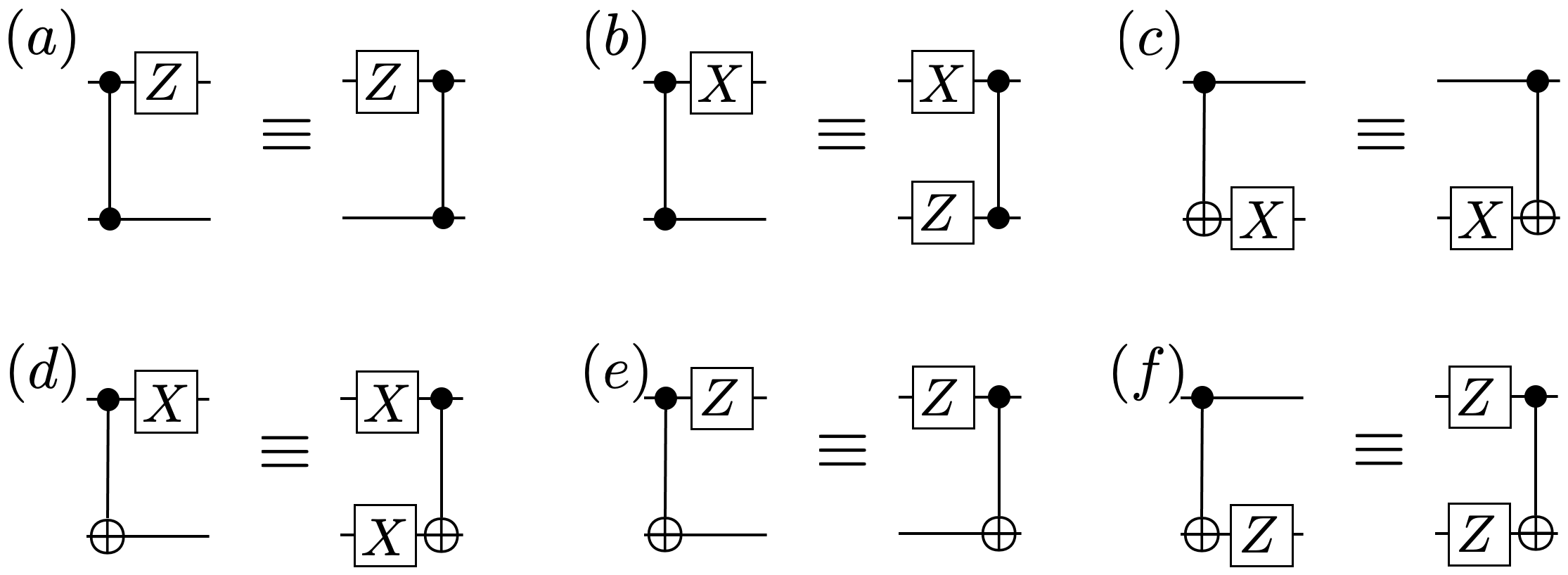}
    \caption{Commutation and composition relations for the two-qubit CZ and CNOT gates with single-qubit $X$ and $Z$ gates.}
   \label{fig:TwoQubitCliffords}
\end{figure}
The following identities for CZ and CNOT are easy to verify by enumerating the action of the gate sequences on the four computational basis states, and summarized in Fig.~\ref{fig:TwoQubitCliffords}:
\begin{enumerate}[label=(\alph*)]
\item CZ commutes with $Z$ gate(s) on either or both of the qubits. This implies: ${\rm CZ}_{1,2} (I_1 Z_2) {\rm CZ}_{1,2} = I_1 Z_2$, ${\rm CZ}_{1,2} (Z_1 I_2) {\rm CZ}_{1,2} = Z_1 I_2$, and similarly, ${\rm CZ}_{1,2} (Z_1 Z_2) {\rm CZ}_{1,2} = Z_1 Z_2$.
\item ${\rm CZ}_{1,2}$ followed by $X$ on qubit 1 is same as applying $X$ on qubit 1 and $Z$ on qubit 2, followed by the ${\rm CZ}_{1,2}$. This implies: ${\rm CZ}_{1,2} (I_1 X_2) {\rm CZ}_{1,2} = Z_1 X_2$, ${\rm CZ}_{1,2} (X_1 I_2) {\rm CZ}_{1,2} = X_1 Z_2$, ${\rm CZ}_{1,2} (X_1 X_2) {\rm CZ}_{1,2} = (Z_1X_1)(X_2Z_2) = Y_1Y_2$. 
\item CNOT commutes with $X$ on the target qubit. This implies: ${\rm CNOT}_{1,2} (I_1 X_2) {\rm CNOT}_{1,2} = I_1 X_2$.
\item ${\rm CNOT}_{1,2}$ followed by $X$ on the control qubit is same as $X$ on both qubits followed by ${\rm CNOT}_{1,2}$. This implies: ${\rm CNOT}_{1,2} (X_1 I_2) {\rm CNOT}_{1,2} =X_1 X_2$.
\item CNOT commutes with $Z$ on the control qubit. This implies: ${\rm CNOT}_{1,2} (Z_1 I_2) {\rm CNOT}_{1,2} = Z_1 I_2$.
\item ${\rm CNOT}_{1,2}$ followed by $Z$ on the target qubit is same as $Z$ on both qubits followed by ${\rm CNOT}_{1,2}$. This implies: ${\rm CNOT}_{1,2} (I_1 Z_2) {\rm CNOT}_{1,2} = Z_1 Z_2$. Also, using this, and (d), ${\rm CNOT}_{1,2} (X_1 Z_2) {\rm CNOT}_{1,2} = (X_1Z_1)(X_2Z_2) = -Y_1Y_2$.
\end{enumerate}

If $\ket{\psi^\prime} = U\ket{\psi}$ for unitary $U$, where the $n$-qubit stabilizer state $\ket{\psi}$ is uniquely described by stabilizer generators $\{M_k\}$, $k = 1, \ldots, n$, the state $\ket{\psi^\prime}$ will be uniquely described by stabilizer generators $\{UM_kU^\dagger\}$. Consider a stabilizer measurement on $\ket{\psi^\prime}$ described by stabilizer operator(s) $\{L_i\}$, $i = 1, \ldots, m$, $m \le n$. The same measurement outcome can be obtained by measuring stabilizer operator(s) $\{U^\dagger L_i U\}$ on the state $\ket{\psi}$, followed by applying $U$ on the post-measurement state obtained. A simple way to see this is to consider the binary-outcome projective measurement discussed in Section~\ref{subsec:measurements} described by stabilizer operator $L = A_+ - A_-$, acting on state $\ket{\psi^\prime}$, which produces one of the two post-measurement states $A_\pm \ket{\psi^\prime}/\sqrt{p_\pm}$ with probabilities $p_\pm = \bra{\psi^\prime} A_\pm \ket{\psi^\prime}$. Since $\ket{\psi^\prime} = U\ket{\psi}$, the same measurement outcome can be obtained by measuring $L^\prime = U^\dagger L U = (U^\dagger A_+ U) - (U^\dagger A_- U)$ on $\ket{\psi}$, which would lead to post-measurement states: $U^\dagger A_\pm U\ket{\psi}/\sqrt{p_\pm} = U^\dagger A_\pm\ket{\psi^\prime}/\sqrt{p_\pm}$ with probabilities $p_\pm = \bra{\psi}U^\dagger A_\pm U\ket{\psi} = \bra{\psi^\prime} A_\pm \ket{\psi^\prime}$. By applying the unitary $U$ to the above post-measurement states, one would obtain the states $A_\pm \ket{\psi^\prime}/\sqrt{p_\pm}$. When transforming a stabilizer measurement with a unitary $U$ applied before it, we will often ignore the last step above, of applying $U$ to the post-measurement state. This is because in most applications, we would be concerned only with reproducing the correct measurement probabilities. Also, most physical measurement devices do not leave behind any post-measurement state, i.e., they are descructive.

In terms of POVM operators, if a $K$-outcome projective measurement is described by operators $\{\Pi_k\}$, $k = 1, \ldots, K$, a unitary $U$ followed by that measurement would be described by projection operators $\{U^\dagger \Pi_k U\}$, followed by applying $U$ to the post-measurement state.

Finally, in TABLE~\ref{tab:CliffordTransforms}, we tabulate all the Clifford transformations of all the Pauli operators. 

\begin{table*}[ht]
    \centering
  \begin{tabular}{ ?P{3cm}|P{1cm}|P{1.5cm}||P{3cm}|P{1cm}|P{1.5cm}||P{3cm}|P{1cm}|P{1.5cm}? } 
\noalign{\hrule height 1pt}\xrowht{10pt}
Clifford Unitary & Input & Output&Clifford Unitary & Input & Output&Clifford Unitary & Input & Output \\
\noalign{\hrule height 1pt}\xrowht{10pt}
\multirow{3}{3em}{$H$} & $X$ & $Z$ & \multirow{6}{3em}{${\rm{CNOT}}_{1,2}$} & $X_1$ & $X_1X_2$ &\multirow{6}{3em}{${\rm{CZ}}_{1,2}$} & $X_1$ & $X_1Z_2$\\ 
& $Y$ & $-Y$ & &$Y_1$ & $Y_1X_2$ && $Y_1$ & $Y_1Z_2$ \\ 
& $Z$ & $X$ & &$Z_1$ & $Z_1$&& $Z_1$ & $Z_1$ \\ 
\cline{1-3}\xrowht{10pt}
\multirow{3}{3em}{$P$} & $X$ & $Y$ && $X_2$ & $X_2$ && $X_2$ & $X_2Z_1$ \\ 
& $Y$ & $-X$ && $Y_2$ & $Z_1Y_2$ && $Y_2$ &  $Z_1Y_2$\\ 
& $Z$ & $Z$ &&  $Z_2$ & $Z_1Z_2$ && $Z_2$ & $Z_2$   \\
\noalign{\hrule height 1pt}
\end{tabular}
  \caption{Here, we tabulate the transformations of Pauli operators under all Clifford unitaries. For example, $HXH^\dagger = Z$, $PYP^\dagger = -X$, ${\rm{CNOT}}_{1,2} \, X_1 {\rm{CNOT}}_{1,2}\,^\dagger = X_1X_2$, and ${\rm{CZ}}_{1,2}\, Y_2\, {\rm{CZ}}_{1,2}^\dagger = Z_1Y_2$.}
  \label{tab:CliffordTransforms}
  \end{table*}

\subsection{Single-qubit stabilizer measurement on a multi-qubit stabilizer state}\label{subsec:StabMeasure}

Recall that an $X$ measurement on a qubit means measuring the qubit in the $\{\ket{\pm}\}$ basis, the eigenstates of the $X$ operator, and making a $Z$ measurement on a qubit is measuring it in the computational basis, $\{\ket{0},\ket{1}\}$, the eigenstates of the $Z$ operator. Let us consider a single-qubit measurement described by the Pauli operator $M$, on an $n$-qubit state $\ket{\psi}$ with stabilizer generators $\mathcal{G}(\ket{\psi}) = \langle S_1, S_2, \dots, S_n\rangle$. Any single-qubit measurement would produce a (probabilistic, in general) binary measurement result $m = \pm 1$. The post-measurement state, also a stabilizer state, is given by $\langle  x_m | \psi \rangle$ (normalized), where $\{\ket{x_m}\}$, $m = \pm 1$, are the eigenstates of $M$. The rules to find the stabilizer generators of the post-measurement state are given below \cite{gottesman1998heisenberg, MercedesThesis} and proved in Appendix~\ref{apx:StabMeasure}. Note that we did not include a subscript in $M$ to denote which qubit it acts on. Examples of $M$ can be: $X_2$, $Z_4$, $Y_3$, etc.
\newline\textbf{(a) If $M$ commutes with all elements of $\mathcal{G}$:} The state $\ket{\psi}$ then is an eigenstate of $M$. Hence, the state and its stabilizer generators remain unchanged. The measurement outcome is $1$ and is deterministic.
\newline\textbf{(b) If $M$ anti-commutes with $\langle S_1, S_2, \dots, S_l\rangle$ and commutes with $\langle S_{l+1}, S_{l+2}, \dots, S_n\rangle$:}
\begin{enumerate}
\item  Since at least one of the generators in $\mathcal{G}(\ket{\psi})$ anti-commutes with $M$, we will get a non-deterministic measurement outcome $m\in \{\pm 1\}$.
\item Because product of any two stabilizers is also a stabilizer, without loss of generality (WLOG), the generator set can be modified to: $\langle S_1, S_1S_2,S_1S_3, \dots, S_1S_l,S_{l+1}, S_{l+2}, \dots, S_n\rangle$. In this new set of stabilizer generators, each generator except for $S_1$ commutes with $M$. 
\item  The measured qubit collapses to the $\pm 1$ eigenstate of $M$ with equal probability. Hence, we will replace the anti-commuting generator $S_1$ with $mM$ with $m\in \{\pm 1\}$ the measurement outcome, to describe the state after measurement. 
\item The remainder of the generators that commute with $M$ are left unchanged.
\item  The generator set after the measurement is performed, is: $\langle mM, S_1S_2,S_1S_3, \dots, S_1S_l,S_{l+1}, S_{l+2}, \dots, S_n\rangle$.
\item Finally, if any of the commuting generators $\langle S_1S_2, \ldots, S_1S_l, S_{l+1}, \ldots, S_n\rangle$ has $M$ in it, the Pauli $M$ in that generator is replaced by the measurement result $m$.
\end{enumerate}
If a succession of $k$ Pauli measurements are performed on $k \le n$ qubits of an $n$-qubit stabilizer state, obtaining measurement results $m_1, \ldots, m_k$, the above procedure of evolving stabilizer generators must be repeated $k$ times. Some examples of evolution of stabilizers of a state under Pauli measurements are discussed in Section~\ref{sec:ClusterStates}.

\subsection{Two-qubit fusion measurements}
\label{sub:stab_fusion}
\begin{figure}
    \centering
\includegraphics[width=0.55\columnwidth]{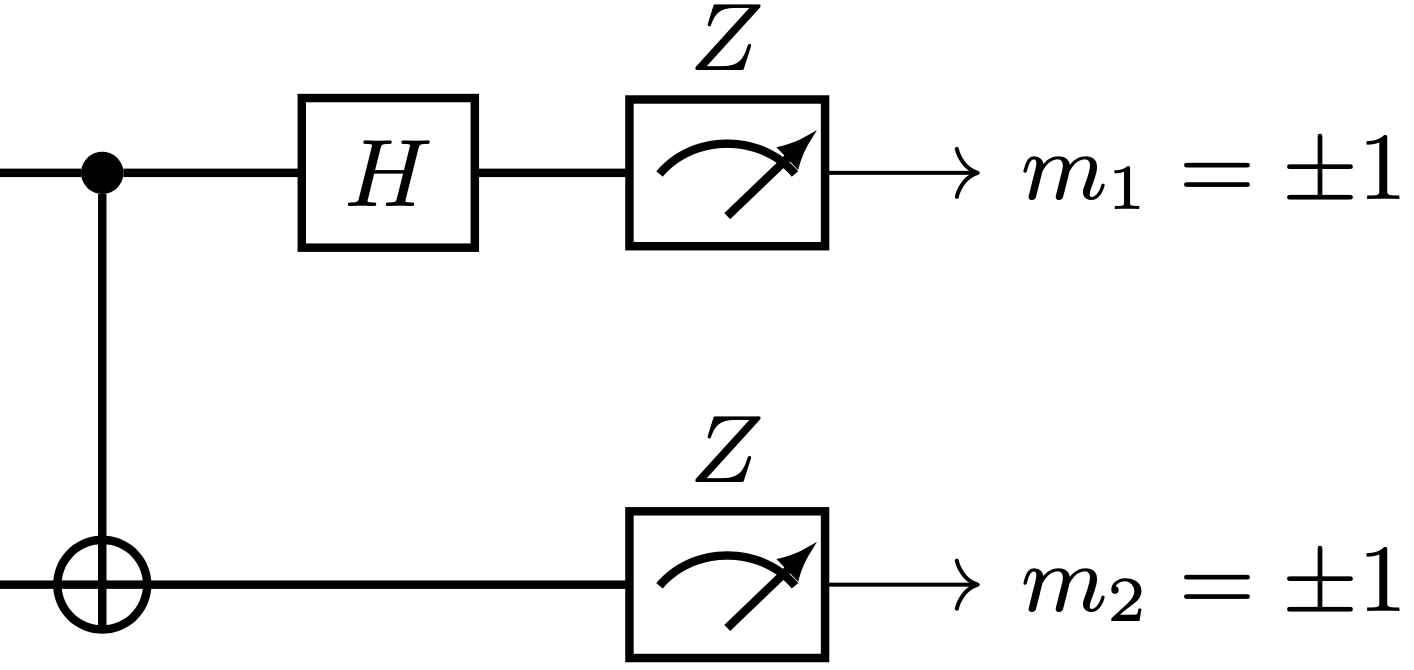}
    \caption{The basic fusion circuit, a von-Neumann projective measurement onto four mutually-orthogonal Bell states.}
   \label{fig:base_fusion}
\end{figure}
In this section, we will discuss two-qubit measurements (and projections) onto four mutually-orthogonal two-qubit Bell-like states (or some subset thereof). We call these~\textit{fusions}. Let us consider the circuit shown in Fig.~\ref{fig:base_fusion}: a ${\rm CNOT}_{1,2}$ followed by $H$ on qubit 1, followed by $Z$ measurements on both qubits. As we discussed in Section~\ref{subsec:measurements}, the two $Z$ measurements is a four-outcome stabilizer measurement described by $\langle Z_1I_2, I_1Z_2\rangle$. At the end of Section~\ref{subsec:UnitaryOps}, we derived that when a unitary $U$ precedes a stabilizer measurement described by stabilizers $\langle M_1,M_2,\dots,M_k\rangle$, the overall measurement can be described by the stabilizers $\langle U^\dagger M_1 U, U^\dagger M_2 U, \dots, U^\dagger M_k U\rangle$. In this case, $U = H_1 \,{\rm CNOT}_{1,2}$. Therefore, $U^\dagger = {\rm CNOT}_{1,2}\,H_1$. Hence, the overall measurement is described by the two stabilizers ${\rm CNOT}_{1,2}H_1(Z_1I_2)H_1 {\rm CNOT}_{1,2}$ and ${\rm CNOT}_{1,2}H_1(I_1Z_2)H_1 {\rm CNOT}_{1,2}$. Using the commutation rules from Section~\ref{subsec:UnitaryOps}, Fig.~\ref{fig:TwoQubitCliffords}, the above stabilizers simplify to: $\langle X_1X_2, Z_1Z_2\rangle $. The circuit in Fig.~\ref{fig:base_fusion} therefore represents a two-qubit von-Neumann projective measurement described by the simultaneous stabilizer subspaces of $\langle X_1X_2, Z_1Z_2\rangle$, $\langle -X_1X_2, Z_1Z_2\rangle $, $\langle X_1X_2, -Z_1Z_2\rangle$ and $\langle -X_1X_2, -Z_1Z_2\rangle$, i.e., the unique states $\ket{\psi^+} = (\ket{00} + \ket{11})/\sqrt{2}$, $\ket{\psi^-} = (\ket{00} - \ket{11})/\sqrt{2}$, $\ket{\phi^+} = (\ket{01} + \ket{10})/\sqrt{2}$ and $\ket{\phi^-} = (\ket{01} - \ket{10})/\sqrt{2}$, respectively. These four mutually-orthogonal states are called {\em Bell states}---maximally-entangled states of two qubits---and this measurement is the Bell State Measurement (BSM). 

\begin{figure}
	\centering
	\begin{quantikz}
	\lstick{$\ket{c}$} & \gate{R_c^{\dagger}} & \ctrl{1} & \gate{H} & \meter{$Z$} \arrow[r] & \rstick{$m_1=\pm1$}\\
	\lstick{$\ket{t}$} & \gate{R_t^{\dagger}} & \targ{} & \qw & \meter{$Z$} \arrow[r] & \rstick{$m_2=\pm1$}\\
	\end{quantikz}
	\caption{Rotated Bell State Measurement (BSM), or a {\em fusion}. $m_1 \in \{\pm 1\}$ and $m_2 \in \{\pm 1\}$ denote the measurement results. This circuit projects qubits $c$ and $t$ on $R_cR_tZ_c^{(1-m_1)/2}X_t^{(1-m_2)/2}\big({\ket{00}_{c,t}+\ket{11}_{c,t}}\big)/{\sqrt{2}}$. $R_c$ and $R_t$ are single qubit Clifford unitary rotations on the control and target qubits, prior to the Bell basis measurement of Fig.~\ref{fig:base_fusion}.}
	\label{fig:fusionQckt}
\end{figure}
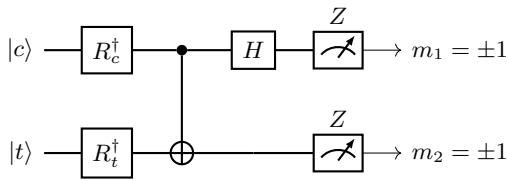
We will term any Clifford-{\em rotated} BSM, as shown in Fig.~\ref{fig:fusionQckt}, a {\em fusion}. The control qubit (denoted $\ket{c}$) undergoes a Clifford unitary $R_c^\dagger$ and the target qubit (denoted $\ket{t}$) undergoes a Clifford unitary $R_t^\dagger$ before undergoing the BSM depicted in Fig.~\ref{fig:base_fusion}. Per the discussion above, the overall measurement on the control and target qubits is now described by the stabilizers $\langle V^\dagger (X_1X_2) V, V^\dagger (Z_1Z_2) V\rangle$, where $V = R_c^\dagger R_t^\dagger$. Using the discussion in Section~\ref{subsec:measurements}, the four POVM operators of this rotated BSM are: $\{V^\dagger\ket{\psi^\pm}\bra{\psi^\pm}V, V^\dagger\ket{\phi^\pm}\bra{\phi^\pm}V\}$.
\begin{figure*}
    \centering
\includegraphics[width=0.9\textwidth]{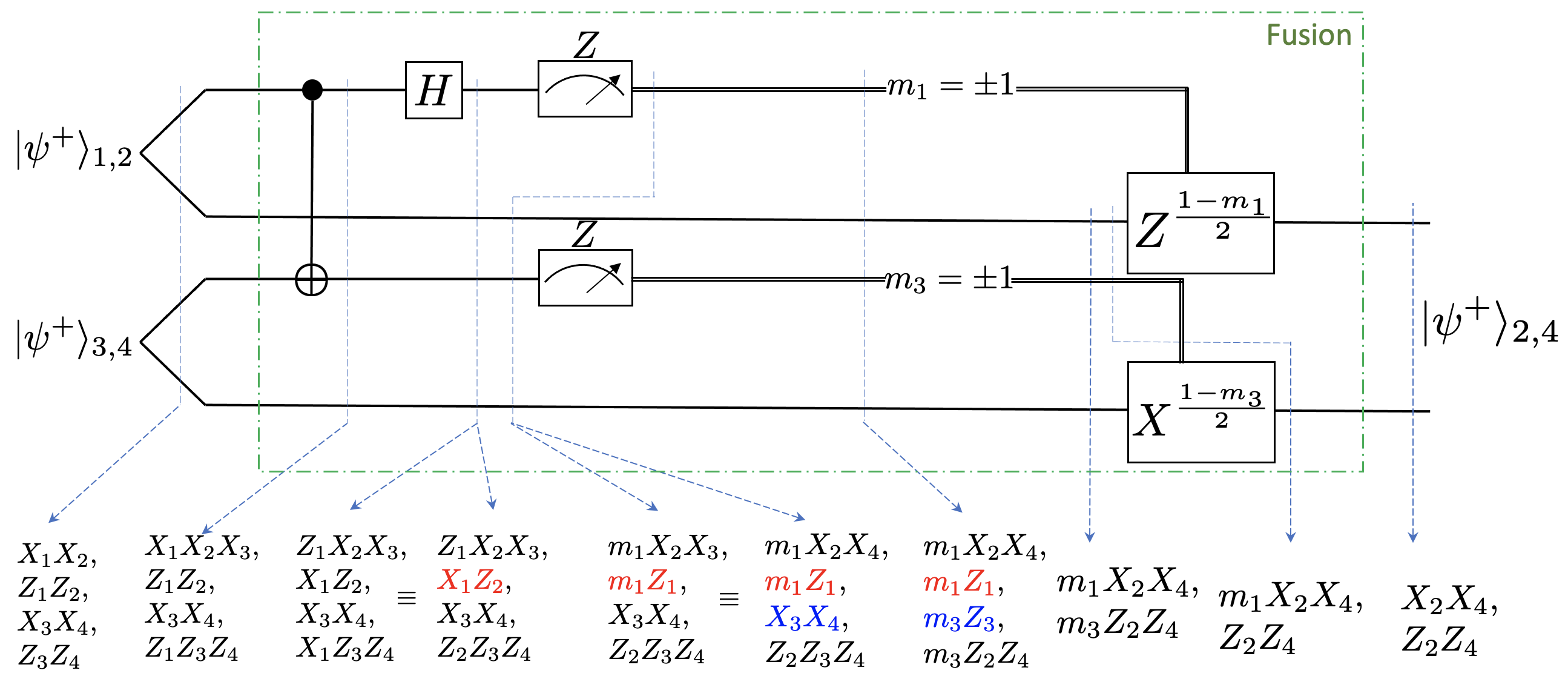}
    \caption{Fusion, with $R_c^\dagger = R_t^\dagger$ both set to identity, on one qubit each of two copies of the $2$-qubit $\ket{\psi^+}$ Bell state, followed by conditional single-qubit Pauli unitaries on the unmeasured qubits, gets a $\ket{\psi^+}$ state on the unmeasured qubits.}
   \label{fig:fusion}
\end{figure*}
To appreciate why this is called a {\em fusion}, let us consider performing the rotated BSM shown in Fig.~\ref{fig:fusionQckt}, with $R_c^\dagger = R_t^\dagger = I$, on qubits $1$ (control) and $3$ (target) of two copies of the Bell state $\ket{\psi^+}_{1,2} \ket{\psi^+}_{3,4}$. It is straightforward to evolve the states, represented by their stabilizer generators, through the unitaries and the measurements---using rules described in Sections~\ref{sec:stabilizeroperators},~\ref{subsec:measurements},~\ref{subsec:UnitaryOps} and~\ref{subsec:StabMeasure} (see Fig.~\ref{fig:fusion})---obtaining the state $\langle m_1X_2X_4, m_3Z_2Z_4\rangle$ conditioned on the measurement outcomes $m_1$ and $m_3$. By applying $I$ or $Z$ to qubit $2$ based on whether $m_1 = 1$ or $-1$, and applying $I$ or $X$ to qubit $4$ based on whether $m_3 = 1$ or $-1$, we obtain the $\ket{\psi^+}_{2,4}$ Bell state among qubits $2$ and $4$. The rotated BSM together with the two conditional single-qubit unitary operations has an action of `fusing' qubits $2$ and $4$ that were originally part of two separate entangled states, into a new entangled state. For this reason, the circuit in Fig.~\ref{fig:fusion}, comprising a rotated BSM and two conditional Paulis, is known as a {\em fusion} measurement. Note that if the $(m_1,m_3)$-dependent Pauli correction are not applied to the output state $\langle m_1X_2X_4, m_3Z_2Z_4 \rangle$, it is a von Neumann projective measurement. If both (or one) of the Pauli corrections are applied, it is a projection onto one (or respectively, two) of the four Bell state(s).

If one wishes to implement a {\em projection} onto a subset, e.g., one or two, of the $4$ Bell states $\{\ket{\psi^\pm},\ket{\phi^\pm}\}$, one can use the same circuit as above, but herald the output based on specific values of $m_1$ and/or $m_3$. In the remainder of this paper, we will simply refer to the rotated BSM as a {\em fusion}, for both cases of when the circuit is used to perform a full Bell State Measurement or a projection, and will implicitly assume that the conditional Pauli corrections have been applied. In this paper, we will study the four types of fusions, shown in Table~\ref{tab:fusion_rules}.

\subsection{Multi-qubit fusion measurement}
\label{subsec:GHZproj1}
\begin{figure}
	\centering
	\begin{quantikz}
		\lstick{$\ket{c}$}  & \ctrl{1} & \ctrl{2} & \gate{H} & \meter{$Z$} \arrow[r] & \rstick{$m_1=\pm1$}\\
		\lstick{$\ket{t_1}$} & \targ{} & \qw & \qw & \meter{$Z$} \arrow[r] & \rstick{$m_2=\pm1$}\\
		\lstick{$\ket{t_2}$} & \qw & \targ{} & \qw & \meter{$Z$} \arrow[r] & \rstick{$m_3=\pm1$}\\
	\end{quantikz}
	\caption{A three-qubit fusion. This circuit projects qubits $c$, $t_1$ and $t_2$ on the $3$-qubit GHZ state $\frac{\ket{000}_{c,t_1,t_2}+\ket{111}_{c,t_1,t_2}}{\sqrt{2}}$, provided all the measurement results, $m_1 = m_2 = m_3 = +1$.} 
	\label{fig:GHZfusionQckt}
\end{figure}
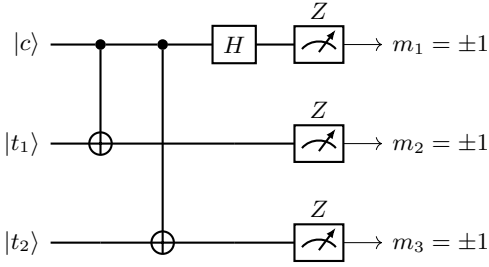

An $n$-fusion, $n \ge 2$ is a projective measurement (or, a projection) onto the $2^n$ mutually-orthogonal maximally-entangled $n$-qubit states (or a subset thereof), known as the GHZ states. We can extend the circuit for two-qubit fusion discussed in Section~\ref{sub:stab_fusion} to realize $n$-fusions. Consider a measurement on three qubits realized by the circuit shown in Fig.~\ref{fig:GHZfusionQckt}. This is a $3$ qubit fusion measurement described by stabilizers: $\langle X_1X_2X_3, Z_1Z_2, Z_1Z_3\rangle$. This can be seen by observing that the independent $Z$ measurements on the three qubits is described by stabilizers: $\langle Z_1I_2I_3, I_1Z_2I_3, I_1I_2Z_3 \rangle$, and then transforming these measurement stabilizers by the two CNOTs, as we did before for two-fusion. In other words, this is a von Neumann projective measurement described by $2^3 = 8$ projection operators, each of which is a rank-one projector onto one of the $8$ three-qubit maximally-entangled states described by stabilizers $\langle\pm X_1X_2X_3, \pm Z_1Z_2, \pm Z_1Z_3\rangle$. Generalization of this circuit to realize an $n$-fusion, described by stabilizers $\langle X_1X_2\dots X_n, Z_1Z_2, Z_1Z_3,\ldots, Z_1Z_n\rangle$ is straightforward. The $n$-qubit GHZ state stabilized by $\langle X_1X_2\dots X_n, Z_1Z_2, Z_1Z_3,\ldots, Z_1Z_n\rangle$, is given by $\frac{\ket{0}^{\otimes n}+\ket{1}^{\otimes n}}{\sqrt{2}}$. Depending upon the measurement outcome, single qubit Pauli operations are applied to the state resulting after measurements to achieve the desired projection. It is simple to create rotated $n$-fusions by applying single qubit unitaries before the CNOTs in Fig.~\ref{fig:GHZfusionQckt}.

Fusions are used to grow multi-qubit stabilizer states, esp. cluster states~\cite{MercedesThesis} (described in Section~\ref{sec:ClusterStates}). Fusions performed at network nodes can be used to efficiently distribute GHZ states among distant users~\cite{patil2020entanglement,patil2021distance}. 

Now, we generalize the rules for the transformation of stabilizers of an $n$-qubit state after an $l$-qubit joint measurement, $l\leq n$, of the stabilizers $M=\langle M_1, M_2,\dots, M_l\rangle$. Let $\mathcal{G} = \langle S_1,S_2,\dots, S_n\rangle$ be the stabilizer generators of the $n$-qubit state that is measured. 
\begin{itemize}
    \item If all elements in $M$ commute with all stabilizer generators in $\mathcal{G}$, then the state remains unchanged after the measurement. 
    \item Otherwise, $\mathcal{G}$ is transformed as follows:
    \begin{enumerate}
        \item Let $S_{M_i}=\{S_{i1},S_{i2},\dots\}$ be the subset of the generators that anti-commute with $M_i$ and let $S'$ be the set of generators s.t. every generator in $S'$ commutes with all elements in $M$. Note that, $\mathcal{G} = (\cup_{i=1}^lS_{M_i})\cup S'$. Modify the stabilizer generators in $\mathcal{G}$ (by replacing a stabilizer generator by the product of that generator with other generators) such that for every $M_i\in M$, there is exactly one anti-commuting generator in $\mathcal{G}$. Equivalently, replace every $S_{M_i}$ with $S'_{M_i}=\{S_{i1},S_{i1}S_{i2},S_{i1}S_{i3},\dots\}$. Now $\mathcal{G} = (\cup_{i=1}^lS'_{M_i})\cup S'$.
        \item Replace the anti-commuting generator $S_{i1}$ in $\mathcal{G}$ with $m_iM_{i}$ $\forall i \in \{1,2,\dots l\}$.  Here, $m_i$ is the measurement outcome. The new set of generators are: $\mathcal{G}' = (\cup_{i=1}^lS''_{M_i})\cup S'$, $S''_{M_i}=\{m_iM_{i},S_{i1}S_{i2},S_{i1}S_{i3},\dots\}$. 
    \end{enumerate}
\end{itemize}

The proof for these rules is described in Appendix~\ref{apx:StabMeasure}.

\textbf{Example 1}---Given two 3-qubit star-shaped cluster states with generators $\mathcal{G} = \langle X_1Z_2Z_3, Z_1X_2, Z_1X_3, X_4Z_5Z_6, Z_4X_5, Z_4X_6\rangle$, perform fusion between qubits 2 and 4, i.e., measure $M=\langle X_2X_4,Z_2Z_4\rangle$. 
\newline
$\mathcal{G}$ contains generators that anti-commute with operators in $M$. 
We can rewrite $\mathcal{G}$ as $\mathcal{G} = \langle X_1Z_2Z_3, Z_1X_2, Z_1X_3, Z_1X_2X_4Z_5Z_6, X_1Z_2Z_3Z_4X_5,$ $ X_1Z_2Z_3Z_4X_6\rangle$ so that the measurement operators, $X_2X_4$ and $Z_2Z_4$ anti-commute with exactly one generator each ($X_1Z_2Z_3, Z_1X_2$, respectively). The post-measurement state then is - $\langle m_1X_2X_4, m_2Z_2Z_4, Z_1X_3, Z_1X_2X_4Z_5Z_6, X_1Z_2Z_3Z_4X_5, $ $X_1Z_2Z_3Z_4X_6\rangle$. Or, equivalently - $\langle m_1X_2X_4, m_2Z_2Z_4, Z_1X_3, Z_1Z_5Z_6, $ $X_1Z_3X_5, X_1Z_3X_6\rangle$.

\textbf{Example 2}---Given three Bell states with generators $\mathcal{G} = \langle X_1X_2,Z_1Z_2,X_3X_4,Z_3Z_4,X_5X_6,Z_5Z_6\rangle$, perform a 3-fusion measurement with $M=\langle X_1X_3X_5, Z_1Z_3, Z_1Z_5\rangle$.
\newline 
Let's rewrite $\mathcal{G} = \langle X_1X_2X_3X_4X_5X_6, Z_1Z_2, X_3X_4, $ $Z_1Z_2Z_3Z_4, X_5X_6, Z_1Z_2Z_5Z_6\rangle$. The generators after the fusion are:  $\mathcal{G}' = \langle X_1X_2X_3X_4X_5X_6,$ $m_1X_1X_3X_5,m_2Z_1Z_3,Z_1Z_2Z_3Z_4,m_3Z_1Z_5,Z_1Z_2Z_5Z_6\rangle$. After simplifying, $\mathcal{G}' = \langle m_1X_1X_3X_5, m_2Z_1Z_3, $ $m_3Z_1Z_5, X_2X_4X_6, Z_2Z_4,Z_2Z_6 \rangle$, which reflects the GHZ-state created between qubits 2, 4, and 6 as a result of the 3-qubit fusion.

\section{Cluster States}
\label{sec:ClusterStates}
Cluster states, or graph states, are highly entangled states used as the building block of measurement-based quantum computing (MBQC)~\cite{raussendorf2001one,raussendorf2003measurement,walther2005experimental,vallone2008active,wang201816,bourassa2021blueprint,fukui2018high} and can themselves act as error-correcting codes to achieve fault-tolerant MBQC~\cite{raussendorf2005long,raussendorf2006fault,raussendorf2007fault,nickerson2018measurement,bolt2016foliated,yao2012experimental,lanyon2013measurement,schlingemann2001quantum,zwerger2014hybrid,schlingemann2001stabilizer}. They can also be used for to distribute shared entanglement in quantum networks~\cite{pant2017allOptical, grassl2002graphs}. 

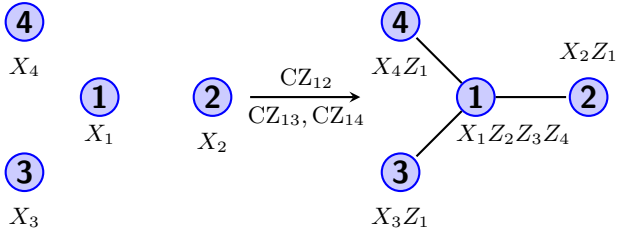
\begin{figure}[h]
		\centering
		\begin{tikzpicture}[shorten >=1pt, auto, node distance=2cm, thick,
		node_style/.style={circle,inner sep=0pt,minimum size=0.5cm,draw=blue,fill=blue!20!,font=\sffamily\large\bfseries},
		node3_style/.style={circle,inner sep=0pt,minimum size=2cm,yscale=.35,draw=black,font=\sffamily\large\bfseries},
		node2_style/.style={circle,dashed,inner sep=0pt,minimum size=0.5cm,draw=blue,fill=blue!20!,font=\sffamily\large\bfseries},
		edge_style/.style={draw=black, thick}
		arrow_style/.style={draw=black,thick,->}]
		
		\node[node_style] (u4) at (-10.5,4) {4};
		\node at (-10.5,3.4) {$X_4$};
		\node[node_style] (u1) at (-9.5,3) {1};
		\node at (-9.5,2.5) {$X_1$};
		\node[node_style] (u3) at (-10.5,2) {3};
		\node at (-10.5,1.4) {$X_3$};	
		\node[node_style] (u2) at (-8,3) {2};
		\node at (-8,2.4) {$X_2$};
		
		\node[node_style] (u4) at (-5.5,4) {4};
		\node at (-5.5,3.4) {$X_4Z_1$};
		\node[node_style] (u1) at (-4.5,3) {1};
		\node at (-4,2.5) {$X_1Z_2Z_3Z_4$};
		\node[node_style] (u3) at (-5.5,2) {3};
		\node at (-5.5,1.4) {$X_3Z_1$};	
		\node[node_style] (u2) at (-3,3) {2};
		\node at (-3,3.6) {$X_2Z_1$};
		\draw[edge_style]  (u1) edge (u2);
		\draw[edge_style]  (u1) edge (u3);
		\draw[edge_style]  (u1) edge (u4);
		\draw [-stealth,thick] (-7.5,3) -- node[above] {$\text{CZ}_{12}$} node[below] {$\text{CZ}_{13},\text{CZ}_{14}$} (-6,3);

		\end{tikzpicture}
		\caption{ Starting with four single qubits in state $\ket{+}$, described by stabilizer generators $\left\langle X_i\right\rangle $, a 4-qubit star-shaped cluster state is created by applying three CZ gates. The resulting state has four stabilizer generators of the form~(\ref{eq:clusterStab}).}

		\label{fig:star}
	\end{figure}

A cluster state is described by a graph $G(V, E)$, of $n = |V|$ vertices. A cluster state $|\psi\rangle_G$ is created by placing a qubit initialized in the $\ket{+}$ state at each vertex $v \in V$, and applying CZ$_{u,v}$ gates for every edge $e \equiv (u,v) \in E$. An example of a $4$-qubit star-topology cluster state is shown in FIG.~\ref{fig:star}. Cluster states are stabilizer states. But not all stabilizer states are cluster states. The stabilizer generators of $|\psi\rangle_G$, $S_i$, $i \in 1, \ldots, n$, where $n = |V|$ is the number of qubits in $|\psi\rangle_G$, are given by:
\begin{align}
\label{eq:clusterStab}
    S_i&=X_i\prod_{j\in \mathcal{N}_i} Z_j,
\end{align}
where $\mathcal{N}_i$ is the neighborhood of vertex $i$ in $G$. Eq. (\ref{eq:clusterStab}) can be derived using the tools developed in Section~\ref{subsec:UnitaryOps}, starting with the stabilizers of $\ket{+}^{\otimes n}$, i.e., $X_1II\ldots I, IX_2I\ldots I, IIX_3\ldots I, \ldots, I\ldots IX_n$, followed by evolving the stabilizers for each CZ gate applied, to create $|\psi\rangle_G$. We would often associate the stabilizer generator $S_i$ with qubit $i$ of the cluster state. 

\subsection{Relationship between cluster state and GHZ state}
\label{sub:cluster_GHZ}

\begin{figure*}[htb]
		\centering
		\begin{tikzpicture}[shorten >=1pt, auto, node distance=2cm, thick,
		node_style/.style={circle,inner sep=0pt,minimum size=0.5cm,draw=blue,fill=blue!20!,font=\sffamily\large\bfseries},
  node4_style/.style={circle,inner sep=0pt,minimum size=0.5cm,draw=red,fill=red!20!,font=\sffamily\large\bfseries},  
  node5_style/.style={circle,dashed,inner sep=0pt,minimum size=0.5cm,draw=red,fill=red!20!,font=\sffamily\large\bfseries},
		node3_style/.style={circle,inner sep=0pt,minimum size=2cm,yscale=.85,draw=black,font=\sffamily\large\bfseries},
		node2_style/.style={circle,dashed,inner sep=0pt,minimum size=0.5cm,draw=blue,fill=blue!20!,font=\sffamily\large\bfseries},
		edge_style/.style={draw=black, thick}
		arrow_style/.style={draw=black,thick,<->}]
            
            \node at (-12-.5-.5,4-.5) {$Z_1Z_2Z_3Z_4$, $X_1X_3$};
            \node at (-12-.5-.5,4-0.5-0.5) {$X_1X_2$, $X_1X_4$};
            \node[node5_style] (u4) at (-13.5,2) {4};
            \node[node5_style] (u4) at (-12.5,2) {2};
            \node[node5_style] (u4) at (-13,2.5) {1};
            \node[node5_style] (u4) at (-13,1.5) {3};
            \node[node3_style] (u4) at (-13,2) {};
            \node at (-12-.5-.5,4) {GHZ state in +/- basis};
		\draw [-stealth,thick] (-10.5,2.75) -- node[above] {$\text{H}_{1}$}  (-11.5,2.75);
            \draw [stealth-,thick] (-10.5,2.75) --  (-11.5,2.75);

		\node[node_style] (u4) at (-10.5,4) {4};
		\node at (-10.5,3.4) {$X_4Z_1$};
		\node[node_style] (u1) at (-9.5,3) {1};
		\node at (-9,3.4) {$X_1Z_2Z_3Z_4$};
		\node[node_style] (u3) at (-10.5,2) {3};
		\node at (-10.5,1.4) {$X_3Z_1$};	
		\node[node_style] (u2) at (-8,3) {2};
		\node at (-8,2.4) {$X_2Z_1$};
            \draw[edge_style]  (u1) edge (u2);
		\draw[edge_style]  (u1) edge (u3);
		\draw[edge_style]  (u1) edge (u4);

		 \node at (-4-.5-1.25,4-.5-0.5) {$X_1X_2Z_3Z_4$};
            \node at (-4-.5-1.25,4-0.5-0.5-0.5) {$Z_1Z_2$};
            \node at (-4-.5-1.25,4-0.5-0.5-0.5-0.5) {$Z_1X_3$};
            \node at (-4-.5-1.25,4-0.5-0.5-0.5-0.5-0.5) {$Z_1X_4$};
            \node at (-4-.5-1.25,4) {Not a GHZ or};
            \node at (-4-.5-1.25,4-0.5) {a cluster state};
		\draw [-stealth,thick] (-7.5,2.75) -- node[above] {$\text{H}_{2}$}  (-6.5,2.75);
  \draw [stealth-,thick] (-7.5,2.75) -- (-6.5,2.75);
  
            \draw [-stealth,thick] (-4.5-0.5,2.75) -- node[above] {$\text{H}_{3}$}  (-4.5+.5,2.75);
            \draw [stealth-,thick] (-4.5-0.5,2.75) --   (-4.5+0.5,2.75);

            \node[node4_style] (u4) at (-3.5,4) {1};
		\node at (-3.5,3.4) {$X_4Z_1$};
		\node[node4_style] (u1) at (-2.5,3) {4};
		\node at (-2,2.5) {$Z_4X_1X_2X_3$};
            \node at (-1.8,2) {Cluster state in};
            \node at (-1.8,1.5) {the +/- basis};
		\node[node4_style] (u3) at (-3.5,2) {3};
		\node at (-3.5,1.4) {$X_4Z_3$};	
		\node[node4_style] (u2) at (-1,3) {2};
		\node at (-1,3.6) {$X_4Z_2$};
		\draw[edge_style]  (u1) edge (u2);
		\draw[edge_style]  (u1) edge (u3);
		\draw[edge_style]  (u1) edge (u4);
            \draw [-stealth,thick] (-.5,-1+4) -- node[above] {$\text{H}_{4}$}  (.5,-1+4);
            \draw [stealth-,thick] (-.5,-1+4) --   (.5,-1+4);

            \node at (-.5+2,-.5+4) {$X_1X_2X_3X_4$, $Z_1Z_3$};
            \node at (-.5+2,-0.5-0.5+4) {$Z_1Z_2$, $Z_1Z_4$};
            \node at (-.5+2,0+4) {GHZ state in 0/1-basis};
            \node[node2_style] (u4) at (-13.5+14.5,2) {4};
            \node[node2_style] (u4) at (-12.5+14.5,2) {2};
            \node[node2_style] (u4) at (-13+14.5,2.5) {1};
            \node[node2_style] (u4) at (-13+14.5,1.5) {3};
            \node[node3_style] (u4) at (-13+14.5,2) {};
            
		\end{tikzpicture}
		\caption{ Applying a Hadamard gate on qubit number $k$ of an $n$-qubit GHZ state in the $(0/1)$ basis (resp. $(+/-)$ basis) results in a star-shaped cluster state in the $(+/-)$ basis (resp. $(0/1)$ basis) with the qubit $k$ at the center of the star, and the remaining $n-1$ qubits as leaf nodes. If Hadamard gates are applying to all the qubits except qubit number $k$, of an $n$-qubit GHZ state in the $(0/1)$ basis (resp. $(+/-)$ basis), one obtains a star-shaped cluster state in the $(0/1)$ basis (resp. $(+/-)$ basis) with the qubit $k$ at the center of the star.}

		\label{fig:GHZ_cluster}
	\end{figure*}
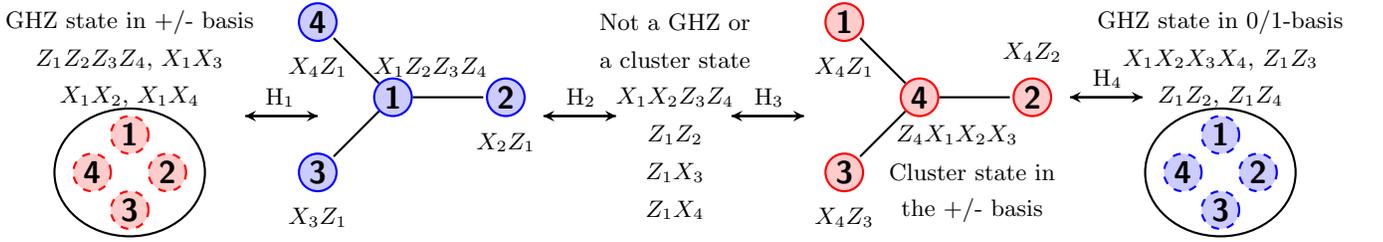

Cluster states are local Clifford (LC) equivalent to stabilizer states \cite{van2004efficient, grassl2002graphs, schlingemann2001stabilizer}. Let us consider one illustrative example of this equivalence, which we will find useful in the subsequent subsections. Consider the $n$-qubit GHZ state in the computational ($0/1$) basis, $|{\text{GHZ}}_{(0/1)}\rangle = \frac{\ket{0}^{\otimes n}+\ket{1}^{\otimes n}}{\sqrt{2}}$ described by stabilizer generators $\langle X_1X_2\dots X_n, Z_1Z_2, Z_1Z_3,\ldots, Z_1Z_n\rangle$. Also consider the $n$-qubit GHZ state in the conjugate ($+/-$) basis, $|{\text{GHZ}}_{(+/-)}\rangle = \frac{\ket{+}^{\otimes n}+\ket{-}^{\otimes n}}{\sqrt{2}}$ described by stabilizer generators $\langle Z_1Z_2\dots Z_n, X_1X_2, X_1X_3,\ldots, X_1X_n\rangle$. Consider the $n$-qubit star-shaped cluster state with node-$k$ at the center (and the other $n-1$ nodes being leaf nodes), $|{\text{star}}_k\rangle$. It is a simple calculation to see that if we apply Hadamard gates on all the qubits except qubit number $k$ of the $n$-GHZ state $|{\text{GHZ}}_{(0/1)}\rangle$, we would obtain $\ket{{\text{star}}_k}$. If we then apply a Hadamard gate on qubit-$k$ as well, we would obtain the state $|{\text{GHZ}}_{(+/-)}\rangle$. Consequently, applying a single Hadamard gate on qubit number $l$ of $|{\text{GHZ}}_{(+/-)}\rangle$ results in $|{\text{star}}_l\rangle$, for any $l$. Finally, the GHZ state $|{\text{GHZ}}_{(0/1)}\rangle$ (which is symmetric with respect to interchanging any pair of qubits) can be converted into $|{\text{star}}_k\rangle$ by applying Hadamard gates on all but the qubit $k$. The aforesaid relationship between GHZ and cluster states is summarized in Figure~\ref{fig:GHZ_cluster}.

In the remainder of this Section, we use the material developed in Section~\ref{subsec:StabMeasure} to work out how the cluster state stabilizers change after undergoing Pauli measurements on a qubit, and a few other important Clifford unitaries and measurements, express the results using graph-theoretic rules, e.g., in the form of addition or deletion of edges, and inverting the neighborhood of the graph at a node. One important thing to remember is that two (parallel) edges between two qubits cancel each other out, as this is equivalent to applying a CZ twice, which results in an Identity operation. As a result, if a graph rule corresponding to any of the stabilizer (gate of measurement) operation we study henceforth asks to add an edge between two qubits that already have an edge, the existing edge must be removed.

\subsection{Pauli Z basis measurement}
\label{sec:PauliZ}

\begin{figure*}[htb]
		\centering
		\begin{tikzpicture}[shorten >=1pt, auto, node distance=2cm, thick,
		node_style/.style={circle,inner sep=0pt,minimum size=0.5cm,draw=blue,fill=blue!20!,font=\sffamily\large\bfseries},
  node4_style/.style={circle,inner sep=0pt,minimum size=0.5cm,draw=red,fill=red!20!,font=\sffamily\large\bfseries},  
  node5_style/.style={circle,dashed,inner sep=0pt,minimum size=0.5cm,draw=red,fill=red!20!,font=\sffamily\large\bfseries},
		node3_style/.style={circle,inner sep=0pt,minimum size=2cm,yscale=.85,draw=black,font=\sffamily\large\bfseries},
		node2_style/.style={circle,dashed,inner sep=0pt,minimum size=0.5cm,draw=blue,fill=blue!20!,font=\sffamily\large\bfseries},
		edge_style/.style={draw=black, thick}
		arrow_style/.style={draw=black,thick,<->}]
         
		\node[node_style] (v1) at (-4,2) {1};
		\node at (-4,1.5) {$X_1Z_2$};
            \node at (-4.25,2.75) {(a)};
		\node[node_style] (v2) at (-2.5,2) {2};
		\node at (-2.5,1.5) {$Z_1X_2Z_3$};
		\node[node_style] (v3) at (-1,2) {3};
		\node at (-1,1.5) {$Z_2X_3Z_4$};
		\node[node_style] (v4) at (0.5,2) {4};
		\node at (0.5,1.5) {$Z_3X_4Z_5$};	
		\node[node_style] (v5) at (2,2) {5};
		\node at (2,1.5) {$Z_4X_5$};

		\draw[edge_style]  (v1) edge (v2);
		\draw[edge_style]  (v2) edge (v3);
		\draw[edge_style]  (v3) edge (v4);
		\draw[edge_style]  (v4) edge (v5);
		
        \node[inner sep=0pt] (meter) at (3,2.25)
    {\begin{quantikz}
		 \meter{$Z_3$}
    \end{quantikz}};

    \node[node_style] (v1) at (4,2) {1};
		\node at (4,1.5) {$X_1Z_2$};
		\node[node_style] (v2) at (5.5,2) {2};
		\node at (5.5,1.5) {$Z_1X_2$};
		\node[node_style] (v3) at (7,2) {3};
		\node at (0+7,1.5) {$Z_3$};
            \node at (3.75,2.75) {(b)};
		\node[node_style] (v4) at (8.5,2) {4};
		\node at (8.5,1.5) {$X_4Z_5$};	
		\node[node_style] (v5) at (10,2) {5};
		\node at (10,1.5) {$Z_4X_5$};

		\draw[edge_style]  (v1) edge (v2);
		\draw[edge_style]  (v4) edge (v5);

		\end{tikzpicture}
		\caption{Final state after qubit $3$ of the $5$-qubit line cluster state is measured in the Pauli $Z$ basis.}

		\label{fig:measureZ_line}
	\end{figure*}
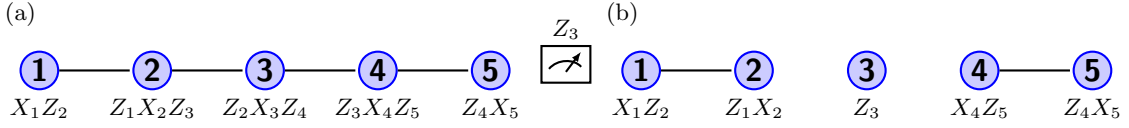

Consider a 5-qubit line cluster state as shown in FIG.~\ref{fig:measureZ_line}(a), and let us say we measure qubit 3 in the Z (computational, or $0/1$) basis. $Z_3$ anti-commutes with $Z_2X_3Z_4$ and commutes with every other stabilizer of the line cluster state. Following the rules in Section~\ref{subsec:StabMeasure}, to evaluate the post-measurement state, we first replace $Z_2X_3Z_4$ with ${m_3}Z_3$, with $m_3\in \{\pm 1\}$ corresponding to the (random) measurement result. The new generators are $\langle X_1Z_2,m_3Z_1X_2,m_3Z_3,m_3X_4Z_5,Z_4X_5\rangle $. As discussed in Section~\ref{subsec:StabMeasure}, if $m_3=-1$, we can flip the signs of the stabilizer generators $m_3Z_3$, $m_3Z_1X_2$ and $m_3X_4Z_5$ by applying a single-qubit $Z$ to qubits $3$, $2$ and $4$ respectively. Therefore, henceforth, we will not bother carrying the $m_3$. The final set of generators after the $Z$ measurement on qubit $3$ is, therefore: $\langle X_1Z_2,Z_1X_2,Z_3,X_4Z_5,Z_4X_5\rangle $. Comparing this to Eq. (\ref{eq:clusterStab}), we observe that it corresponds to two $2$-qubit line-cluster states and a detached qubit $3$ in state $\ket{0}$ (see FIG. \ref{fig:measureZ_line}(b)).

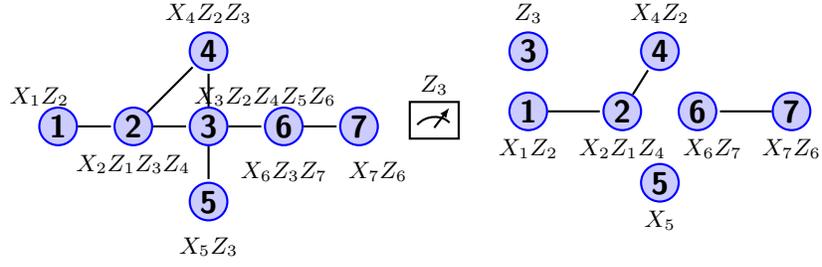
\begin{figure*}[htb]
		\centering
		\begin{tikzpicture}[shorten >=1pt, auto, node distance=2cm, thick,
		node_style/.style={circle,inner sep=0pt,minimum size=0.5cm,draw=blue,fill=blue!20!,font=\sffamily\large\bfseries},
  node4_style/.style={circle,inner sep=0pt,minimum size=0.5cm,draw=red,fill=red!20!,font=\sffamily\large\bfseries},  
  node5_style/.style={circle,dashed,inner sep=0pt,minimum size=0.5cm,draw=red,fill=red!20!,font=\sffamily\large\bfseries},
		node3_style/.style={circle,inner sep=0pt,minimum size=2cm,yscale=.85,draw=black,font=\sffamily\large\bfseries},
		node2_style/.style={circle,dashed,inner sep=0pt,minimum size=0.5cm,draw=blue,fill=blue!20!,font=\sffamily\large\bfseries},
		edge_style/.style={draw=black, thick}
		arrow_style/.style={draw=black,thick,<->}]
         
		\node[node_style] (u2) at (-12,3-0.5) {2};
		\node at (-12,2.5-0.5) {$X_2Z_1Z_3Z_4$};
            \node[node_style] (u1) at (-13,3-0.5) {1};
		\node at (-13-0.25,3.4-0.5) {$X_1Z_2$};
            \node[node_style] (u4) at (-11,4-0.5) {4};
		\node at (-11,4) {$X_4Z_2Z_3$};
		\node[node_style] (u3) at (-11,3-0.5) {3};
		\node at (-10.25,3.4-0.5) {$X_3Z_2Z_4Z_5Z_6$};
		\node[node_style] (u5) at (-11,2-0.5) {5};
		\node at (-11,1.4-0.5) {$X_5Z_3$};	
		\node[node_style] (u6) at (-10,3-0.5) {6};
		\node at (-10,2.4-0.5) {$X_6Z_3Z_7$};
            \node[node_style] (u7) at (-9,3-0.5) {7};
		\node at (-8.75,2.4-0.5) {$X_7Z_6$};
            \draw[edge_style]  (u1) edge (u2);
		\draw[edge_style]  (u2) edge (u3);
  \draw[edge_style]  (u2) edge (u4);
		\draw[edge_style]  (u3) edge (u4);
            \draw[edge_style]  (u3) edge (u5);
            \draw[edge_style]  (u3) edge (u6);
            \draw[edge_style]  (u6) edge (u7);

            \node[node_style] (u1) at (-7+0.25,2+1-0.3) {1};
             \node at (-7+0.25,2+1-0.3-0.5) {$X_1Z_2$};
             \node[node_style] (u7) at (-3.25,2+1-0.3) {7};
             \node at (-3.25,2+1-0.3-0.5) {$X_7Z_6$};
            \node[node_style] (u3) at (-6.75,3.5) {3};
            \node at (-7+0.25,4) {$Z_3$};
		\node[node_style] (u4) at (-13.5+8,2+1-0.3) {2};
  \node at (-13.5+8,2+1-0.8) {$X_2Z_1Z_4$};
            \node[node_style] (u2) at (-12.5+8,2+1-0.3) {6};
            \node at (-12.5+8.2,2+1-0.8) {$X_6Z_7$};
            \node[node_style] (u5) at (-13+8,3.5) {4};
            \node at (-13+8,4) {$X_4Z_2$};
            \node[node_style] (u6) at (-13+8,1.75) {5};
            \node at (-13+8,1.25) {$X_5$};
        
            \draw[edge_style]  (u1) edge (u4);
            \draw[edge_style]  (u4) edge (u5);
		\draw[edge_style]  (u7) edge (u2);
        \node[inner sep=0pt] (meter) at (-7-1,2.75)
    {\begin{quantikz}
		 \meter{$Z_3$}
    \end{quantikz}};

		\end{tikzpicture}
		\caption{A single-qubit $Z$ measurement (performed on qubit $3$ in the example shown) can break up a cluster state into disjointed smaller clusters.}

		\label{fig:measureZ_plus_edge}
	\end{figure*}

\begin{figure}[h]
		\centering
		
	\begin{tikzpicture}[darkstyle/.style={circle,inner sep=0pt,minimum size=0.4cm,draw=blue,fill=blue!20!,font=\sffamily\small\bfseries},
 edge_style/.style={draw=black,  thick}]
 
 \draw [step=0.75cm,thick] (0,0) grid (3,3);

  \foreach \x in {0,0.75,...,3}
    \foreach \y in {0,0.75,...,3} 
       {\pgfmathtruncatemacro{\label}{4*\x/3 - 5 *  4*\y/3 +21}
       \node [darkstyle]  (\x\y) at (\x,\y) {\label};} 
 ;

 \draw [step=0.75cm,thick] (0+4.5,0) grid (3+4.5,.75);
  \draw [step=0.75cm,thick] (4.5,2.25) grid (3+4.5,3);
             \draw (4.5,0) --   (4.5,3);
             \draw (4.5+2.25,0) --   (4.5+2.25,3);
    \draw [step=0.75cm, thick] (4.5,0) grid (.75+4.5,3);
     \draw [step=0.75cm, thick] (4.5+2.25,0) grid (3+4.5,3);

  \foreach \x in {0,0.75,...,3}
    \foreach \y in {0,0.75,...,3} 
       {\pgfmathtruncatemacro{\label}{4*\x/3 - 5 *  4*\y/3 +21}
       \node [darkstyle]  (\x\y) at (\x+4.5,\y) {\label};} 
 ;
      \node[inner sep=0pt] (meter) at (3.7,1.75)
    {\begin{quantikz}
		 \meter{$Z_{13}$}
    \end{quantikz}};
		\end{tikzpicture}
		\caption{$Z$ measurement on a qubit of a square grid cluster leaves a square hole of linear dimension equaling two edges.}
		\label{fig:measureZ_squaregrid}
	\end{figure}
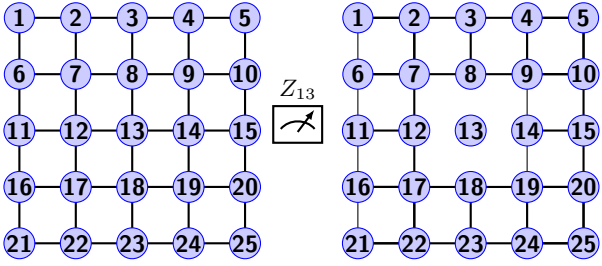

Let us consider measuring $Z$ on node $u$ of a general cluster state corresponding to the graph $G(V,E)$ resulting in a (random) measurement result $m_u$. $Z_u$ will commute with every stabilizer generator of the cluster state except for the stabilizer corresponding to node $u$, i.e., $X_u\Pi_{v \in {\cal N}_u}Z_v$, where ${\cal N}_u$ are nodes in the neighborhood of $u$ in the cluster state. This is because: (1) the stabilizer generator corresponding to any node $v \in {\cal N}_u$, i.e., $X_v\Pi_{w \in {\cal N}_v}Z_w$ commutes with $Z_u$, since the only qubit-$u$ Pauli literal in the set $w \in {\cal N}_v$ is $Z_u$; and (2) none of the other stabilizer generators have any  qubit-$u$ Pauli literal, so they trivially commute with $Z_u$. Applying the rules of Section~\ref{subsec:StabMeasure}, and ignoring the unimportant sign of $m_u$, the stabilizer generators of $G$ are modified in the following way, as a result of the $Z_u$ measurement: 
\begin{itemize}
\item The (now detached) qubit-$u$ stabilizer becomes $Z_u$,
\item The stabilizer generator corresponding to any node $v$ in $G$ that was a neighbor of $u$ (i.e., $v \in {\cal N}_u$), viz., $X_v\Pi_{w \in {\cal N}_v}Z_w$ transforms to $X_v\Pi_{\left\{w \in {\cal N}_v, w \ne u\right\}}Z_w$,
\item All other stabilizer generators remain unchanged.
\end{itemize}
{\em{Compact graphical rule}}---The effect of performing a $Z$ measurement on qubit $u$ of a cluster state removes the qubit from the cluster state, and all the edges it was connected to~\cite{dahlberg2018transforming}. We show two examples in FIG.~\ref{fig:measureZ_plus_edge} and FIG.~\ref{fig:measureZ_squaregrid}.

\subsection{Local complementation}
\label{subsec:LoComp}
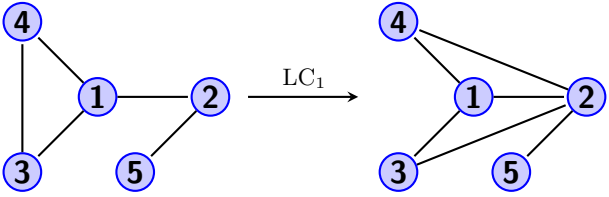
\begin{figure}[h]
	\centering
	\begin{tikzpicture}[shorten >=1pt, auto, node distance=2cm, thick,
		node_style/.style={circle,inner sep=0pt,minimum size=0.5cm,draw=blue,fill=blue!20!,font=\sffamily\large\bfseries},
		node3_style/.style={circle,inner sep=0pt,minimum size=2cm,yscale=.35,draw=black,font=\sffamily\large\bfseries},
		node2_style/.style={circle,dashed,inner sep=0pt,minimum size=0.5cm,draw=blue,fill=blue!20!,font=\sffamily\large\bfseries},
		edge_style/.style={draw=black, thick}
		arrow_style/.style={draw=black,thick,->}]
		
		\node[node_style] (u4) at (-10.5,4) {4};
		\node[node_style] (u1) at (-9.5,3) {1};
		\node[node_style] (u3) at (-10.5,2) {3};
		\node[node_style] (u2) at (-8,3) {2};
		\node[node_style] (u5) at (-9,2) {5};
		\draw[edge_style]  (u1) edge (u2);
		\draw[edge_style]  (u1) edge (u3);
		\draw[edge_style]  (u1) edge (u4);
		\draw[edge_style]  (u3) edge (u4);
		\draw[edge_style]  (u2) edge (u5);
		
		\node[node_style] (u4) at (-5.5,4) {4};
		\node[node_style] (u1) at (-4.5,3) {1};
		\node[node_style] (u3) at (-5.5,2) {3};
		\node[node_style] (u2) at (-3,3) {2};
		\node[node_style] (u5) at (-4,2) {5};
		\draw[edge_style]  (u1) edge (u2);
		\draw[edge_style]  (u1) edge (u3);
		\draw[edge_style]  (u1) edge (u4);
		\draw[edge_style]  (u3) edge (u2);
		\draw[edge_style]  (u2) edge (u4);
		\draw[edge_style]  (u2) edge (u5);
		\draw [-stealth,thick] (-7.5,3) --  node[above] {LC$_1$}(-6,3);

		\end{tikzpicture}

	\caption{Local complementation on qubit 1 deletes the existing edges between every pair of neighbors of qubit 1 and adds new edges if there aren't any.}
	\label{fig:localComp}
\end{figure}

Local complementation plays a key role in determining local Clifford (LC) equivalence of cluster states. Two cluster states are LC equivalent if and only if there exists a sequence of local complementations that converts the graph corresponding to one state to that of the other~\cite{van2003graphical}. The unitary to perform local complementation of qubit $k$ of a cluster state,
\begin{equation}
	{\text{LC}}_{k} = e^{-i\frac{\pi}{4}X_k}\prod_{j \in \mathcal{N}_k}e^{i\frac{\pi}{4}Z_j},
	\label{eq:LocalComp}
\end{equation}
where $\mathcal{N}_k$ is the neighborhood of $k$ ~\cite{Grosshans2019distributing}. For a given cluster state, performing local complementation on a qubit inverts the local neighborhood of that qubit as shown in FIG.~\ref{fig:localComp}. In other words, for each $\binom{|\mathcal{N}_k|}{2}$ pairs of neighbors of $k$, if an edge existed we delete the edge, and if there was no edge we draw one.

\subsection{Pauli X basis measurement}
\label{sec:PauliX}

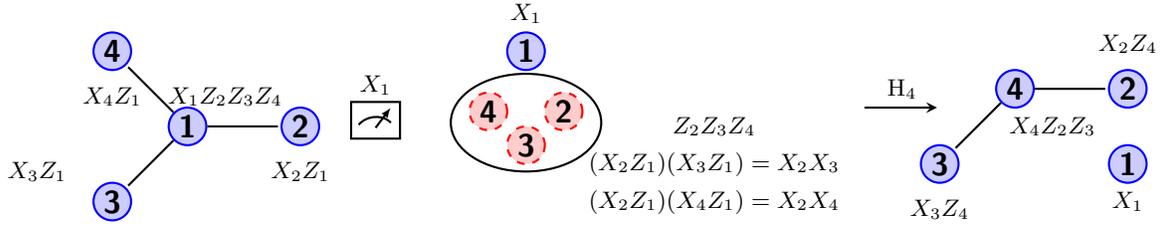
\begin{figure*}[htb]
		\centering
		\begin{tikzpicture}[shorten >=1pt, auto, node distance=2cm, thick,
		node_style/.style={circle,inner sep=0pt,minimum size=0.5cm,draw=blue,fill=blue!20!,font=\sffamily\large\bfseries},
  node4_style/.style={circle,inner sep=0pt,minimum size=0.5cm,draw=red,fill=red!20!,font=\sffamily\large\bfseries},  
  node5_style/.style={circle,dashed,inner sep=0pt,minimum size=0.5cm,draw=red,fill=red!20!,font=\sffamily\large\bfseries},
		node3_style/.style={circle,inner sep=0pt,minimum size=2cm,yscale=.65,draw=black,font=\sffamily\large\bfseries},
		node2_style/.style={circle,dashed,inner sep=0pt,minimum size=0.5cm,draw=blue,fill=blue!20!,font=\sffamily\large\bfseries},
		edge_style/.style={draw=black, thick}
		arrow_style/.style={draw=black,thick,<->}]
         
		\node[node_style] (u4) at (-10.5,4-0.5) {4};
		\node at (-10.5,3.4-0.5) {$X_4Z_1$};
		\node[node_style] (u1) at (-9.5,3-0.5) {1};
		\node at (-9,3.4-0.5) {$X_1Z_2Z_3Z_4$};
		\node[node_style] (u3) at (-10.5,2-0.5) {3};
		\node at (-11.5,2.4-0.5) {$X_3Z_1$};	
		\node[node_style] (u2) at (-8,3-0.5) {2};
		\node at (-8,2.4-0.5) {$X_2Z_1$};
            \draw[edge_style]  (u1) edge (u2);
		\draw[edge_style]  (u1) edge (u3);
		\draw[edge_style]  (u1) edge (u4);

            \node[node_style] (u2) at (-5,3.5) {1};
            \node at (-5,4) {$X_1$};
		\node[node5_style] (u4) at (-13.5+8,2+1-0.3) {4};
            \node[node5_style] (u4) at (-12.5+8,2+1-0.3) {2};
            \node[node5_style] (u4) at (-13+8,1.5+1-0.25) {3};
            \node[node3_style] (u4) at (-13+8,2+1-0.45) {};
            \node at (-5+3-0.5,0.5+2) {$Z_2Z_3Z_4$};
            
            \node at (-5+3-0.5,2) {$(X_2Z_1)(X_3Z_1) = X_2X_3$};
            \node at (-5+3-0.5,1.5) {$(X_2Z_1)(X_4Z_1) = X_2X_4$};
  \node[inner sep=0pt] (meter) at (-7,2.75)
    {\begin{quantikz}
		 \meter{$X_1$}
    \end{quantikz}};
  
            \draw [-stealth,thick] (-4.5+1+3,2.75) -- node[above] {$\text{H}_{4}$}  (-4.5+2+3,2.75);

		\node[node_style] (u1) at (-2.5+1+3,3) {4};
		\node at (-2+1+3,2.5) {$X_4Z_2Z_3$};
		\node[node_style] (u3) at (-3.5+1+3,2) {3};
		\node at (-3.5+1+3,1.4) {$X_3Z_4$};	
		\node[node_style] (u2) at (-1+1+3,3) {2};
  \node[node_style] (u4) at (-1+1+3,2) {1};
		\node at (-1+1+3,3.6) {$X_2Z_4$};
  \node at (-1+1+3,1.5) {$X_1$};
		\draw[edge_style]  (u1) edge (u2);
		\draw[edge_style]  (u1) edge (u3);

		\end{tikzpicture}
		\caption{Measuring $X$ at the root node of an $n+1$ qubit star-shaped cluster state, i.e., of degree-$n$, results in an $n$-qubit GHZ state in the $+/-$ basis, with the measured qubit detached. Performing a Hadamard gate on any of the $n$ qubits of the GHZ state, say qubit-$u$, results in an $n$ qubit star-shaped cluster state (degree $n-1$) in the $0/1$ basis, with qubit $u$ as its root node.}

		\label{fig:measureX_star}
	\end{figure*}

Let us begin by analyzing the effect of an $X$ measurement at the root node of a star-shaped cluster state. See FIG.~\ref{fig:measureX_star} as an example of a degree-$3$ star graph where we perform $X$ measurement on qubit-$1$, the root node, obtaining an outcome $m_1$. Applying the rules of Section~\ref{subsec:StabMeasure}, we find that $X_1$ commutes with the root-node (qubit-$1$) stabilizer $X_1Z_2Z_3Z_4$, and anti-commutes with every other stabilizer generator. We now must turn all but one stabilizer to commute with $X_1$. Let us (arbitrarily) pick the qubit-$2$ stabilizer $X_2Z_1$ to be that stabilizer. We multiply every other non-commuting stabilizer (i.e., $X_3Z_1$ and $X_4Z_1$) with $X_2Z_1$, obtaining generators: $\left\langle X_2Z_1, X_1Z_2Z_3Z_4, X_2X_3, X_2X_4\right\rangle $, where now only the first generator anti-commutes with $X_1$. At this point, per the rule in Section~\ref{subsec:StabMeasure}, measuring $X_1$ corresponds to: (1) replacing $X_2Z_1$ by $m_1X_1$ and ignoring the sign of $m_1$ to re-write it as $X_1$ (this is the new generator corresponding to the now-detached qubit-$1$ in the $\ket{+}$ state), (2) and replacing $X_1Z_2Z_3Z_4$ by $m_1Z_2Z_3Z_4$, which we re-write as $Z_2Z_3Z_4$ by ignoring the sign of $m_1$ as usual. Therefore, the final set of stabilizers after the measurement are: $\left\langle X_1, Z_2Z_3Z_4, X_2X_3, X_2X_4\right\rangle $. We recognize that the post-measurement state of the original leaf-qubits of the star-cluster, qubits $2$, $3$ and $4$, from Section~\ref{sub:cluster_GHZ}, as the $3$-qubit GHZ state in the $+/-$ basis, i.e., $(\ket{+++}+\ket{---})/\sqrt{2}$. As discussed in Section~\ref{sub:cluster_GHZ}, we can turn this into a cluster state (in the $0/1$ basis) by performing a Hadamard gate on any of the qubits $2$, $3$ or $4$. In general, it is simple to see that measuring $X$ at the degree-$n$ root node, say qubit-$v$, of an $n+1$ qubit star-shaped cluster state results in an $n$-qubit GHZ state in the $+/-$ basis, with qubit-$v$ getting detached. Thereafter, performing a Hadamard gate on any of the $n$ qubits of that GHZ state, say qubit-$u$, results in an $n$ qubit star-shaped cluster state (degree $n-1$), with qubit $u$ as its root (see FIG.~\ref{fig:measureX_star} for an illustration of the $n=3$ case).

Now, let us consider performing an $X$ measurement on any (say degree-$d$) qubit $u$ of a cluster state described by graph $G$. It results in all the $d$ neighbors of $u$ in $G$ to turn into a $d$-qubit $+/-$ basis GHZ state as above. But all the original edges incident on those $d$ neighbors of $u$ (both from other nodes in $G$, and any edges between those $d$ nodes) are preserved. We can think of the GHZ state in the $+/-$ basis among the $d$ erstwhile neighbors of $u$, $(\ket{+}^{\otimes d}+\ket{-}^{\otimes d})/\sqrt{2}$, as a {\em redundantly-encoded qubit} (term first introduced in Ref.~\cite{browne2005Fusion}), i.e., a logical qubit made up of $d$ physical qubits in the $+/-$ basis. The edges incident on the qubits of that GHZ state represent CZ gates (see FIG.~\ref{fig:measureX_line}(b) for an example). 

This overall post-measurement state, therefore, is neither a GHZ state nor a cluster state. But, we know that performing a Hadamard gate on any of those $d$ qubits of the GHZ state portion would turn that $d$-qubit $+/-$ basis GHZ state into a degree $d-1$ star cluster state. Therefore, if we perform a Hadamard gate on any of those $d$ qubits of the GHZ state portion of the post-$X_u$-measurement state, say qubit $v\in \mathcal{N}_u$, we retrieve a cluster state described by, say $G'$. $G'$ can be obtained by modifying the edges of $G$ using the following steps. In these steps, the $\mathcal{N}_i$ refers to the neighborhood of qubit $i$ in the pre-measurement cluster state.
\begin{enumerate}
        \item Delete all the edges to $u$.
		\item Draw edges from $v$ to all qubits in the set $\mathcal{N}_u\setminus (\mathcal{N}_v\cup\{v\})$. This step mimics the creation of the degree $d-1$ star cluster state.

		\item $\forall i \in \mathcal{N}_{vu}$ s.t. $\mathcal{N}_{vu} = \mathcal{N}_v\setminus\mathcal{N}_u$ and $\forall j \in \mathcal{N}_{u}$, invert the $(i,j)$ edge, i.e., invert all the edges from qubits that are neighbors if $v$ but not $u$ to all the neighbors of $u$, including $v$. Note that, this rule deletes all the $(i,v)$ edges.
		\item $\forall i \in  \mathcal{N}_v\cap\mathcal{N}_u$ and $\forall j \in  \mathcal{N}_u\setminus(\mathcal{N}_v\cup\{v\})$, invert the $(i,j)$ edge. 
  \item All other edges remain unchanged.
\end{enumerate}

Let us now consider a few illustrative examples.

	
{\textbf{Example 1}}---Let us consider measuring qubit $3$ of the $5$-qubit line cluster state shown in FIG.~\ref{fig:measureX_line}(a) in the $X$-basis. The post-measurement state is depicted in FIG.~\ref{fig:measureX_line}(b). As a result of the $X$ measurement on qubit $3$, qubit $3$ gets detached, and qubits $2$ and $4$ turn into a $+/-$ basis GHZ state, the redundantly encoded qubit, described by stabilizer generators $\left\langle Z_2Z_4, X_2X_4\right\rangle $, while retaining the original two {\em edges} that qubits $2$ and $4$ had in the line cluster, i.e., (1) a CZ between qubit $2$ of the GHZ state with the qubit $1$ (originally in the $\ket{+}$ state, the CZ thereby turning the qubit $1$'s stabilizer generator from $X_1$ to $X_1Z_2$ and the stabilizer generator $X_2X_4$ of the GHZ state into $Z_1X_2X_4$); and then (2) the CZ between qubit $4$ of the GHZ state with the qubit $5$ (originally in the $\ket{+}$ state, the CZ thereby turning the qubit $5$'s stabilizer generator from $X_5$ to $X_5Z_4$, and turning the aforesaid stabilizer generator $Z_1X_2X_4$ into $Z_1X_2X_4Z_5$). The stabilizer generators of the post-measurement stabilizer state therefore are: $\left\langle X_3, X_1Z_2, Z_2Z_4, Z_1X_2X_4Z_5, Z_4X_5\right\rangle $. This state is depicted in FIG.~\ref{fig:measureX_line}(b) using our newly-introduced pictorial representation of a $+/-$ basis GHZ state, some of those qubits are also connected via cluster-state-like edges to other qubits. 

To convert this post-measurement stabilizer state in FIG.~\ref{fig:measureX_line}(b) to a cluster state, as discussed above, we can apply a Hadamard gate on either qubit $2$ or on qubit $4$. Let us consider applying a Hadamard on qubit $2$ (see FIG.~\ref{fig:measureX_line}(c)). This results in the two-qubit $+/-$ basis GHZ state among qubits $2$ and $4$ to turn into a cluster state with qubit $2$ at the root node. Per the rule described above, the ($4$,$5$) edge is retained, the original edge incident on qubit $2$ (i.e., from $1$) is deleted, and a new edge is formed from $1$ to $4$, resulting in the cluster state shown in FIG.~\ref{fig:measureX_line}(c). It is simple to also derive the above by transforming each of the stabilizer generators of the post-measurement stabilizer state in FIG.~\ref{fig:measureX_line}(b) under $H_2$, and observing that the stabilizer generators of new state are that of the cluster state shown in FIG.~\ref{fig:measureX_line}(c).
 \begin{figure*}[htb]
		\centering
		\begin{tikzpicture}[shorten >=1pt, auto, node distance=2cm, thick,
		node_style/.style={circle,inner sep=0pt,minimum size=0.5cm,draw=blue,fill=blue!20!,font=\sffamily\large\bfseries},
  node4_style/.style={circle,inner sep=0pt,minimum size=0.5cm,draw=red,fill=red!20!,font=\sffamily\large\bfseries},  
  node5_style/.style={circle,dashed,inner sep=0pt,minimum size=0.5cm,draw=red,fill=red!20!,font=\sffamily\large\bfseries},
		node3_style/.style={circle,inner sep=0pt,minimum size=2cm,yscale=.35,draw=black,font=\sffamily\large\bfseries},
		node2_style/.style={circle,dashed,inner sep=0pt,minimum size=0.5cm,draw=blue,fill=blue!20!,font=\sffamily\large\bfseries},
		edge_style/.style={draw=black, thick}
		arrow_style/.style={draw=black,thick,<->}]
         
		\node[node_style] (v1) at (-4,2) {1};
		\node at (-4,1.5) {$X_1Z_2$};
  \node at (-4.25,2.75) {(a)};
		\node[node_style] (v2) at (-3,2) {2};
		\node at (-3,2.5) {$Z_1X_2Z_3$};
		\node[node_style] (v3) at (-2,2) {3};
		\node at (-2,1.5) {$Z_2X_3Z_4$};
		\node[node_style] (v4) at (-1,2) {4};
		\node at (-1,2.5) {$Z_3X_4Z_5$};	
		\node[node_style] (v5) at (0,2) {5};
		\node at (0,1.5) {$Z_4X_5$};

		\draw[edge_style]  (v1) edge (v2);
		\draw[edge_style]  (v2) edge (v3);
		\draw[edge_style]  (v3) edge (v4);
		\draw[edge_style]  (v4) edge (v5);
		
        \node[inner sep=0pt] (meter) at (1,2.25)
    {\begin{quantikz}
		 \meter{$X_3$}
    \end{quantikz}};

   \node[node_style] (v1) at (2,2) {3};
		\node at (2,1.4) {$X_3$};
		\node[node_style] (v2) at (3,2) {1};
		\node at (3,1.4) {$X_1Z_2$};
  \node at (1.75,2.75) {(b)};
		\node[node5_style] (v3) at (4,2) {2};
		\node at (4.5,1.4) {$Z_2Z_4$,};
		\node[node5_style] (v4) at (5,2) {4};
		\node[node3_style] (v) at (4.5,2) {};
		\node at (4.5,0.95) {$Z_1X_2X_4Z_5$};	
		\node[node_style] (v5) at (6,2) {5};
		\node at (6,1.4) {$Z_4X_5$};
		
		\draw[edge_style]  (v2) edge (v3);
		\draw[edge_style]  (v4) edge (v5);

            \draw [-stealth,thick] (6.5,2) -- node[above] {$\text{H}_{2}$} (7.5,2);
         \node[node_style] (v1) at (-4+12,2) {3};
		\node at (-4+12,1.4) {$X_3$};
  \node at (-4.25+12,2.75) {(c)};
		\node[node_style] (v2) at (-2.5+12,2) {1};
		\node at (-2.5+12,1.4) {$X_1Z_2$};
		\node[node_style] (v3) at (-1+12,3) {2};
		\node at (-1+12,3.5) {$X_2Z_4$};
		\node[node_style] (v4) at (-1+12,2) {4};
		\node at (-1+12,1.4) {$Z_1Z_2X_4Z_5$};	
		\node[node_style] (v5) at (0.5+12,2) {5};
		\node at (0.5+12,1.4) {$Z_4X_5$};
		
		\draw[edge_style]  (v2) edge (v4);
		\draw[edge_style]  (v3) edge (v4);
		
		\draw[edge_style]  (v4) edge (v5);
		\end{tikzpicture}
		\caption{Pauli-X measurement on  qubit $3$ of the $5$-qubit line cluster state puts its neighbors, qubits 2 and 4 in a redundantly encoded state. The post-measurement state can be converted into a cluster state by applying Hadamard on one of the neighbors. The state after $H_2$ is shown.}

		\label{fig:measureX_line}
	\end{figure*}
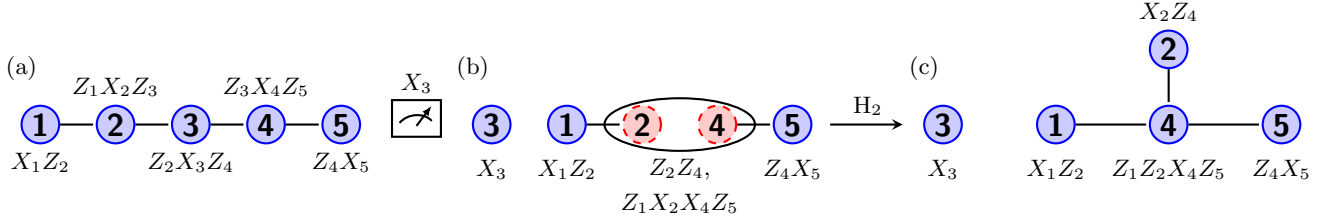
	
 \begin{figure*}[htb]
		\centering
		\begin{tikzpicture}[shorten >=1pt, auto, node distance=2cm, thick,
		node_style/.style={circle,inner sep=0pt,minimum size=0.5cm,draw=blue,fill=blue!20!,font=\sffamily\large\bfseries},
  node4_style/.style={circle,inner sep=0pt,minimum size=0.5cm,draw=red,fill=red!20!,font=\sffamily\large\bfseries},  
  node5_style/.style={circle,dashed,inner sep=0pt,minimum size=0.5cm,draw=red,fill=red!20!,font=\sffamily\large\bfseries},
		node3_style/.style={circle,inner sep=0pt,minimum size=2cm,yscale=.85,draw=black,font=\sffamily\large\bfseries},
		node2_style/.style={circle,dashed,inner sep=0pt,minimum size=0.5cm,draw=blue,fill=blue!20!,font=\sffamily\large\bfseries},
		edge_style/.style={draw=black, thick}
		arrow_style/.style={draw=black,thick,<->}]
         
		\node[node_style] (u2) at (-12,3-0.5) {2};
		\node at (-12,3.4-0.5) {$X_2Z_1Z_3$};
            \node[node_style] (u1) at (-13,3-0.5) {1};
		\node at (-13-0.25,3.4-0.5) {$X_1Z_2$};
            \node[node_style] (u4) at (-11,4-0.5) {4};
		\node at (-11,4) {$X_4Z_3$};
		\node[node_style] (u3) at (-11,3-0.5) {3};
		\node at (-10.25,3.4-0.5) {$X_3Z_2Z_4Z_5Z_6$};
		\node[node_style] (u5) at (-11,2-0.5) {5};
		\node at (-11,1.4-0.5) {$X_5Z_3$};	
		\node[node_style] (u6) at (-10,3-0.5) {6};
		\node at (-10,2.4-0.5) {$X_6Z_3Z_7$};
            \node[node_style] (u7) at (-9,3-0.5) {7};
		\node at (-8.75,2.4-0.5) {$X_7Z_6$};
            \draw[edge_style]  (u1) edge (u2);
		\draw[edge_style]  (u2) edge (u3);
		\draw[edge_style]  (u3) edge (u4);
            \draw[edge_style]  (u3) edge (u5);
            \draw[edge_style]  (u3) edge (u6);
            \draw[edge_style]  (u6) edge (u7);

            \node[node_style] (u1) at (-7+0.25,2+1-0.3) {1};
             \node at (-7+0.25,2+1-0.3-0.5) {$X_1Z_2$};
             \node[node_style] (u7) at (-3.25,2+1-0.3) {7};
             \node at (-3.25,2+1-0.3-0.5) {$X_7Z_6$};
            \node[node_style] (u3) at (-6.75,3.5) {3};
            \node at (-7+0.25,4) {$X_3$};
		\node[node5_style] (u4) at (-13.5+8,2+1-0.3) {2};
            \node[node5_style] (u2) at (-12.5+8,2+1-0.3) {6};
            \node[node5_style] (u5) at (-13+8,3.2) {4};
            \node[node5_style] (u6) at (-13+8,1.5+1-0.25) {5};
            \node[node3_style] (u) at (-13+8,2+1-0.25) {};
            \node at (-5,1.5) {$Z_2Z_4Z_5Z_6, X_2X_4Z_1$};
            \node at (-5,1) {$X_2X_5Z_1$, $X_2X_6Z_1Z_7$};
            \draw[edge_style]  (u1) edge (u4);
		\draw[edge_style]  (u7) edge (u2);
        \node[inner sep=0pt] (meter) at (-7-1,2.75)
    {\begin{quantikz}
		 \meter{$X_3$}
    \end{quantikz}};

            \draw [-stealth,thick] (-4.5+2-0.25,2.75) -- node[left] {$\text{H}_{2}$}  (-4.5+3,3.75);
             \draw [-stealth,thick] (-4.25+2-0.5,2.75) -- node[left] {$\text{H}_{4}$}  (-4.5+3,1.75);
           
            \node[node_style] (u2) at (-2+2,2.5+1.5) {4};
		\node at (-2.5+2,3.4-0.5+1.5) {$X_2Z_1Z_3$};
            \node[node_style] (u1) at (-1.5+2,4-0.5+1.5) {1};
		\node at (.5,5.5) {$X_1Z_4Z_5Z_6$};
            \node[node_style] (u4) at (-.5+2,4+1) {5};
		\node at (2.5,4+1) {$X_5Z_1Z_2$};
		\node[node_style] (u3) at (-1+2,3-0.5+1.5) {2};
		\node at (-1+2,2.5-0.5+1.5) {$X_2Z_4Z_5Z_6$};
		\node[node_style] (u5) at (-1,3.5+1.5) {3};
		\node at (-1,4+1.5) {$X_3$};	
		\node[node_style] (u6) at (-0+2,3-0.5+1.5) {6};
		\node at (-0+2.2,2.5-0.5+2.5) {$X_6Z_1Z_2Z_7$};
            \node[node_style] (u7) at (1+2,3-0.5+1.5) {7};
		\node at (-8.75+12,2.5-0.5+1.5) {$X_7Z_6$};
            \draw[edge_style]  (u1) edge (u2);
            \draw[edge_style]  (u1) edge (u4);
            \draw[edge_style]  (u1) edge (u6);
		\draw[edge_style]  (u2) edge (u3);
		\draw[edge_style]  (u3) edge (u4);
            \draw[edge_style]  (u3) edge (u6);
            \draw[edge_style]  (u6) edge (u7);

             \node[node_style] (u2) at (-2+2,2.5-0.75) {2};
		\node at (-2+2,3.4-0.5-0.75) {$X_2Z_1Z_4$};
            \node[node_style] (u1) at (-3+2,3-0.5-0.75) {1};
		\node at (-3-0.25+2,3-1-0.75) {$X_1Z_2$};
            \node[node_style] (u4) at (-1+2,4-0.5+1.5-2.25) {5};
		\node at (-1+2.75,4-0.5+1.5-2.25) {$X_5Z_4$};
		\node[node_style] (u3) at (-1+2,3-0.5+1.5-2.25) {4};
		\node at (1,3.2-0.5+1.25-2.75) {$X_4Z_2Z_5Z_6$};
		\node[node_style] (u5) at (-1,3.5+1.5-2.25) {3};
		\node at (-1,4+1.5-2.25) {$X_3$};	
		\node[node_style] (u6) at (-0+2,3-0.5+1.5-2.25) {6};
		\node at (-0+2,3+1.5-2.25) {$X_6Z_4Z_7$};
            \node[node_style] (u7) at (1+2,3-0.5+1.5-2.25) {7};
		\node at (-8.75+12,2.5-0.5+1.5-2.25) {$X_7Z_6$};
            \draw[edge_style]  (u1) edge (u2);
		\draw[edge_style]  (u2) edge (u3);
		\draw[edge_style]  (u3) edge (u4);
            \draw[edge_style]  (u3) edge (u6);
            \draw[edge_style]  (u6) edge (u7);

		\end{tikzpicture}
		\caption{An example of an $X$ measurement performed on a qubit ($3$) of a tree cluster state. The post-measurement stabilizer state is depicted using our notation of a $+/-$ basis GHZ state (the redundantly encoded logical qubit) with the original cluster-state edges incident on qubits of that GHZ state. On the right, we show two examples of how performing a Hadamard gate on any one of the qubits within the GHZ state portion retrieves an overall cluster state.}

		\label{fig:measureX_plus}
	\end{figure*}
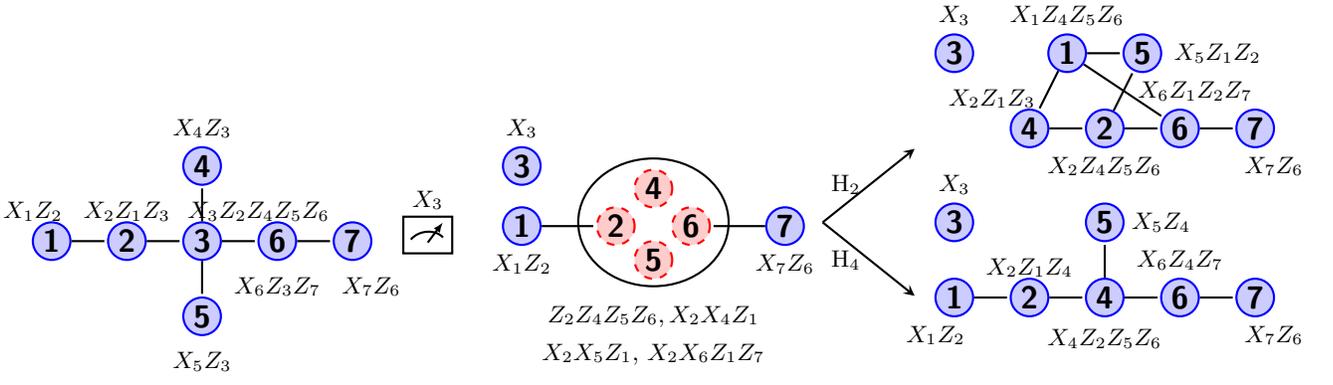

{\textbf{Example 2}}---Consider making an $X$ measurement on qubit $3$ of the tree cluster shown in FIG.~\ref{fig:measureX_plus}. The post-measurement stabilizer state is an isolated qubit $3$, a $4$-qubit $+/-$ basis GHZ state among qubits $2$, $4$, $5$ and $6$, with the original ($1$,$2$) and ($6$,$7$) edges retained. We can now turn this state into a cluster state by performing a Hadamard on any one of the qubits $2$, $4$, $5$ and $6$. Two examples of the cluster state obtained by performing Hadamard on qubit $2$ or on qubit $4$ are shown in FIG.~\ref{fig:measureX_plus}. Performing $H_4$ creates a $4$-qubit star cluster with $4$ as root node, with the original ($1$,$2$) and ($6$,$7$) edges retained. Performing $H_2$ on the post-measurement state creates a $4$-qubit star cluster with $2$ as root node, with the original ($6$,$7$) edge retained, while the ($1$,$2$) edge is deleted with $1$ inheriting new edges to qubits $4$, $5$ and $6$.

 \begin{figure*}[htb]
\centering
\begin{tikzpicture}[shorten >=1pt, auto, node distance=2cm, thick,
node_style/.style={circle,inner sep=0pt,minimum size=0.5cm,draw=blue,fill=blue!20!,font=\sffamily\large\bfseries},
  node4_style/.style={circle,inner sep=0pt,minimum size=0.5cm,draw=red,fill=red!20!,font=\sffamily\large\bfseries},  
  node5_style/.style={circle,dashed,inner sep=0pt,minimum size=0.5cm,draw=red,fill=red!20!,font=\sffamily\large\bfseries},
node3_style/.style={circle,inner sep=0pt,minimum size=2cm,yscale=.85,draw=black,font=\sffamily\large\bfseries},
node2_style/.style={circle,dashed,inner sep=0pt,minimum size=0.5cm,draw=blue,fill=blue!20!,font=\sffamily\large\bfseries},
edge_style/.style={draw=black, thick}
arrow_style/.style={draw=black,thick,<->}]
         
\node[node_style] (u2) at (-12,3-0.5) {2};
\node at (-12,2.5-0.5) {$X_2Z_1Z_3Z_4$};
            \node[node_style] (u1) at (-13,3-0.5) {1};
\node at (-13-0.25,3.4-0.5) {$X_1Z_2$};
            \node[node_style] (u4) at (-11,4-0.5) {4};
\node at (-11,4) {$X_4Z_2Z_3$};
\node[node_style] (u3) at (-11,3-0.5) {3};
\node at (-10.25,3.4-0.5) {$X_3Z_2Z_4Z_5Z_6$};
\node[node_style] (u5) at (-11,2-0.5) {5};
\node at (-11,1.4-0.5) {$X_5Z_3$}; 
\node[node_style] (u6) at (-10,3-0.5) {6};
\node at (-10,2.4-0.5) {$X_6Z_3Z_7$};
            \node[node_style] (u7) at (-9,3-0.5) {7};
\node at (-8.75,2.4-0.5) {$X_7Z_6$};
            \draw[edge_style]  (u1) edge (u2);
\draw[edge_style]  (u2) edge (u3);
  \draw[edge_style]  (u2) edge (u4);
\draw[edge_style]  (u3) edge (u4);
            \draw[edge_style]  (u3) edge (u5);
            \draw[edge_style]  (u3) edge (u6);
            \draw[edge_style]  (u6) edge (u7);

            \node[node_style] (u1) at (-7+0.25,2+1-0.3) {1};
             \node at (-7+0.25,2+1-0.3-0.5) {$X_1Z_2$};
             \node[node_style] (u7) at (-3.25,2+1-0.3) {7};
             \node at (-3.25,2+1-0.3-0.5) {$X_7Z_6$};
            \node[node_style] (u3) at (-6.75,3.5) {3};
            \node at (-7+0.25,4) {$X_3$};
\node[node5_style] (u4) at (-13.5+8,2+1-0.3) {2};
            \node[node5_style] (u2) at (-12.5+8,2+1-0.3) {6};
            \node[node5_style] (u5) at (-13+8,3.2) {4};
            \node[node5_style] (u6) at (-13+8,1.5+1-0.25) {5};
            \node[node3_style] (u) at (-13+8,2+1-0.25) {};
            \node at (-5,1.5) {$Z_2Z_4Z_5Z_6,X_4X_5Z_2$};
            \node at (-5,1) {$ X_2X_5Z_1Z_4$, $X_5X_6Z_7$};
            \draw[edge_style]  (u1) edge (u4);
            \draw[edge_style]  (u4) edge (u5);
\draw[edge_style]  (u7) edge (u2);
\node[inner sep=0pt] (meter) at (-7-1,2.75)
    {\begin{quantikz}
		 \meter{$X_3$}
    \end{quantikz}};

            \draw [-stealth,thick] (-4.5+2-0.25,2.75) -- node[left] {$\text{H}_{4}$}  (-4.5+3,3.75);
             \draw [-stealth,thick] (-4.25+2-0.5,2.75) -- node[left] {$\text{H}_{2}$}  (-4.5+3,1.75);
             \draw (-4.25+2-0.5,2.75) --   (-4.25+1.5,-1);
             \draw [-stealth,thick] (-4.25+1.5,-1) -- node[above] {$\text{H}_{5}$}  (-4.5+3,-1);
              \draw [-stealth,thick] (-4.25+1.5,-1) -- node[above] {$\text{H}_{6}$}  (-4.,-1);
           
            \node[node_style] (u2) at (-2+2,2.5+1.5) {5};
\node at (-2.5+2,3.4-0.5+1.5) {$X_5Z_2Z_4$};
            \node[node_style] (u1) at (-1.5+2,4-0.5+1.5) {2};
\node at (.5,5.5) {$X_2Z_1Z_4Z_5Z_6$};
            \node[node_style] (u4) at (-.5+2,4+1) {1};
\node at (2.5,4+1) {$X_1Z_2$};
\node[node_style] (u3) at (-1+2,3-0.5+1.5) {4};
\node at (-1+2,2.5-0.5+1.5) {$X_4Z_2Z_5Z_6$};
\node[node_style] (u5) at (-1,3.5+1.5) {3};
\node at (-1,4+1.5) {$X_3$}; 
\node[node_style] (u6) at (-0+2,3-0.5+1.5) {6};
\node at (-0+2.2,2.5-0.5+2.5) {$X_6Z_4Z_2Z_7$};
            \node[node_style] (u7) at (1+2,3-0.5+1.5) {7};
\node at (-8.75+12,2.5-0.5+1.5) {$X_7Z_6$};
            \draw[edge_style]  (u1) edge (u2);
            \draw[edge_style]  (u1) edge (u4);
            \draw[edge_style]  (u1) edge (u6);
\draw[edge_style]  (u2) edge (u3);
\draw[edge_style]  (u3) edge (u1);
            \draw[edge_style]  (u3) edge (u6);
            \draw[edge_style]  (u6) edge (u7);

             \node[node_style] (u2) at (-2+2,2.5-0.75-0.5) {5};
\node at (-2.25+2,3.4-0.5-0.75-0.5) {$X_5Z_1Z_2Z_4$};
            \node[node_style] (u1) at (1,3-0.5-0.75-0.5) {4};
\node at (1,3-1-0.75-0.5) {$X_4Z_1Z_2Z_5Z_6$};
            \node[node_style] (u4) at (-1+2.5,4-0.5+1.5-2.25-0.5) {1};
\node at (-1+2.75+0.25,4+1.5-2.25-0.5) {$X_1Z_4Z_5Z_6$};
\node[node_style] (u3) at (-1.5+2,4-0.5+1.5-2.25-0.5) {2};
\node at (0.25,4-0.5+1.5-2.25) {$X_2Z_4Z_5Z_6$};
\node[node_style] (u5) at (-1,3.5+1.5-2.25-0.5) {3};
\node at (-1,4+1.5-2.25-0.5) {$X_3$}; 
\node[node_style] (u6) at (-0+2,3-0.5+1.5-2.25-0.5) {6};
\node at (-0+2.75,3+1.5-2.25-0.5) {$X_6Z_1Z_2Z_4Z_7$};
            \node[node_style] (u7) at (1+2,3-0.5+1.5-2.25-0.5) {7};
\node at (-8.75+12,2.5-0.5+1.5-2.25-0.5) {$X_7Z_6$};
            \draw[edge_style]  (u1) edge (u2);
            \draw[edge_style]  (u1) edge (u6);
            \draw[edge_style]  (u1) edge (u3);
\draw[edge_style]  (u2) edge (u3);
\draw[edge_style]  (u1) edge (u4);
            \draw[edge_style]  (u4) edge (u2);
            \draw[edge_style]  (u4) edge (u6);
            \draw[edge_style]  (u3) edge (u6);
            \draw[edge_style]  (u6) edge (u7);
  
\node[node_style] (u2) at (-2+2,2.5-0.75-2.75) {2};
\node at (-2+2,3.4-0.5-0.75-2.75) {$X_2Z_1Z_4Z_5$};
            \node[node_style] (u1) at (-3+2,3-0.5-0.75-2.75) {1};
\node at (-3-0.25+2,3-1-0.75-2.75) {$X_1Z_2$};
            \node[node_style] (u4) at (-1+2,4-0.5+1.5-2.25-2.75) {4};
\node at (-1+2.75,4-0.5+1.5-2.25-2.75) {$X_4Z_5$};
\node[node_style] (u3) at (-1+2,3-0.5+1.5-2.25-2.75) {5};
\node at (1,3.2-0.5+1.25-2.75-2.75) {$X_5Z_2Z_4Z_6$};
\node[node_style] (u5) at (-1,3.5+1.5-2.25-2.75) {3};
\node at (-1,4+1.5-2.25-2.75) {$X_3$}; 
\node[node_style] (u6) at (-0+2,3-0.5+1.5-2.25-2.75) {6};
\node at (-0+2,3+1.5-2.25-2.75) {$X_6Z_5Z_7$};
            \node[node_style] (u7) at (1+2,3-0.5+1.5-2.25-2.75) {7};
\node at (-8.75+12,2.5-0.5+1.5-2.25-2.75) {$X_7Z_6$};
            \draw[edge_style]  (u1) edge (u2);
\draw[edge_style]  (u2) edge (u3);
\draw[edge_style]  (u3) edge (u4);
            \draw[edge_style]  (u4) edge (u2);
            \draw[edge_style]  (u3) edge (u6);
            \draw[edge_style]  (u6) edge (u7);

 \node[node_style] (u2) at (-1+2-8,2.5-0.75-2.75) {2};
\node at (-7,3-1-0.75-2.75) {$X_2Z_1Z_4Z_6$};
            \node[node_style] (u1) at (-1.25+1-8,3-0.5-0.75-2.75) {1};
\node at (-8.25,3-1-0.75-2.75) {$X_1Z_2$};
            \node[node_style] (u4) at (2-8,4-0.5+1.5-2.25-2.75) {4};
\node at (-5,4-0.5+1.5-2.25-2.75) {$X_4Z_2Z_6$};
\node[node_style] (u3) at (2-8,3-0.5+1.5-2.25-2.75-1) {5};
\node at (2-8,3-0.5+1.5-2.75-2.75-1) {$X_5Z_6$};
\node[node_style] (u5) at (-8,3.5+1.5-2.25-2.75) {3};
\node at (-7.5,3.5+1.5-2.25-2.75) {$X_3$}; 
\node[node_style] (u6) at (-7+1,3-0.5+1.5-2.25-2.75) {6};
\node at (-6+.3,3+1.5-2.25-2.75) {$X_6Z_2Z_5Z_7$};
            \node[node_style] (u7) at (1+2-8,3-0.5+1.5-2.25-2.75) {7};
\node at (1+2-8,3-0.5+1-2.25-2.75) {$X_7Z_6$};
            \draw[edge_style]  (u1) edge (u2);
\draw[edge_style]  (u2) edge (u6);
\draw[edge_style]  (u6) edge (u4);
            \draw[edge_style]  (u4) edge (u2);
            \draw[edge_style]  (u3) edge (u6);
            \draw[edge_style]  (u6) edge (u7);

\end{tikzpicture}
\caption{An example of an $X$ measurement performed on a qubit of a cluster state where there is a direct edge between two neighboring qubits of the qubit on which the $X$ measurement is performed.}

\label{fig:measureX_plus_edge}
\end{figure*}
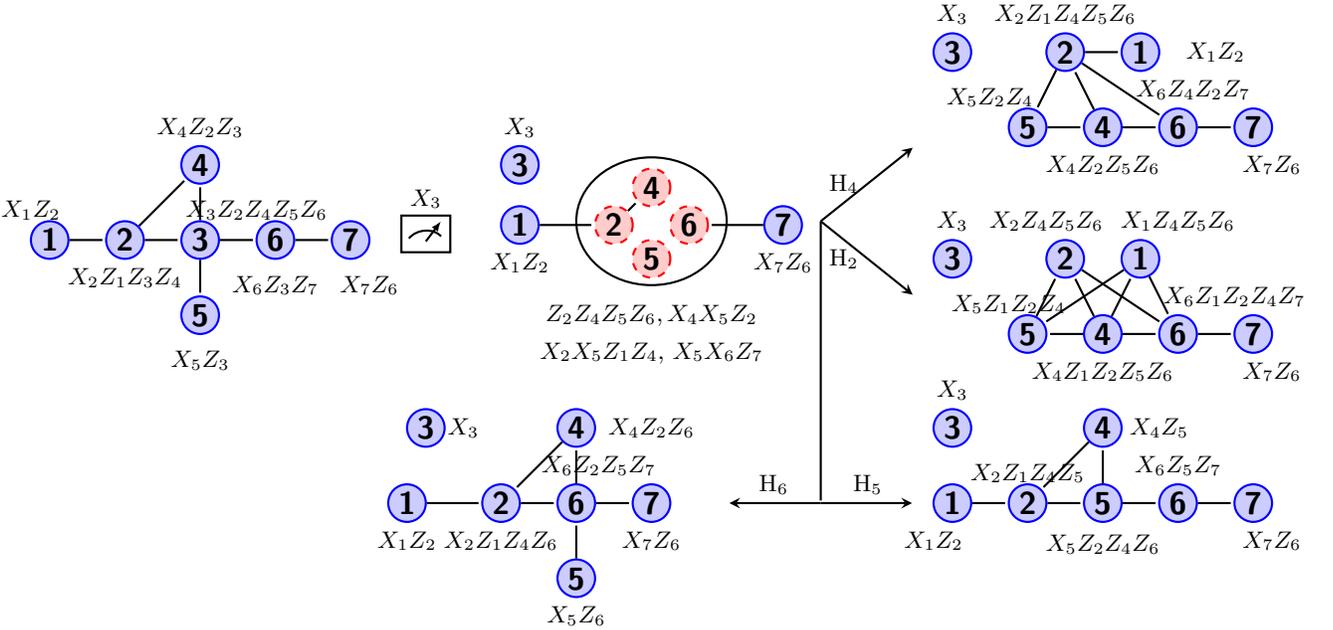

{\textbf{Example 3}}---Consider making an $X$ measurement on qubit $3$ of the cluster shown in FIG.~\ref{fig:measureX_plus_edge}. The only difference with Example 2 is an additional ($2$,$4$) edge. Notice that this edge is retained in the post-measurement stabilizer state, {\em within} the GHZ state portion. FIG.~\ref{fig:measureX_plus_edge} shows all the four possible choices of single-qubit Hadamard gates on the GHZ qubits to turn the post-measurement stabilizer state into a cluster state. We will discuss the $H_2$ case, and leave the remaining for the reader to work through, with FIG.~\ref{fig:measureX_plus_edge} providing visual aid. The $H_2$ converts the GHZ state into a star cluster with qubit $2$ as root node. The ($6$,$7$) edge is retained. Qubit $2$, on which the Hadamard was performed, had two former edges, one to qubit $1$ and the other to qubit $4$. As for the ($1$,$2$) edge, it is removed, and three new edges from qubit $1$ to all the non-qubit-$2$ qubits of the GHZ state, i.e., qubits $4$, $5$ and $6$ are drawn. Similarly, the ($4$,$2$) edge within the GHZ portion of the post-measurement stabilizer state is removed (note that a new ($4$,$2$) edge was formed as a result of the conversion of the GHZ state to a star cluster, which is unaffected), and new edges are drawn from qubit $4$ to all the non-qubit-$2$ qubits of the GHZ state (which is qubits $4$, $5$ and $6$). The ($4$,$4$) self loop can be ignored, and two new edges ($4$,$5$) and ($4$,$6$) are drawn.

\begin{figure*}[htb]
	\centering
	\begin{tikzpicture}[shorten >=1pt, auto, node distance=2cm, thick,
		node_style/.style={circle,inner sep=0pt,minimum size=0.5cm,draw=blue,fill=blue!20!,font=\sffamily\large\bfseries},
		node4_style/.style={circle,inner sep=0pt,minimum size=0.5cm,draw=black,font=\sffamily\large\bfseries},  
		node5_style/.style={circle,dashed,inner sep=0pt,minimum size=0.5cm,draw=red,fill=red!20!,font=\sffamily\large\bfseries},
		node3_style/.style={circle,inner sep=0pt,minimum size=2cm,yscale=.85,draw=black,font=\sffamily\large\bfseries},
		node2_style/.style={circle,dashed,inner sep=0pt,minimum size=0.5cm,draw=blue,fill=blue!20!,font=\sffamily\large\bfseries},
		edge_style/.style={draw=black, thick}
		arrow_style/.style={draw=black,thick,<->}]
		
		\node[node_style] (u2) at (-12,3-0.5) {2};
		\node at (-12,2.5-0.5) {$X_2Z_1Z_3Z_4Z_5$};
		\node[node_style] (u1) at (-13,3-0.5) {1};
		\node at (-13-0.25,3.4-0.5) {$X_1Z_2Z_4$};
		\node[node_style] (u4) at (-11,4-0.5) {4};
		\node at (-11,4) {$X_4Z_2Z_3Z_6$};
		\node[node_style] (u3) at (-11,3-0.5) {3};
		\node at (-10.25,3.4-0.5) {$X_3Z_2Z_4Z_5Z_6$};
		\node[node_style] (u5) at (-11,2-0.5) {5};
		\node at (-11,1.4-0.5) {$X_5Z_2Z_3$}; 
		\node[node_style] (u6) at (-10,3-0.5) {6};
		\node at (-10,2.4-0.5) {$X_6Z_3Z_4Z_7$};
		\node[node_style] (u7) at (-9,3-0.5) {7};
		\node at (-8.75,2.4-0.5) {$X_7Z_6$};
		\draw[edge_style]  (u1) edge (u2);
		\draw[edge_style]  (u2) edge (u3);
		\draw[edge_style]  (u2) edge (u4);
		\draw[edge_style]  (u2) edge (u5);
		\draw[edge_style]  (u1) edge (u4);
		\draw[edge_style]  (u6) edge (u4);
		\draw[edge_style]  (u3) edge (u4);
		\draw[edge_style]  (u3) edge (u5);
		\draw[edge_style]  (u3) edge (u6);
		\draw[edge_style]  (u6) edge (u7);

		\node[node_style] (u1) at (-7+0.25,2+1-0.3) {1};
		\node at (-7+0.25,2+1-0.3-0.5) {$X_1Z_2$};
		\node[node_style] (u7) at (-3.25,2+1-0.3) {7};
		\node at (-3.25,2+1-0.3-0.5) {$X_7Z_6$};
		\node[node_style] (u3) at (-6.75,3.5) {3};
		\node at (-7+0.25,4) {$X_3$};
		\node[node5_style] (u4) at (-13.5+8,2+1-0.3) {2};
		\node[node5_style] (u2) at (-12.5+8,2+1-0.3) {6};
		\node[node5_style] (u5) at (-13+8,3.2) {4};
		\node[node5_style] (u6) at (-13+8,1.5+1-0.25) {5};
		\node[node3_style] (u) at (-13+8,2+1-0.25) {};
		\node at (-5,1.5) {$Z_2Z_4Z_5Z_6,X_4X_5Z_1Z_6$};
		\node at (-5,1) {$ Y_2Y_5Z_1Z_4$, $X_5X_6Z_2Z_4Z_7$};
		\draw[edge_style]  (u1) edge (u4);
		\draw[edge_style]  (u4) edge (u5);
		\draw[edge_style]  (u1) edge (u5);
		\draw[edge_style]  (u4) edge (u6);
		\draw[edge_style]  (u2) edge (u5);
		\draw[edge_style]  (u7) edge (u2);
  \node[inner sep=0pt] (meter) at (-6.75-1.25,2.75)
    {\begin{quantikz}
		 \meter{$X_3$}
    \end{quantikz}};

		\draw [-stealth,thick] (-4.5+2-0.25,2.75) -- node[above] {$\text{H}_{2}$}
		node[below] {Steps 1-2}(-4.5+3.25,2.75);
		\draw [-stealth,thick] (1,2.75-1.5) -- node[right] {Step 3}  (1,1.75-1.5);
		
		\draw [-stealth,thick] (-4.25+3,-1) -- node[above] {Steps 4-5}  (-4.5+1.5,-1);
		
		\node[node4_style] (u2) at (-12+12.5,3-0.5) {2};
		\node[node4_style] (u1) at (-13+12.5,3-0.5) {1};
		\node[node4_style] (u4) at (-11+11.5,4-0.5) {4};
		\node[node_style] (u3) at (-11+13.5,3+.5) {3};
		\node[node4_style] (u5) at (-11+11.5,2-0.5) {5};
		
		\node[node4_style] (u6) at (-10+11.5,3-0.5) {6};
		
		\node[node4_style] (u7) at (-9+11.5,3-0.5) {7};
		
		\draw[edge_style]  (u1) edge (u2);
		
		\draw[edge_style]  (u2) edge (u4);
		\draw[edge_style]  (u2) edge (u5);
		\draw[edge_style]  (u1) edge (u4);
		\draw[edge_style]  (u6) edge (u4);
		\draw[edge_style]  (u2) edge (u4);
		\draw[edge_style]  (u2) edge (u5);
		\draw[edge_style]  (u2) edge (u6);
		\draw[edge_style]  (u6) edge (u7);
		
		\node[node4_style] (u2) at (-12+12.5,3-0.5-3.5) {2};
		\node[node4_style] (u1) at (-13+12.5,3-0.5-3.5) {1};
		\node[node4_style] (u4) at (-11+12.5,3-0.5-3.5) {4};
		\node[node_style] (u3) at (-11+13.5,3+.5-3.5) {3};
		\node[node4_style] (u5) at (-11+11.5,2-0.5-3.5) {5};
		
		\node[node4_style] (u6) at (-10+10.5,4-0.5-3.5) {6};
		
		\node[node4_style] (u7) at (-9+10.5,4-0.5-3.5) {7};

		\draw[edge_style]  (u2) edge (u4);
		\draw[edge_style]  (u2) edge (u5);
		\draw[edge_style]  (u1) edge (u5);
		\draw[edge_style]  (u1) edge (u6);
		\draw[edge_style]  (u6) edge (u4);
		
		\draw[edge_style]  (u2) edge (u5);
		\draw[edge_style]  (u2) edge (u6);
		\draw[edge_style]  (u6) edge (u7);

		\node[node_style] (u2) at (-12+12.5-6,3-0.5-3.5) {2};
            \node at (-12+12.75-6.,3-1-3.5) {$X_2Z_4Z_5Z_6$};
		\node[node_style] (u1) at (-13+12.5-6,3-0.5-3.5) {1};
            \node at (-13+12-6.4,-1) {$X_1Z_5Z_6$};
		\node[node_style] (u4) at (-11+12.5-6,3-0.5-3.5) {4};
            \node at (-11+13.25-6,3-1-3) {$X_4Z_2$};
		\node[node_style] (u3) at (-11+13.5-6,3+.5-3.5) {3};
  \node at (-11+13.5-6,3-3.5) {$X_3$};
		\node[node_style] (u5) at (-11+10.5-6,2-0.5-3.5) {5};
            \node at (-11+10-6.5,2-0.5-3.5) {$X_5Z_1Z_2Z_6$};
		\node[node_style] (u6) at (-10+10.5-6,4-0.5-3.5) {6};
  \node at (-10+10-6.75,4-0.5-3.5) {$X_6Z_1Z_2Z_5Z_7$};
		\node[node_style] (u7) at (-9+10.5-6,4-0.5-3.5) {7};
  \node at (-9+10.5-6,3-0.5-3) {$X_7Z_6$};
		
		\draw[edge_style]  (u2) edge (u4);
		\draw[edge_style]  (u2) edge (u5);
		\draw[edge_style]  (u1) edge (u5);
		\draw[edge_style]  (u1) edge (u6);
		\draw[edge_style]  (u6) edge (u5);
		
		\draw[edge_style]  (u2) edge (u5);
		\draw[edge_style]  (u2) edge (u6);
		\draw[edge_style]  (u6) edge (u7);       
	\end{tikzpicture}
	\caption{An example action of $X$ measurement on the qubit $3$ of a well-connected cluster state. Here, we work out the final cluster state, step by step following the graph rules described in this section, but only for the case when the Hadamard gate (after the $X_3$ measurement) is performed on qubit $2$. Here we use white circles to denote the intermediate steps.}
	\label{fig:measureX_plus_doubleedge}
\end{figure*}
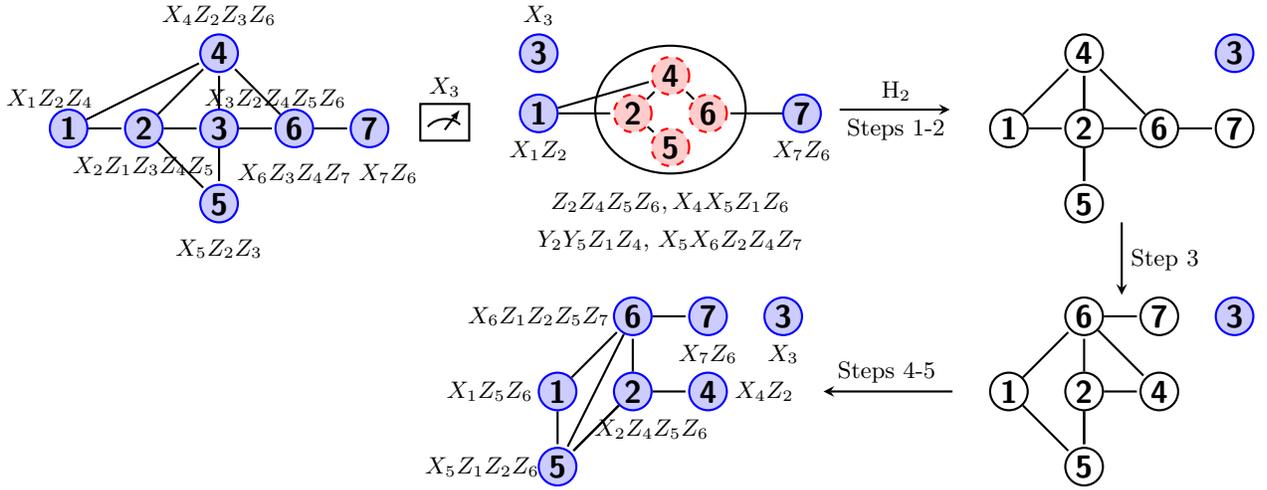

{\textbf{Example 4}}---We consider an example (See FIG.~\ref{fig:measureX_plus_doubleedge}) where a double-edge cancellation occurs. We perform an $X$ measurement on qubit $3$, followed by the Hadamard $H_2$. This time, we show the step-by-step arrival at the cluster state produced after the Hadamard $H_2$ is done.


 \begin{figure}[h]
		\centering
		
	\begin{tikzpicture}[darkstyle/.style={circle,inner sep=0pt,minimum size=0.4cm,draw=blue,fill=blue!20!,font=\sffamily\small\bfseries},
 edge_style/.style={draw=black,  thick}]
 
 \draw [step=0.75cm,thick] (0,0) grid (3,3);

  \foreach \x in {0,0.75,...,3}
    \foreach \y in {0,0.75,...,3} 
       {\pgfmathtruncatemacro{\label}{4*\x/3 - 5 *  4*\y/3 +21}
       \node [darkstyle]  (\x\y) at (\x,\y) {\label};} 
 ;

 \draw [step=0.75cm,thick] (0+4.5,0) grid (1.5+4.5,.75);
 \draw [step=0.75cm,thick] (3+4.5,0) grid (3.75+4.5,.75);
  \draw [thick](4.5,1.5) --   (7.5+0.75,1.5);
  \draw [thick](4.5,3) --   (7.5+0.75,3);
             \draw [thick](4.5,0) --   (4.5,3);
             \draw [thick](4.5+3.75,0) --   (4.5+3.75,3);
              \draw [thick](4.5+3,0) --   (4.5+3,1.5);

  \foreach \x in {0,0.75}
    \foreach \y in {0,0.75,...,3} 
       {\pgfmathtruncatemacro{\label}{4*\x/3 - 5 *  4*\y/3 +21}
       \node [darkstyle]  (\label) at (\x+4.5,\y) {\label};} 
 ;
  \foreach \y in {0,0.75,...,3}
    \foreach \x in {2.25+0.75,3+0.75} 
       {\pgfmathtruncatemacro{\label}{4*(\x-0.75)/3 - 5 *  4*\y/3 +21}
       \node [darkstyle]  (\label) at (\x+4.5,\y) {\label};} 
 ;
 \foreach \y in {0,0.75}
    \foreach \x in {1.5} 
       {\pgfmathtruncatemacro{\label}{4*\x/3 - 5 *  4*\y/3 +21}
       \node [darkstyle]  (\label) at (\x+4.5,\y) {\label};} 
 ;
 \foreach \y in {1.5}
    \foreach \x in {2} 
       {\pgfmathtruncatemacro{\label}{4*\x/3 - 5 *  4*(\y+0.75)/3 +20}
       \node [darkstyle]  (8) at (\x+4.5,\y) {8};} 
        \foreach \y in {2.54}
    \foreach \x in {2.4} 
       {\pgfmathtruncatemacro{\label}{4*\x/3 - 5 *  4*(\y+0.75)/3 +20}
       \node [darkstyle]  (13) at (\x+4.5,\y) {13};} 
 ;
 \foreach \y in {3}
    \foreach \x in {1.5} 
       {\pgfmathtruncatemacro{\label}{4*\x/3 - 5 *  4*\y/3 +21}
       \node [darkstyle]  (\label) at (\x+4.5,\y) {\label};} 
 ;
 
  \node[inner sep=0pt] (meter) at (3.75,1.75)
    {\begin{quantikz}
		 \meter{$X_{13}$}
    \end{quantikz}};
    \node at (3.8,1) {$H_8$};
      \draw[edge_style] (3) edge (12);
      \draw[edge_style] (3) edge (14);
      \draw[edge_style] (3) edge (18);
      \draw[edge_style] (19) edge (18);
      \draw[edge_style] (23) edge (24);
      \draw[edge_style] (9) edge (12);
       \draw[edge_style] (9) edge (18);
       \draw[edge_style] (7) edge (2);
       \draw[edge_style] (8) edge (18);
       \draw[edge_style] (4) edge (9);
       \draw[edge_style] (6) edge (7);
       \draw[edge_style] (7) edge (18);
       \draw[edge_style] (7) edge (14);
      
		\end{tikzpicture}
		\caption{The action of an $X$ measurement on a single qubit of a square grid cluster state leaves a highly-connected distortion in the immediate neighborhood of the measured qubit.}
		\label{fig:measureX_squaregrid}
	\end{figure}
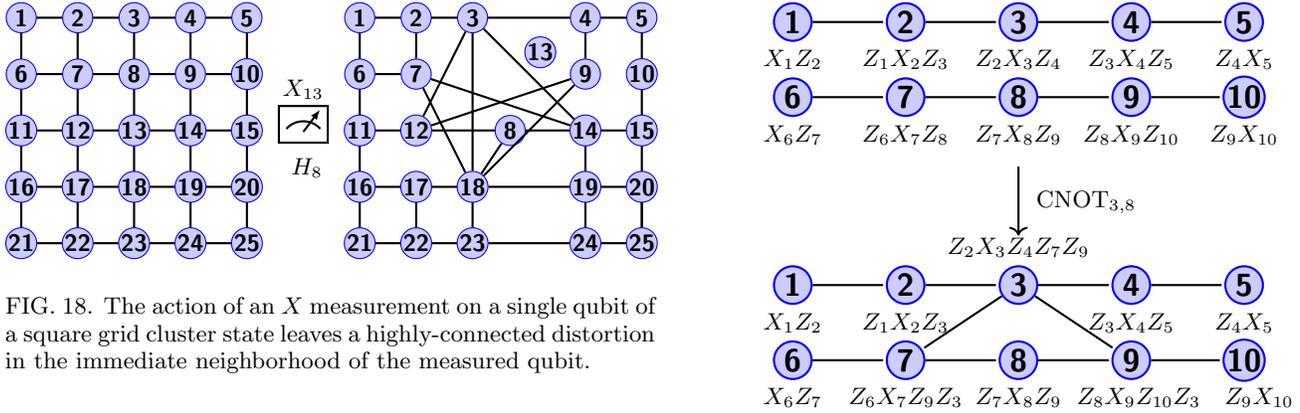

{\textbf{Example 5}}---Finally, we show an example where we depict the distortion caused due to performing an $X$ measurement on a qubit in a square-grid cluster state, followed by a Hadamard on one of its neighbors. See FIG.~\ref{fig:measureX_squaregrid} for an illustration.
	
{\em{Compact graphical rule}}---The four-step graphical rule we have been following for the above examples, can be condensed into a compact expression. If qubit $v\in V$ of cluster state represented by $G(V,E)$ is measured in the $X$ basis, followed by Hadamard $H_u$, where $u$ is a neighbor of $v$ in $G$, then the resulting cluster state is given by the graph $G' = (((G.v).u).v)\backslash\{v\}$~\cite{dahlberg2018transforming}. Recall that $G.x$ denotes the local complementation of graph $G$ at node $x$ (See Section~\ref{subsec:LoComp}).

\subsection{Two adjacent Pauli X measurements}  
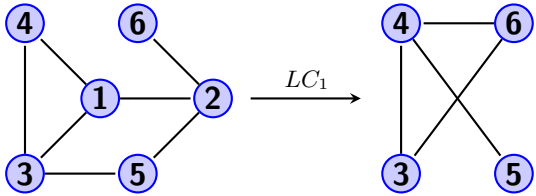
\begin{figure}[h]
	\centering
	\begin{tikzpicture}[shorten >=1pt, auto, node distance=2cm, thick,
		node_style/.style={circle,inner sep=0pt,minimum size=0.5cm,draw=blue,fill=blue!20!,font=\sffamily\large\bfseries},
		node3_style/.style={circle,inner sep=0pt,minimum size=2cm,yscale=.35,draw=black,font=\sffamily\large\bfseries},
		node2_style/.style={circle,dashed,inner sep=0pt,minimum size=0.5cm,draw=blue,fill=blue!20!,font=\sffamily\large\bfseries},
		edge_style/.style={draw=black, thick}
		arrow_style/.style={draw=black,thick,->}]
		
		\node[node_style] (u4) at (-10.5,4) {4};
		\node[node_style] (u1) at (-9.5,3) {1};
		\node[node_style] (u3) at (-10.5,2) {3};
		\node[node_style] (u2) at (-8,3) {2};
		\node[node_style] (u5) at (-9,2) {5};
		\node[node_style] (u6) at (-9,4) {6};
		\draw[edge_style]  (u1) edge (u2);
		\draw[edge_style]  (u1) edge (u3);
		\draw[edge_style]  (u1) edge (u4);
		\draw[edge_style]  (u3) edge (u4);
		\draw[edge_style]  (u2) edge (u5);
		\draw[edge_style]  (u2) edge (u6);
		\draw[edge_style]  (u3) edge (u5);
		
		\node[node_style] (u4) at (-5.5,4) {4};
		\node[node_style] (u3) at (-5.5,2) {3};
		\node[node_style] (u5) at (-4,2) {5};
		\node[node_style] (u6) at (-4,4) {6};
		\draw[edge_style]  (u4) edge (u5);
		\draw[edge_style]  (u6) edge (u3);
		\draw[edge_style]  (u6) edge (u4);
		\draw[edge_style]  (u3) edge (u4);
		\draw [-stealth,thick] (-7.5,3) --  node[above]{$LC_1$}(-6,3);
		
		\end{tikzpicture}

	\caption{Pauli-X basis measurements on the neighboring qubits $1$ and $2$: connects neighbors of qubit $1$ to the neighbors of qubit $2$ if they aren't already connected, and vice versa; if they are connected, then deletes the corresponding edge (e.g., qubits $3$ and $5$), and vice versa.}
	\label{fig:TwoX}
\end{figure}

The result of performing $X$ measurements on two neighboring qubits of a cluster state (i.e., ones that share an edge) can be worked out by sequentially performing the two $X$ measurements using the rules described in the previous subsection. When two adjacent $X$ measurements are performed say on qubits $u$ and $v$ that share an edge, (1) both $u$ and $v$ detach themselves, (2) all edges $(u, w)$ where $w \in {\cal N}_u, w \ne v$, and all edges $(v, s)$ where $s \in {\cal N}_v, s \ne u$ are inverted (see FIG.~\ref{fig:TwoX} for an example). This operation is used for counterfactual error correction when {\em attaching} a tree code to a photon in a cluster state that needs to be protected from unheralded photon loss~\cite{varnava2006loss}, and also to create highly connected cluster states from sparsely connected states. Note that the final state is a cluster state and, unlike the case of a single-qubit $X$ measurement, no additional Hadamard gates are necessary to turn the post-measurement state into a cluster state. 

\subsection{Controlled NOT gate} 
\label{sub:clusterCNOT}
As discussed in Section~\ref{subsec:UnitaryOps} and FIG.~\ref{fig:TwoQubitCliffords}, a CNOT changes the $X$ stabilizer of the control qubit as, $X_c\rightarrow X_cX_t$ and the $Z$ stabilizer of the target qubit as, $Z_t\rightarrow Z_cZ_t$. Hence, doing a CNOT between any two qubits of a cluster state (either one connected cluster state, or each of the two qubits in a cluster state disconnected from one another) inverts the edges connecting the control qubit to all the neighbors of the target qubit, i.e., adds an edge if there is none and removes any pre-existing edge, as shown in the example depicted in FIG.~\ref{fig:CNOT}.
\begin{figure}[h]
	\centering
	\begin{tikzpicture}[shorten >=1pt, auto, node distance=2cm,   thick,
	node_style/.style={circle,inner sep=0pt,minimum size=0.5cm,draw=blue,fill=blue!20!,font=\sffamily\large\bfseries},
	edge_style/.style={draw=black,   thick}
	arrow_style/.style={thick,->}
			]
	
	\node[node_style] (v1) at (-4,2) {1};
	\node at (-4,1.5) {$X_1Z_2$};
	\node[node_style] (v2) at (-2.5,2) {2};
	\node at (-2.5,1.5) {$Z_1X_2Z_3$};
	\node[node_style] (v3) at (-1,2) {3};
	\node at (-1,1.5) {$Z_2X_3Z_4$};
	\node[node_style] (v4) at (0.5,2) {4};
	\node at (0.5,1.5) {$Z_3X_4Z_5$};	
	\node[node_style] (v5) at (2,2) {5};
	\node at (2,1.5) {$Z_4X_5$};

	\draw[edge_style]  (v1) edge (v2);
	\draw[edge_style]  (v2) edge (v3);
	\draw[edge_style]  (v3) edge (v4);
	
	\draw[edge_style]  (v4) edge (v5);

	
	\node[node_style] (v6) at (-4,2-1) {6};
	\node at (-4,1.5-1) {$X_6Z_7$};
	\node[node_style] (v7) at (-2.5,2-1) {7};
	\node at (-2.5,1.5-1) {$Z_6X_7Z_8$};
	\node[node_style] (v8) at (-1,2-1) {8};
	\node at (-1,1.5-1) {$Z_7X_8Z_9$};
	\node[node_style] (v9) at (0.5,2-1) {9};
	\node at (0.5,1.5-1) {$Z_8X_9Z_{10}$};	
	\node[node_style] (v10) at (2,2-1) {10};
	\node at (2,1.5-1) {$Z_9X_{10}$};

	\draw[edge_style]  (v6) edge (v7);
	\draw[edge_style]  (v7) edge (v8);
	\draw[edge_style]  (v8) edge (v9);
	
	\draw[edge_style]  (v9) edge (v10);
	
	\node (ct) at (-1,0.2) {};
	\node (cb) at (-1,-1) {};
	\node (cn) at (-0.1,-0.4) {${\rm{CNOT}}_{3,8}$};
	\draw[->, thick] (ct) to (cb);
	\node[node_style] (v1) at (-4,2-3.5) {1};
	\node at (-4,1.5-3.5) {$X_1Z_2$};
	\node[node_style] (v2) at (-2.5,2-3.5) {2};
	\node at (-2.5,1.5-3.5) {$Z_1X_2Z_3$};
	\node[node_style] (v3) at (-1,2-3.5) {3};
	\node at (-1,1.5-2.5) {$Z_2X_3Z_4Z_7Z_9$};
	\node[node_style] (v4) at (0.5,2-3.5) {4};
	\node at (0.5,1.5-3.5) {$Z_3X_4Z_5$};	
	\node[node_style] (v5) at (2,2-3.5) {5};
	\node at (2,1.5-3.5) {$Z_4X_5$};

	\draw[edge_style]  (v1) edge (v2);
	\draw[edge_style]  (v2) edge (v3);
	\draw[edge_style]  (v3) edge (v4);
	
	\draw[edge_style]  (v4) edge (v5);

	
	\node[node_style] (v6) at (-4,2-1-3.5) {6};
	\node at (-4,1.5-1-3.5) {$X_6Z_7$};
	\node[node_style] (v7) at (-2.5,2-1-3.5) {7};
	\node at (-2.5,1.5-1-3.5) {$Z_6X_7Z_9Z_3$};
	\node[node_style] (v8) at (-1,2-1-3.5) {8};
	\node at (-1,1.5-1-3.5) {$Z_7X_8Z_9$};
	\node[node_style] (v9) at (0.5,2-1-3.5) {9};
	\node at (0.6,1.5-1-3.5) {$Z_8X_9Z_{10}Z_3$};	
	\node[node_style] (v10) at (2,2-1-3.5) {10};
	\node at (2.2,1.5-1-3.5) {$Z_9X_{10}$};

	\draw[edge_style]  (v6) edge (v7);
	\draw[edge_style]  (v7) edge (v8);
	\draw[edge_style]  (v8) edge (v9);
	\draw[edge_style]  (v9) edge (v10);
	\draw[edge_style]  (v3) edge (v7);
	\draw[edge_style]  (v3) edge (v9);
	
	\end{tikzpicture}
	\caption{Performing CNOT between qubit 3 (control) and qubit 8 (target) inverts the edges between all the neighbors of the target qubit and the control qubit. In this example, two new edges ($3$,$7$) and ($3$,$9$) are created.}
	\label{fig:CNOT}
\end{figure}

\subsection{Pauli Y basis measurement}


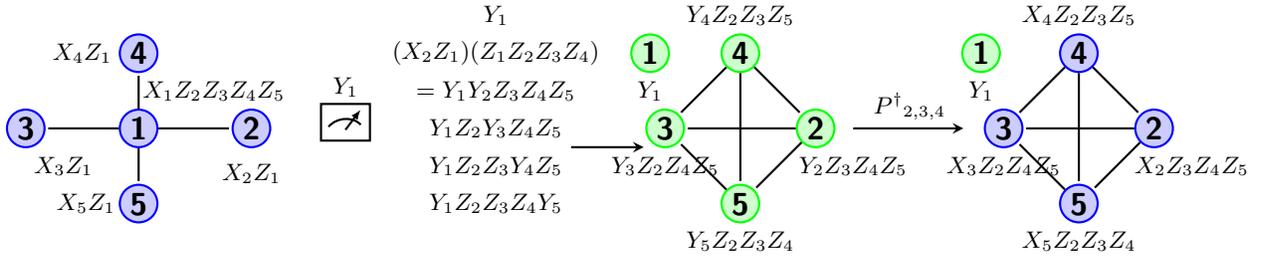
\begin{figure*}[htb]
		\centering
		\begin{tikzpicture}[shorten >=1pt, auto, node distance=2cm, thick,
		node_style/.style={circle,inner sep=0pt,minimum size=0.5cm,draw=blue,fill=blue!20!,font=\sffamily\large\bfseries},
  node4_style/.style={circle,inner sep=0pt,minimum size=0.5cm,draw=red,fill=red!20!,font=\sffamily\large\bfseries},  
  node5_style/.style={circle,inner sep=0pt,minimum size=0.5cm,draw=green,fill=green!20!,font=\sffamily\large\bfseries},
		node3_style/.style={circle,inner sep=0pt,minimum size=2cm,yscale=.65,draw=black,font=\sffamily\large\bfseries},
		node2_style/.style={circle,dashed,inner sep=0pt,minimum size=0.5cm,draw=blue,fill=blue!20!,font=\sffamily\large\bfseries},
		edge_style/.style={draw=black, thick}
		arrow_style/.style={draw=black,thick,<->}]
         
		\node[node_style] (u4) at (-9.5,3.5) {4};
		\node at (-10.25,3.5) {$X_4Z_1$};
        \node[node_style] (u5) at (-9.5,1.5) {5};
		\node at (-10.2,1.5) {$X_5Z_1$};
		\node[node_style] (u1) at (-9.5,3-0.5) {1};
		\node at (-8.5,3.) {$X_1Z_2Z_3Z_4Z_5$};
		\node[node_style] (u3) at (-11,2.5) {3};
		\node at (-10.5,2) {$X_3Z_1$};	
		\node[node_style] (u2) at (-8,3-0.5) {2};
		\node at (-8,2.4-0.5) {$X_2Z_1$};
            \draw[edge_style]  (u1) edge (u2);
		\draw[edge_style]  (u1) edge (u3);
		\draw[edge_style]  (u1) edge (u4);
  \draw[edge_style]  (u1) edge (u5);
  
             \node at (-5+3-0.25-3+1-0.5,3+0.5+0.5) {$Y_1$}; 
             \node at (-5+3-0.25-3+1-0.5,0.5+2.5+0.5) {$(X_2Z_1)(Z_1Z_2Z_3Z_4)$};
            \node at (-5+3-0.25-3+1-0.5,0.5+2+0.5) {$=Y_1Y_2Z_3Z_4Z_5$};
            
            \node at (-5+3-0.25-3+1-0.5,2+0.5) {$Y_1Z_2Y_3Z_4Z_5$};
            \node at (-5+3-0.25-3+1-0.5,1.5+0.5) {$Y_1Z_2Z_3Y_4Z_5$};
            \node at (-5+3-0.25-3+1-0.5,1.5) {$Y_1Z_2Z_3Z_4Y_5$};
		\draw [-stealth,thick] (-7.5+3.25+1-0.5,2.75-0.5) --   (-6.5+3.25+1-0.5,2.75-0.5);
  
            \node[node5_style] (u1) at (-2+.5,3.5) {4};
		\node at (-2+.5,4) {$Y_4Z_2Z_3Z_5$};
 \node[node5_style] (u5) at (-2+.5,1.5) {5};
		\node at (-2+.5,1) {$Y_5Z_2Z_3Z_4$};
		\node[node5_style] (u3) at (-3+.5,2.5) {3};
		\node[node5_style] (u4) at (-3+.3,3.5) {1};
		\node at (-3+.3,3) {$Y_1$};
		\node at (-3+.5,1.5+.5) {$Y_3Z_2Z_4Z_5$};	
		\node[node5_style] (u2) at (-1+.5,2.5) {2};
		\node at (-1+1,1.5+.5) {$Y_2Z_3Z_4Z_5$};
		\draw[edge_style]  (u1) edge (u2);
		\draw[edge_style]  (u1) edge (u3);
            \draw[edge_style]  (u2) edge (u3);
            \draw[edge_style]  (u1) edge (u5);
		\draw[edge_style]  (u5) edge (u3);
            \draw[edge_style]  (u2) edge (u5);
        
  \node[inner sep=0pt] (meter) at (-6.25-.5,2.75)
    {\begin{quantikz}
		 \meter{$Y_1$}
    \end{quantikz}};
  
            \draw [-stealth,thick] (-4+0.5+3.5,2.75-0.25) -- node[above] {$\text{$P^{\dagger}$}_{2,3,4}$}  (-4+2+3.5,2.75-0.25);
           
\node[node_style] (u1) at (-2+.5+4.5,3.5) {4};
		\node at (-2+.5+4.5,4) {$X_4Z_2Z_3Z_5$};
 \node[node_style] (u5) at (-2+.5+4.5,1.5) {5};
		\node at (-2+.5+4.5,1) {$X_5Z_2Z_3Z_4$};
		\node[node_style] (u3) at (-3+.5+4.5,2.5) {3};
		\node at (-3+.5+4.5,1.5+.5) {$X_3Z_2Z_4Z_5$};
		\node[node5_style] (u4) at (-3.5+1+3+1.2,3.5) {1};
		\node at (-3.5+1+3+1.2,3) {$Y_1$};	
		\node[node_style] (u2) at (-1+.5+4.5,2.5) {2};
		\node at (-1+1+4.5,1.5+.5) {$X_2Z_3Z_4Z_5$};
		\draw[edge_style]  (u1) edge (u2);
		\draw[edge_style]  (u1) edge (u3);
            \draw[edge_style]  (u2) edge (u3);
            \draw[edge_style]  (u1) edge (u5);
		\draw[edge_style]  (u5) edge (u3);
            \draw[edge_style]  (u2) edge (u5);
             
		\end{tikzpicture}
		\caption{Here, green circles denote cluster state qubits with the Phase gate applied to them. }

		\label{fig:measureY_star}
	\end{figure*}

Let us begin by analyzing the effect of a $Y$ measurement at the root node (say, qubit $1$) of an $n$ qubit star cluster obtaining outcome $m_1$, where the $n-1$ leaf qubits are numbered $2, 3, \ldots, n$. The stabilizer generators are: $\left\langle X_1Z_2 \ldots Z_n, X_2Z_1, X_3Z_1, \ldots, X_{n}Z_1\right\rangle $. Each of the stabilizers anti-commute with $Y_1$. We first multiply each of the stabilizers except the first one, with the first one, to get the transformed set: $\left\langle X_1Z_2 \ldots Z_n\right.$, $Y_1Y_2\Pi_{i \in \left\{1, \ldots, n\right\}, i \ne 2}Z_i$, $Y_1Y_3\Pi_{i \in \left\{1, \ldots, n\right\}, i \ne 3}Z_i$, $\ldots$, $\left. Y_1Y_n\Pi_{i \in \left\{1, \ldots, n\right\}, i \ne n}Z_i\right\rangle $. Each of the generators in this new set commutes with $Y_1$ except for the first generator. Now, we can apply the rules from Section~\ref{subsec:StabMeasure} to evaluate the post-measurement stabilizer state, whose stabilizer generators are given by (ignoring $m_1$ as usual): $\left\langle Y_1, Y_2\Pi_{i \in \left\{1, \ldots, n\right\}, i \ne 2}Z_i, \ldots, Y_n\Pi_{i \in \left\{1, \ldots, n\right\}, i \ne n}Z_i\right\rangle $. This post-measurement stabilizer state is clearly not a cluster state. However, noting that $P^\dagger Y P = X$, we can apply $P^\dagger$ on each of the qubits $2, \ldots, n$, to obtain the clique-topology (fully-connected) $n-1$ qubit cluster state among qubits $(2, \ldots, n)$ given by stabilizer generators: $\left\langle X_2\Pi_{i \in \left\{1, \ldots, n\right\}, i \ne 2}Z_i, \ldots, X_n\Pi_{i \in \left\{1, \ldots, n\right\}, i \ne n}Z_i\right\rangle $. The measured qubit $1$ is detached from the clique, in the $+1$ eigenstate of $Y_1$, i.e., $(\ket{0} + i\ket{1})/\sqrt{2}$. FIG.~\ref{fig:measureY_star} shows the above for the $n=5$ case, as an example illustration.

{\em{Compact graphical rule}}---Note that transforming an $n$-star into an $n-1$ clique can be interpreted as applying local complementation (LC) on the star's root node, followed by deleting that (isolated) root node. If a $Y$ measurement is performed on a qubit $v$ of a general cluster state described by graph $G$, it is simple to reason that every manipulation we did above to uncover the effect of a $Y$ measurement on the root node of a star cluster will apply verbatim, not affecting the remainder of the topology among the qubits connected to $v$ in $G$ via direct edges. Therefore, the action of the $Y$ measurement on the qubit $v$ in $G$, followed by application of $P^\dagger$ to all neighbors of $v$, i.e., on qubits $u \in {\mathcal N}_v$, is given by a new cluster state described by the graph, $G' = (G.v)\backslash\{v\}$  ~\cite{dahlberg2018transforming}.


In hardware platforms such as the linear optics, where the entangling gates such as CZ and CNOT are inherently probabilistic (see Section~\ref{sec:LO}), generating a highly connected cluster state such as a clique starting with single photons becomes very expensive. Instead, one could prepare an easier-to-realize $n$ qubit star cluster, followed by a single $Y$ measurement, and $n-1$ single-qubit $P^\dagger$ unitaries (which are realizable deterministally) to create an $n-1$ clique. If all the subsequent unitaries and measurements in the application at hand are Clifford, one doesn't even need to perform the $P^\dagger$ unitaries, but simply account for them during the final measurements of the qubits. This method was proposed in the preparation of the error-corrected repeater state for an all-optical quantum repeater scheme~\cite{pant2017allOptical}.

\begin{figure*}
		\centering
		\begin{tikzpicture}[shorten >=1pt, auto, node distance=2cm, thick,
		node_style/.style={circle,inner sep=0pt,minimum size=0.5cm,draw=blue,fill=blue!20!,font=\sffamily\large\bfseries},
  node4_style/.style={circle,inner sep=0pt,minimum size=0.5cm,draw=red,fill=red!20!,font=\sffamily\large\bfseries},  
  node5_style/.style={circle,inner sep=0pt,minimum size=0.5cm,draw=green,fill=green!20!,font=\sffamily\large\bfseries},
		node3_style/.style={circle,inner sep=0pt,minimum size=2cm,yscale=.85,draw=black,font=\sffamily\large\bfseries},
		node2_style/.style={circle,dashed,inner sep=0pt,minimum size=0.5cm,draw=blue,fill=blue!20!,font=\sffamily\large\bfseries},
		edge_style/.style={draw=black, thick}
		arrow_style/.style={draw=black,thick,<->}]
         
		\node[node_style] (u2) at (-12,3-0.5) {2};
		\node at (-12,3.4-0.5) {$X_2Z_1Z_3$};
            \node[node_style] (u1) at (-13,3-0.5) {1};
		\node at (-13-0.25,3.4-0.5) {$X_1Z_2$};
            \node[node_style] (u4) at (-11,4-0.5) {4};
		\node at (-11,4) {$X_4Z_3$};
		\node[node_style] (u3) at (-11,3-0.5) {3};
		\node at (-10.25,3.4-0.5) {$X_3Z_2Z_4Z_5Z_6$};
		\node[node_style] (u5) at (-11,2-0.5) {5};
		\node at (-11,1.4-0.5) {$X_5Z_3$};	
		\node[node_style] (u6) at (-10,3-0.5) {6};
		\node at (-10,2.4-0.5) {$X_6Z_3Z_7$};
            \node[node_style] (u7) at (-9,3-0.5) {7};
		\node at (-8.75,2.4-0.5) {$X_7Z_6$};
            \draw[edge_style]  (u1) edge (u2);
		\draw[edge_style]  (u2) edge (u3);
		\draw[edge_style]  (u3) edge (u4);
            \draw[edge_style]  (u3) edge (u5);
            \draw[edge_style]  (u3) edge (u6);
            \draw[edge_style]  (u6) edge (u7);

            \node[node_style] (u1) at (-7+0.25,2) {1};
             \node at (-7+0.25,1.5) {$X_1Z_2$};
             \node[node_style] (u7) at (-3.5,2+1-0.3) {7};
             \node at (-3.25+.5,2+2-0.5-.75) {$X_7Z_6$};
            \node[node5_style] (u3) at (-2.75,3.75) {3};
            \node at (-5+2.25,4.25) {$Y_3$};
		\node[node5_style] (u4) at (-5.75,2+1-0.3) {2};
            \node[node5_style] (u2) at (-4.5,2+1-0.3) {6};
            \node[node5_style] (u5) at (-6,4) {4};
            \node[node5_style] (u6) at (-13+9-.25,4) {5};
        
            \node at (-6,4.5) {$Y_4Z_2Z_5Z_6$};
             \node at (-4,4.5) {$Y_5Z_4Z_2Z_6$};
            \node at (-13.5+7.5,2+2-0.9) {$Y_2Z_1Z_4Z_5Z_6$};
             \node at (-12.5+8.5,2.2) { $Y_6Z_2Z_4Z_5Z_7$};
            \draw[edge_style]  (u1) edge (u4);
		\draw[edge_style]  (u7) edge (u2);
  \draw[edge_style]  (u5) edge (u4);
  \draw[edge_style]  (u2) edge (u4);
  \draw[edge_style]  (u6) edge (u4);
  \draw[edge_style]  (u5) edge (u6);
  \draw[edge_style]  (u5) edge (u2);
  \draw[edge_style]  (u6) edge (u2);
  \node[inner sep=0pt] (meter) at (-6.75-1.25,2.75+.25)
    {\begin{quantikz}
		 \meter{$Y_3$}
    \end{quantikz}};

           \draw [-stealth,thick] (-1.75-.5,2.75) -- node[above] {$P^{\dagger}_{2,4,5,6}$}  (-.5-.5,2.75);

           \node[node_style] (u1) at (-7+0.25+7-1,2) {1};
             \node at (-7+0.25+7-1,1.5) {$X_1Z_2$};
             \node[node_style] (u7) at (-3.5+7-1,2+1-0.3) {7};
             \node at (-3.25+.5+7-1,2+2-0.5-.75) {$X_7Z_6$};
            \node[node5_style] (u3) at (-2.75+7-1,3.75) {3};
            \node at (-5+2.25+7-1,4.25) {$Y_3$};
		\node[node_style] (u4) at (-5.75+7-1,2+1-0.3) {2};
            \node[node_style] (u2) at (-4.5+7-1,2+1-0.3) {6};
            \node[node_style] (u5) at (-6+7-1,4) {4};
            \node[node_style] (u6) at (-13+9-.25+7-1,4) {5};
        
            \node at (-6+7-1,4.5) {$Y_4Z_2Z_5Z_6$};
             \node at (-4+7-1,4.5) {$Y_5Z_4Z_2Z_6$};
            \node at (-13.5+7.5+7-1,2+2-0.9) {$Y_2Z_1Z_4Z_5Z_6$};
             \node at (-12.5+8.5+7-1,2.2) { $Y_6Z_2Z_4Z_5Z_7$};
            \draw[edge_style]  (u1) edge (u4);
		\draw[edge_style]  (u7) edge (u2);
  \draw[edge_style]  (u5) edge (u4);
  \draw[edge_style]  (u2) edge (u4);
  \draw[edge_style]  (u6) edge (u4);
  \draw[edge_style]  (u5) edge (u6);
  \draw[edge_style]  (u5) edge (u2);
  \draw[edge_style]  (u6) edge (u2);
  
		\end{tikzpicture}
		\caption{A $Y$ measurement is performed on qubit $3$ of the tree cluster shown on the left. The immediate neighborhood of $3$ after the $Y$ measurement (and subsequent application of $P^\dagger$ gates) turns into a clique, while retaining external edges $(2,1)$ and $(6,7)$. The measured qubit $3$ is detached as expected.}

		\label{fig:measureY_plus}
	\end{figure*}
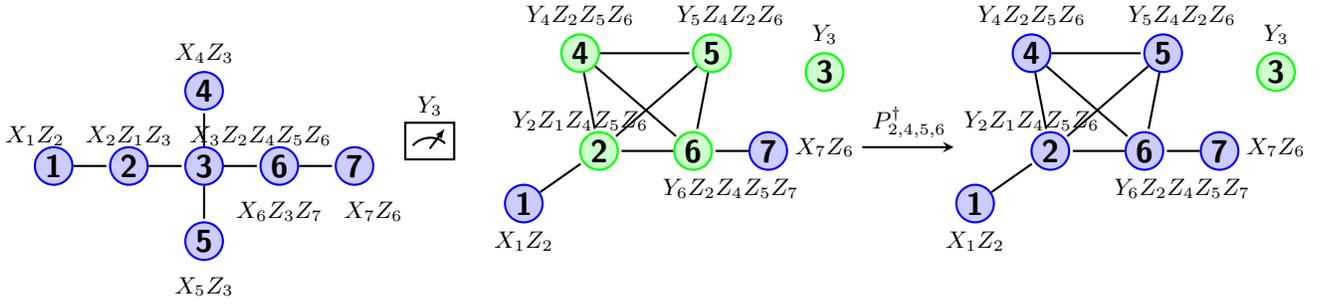


 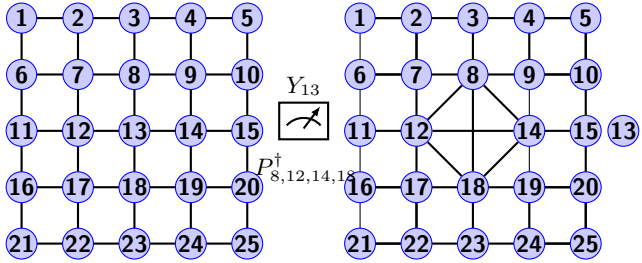
\begin{figure}[h]
		\centering
		
	\begin{tikzpicture}[darkstyle/.style={circle,inner sep=0pt,minimum size=0.4cm,draw=blue,fill=blue!20!,font=\sffamily\small\bfseries},
 edge_style/.style={draw=black,  thick}]
 
 \draw [step=0.75cm,thick] (0,0) grid (3,3);

  \foreach \x in {0,0.75,...,3}
    \foreach \y in {0,0.75,...,3} 
       {\pgfmathtruncatemacro{\label}{4*\x/3 - 5 *  4*\y/3 +21}
       \node [darkstyle]  (\x\y) at (\x,\y) {\label};} 
 ;

 \draw [step=0.75cm,thick] (0+4.5,0) grid (3+4.5,.75);
  \draw [step=0.75cm,thick] (4.5,2.25) grid (3+4.5,3);
             \draw (4.5,0) --   (4.5,3);
             \draw (4.5+2.25,0) --   (4.5+2.25,3);
    \draw [step=0.75cm, thick] (4.5,0) grid (.75+4.5,3);
     \draw [step=0.75cm, thick] (4.5+2.25,0) grid (3+4.5,3);

  \foreach \x in {0,0.75,...,3}
    \foreach \y in {0,0.75} 
       {\pgfmathtruncatemacro{\label}{4*\x/3 - 5 *  4*\y/3 +21}
       \node [darkstyle]  (\label) at (\x+4.5,\y) {\label};} 
 ;
 \foreach \x in {0,0.75,...,3}
    \foreach \y in {2.25,3} 
       {\pgfmathtruncatemacro{\label}{4*\x/3 - 5 *  4*\y/3 +21}
       \node [darkstyle]  (\label) at (\x+4.5,\y) {\label};} 
 ;
 \foreach \x in {0,0.75}
    \foreach \y in {1.5} 
       {\pgfmathtruncatemacro{\label}{4*\x/3 - 5 *  4*\y/3 +21}
       \node [darkstyle]  (\label) at (\x+4.5,\y) {\label};} 
 ;
 \foreach \x in {02.25,3}
    \foreach \y in {1.5} 
       {\pgfmathtruncatemacro{\label}{4*\x/3 - 5 *  4*\y/3 +21}
       \node [darkstyle]  (\label) at (\x+4.5,\y) {\label};} 
 ;
 \node[darkstyle] (13) at (8,1.5) {13};
  \node[inner sep=0pt] (meter) at (3.75,1.8)
    {\begin{quantikz}
		 \meter{$Y_{13}$}
    \end{quantikz}};
    \node at (3.78,1) {$P^{\dagger}_{8,12,14,18}$};
       \draw[edge_style] (8) edge (18);
        \draw[edge_style] (8) edge (12);
         \draw[edge_style] (8) edge (14);
         \draw[edge_style] (12) edge (14);
        \draw[edge_style] (18) edge (12);
         \draw[edge_style] (18) edge (14);
         
		\end{tikzpicture}
		\caption{The action of a single-qubit $Y$ measurement on a qubit of a square grid cluster state.}
		\label{fig:measureY_squaregrid}
	\end{figure}

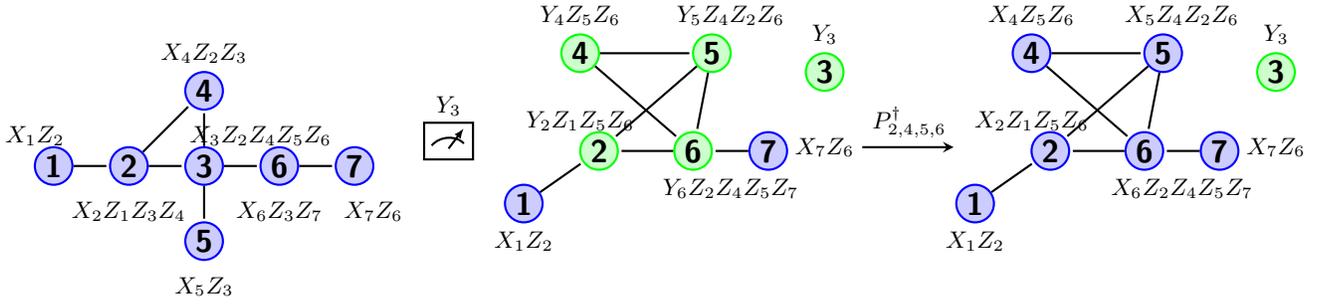
\begin{figure*}[htb]

		\centering
		\begin{tikzpicture}[shorten >=1pt, auto, node distance=2cm, thick,
		node_style/.style={circle,inner sep=0pt,minimum size=0.5cm,draw=blue,fill=blue!20!,font=\sffamily\large\bfseries},
  node4_style/.style={circle,inner sep=0pt,minimum size=0.5cm,draw=red,fill=red!20!,font=\sffamily\large\bfseries},  
  node5_style/.style={circle,inner sep=0pt,minimum size=0.5cm,draw=green,fill=green!20!,font=\sffamily\large\bfseries},
		node3_style/.style={circle,inner sep=0pt,minimum size=2cm,yscale=.85,draw=black,font=\sffamily\large\bfseries},
		node2_style/.style={circle,dashed,inner sep=0pt,minimum size=0.5cm,draw=blue,fill=blue!20!,font=\sffamily\large\bfseries},
		edge_style/.style={draw=black, thick}
		arrow_style/.style={draw=black,thick,<->}]
         
		\node[node_style] (u2) at (-12,3-0.5) {2};
		\node at (-12,3.4-1.5) {$X_2Z_1Z_3Z_4$};
            \node[node_style] (u1) at (-13,3-0.5) {1};
		\node at (-13-0.25,3.4-0.5) {$X_1Z_2$};
            \node[node_style] (u4) at (-11,4-0.5) {4};
		\node at (-11,4) {$X_4Z_2Z_3$};
		\node[node_style] (u3) at (-11,3-0.5) {3};
		\node at (-10.25,3.4-0.5) {$X_3Z_2Z_4Z_5Z_6$};
		\node[node_style] (u5) at (-11,2-0.5) {5};
		\node at (-11,1.4-0.5) {$X_5Z_3$};	
		\node[node_style] (u6) at (-10,3-0.5) {6};
		\node at (-10,2.4-0.5) {$X_6Z_3Z_7$};
            \node[node_style] (u7) at (-9,3-0.5) {7};
		\node at (-8.75,2.4-0.5) {$X_7Z_6$};
            \draw[edge_style]  (u1) edge (u2);
		\draw[edge_style]  (u2) edge (u3);
  \draw[edge_style]  (u2) edge (u4);
		\draw[edge_style]  (u3) edge (u4);
            \draw[edge_style]  (u3) edge (u5);
            \draw[edge_style]  (u3) edge (u6);
            \draw[edge_style]  (u6) edge (u7);

            \node[node_style] (u1) at (-7+0.25,2) {1};
             \node at (-7+0.25,1.5) {$X_1Z_2$};
             \node[node_style] (u7) at (-3.5,2+1-0.3) {7};
             \node at (-3.25+.5,2+2-0.5-.75) {$X_7Z_6$};
            \node[node5_style] (u3) at (-2.75,3.75) {3};
            \node at (-5+2.25,4.25) {$Y_3$};
		\node[node5_style] (u4) at (-5.75,2+1-0.3) {2};
            \node[node5_style] (u2) at (-4.5,2+1-0.3) {6};
            \node[node5_style] (u5) at (-6,4) {4};
            \node[node5_style] (u6) at (-13+9-.25,4) {5};
        
            \node at (-6,4.5) {$Y_4Z_5Z_6$};
             \node at (-4,4.5) {$Y_5Z_4Z_2Z_6$};
            \node at (-13.5+7.5,2+2-0.9) {$Y_2Z_1Z_5Z_6$};
             \node at (-12.5+8.5,2.2) { $Y_6Z_2Z_4Z_5Z_7$};
            \draw[edge_style]  (u1) edge (u4);
		\draw[edge_style]  (u7) edge (u2);
  \draw[edge_style]  (u2) edge (u4);
  \draw[edge_style]  (u6) edge (u4);
  \draw[edge_style]  (u5) edge (u6);
  \draw[edge_style]  (u5) edge (u2);
  \draw[edge_style]  (u6) edge (u2);
  \node[inner sep=0pt] (meter) at (-6.75-1,2.75+.25)
    {\begin{quantikz}
		 \meter{$Y_3$}
    \end{quantikz}};

           \draw [-stealth,thick] (-1.75-.5,2.75) -- node[above] {$P^{\dagger}_{2,4,5,6}$}  (-.5-.5,2.75);

           \node[node_style] (u1) at (-7+0.25+7-1,2) {1};
             \node at (-7+0.25+7-1,1.5) {$X_1Z_2$};
             \node[node_style] (u7) at (-3.5+7-1,2+1-0.3) {7};
             \node at (-3.25+.5+7-1,2+2-0.5-.75) {$X_7Z_6$};
            \node[node5_style] (u3) at (-2.75+7-1,3.75) {3};
            \node at (-5+2.25+7-1,4.25) {$Y_3$};
		\node[node_style] (u4) at (-5.75+7-1,2+1-0.3) {2};
            \node[node_style] (u2) at (-4.5+7-1,2+1-0.3) {6};
            \node[node_style] (u5) at (-6+7-1,4) {4};
            \node[node_style] (u6) at (-13+9-.25+7-1,4) {5};
        
            \node at (-6+7-1,4.5) {$X_4Z_5Z_6$};
             \node at (-4+7-1,4.5) {$X_5Z_4Z_2Z_6$};
            \node at (-13.5+7.5+7-1,2+2-0.9) {$X_2Z_1Z_5Z_6$};
             \node at (-12.5+8.5+7-1,2.2) { $X_6Z_2Z_4Z_5Z_7$};
            \draw[edge_style]  (u1) edge (u4);
		\draw[edge_style]  (u7) edge (u2);
  \draw[edge_style]  (u2) edge (u4);
  \draw[edge_style]  (u6) edge (u4);
  \draw[edge_style]  (u5) edge (u6);
  \draw[edge_style]  (u5) edge (u2);
  \draw[edge_style]  (u6) edge (u2);
  
		\end{tikzpicture}
		\caption{A $Y$ measurement is performed on qubit $3$ of cluster shown on the left. The immediate neighborhood of qubit $3$ after the $Y$ measurement (and subsequent application of $P^\dagger$ gates) turns into a clique, while retaining all other previous edges, i,e., external edges $(2,1)$ and $(6,7)$, and the internal edge $(4,2)$. Since there are two (parallel) $(4,2)$ edges (one from before, and the new one that is part of the clique), this edge is deleted from the final cluster state. Qubit $3$ gets detached as expected.}

		\label{fig:measureY_4}
	\end{figure*}

FIGs.~\ref{fig:measureY_plus}, ~\ref{fig:measureY_4} and ~\ref{fig:measureY_squaregrid} show some illustrative examples of the action of performing a single-qubit $Y$ measurement at a node of a cluster state, followed by applying $P^\dagger$ to the neighbors of that node. 

\subsection{Self loops}\label{sec:self_loops}
If a qubit $u$ in a cluster state has a self loop, the generator of that qubit changes from $X_u\prod_{j\in \mathcal{N}_u}Z_j$ to $Y_u\prod_{j\in \mathcal{N}_u}Z_j$. As we saw before, an application of $P^\dagger$ to that qubit will change this stabilizer generator back to $X_u\prod_{j\in \mathcal{N}_u}Z_j$, since $P^\dagger Y P = X$, while leaving all the stabilizer generators involving $Z_u$ unchanged, since $P^\dagger Z P = Z$. Because of this reason, we can safe `ignore' self-loops in cluster states if they appear when we apply the graphical rules discussed in this paper (in a similar spirit as us ignoring the minus signs in our Pauli measurement examples in calculating the post-measurement states), as long as we keep track of the accumulated phase gates.

We summarize the graphical rules corresponding to all the operations studied above, both unitaries and measurements, in TABLE~\ref{tab:graph_rules}. Also see~\cite{dahlberg2018transforming}.
\begin{table*}[ht]
    \centering
  \begin{tabular}{?P{10cm}|P{7cm}?}
      \hline\xrowht{10pt}
     \textbf{Clifford operation} & \textbf{Graph rule} \\
    \noalign{\hrule height 1pt}\xrowht{20pt}
     Pauli Z-basis measurement on $v\in V$ & $G' = G\backslash\{v\}$ \\
    \hline\xrowht{20pt}
    Pauli-X measurement on $v\in V$ followed by unitary $H_u$, $u\in \mathcal{N}_v$  & $G' = (((G.v).u).v)\backslash\{v\}$ \\
    \hline\xrowht{20pt}
   Pauli-Y measurement on $v\in V$ followed by unitary $P_u^\dagger$, $ u \in \mathcal{N}_v$ & $G' = (G.v)\backslash\{v\}$\\
    \hline\xrowht{20pt}
    CNOT$_{c,t}$ (qubits $c$ and $t$ may not belong to a connected cluster state) & Inverts the edges between $\mathcal{N}_t$ and $c$ \\
    \hline\xrowht{20pt}
    Pauli-X measurements on two neighboring qubits $u,v \in V$  & Inverts the edges between $\mathcal{N}_u$ and $\mathcal{N}_v$ \\
    \noalign{\hrule height 1pt}
  \end{tabular}
\caption{This table summarizes the graph rules for various Clifford operators studied in Section~\ref{sec:ClusterStates}, applied on a cluster state given by the graph $G(V,E)$. For $x\in V$, the neighborhood of $x$ is denoted by $\mathcal{N}_x$, $G\backslash\{x\}$ denotes the graph obtained by deleting the node $x \in V$ (and edges incident on it), and $G.x$ denotes local complementation w.r.t. $x$.} 
    \label{tab:graph_rules}
\end{table*}

\subsection{Two-qubit fusion measurement}
\label{subsec:fusions}
As discussed in Section~\ref{sub:stab_fusion}, two qubit fusions are projective measurements (resp., projections) on unitarily rotated Bell bases (resp., states). The fusion operation has become the fundamental building block for photonic quantum computation. One dimensional or linear cluster states cannot support universal measurement based quantum computing (MBQC) as they cannot be used to implement two qubit gates~\cite{raussendorf2001one}. As a result, it becomes essential to be able to grow cluster states into higher dimensions starting with single qubits or small cluster states. Two-qubit fusions are essential for creating two or higher dimensional cluster states by stitching together smaller cluster states, especially for linear optical hardware \cite{browne2005Fusion,MercedesThesis,pant2019percolation}. Refs.~\cite{seshadreesan2021coherent,fukui2018high} study the use of fusion operations to grow Gottesman-Kitaev-Preskill (GKP) basis cluster states. Ref.~\cite{seshadreesan2021coherent} found that performing fusions on finitely squeezed GKP cluster states gives rise to correlated noise errors in the fused cluster state which propagate only up to second-nearest neighbors of the fused qubits. Recently, an architecture was developed where fault-tolerant quantum computation is encoded into a sequence of two-qubit fusion outcomes performed on pairs of photonic qubits of distinct small (e.g., $6$-qubit) cluster states. This architecture, known as Fusion Based Quantum Computation (FBQC) results in a reduction in errors (compared to MBQC) due to photon loss  errors~\cite{FBQC}. 

Each Clifford-rotated two-qubit fusion joins (two originally disconnected) cluster states, one photon each of which are fused, in a particular fashion, depending upon the rotations $R_c$ and $R_t$ in FIG.~\ref{fig:fusionQckt}. As a result, each rotated fusion has specific graph rules to track the fused cluster state. The action of any Clifford-rotated two-qubit fusion, on the cluster state(s) of whose a pair of photons are being fused, can in principle be evaluated using the rules we described in the above subsections. This is because, a fusion can be expressed as a sequence of single-qubit Clifford unitaries, a CZ gate, and single-qubit Pauli measurements. Changing the unitary operators to rotate the projector changes the graph rules. Instead of individually deriving these graph rules, we leave the derivation for the reader, but tabulate the graph rules for four types of Clifford-rotated fusions in TABLE~\ref{tab:fusion_rules}. Note that the state created using a fusion is a stabilizer state but may not be a cluster state. It can be converted to a cluster state using single qubit gate(s), e.g. Hadamard or Phase gate. The graph rules described in TABLE~\ref{tab:fusion_rules} are to obtain the cluster state after the application of any single-qubit gates (if required). The derivation of these rules is shown in Appendix~\ref{apx:fusion}. Examples of these fusions are shown in FIG.~\ref{fig:chp_fusion}. 
\begin{table*}[ht]
    \centering
  \begin{tabular}{?p{0.5cm}|p{.75cm}?p{1.8cm}|p{1.8cm}|p{4.2cm}?p{1.5cm}|p{1.8cm}| p{4.2cm}?}
    \noalign{\hrule height 1pt}
    \multicolumn{2}{?c?}{} & 
      \multicolumn{3}{c?}{\textbf{Success}} &
      \multicolumn{3}{c?}{\textbf{Failure}} \\
      \hline\xrowht{10pt}
     \centering$R_c$ & \centering $R_t$ & \centering Project on & \centering Unitary & \centering Graph rule & \centering  Project on & \centering Unitary & Graph rule \\
    \noalign{\hrule height 1pt}\xrowht{20pt}
     \centering$H$ &\centering $H$ & \centering$\frac{\bra{++}\pm\bra{--}}{\sqrt{2}}$ & \centering $H_u$ s.t. $u\in \mathcal{N}_c\cup\mathcal{N}_t$ & For a $v \in \{c,t\}$, $w\in\{c,t\}\setminus\{v\}$ and $u\in \mathcal{N}_v$, perform $(((G.v).u).v)\backslash\{v,w\}$. Invert the edges from $\mathcal{N}_w$ to $\mathcal{N}_u \cup \{u\} $.  & $\bra{+-}$ or $\bra{-+}$ & \centering $H_{u,v}$ s.t. $u\in \mathcal{N}_c, v\in\mathcal{N}_t$ &Pauli-X basis measurements on $c$ and $t$, followed by Hadamards on one of their neighbors. \\
    \hline\xrowht{20pt}
     \centering $I$ &\centering $I$ & \centering$\frac{\bra{00}\pm\bra{11}}{\sqrt{2}}$ & \centering $H_u$ s.t. $u\in \mathcal{N}_c\cup\mathcal{N}_t$ & For a $v \in \{c,t\}$, $w\in\{c,t\}\setminus\{v\}$ and $u\in \mathcal{N}_v$, perform $(((G.v).u).v)\backslash\{v,w\}$. Invert the edges from $\mathcal{N}_w$ to $\mathcal{N}_u \cup \{u\} $. & $\bra{10}$ or $\bra{01}$ & \centering Not needed & $G\setminus \{c,t\}$ or Pauli-Z basis measurements on $c$ and $t$ \\
    \hline\xrowht{20pt}
      \centering $H$ &\centering $I$ & \centering$\frac{\bra{+0}\pm\bra{-1}}{\sqrt{2}}$ & \centering Not needed & Remove $c,t$ from the
      cluster.
      Invert the edges from $\mathcal{N}_c$ to $\mathcal{N}_t$.
      & $\bra{+1}$ or $\bra{-0}$ & \centering $H_u$ s.t. $u\in \mathcal{N}_c$ &  For a $u\in \mathcal{N}_c$, perform $(((G.c).u).c)\backslash\{c,t\}$, i.e. Pauli-X measurement of $c$ followed by $H_u$ and Pauli-Z measurement on $t$. 
 \\
    \hline\xrowht{20pt}
    \centering $I$ & \centering $PH$ &  \centering$\frac{\bra{0(-i)}\pm\bra{1(+i)}}{\sqrt{2}}$ $\ket{\pm i}$ : Eigenstates of Pauli-Y & \centering $P^{\dagger}_{\mathcal{N}_c}$& $(((G.c)\backslash\{c,t\}$. Invert the edges from $\mathcal{N}_c$ to $\mathcal{N}_t$.
 & $\bra{0(+i)}$ or $\bra{1(-i)}$& \centering $P^{\dagger}_{\mathcal{N}_t}$ & $(((G.t)\backslash\{c,t\}$, i.e., Pauli Y measurement on $t$ followed by $P^{\dagger}_{\mathcal{N}_t}$ and Pauli Z measurement on $c$.\\
    \noalign{\hrule height 1pt}
  \end{tabular}
\caption{For the fusion operation corresponding to applying the unitary operators $R_c^{\dagger}$ and $R_t^{\dagger}$ applied to the control and target qubits respectively of the BSM projector $\frac{\bra{00}+\bra{11}}{\sqrt{2}}$, the corresponding projection and rules to evolve the original cluster state when the linear optical (LO) fusion circuit either succeeds or fails, are summarized in this Table. Only failed fusion 2 and successful fusion 3 result in a cluster state when the input states are cluster states; all other fusions need local unitaries (as listed in the column ``Unitary") to convert the post-fusion state into a cluster state. Note that, $G$ is the collective graph of the two input cluster states and $\mathcal{N}_i$ refers to the neighborhood of qubit $i$ in $G$.}
\label{tab:fusion_rules}
\end{table*}

\begin{figure*}[htb]
\centering
\begin{tikzpicture}[shorten >=1pt, auto, node distance=2cm, thick,
node_style/.style={circle,inner sep=0pt,minimum size=0.4cm,draw=blue,fill=blue!20!,font=\sffamily\bfseries},
edge_style/.style={draw=black, thick}
arrow_style/.style={draw=black,thick,<->},
node3_style/.style={circle,inner sep=0pt,minimum size=.6cm,yscale=1.6,fill=orange!30!,draw=black,font=\sffamily\large\bfseries}]

            \node[node3_style] (u) at (0-0.5,0-3.75) {};
            \node[node_style] (u1) at (0-0.5,0-4) {1};
            \node[node_style] (u2) at (-.75-0.5,-1-4) {2};
            \node[node_style] (u3) at (0-0.5,-1-4) {3};
            \node[node_style] (u4) at (0.75-0.5,-1-4) {4};

            \draw[edge_style]  (u1) edge (u2);
            \draw[edge_style]  (u1) edge (u3);
            \draw[edge_style]  (u1) edge (u4);

             \node[node_style] (u5) at (1.5-.75-0.5,-1-1.5) {5};
            \node[node_style] (u6) at (1.5-.75-.75-0.5,-3.5) {6};
            \node[node_style] (u7) at (1.50-.75-0.5,-3.5) {7};
            \node[node_style] (u8) at (1.5-0.5,0-3.5) {8};

            \draw[edge_style]  (u5) edge (u6);
            \draw[edge_style]  (u5) edge (u7);
            \draw[edge_style]  (u5) edge (u8);

\draw  (-1.5,-0.5) -- node[above] {Fusion 1} (-3,-.5);
\draw (-3,-.5) -- (-3,1);
\draw (-3,-.5) -- (-3,-2);
\draw [-stealth,thick]  (-3,1) -- node[above]{Success} node[below] {$\frac{\bra{++}\pm\bra{--}}{\sqrt{2}}$}(-4.5,1);
\draw [-stealth,thick]  (-3,-2) -- node[above]{Failure} (-4.5,-2);
\node  at (-3.25,-2.25) {$\bra{+-}$ or $\bra{-+}$};
            \node (H2) at (-5,-1.5+.5) {$H_2$};
            \node[node_style] (u1) at (-5.5,-3.5+.5) {1};
            \node[node_style] (u1) at (-5,-3.5+.5) {5};
            \node[node_style] (u1) at (-4.5,-3.5+.5) {6};
            \node[node_style] (u1) at (-5.5,-4+.5) {7};
            \node[node_style] (u1) at (-4.5,-4+.5) {8};
            \node[node_style] (u2) at (-5,-2+.5) {2};
            \node[node_style] (u3) at (-5.5,-3+.5) {3};
            \node[node_style] (u4) at (-4.5,-3+.5) {4};
            \draw[edge_style]  (u2) edge (u3);
            \draw[edge_style]  (u2) edge (u4);

            \node (H2) at (-5-1.75,-1.5+.5) {$H_3$};
            \node[node_style] (u1) at (-5.5-1.5,-3.5+.5) {1};
            \node[node_style] (u1) at (-5-1.5,-3.5+.5) {5};
            \node[node_style] (u1) at (-4.5-1.5,-3.5+.5) {6};
            \node[node_style] (u1) at (-5.5-1.5,-4+.5) {7};
            \node[node_style] (u1) at (-4.5-1.5,-4+.5) {8};
            \node[node_style] (u2) at (-5-1.5,-2+.5) {3};
            \node[node_style] (u3) at (-5.5-1.5,-3+.5) {2};
            \node[node_style] (u4) at (-4.5-1.5,-3+.5) {4};
            \draw[edge_style]  (u2) edge (u3);
            \draw[edge_style]  (u2) edge (u4);

            \node (H2) at (-5-1.5-1.5,-1.5+.5) {$H_4$};
            \node[node_style] (u1) at (-5.5-1.5-1.5,-3.5+.5) {1};
            \node[node_style] (u1) at (-5-1.5-1.5,-3.5+.5) {5};
            \node[node_style] (u1) at (-4.5-1.5-1.5,-3.5+.5) {6};
            \node[node_style] (u1) at (-5.5-1.5-1.5,-4+.5) {7};
            \node[node_style] (u1) at (-4.5-1.5,-4+.5) {8};
            \node[node_style] (u2) at (-5-1.5-1.5,-2+.5) {4};
            \node[node_style] (u3) at (-5.5-1.5-1.5,-3+.5) {2};
            \node[node_style] (u4) at (-4.5-1.5-1.5,-3+.5) {3};
            \draw[edge_style]  (u2) edge (u3);
            \draw[edge_style]  (u2) edge (u4);

            \node (H2) at (-5,0-.5) {$H_2$};
            \node[node_style] (u1) at (-5.5,1.75-.5) {1};
            \node[node_style] (u5) at (-5,1.75-.5) {5};
            \node[node_style] (u1) at (-5.5,1.25-.5) {6};
            \node[node_style] (u7) at (-4.5,2.25-.5) {7};
            \node[node_style] (u8) at (-5.5,2.25-.5) {8};
            \node[node_style] (u2) at (-5,1-.5) {2};
            \node[node_style] (u3) at (-4.5,.5-.5) {3};
            \node[node_style] (u4) at (-5.5,.5-.5) {4};
            \draw[edge_style]  (u2) edge (u3);
            \draw[edge_style]  (u2) edge (u4);
            \draw[edge_style]  (u5) edge (u8);
            \draw[edge_style]  (u5) edge (u7);
            \draw[edge_style]  (u5) edge (u2);

             \node (H2) at (-5-1.75,0-.5) {$H_3$};
            \node[node_style] (u1) at (-5.5-1.5,1.75-.5) {1};
            \node[node_style] (u5) at (-5-1.5,1.75-.5) {5};
            \node[node_style] (u1) at (-5.5-1.5,1.25-.5) {6};
            \node[node_style] (u7) at (-4.5-1.5,2.25-.5) {7};
            \node[node_style] (u8) at (-5.5-1.5,2.25-.5) {8};
            \node[node_style] (u2) at (-5-1.5,1-.5) {3};
            \node[node_style] (u3) at (-4.5-1.5,.5-.5) {2};
            \node[node_style] (u4) at (-5.5-1.5,.5-.5) {4};
            \draw[edge_style]  (u2) edge (u3);
            \draw[edge_style]  (u2) edge (u4);
            \draw[edge_style]  (u5) edge (u8);
            \draw[edge_style]  (u5) edge (u7);
            \draw[edge_style]  (u5) edge (u2);

            \node (H2) at (-5-1.75-1.75,0-.5) {$H_4$};
            \node[node_style] (u1) at (-5.5-1.5-1.5,1.75-.5) {1};
            \node[node_style] (u5) at (-5-1.5-1.5,1.75-.5) {5};
            \node[node_style] (u1) at (-5.5-1.5-1.5,1.25-.5) {6};
            \node[node_style] (u7) at (-4.5-1.5-1.5,2.25-.5) {7};
            \node[node_style] (u8) at (-5.5-1.5-1.5,2.25-.5) {8};
            \node[node_style] (u2) at (-5-1.5-1.5,1-.5) {4};
            \node[node_style] (u3) at (-4.5-1.5-1.5,.5-.5) {2};
            \node[node_style] (u4) at (-5.5-1.5-1.5,.5-.5) {3};
            \draw[edge_style]  (u2) edge (u3);
            \draw[edge_style]  (u2) edge (u4);
            \draw[edge_style]  (u5) edge (u8);
            \draw[edge_style]  (u5) edge (u7);
            \draw[edge_style]  (u5) edge (u2);

            \node (H2) at (-5-1.5-1.5-1.75,0-.5) {$H_5$};
            \node[node_style] (u1) at (-5.5-1.5-1.5-1.75,.5-.5) {1};
            \node[node_style] (u5) at (-5.5-1.5-1.5-1.75,1-.5) {5};
            \node[node_style] (u1) at (-4.5-1.5-1.5-1.75,.5-.5) {6};
            \node[node_style] (u7) at (-5.5-1.5-1.5-1.75,1.625-.5) {7};
            \node[node_style] (u8) at (-5.5-1.5-1.5-1.75,2.25-.5) {8};
            \node[node_style] (u2) at (-4.5-1.5-1.5-1.75,2.25-.5) {4};
            \node[node_style] (u3) at (-4.5-1.5-1.5-1.75,1-.5) {2};
            \node[node_style] (u4) at (-4.5-1.5-1.5-1.75,1.625-.5) {3};
            \draw[edge_style]  (u2) edge (u7);
            \draw[edge_style]  (u2) edge (u8);
            \draw[edge_style]  (u5) edge (u3);
            \draw[edge_style]  (u5) edge (u4);
            \draw[edge_style]  (u7) edge (u3);
            \draw[edge_style]  (u8) edge (u4);
            \draw[edge_style]  (u8) edge (u3);
            \draw[edge_style]  (u7) edge (u4);
            \draw[edge_style]  (u5) edge (u2);

\draw  (-1.5,-0.5-6.5) -- node[above] {Fusion 2} (-3,-.5-6.5);
\draw (-3,-.5-6) -- (-3,1-6.5);
\draw (-3,-.5-6) -- (-3,-2-6.5);
\draw [-stealth,thick]  (-3,1-6.5) -- node[above]{Success}node[below] {$\frac{\bra{00}\pm\bra{11}}{\sqrt{2}}$} (-4.5,1-6.5);
\draw [-stealth,thick]  (-3,-2-6.5) -- node[above]{Failure} (-4.5,-2-6.5);
\node  at (-3.5,-8.75) {$\bra{01}$ or $\bra{10}$};

            \node[node_style] (u1) at (-5.5-.25,-3.5+.5-6.5) {1};
            \node[node_style] (u1) at (-5-.25,-3.5+.5-6.5) {2};
            \node[node_style] (u1) at (-4.5-.25,-3.5+.5-6.5) {3};
            \node[node_style] (u1) at (-5.5-.25,-4+.5-6.5) {4};
            \node[node_style] (u1) at (-4.5-.25,-4+.5-6.5) {6};
            \node[node_style] (u2) at (-5-.25,-2+.5-6.5) {5};
            \node[node_style] (u3) at (-5.5-.25,-3+.5-6.5) {7};
            \node[node_style] (u4) at (-4.5-.25,-3+.5-6.5) {8};
            \draw[edge_style]  (u2) edge (u3);
            \draw[edge_style]  (u2) edge (u4);

            \node (H2) at (-5,0-.5-6.5) {$H_2$};
            \node[node_style] (u1) at (-5.5,1.75-.5-6.5) {1};
            \node[node_style] (u5) at (-5,1.75-.5-6.5) {5};
            \node[node_style] (u1) at (-5.5,1.25-.5-6.5) {6};
            \node[node_style] (u7) at (-4.5,2.25-.5-6.5) {7};
            \node[node_style] (u8) at (-5.5,2.25-.5-6.5) {8};
            \node[node_style] (u2) at (-5,1-.5-6.5) {2};
            \node[node_style] (u3) at (-4.5,.5-.5-6.5) {3};
            \node[node_style] (u4) at (-5.5,.5-.5-6.5) {4};
            \draw[edge_style]  (u2) edge (u3);
            \draw[edge_style]  (u2) edge (u4);
            \draw[edge_style]  (u5) edge (u8);
            \draw[edge_style]  (u5) edge (u7);
            \draw[edge_style]  (u5) edge (u2);

             \node (H2) at (-5-1.5,0-.5-6.5) {$H_3$};
            \node[node_style] (u1) at (-5.5-1.5,1.75-.5-6.5) {1};
            \node[node_style] (u5) at (-5-1.5,1.75-.5-6.5) {5};
            \node[node_style] (u1) at (-5.5-1.5,1.25-.5-6.5) {6};
            \node[node_style] (u7) at (-4.5-1.5,2.25-.5-6.5) {7};
            \node[node_style] (u8) at (-5.5-1.5,2.25-.5-6.5) {8};
            \node[node_style] (u2) at (-5-1.5,1-.5-6.5) {3};
            \node[node_style] (u3) at (-4.5-1.5,.5-.5-6.5) {2};
            \node[node_style] (u4) at (-5.5-1.5,.5-.5-6.5) {4};
            \draw[edge_style]  (u2) edge (u3);
            \draw[edge_style]  (u2) edge (u4);
            \draw[edge_style]  (u5) edge (u8);
            \draw[edge_style]  (u5) edge (u7);
            \draw[edge_style]  (u5) edge (u2);

            \node (H2) at (-5-1.5-1.5,0-.5-6.5) {$H_4$};
            \node[node_style] (u1) at (-5.5-1.5-1.5,1.75-.5-6.5) {1};
            \node[node_style] (u5) at (-5-1.5-1.5,1.75-.5-6.5) {5};
            \node[node_style] (u1) at (-5.5-1.5-1.5,1.25-.5-6.5) {6};
            \node[node_style] (u7) at (-4.5-1.5-1.5,2.25-.5-6.5) {7};
            \node[node_style] (u8) at (-5.5-1.5-1.5,2.25-.5-6.5) {8};
            \node[node_style] (u2) at (-5-1.5-1.5,1-.5-6.5) {4};
            \node[node_style] (u3) at (-4.5-1.5-1.5,.5-.5-6.5) {2};
            \node[node_style] (u4) at (-5.5-1.5-1.5,.5-.5-6.5) {3};
            \draw[edge_style]  (u2) edge (u3);
            \draw[edge_style]  (u2) edge (u4);
            \draw[edge_style]  (u5) edge (u8);
            \draw[edge_style]  (u5) edge (u7);
            \draw[edge_style]  (u5) edge (u2);

            \node (H2) at (-5-1.5-1.5-1.75,0-.5-6.5) {$H_5$};
            \node[node_style] (u1) at (-5.5-1.5-1.5-1.75,.5-.5-6.5) {1};
            \node[node_style] (u5) at (-5.5-1.5-1.5-1.75,1-.5-6.5) {5};
            \node[node_style] (u1) at (-4.5-1.5-1.5-1.75,.5-.5-6.5) {6};
            \node[node_style] (u7) at (-5.5-1.5-1.5-1.75,1.625-.5-6.5) {7};
            \node[node_style] (u8) at (-5.5-1.5-1.5-1.75,2.25-.5-6.5) {8};
            \node[node_style] (u2) at (-4.5-1.5-1.5-1.75,2.25-.5-6.5) {4};
            \node[node_style] (u3) at (-4.5-1.5-1.5-1.75,1-.5-6.5) {2};
            \node[node_style] (u4) at (-4.5-1.5-1.5-1.75,1.625-.5-6.5) {3};
            \draw[edge_style]  (u2) edge (u7);
            \draw[edge_style]  (u2) edge (u8);
            \draw[edge_style]  (u5) edge (u3);
            \draw[edge_style]  (u5) edge (u4);
            \draw[edge_style]  (u7) edge (u3);
            \draw[edge_style]  (u8) edge (u4);
            \draw[edge_style]  (u8) edge (u3);
            \draw[edge_style]  (u7) edge (u4);
            \draw[edge_style]  (u5) edge (u2);

\draw  (2-0.5,-0.5) -- node[above] {Fusion 3} (3,-.5);
\draw (3,-.5) -- (3,1);
\draw (3,-.5) -- (3,-2);
\draw [-stealth,thick]  (3,1) -- node[above]{Success} node[below] {$\frac{\bra{+0}\pm\bra{-1}}{\sqrt{2}}$}(4.5,1);
\draw [-stealth,thick]  (3,-2) -- node[above]{Failure} (4.5,-2);
\node  at (3.5,-2.25) {$\bra{+1}$ or $\bra{-0}$};

            \node[node_style] (u1) at (-5.5+10.5,1.75-.5) {1};
            \node[node_style] (u5) at (-5+11,1.75-.5-.125) {5};
            \node[node_style] (u1) at (-5.5+10.5,1.25-.5) {6};
            \node[node_style] (u7) at (-4.5+11,2.25-.5) {7};
            \node[node_style] (u8) at (-5.5+11,2.25-.5) {8};
            \node[node_style] (u2) at (-5+11,1-.75) {2};
            \node[node_style] (u3) at (-4.5+11,.25) {3};
            \node[node_style] (u4) at (-5.5+11,.25) {4};
            \draw[edge_style]  (u5) edge (u3);
            \draw[edge_style]  (u5) edge (u4);
            \draw[edge_style]  (u5) edge (u8);
            \draw[edge_style]  (u5) edge (u7);
            \draw[edge_style]  (u5) edge (u2);

            \node (H2) at (5,-3.25) {$H_2$};
            \node[node_style] (u1) at (4.75,-3.5+.75) {1};
            \node[node_style] (u5) at (5.25,-2+.5) {5};
            \node[node_style] (u1) at (5.25,-3.5+.75) {6};
            \node[node_style] (u7) at (5.25,-2.25) {7};
            \node[node_style] (u8) at (5.25,-.75) {8};
            \node[node_style] (u2) at (4.75,-2+.5) {2};
            \node[node_style] (u3) at (4.75,-2.75+.5) {3};
            \node[node_style] (u4) at (4.75,-1.25+.5) {4};
            \draw[edge_style]  (u2) edge (u3);
            \draw[edge_style]  (u2) edge (u4);
             \draw[edge_style]  (u5) edge (u7);
            \draw[edge_style]  (u5) edge (u8);
            
            \node (H2) at (5+1.25,-3.25) {$H_3$};
            \node[node_style] (u1) at (4.75+1.25,-3.5+.75) {1};
            \node[node_style] (u5) at (5.25+1.25,-2+.5) {5};
            \node[node_style] (u1) at (5.25+1.25,-3.5+.75) {6};
            \node[node_style] (u7) at (5.25+1.25,-2.25) {7};
            \node[node_style] (u8) at (5.25+1.25,-.75) {8};
            \node[node_style] (u2) at (4.75+1.25,-2+.5) {3};
            \node[node_style] (u3) at (4.75+1.25,-2.75+.5) {2};
            \node[node_style] (u4) at (4.75+1.25,-1.25+.5) {4};
            \draw[edge_style]  (u2) edge (u3);
            \draw[edge_style]  (u2) edge (u4);
             \draw[edge_style]  (u5) edge (u7);
            \draw[edge_style]  (u5) edge (u8);

            \node (H2) at (5+1.25*2,-3.25) {$H_3$};
            \node[node_style] (u1) at (4.75+1.25*2,-3.5+.75) {1};
            \node[node_style] (u5) at (5.25+1.25*2,-2+.5) {5};
            \node[node_style] (u1) at (5.25+1.25*2,-3.5+.75) {6};
            \node[node_style] (u7) at (5.25+1.25*2,-2.25) {7};
            \node[node_style] (u8) at (5.25+1.25*2,-.75) {8};
            \node[node_style] (u2) at (4.75+1.25*2,-2+.5) {3};
            \node[node_style] (u3) at (4.75+1.25*2,-2.75+.5) {2};
            \node[node_style] (u4) at (4.75+1.25*2,-1.25+.5) {4};
            \draw[edge_style]  (u2) edge (u3);
            \draw[edge_style]  (u2) edge (u4);
             \draw[edge_style]  (u5) edge (u7);
            \draw[edge_style]  (u5) edge (u8);

\draw  (2-0.5,-0.5-6.5) -- node[above] {Fusion 4} (3,-.5-6.5);
\draw (3,-.5-6.5) -- (3,1-6.5);
\draw (3,-.5-6.5) -- (3,-2-6.5);
\draw [-stealth,thick]  (3,1-6.5) -- node[above]{Success}node[below]{$\frac{\bra{0(-i)}\pm\bra{1i}}{\sqrt{2}}$} (4.5,1-6.5);
\draw [-stealth,thick]  (3,-2-6.5) -- node[above]{Failure}node[below]{$\bra{0i}$ or $\bra{1(-i)}$} (4.5,-2-6.5);
    \node[node_style] (u1) at (-5.5+10.5,1.75-.5-6.5) {1};
            \node[node_style] (u5) at (-5+11,1.75-.5-.125-6.5) {5};
            \node[node_style] (u1) at (-5.5+10.5,1.25-.5-6.5) {6};
            \node[node_style] (u7) at (-4.5+11,2.25-.5-6.5) {7};
            \node[node_style] (u8) at (-5.5+11,2.25-.5-6.5) {8};
            \node[node_style] (u2) at (-5+12,1.75-.5-.125-6.5) {2};
            \node[node_style] (u3) at (-4.5+11,.25-6.5) {3};
            \node[node_style] (u4) at (-5.5+11,.25-6.5) {4};
            \draw[edge_style]  (u5) edge (u3);
            \draw[edge_style]  (u5) edge (u4);
            \draw[edge_style]  (u5) edge (u8);
            \draw[edge_style]  (u5) edge (u7);
            \draw[edge_style]  (u5) edge (u2);
             \draw[edge_style]  (u2) edge (u3);
            \draw[edge_style]  (u2) edge (u4); 
            \draw[edge_style]  (u3) edge (u4);
            \node at (-5+11,.25-7) {$P^{\dagger}_{2,3,4}$};
            \node at (-5+11,-2+.5-6) {$P^{\dagger}_{5}$};

        \node[node_style] (u1) at (-5.5+11,-3.5+.5-6.5) {1};
            \node[node_style] (u1) at (-5+11,-3.5+.5-6.5) {2};
            \node[node_style] (u1) at (-4.5+11,-3.5+.5-6.5) {3};
            \node[node_style] (u1) at (-5.5+11,-4+.5-6.5) {4};
            \node[node_style] (u1) at (-4.5+11,-4+.5-6.5) {6};
            \node[node_style] (u2) at (-5+11,-2+.5-6.5) {5};
            \node[node_style] (u3) at (-5.5+11,-3+.5-6.5) {7};
            \node[node_style] (u4) at (-4.5+11,-3+.5-6.5) {8};
            \draw[edge_style]  (u2) edge (u3);
            \draw[edge_style]  (u2) edge (u4);
  
\end{tikzpicture}
\caption{The cluster state(s) resulting from successes and failures of the four different fusions when performed on qubits 1 and 6 of the two degree-3 star states, followed by local unitaries (if any). Note that, qubits 1 and 6 are the control and target qubits, respectively.}

\label{fig:fusion_stars}
\end{figure*}
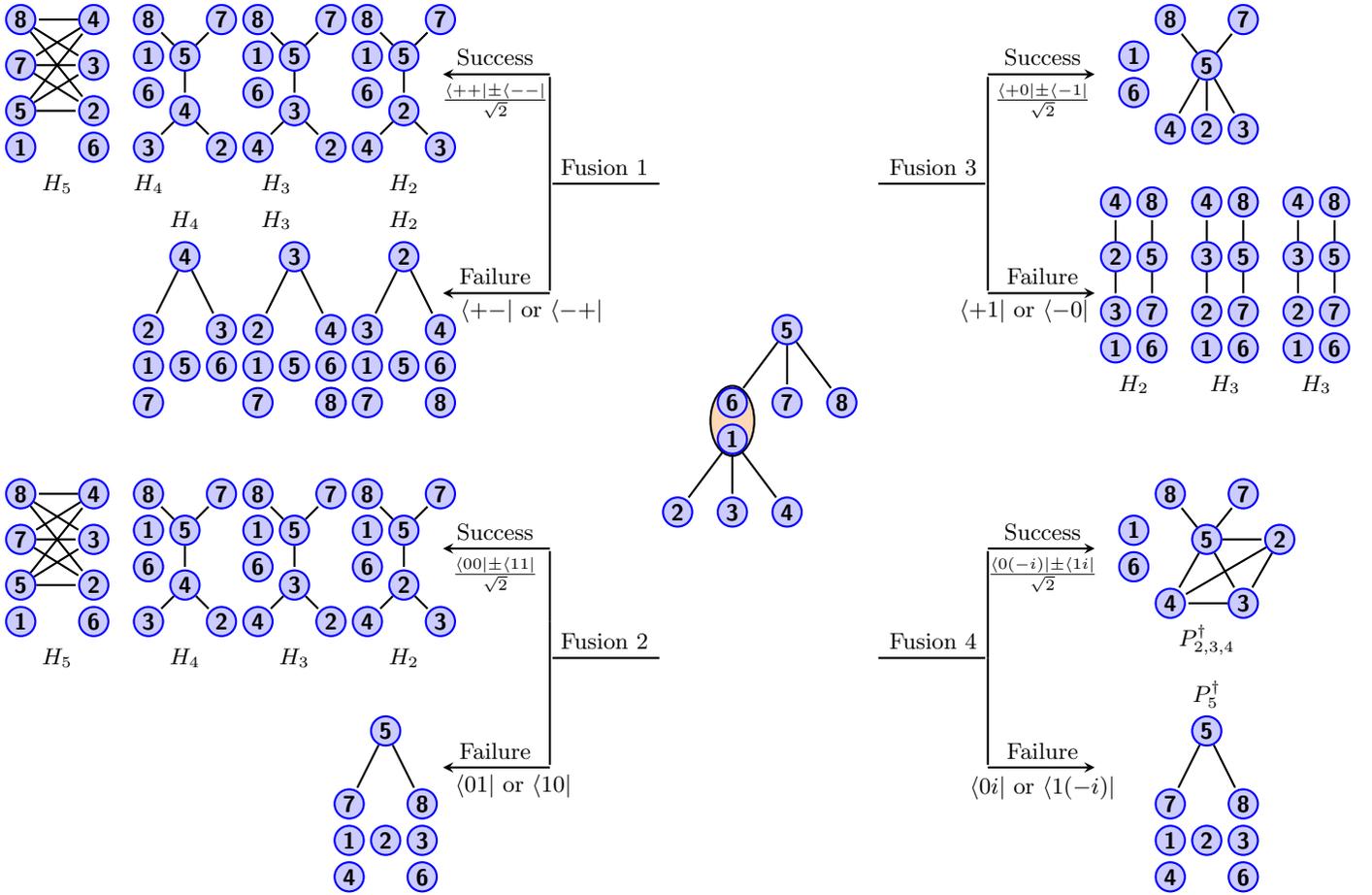

We observe that the graph rule for the $n$-qubit fusion that projects onto the GHZ state $\frac{\ket{0}^{\otimes n}\pm\ket{1}^{\otimes n}}{\sqrt{2}}$ is a generalization of the graph rule for the two-qubit fusion that projects onto $\frac{\ket{00}\pm\ket{11}}{\sqrt{2}}$. If an $n$-qubit fusion is performed on a group of $n$ qubits, each from one of $n$ cluster states collectively described using graph $G$, the state after the fusion is not a cluster state. It can be converted to a cluster state by applying a Hadamard gate on any one of the neighbors of the $n$ measured qubits. If the Hadamard is applied on qubit $j$ in $\mathcal{N}_i$, where $i$ is one of the measured qubits and $\mathcal{N}_i$ is the neighborhood of $i$ in $G$, to obtain the graph describing the post-measurement cluster state:
\begin{itemize}
    \item first perform $(((G.i).j).i)$.
    \item Connect the qubit $j$ and its neighborhood $\mathcal{N}_j$ to neighbors of all measured qubits except $i$.
    \item Delete the $n$ measured qubits from $G$.
\end{itemize}
The rules for the 3-qubit version of this fusion are derived in Appendix~\ref{apx:3fusion} and can be worked out in the same fashion for 4 or higher-qubit fusions.

\section{Stabilizer Simulator}
\label{sec:sta_sim}

The Gottesman-Knill theorem states that a quantum circuit that uses qubits initialized in the computational basis, Clifford unitaries (Hadamard, CNOT, and Phase gates), and Pauli basis measurements can be efficiently simulated on a classical computer~\cite{gottesman1998heisenberg}. Ref.~\cite{aaronson2004improved} improves the algorithm of classical simulation of stabilizer circuits by using twice as many classical bits to encode the quantum state. Based on this method, they have developed a C program called CHP (CNOT-Hadamard-Phase).  They have extended the simulation algorithm to mixed states, non-stabilizer states, and non-stabilizer gates. 

The organization of this Section is as follows: 
\begin{enumerate}
\item We will first review the efficient stabilizer simulation algorithm from~\cite{aaronson2004improved} and explain it by relating the algorithmic steps to the material we covered in Section~\ref{sec:stab_form}. 
\item Thereafter, we will describe our new procedure to convert any Clifford unitary into a series of {\em tableau} operations using Karnaugh maps, a method used to simplify Boolean expressions. 
\item Ref.~\cite{aaronson2004improved} discusses measurement only in the Pauli-Z basis. Although other Pauli measurements can be implemented by first rotating the qubit appropriately and then performing a measurement in the Pauli-Z basis, we explicitly write down the rules for the other two Pauli measurements. 
\item Finally, we discuss the MATLAB program we have developed using the aforesaid method to simulate and visualize various cluster state operations.
\end{enumerate}

\subsection{Destabilizers}
An $n$-qubit stabilizer state has $n$ mutually independent stabilizer generators. These generators can simulate Clifford unitaries in $O(n)$ time. However, updating these generators after stabilizer measurements is more expensive, taking $O(n^3)$ steps. Ref.~\cite{aaronson2004improved} introduced the concept of {\em destabilizers}. Destabilizers, like stabilizers, are Pauli operators and generate the entire Pauli group. Using destabilizers improves the complexity of handling stabilizer quantum measurements to $O(n^2)$ as it doesn't require Gaussian elimination to perform measurements. However, it doubles the number of classical bits used to encode a quantum state, as we discuss below. 

Every stabilizer generator has its corresponding destabilizer (generator). Hence, an $n$-qubit state has $n$ destabilizers. There are some additional properties destabilizers need to satisfy for a given stabilizer state:
\begin{enumerate}
\item Every destabilizer commutes with every other destabilizer.
\item Every destabilizer anti-commutes with its {\em corresponding} stabilizer and commutes with every other stabilizer.
\end{enumerate}

\subsection{Encoding stabilizer states as a tableau}
\label{sub:tableau}
\cite{aaronson2004improved} uses a \textit{tableau} to describe the stabilizers and destabilizers of a stabilizer state. First, they encode every single qubit Pauli operator in 2-bit string as shown in (\ref{eq:TabEnc}). Let's call the first bit the $z$ bit and the second bit as the $x$ bit.
\begin{align}
\label{eq:TabEnc}
    \begin{split}
        &\hspace{1.6em}x\hspace{1em}z\\
	    I&\rightarrow 0\hspace{1em} 0\\
	    Z&\rightarrow 0\hspace{1em}1\\
	    X&\rightarrow 1\hspace{1em}0\\
	    Y&\rightarrow 1\hspace{1em}1\\
    \end{split}
\end{align}
 Every stabilizer and destabilizer of a state is then encoded in bit strings with one additional bit $r$ for the phase such that $r=0$ means positive phase and $r=1$ means negative phase. A Pauli operator comprises 2 bits, and there are $n$ Pauli operators in a stabilizer. As a result, each stabilizer can be written as a $(2n+1)$ bit operator, where the additional bit corresponds to the phase of the operator. Hence, $n(2n+1)$ bits are sufficient to describe the state of $n$-qubits completely. Using destabilizers doubles the number of bits used $(2n(2n+1))$ in exchange for the speedup in simulating quantum measurements on a classical machine.
 
For an $n$-qubit state, the tableau is an $2n\times(2n+1)$ matrix. These stabilizers and destabilizers are arranged in the tableau such that (See Eq.~(\ref{eq:tableau})):
\begin{itemize}
    \item The first $n$ rows correspond to the $n$ destabilizers and the next $n$ rows are the stabilizers.
    \item The $i$-th and $(i+n)$-th columns correspond to $x$-bit and $z$-bit of the encoded Pauli operator of the $i$-th qubit, respectively. 
    \item The last column is reserved for the phases of both stabilizers and destabilizers.
    \item The $x$ bits and the $z$ bits occupy the first $n$ and $n+1\textsuperscript{th}$ to $2n\textsuperscript{th}$ columns of the tableau, respectively.
\end{itemize}

\begin{align}
\label{eq:tableau}
\left(
\begin{array}{ccc|ccc|c}
	x_{11}&  \hdots & 	x_{1n} & z_{11}   &\hdots &	z_{1n}  & r_1 \\
	\vdots& \ddots  & \vdots   & \vdots   &\ddots &\vdots   & \vdots\\
	x_{n1}& \hdots  & 	x_{nn} & z_{n1}   &\hdots &	z_{nn}  & r_n \\
	\hline 
	x_{(n+1)1}&  \hdots & 	x_{(n+1)n} & z_{(n+1)1}   &\hdots &	z_{(n+1)n} & r_{n+1} \\
	\vdots& \ddots  & \vdots   & \vdots   &\ddots &\vdots  & \vdots\\
    x_{2n1}& \hdots  & 	x_{2nn} & z_{2n1}   &\hdots &	z_{2nn} & r_{2n} 
\end{array}
\right).
\end{align}

As an example, the stabilizer generators and corresponding destabilizers of the state $\frac{\ket{+0}_{1,2}-\ket{-1}_{1,2}}{\sqrt{2}}$ are $\{-Z_1X_2, X_1Z_2\}$ and $\{X_1I_2,I_1X_2\}$. The tableau for this state is:

\begin{equation}
\begin{array}{c}
X_1I_2 \rightarrow\\ 
I_1X_2 \rightarrow\\
-\textcolor{blue}{Z_1}\textcolor{red}{X_2} \rightarrow\\ 
X_1Z_2 \rightarrow
\end{array}\left(
\begin{array}{cc|cc|c}
1&0&0&0&0\\
0&1&0&0&0\\
\hline
	\textcolor{blue}{0}&  	\textcolor{red}{1} & 	\textcolor{blue}{1}   &	\textcolor{red}{0} & 1 \\
	1& 0 &0   &1 & 0\\
\end{array}
\right).
\end{equation}
Consider $-Z_1X_2$, $Z_1\rightarrow 01$, hence the first and third columns of the row corresponding to $-Z_1X_2$ make up $01$. Similarly, the second and fourth columns that correspond to qubit 2 have $10$ in them. The negative phase makes the fifth entry 1.

Note that, for cluster states, the stabilizer $z$ sub-matrix of the tableau corresponds to the adjacency matrix of the cluster state and the stabilizer $x$ sub-matrix has full rank.  

\subsection{Unitaries}
In this subsection, we discuss the recipe to convert any Clifford unitary operation to a series of row and column operations on the tableau. If a Clifford unitary is applied to a Pauli unitary encoded as $xz$ and phase $r$ to get $x'z'$ and phase $r'$, the corresponding operations are calculated by writing Kaurnaugh maps for $x'$, $z'$ and $r'$. For example, let's consider the Phase gate $P$. It transforms Paulis according to Eq. (\ref{eq:stabSimPhase}). 
\begin{align}
\label{eq:stabSimPhase}
    \begin{split}
    	x_{ij}z_{ij}         &\xrightarrow{\text{Clifford}}x'_{ij}z'_{ij}\\
	    r_{ij} &\xrightarrow{\text{Clifford}}r'_{ij}
    \end{split}
\end{align} 
\begin{align*}
&\hspace{7.6em}xzr\hspace{1.4em}x'z'r'\\
    I &\xrightarrow{S}I \hspace{2em} \equiv \hspace{2em} 000\xrightarrow{S}000 \\
	    X & \xrightarrow{S}Y  \hspace{1.8em} \equiv \hspace{2em} 100\xrightarrow{S}110 \\
	    Y& \xrightarrow{S}-X  \hspace{0.9em} \equiv \hspace{2em}  110\xrightarrow{S}101 \\
	    Z& \xrightarrow{S}Z  \hspace{1.8em} \equiv \hspace{2em} 100\xrightarrow{S}100\\
\end{align*}

The Boolean expressions to calculate $x'$, $z'$, $r'$ from $x$, $z$, $r$ are derived using Kaurnaugh maps (K-maps). K-maps are used in Boolean algebra to simplify the logical expressions which in turn has applications in digital logic used in classical computers to find reduced equivalent circuit~\cite{karnaugh1953map}. 

The K-maps for the $P$ gate acting on a single qubit are shown in FIG. \ref{fig:K-map}. For the $P$ gate, using the rules of Boolean algebra and the K-maps, 
\begin{align*}
    x' &= x\\
    z' &= \bar{x}.z+x.\bar{z} = x \oplus z\\
    r' &= r+ x.z
\end{align*}
Here, $\oplus$ denotes the Boolean exclusive-OR operation, equivalent to mod 2 addition of bits. Note that, because the acquired phase is due to the Clifford operation multiples existing phase, we add the Boolean expression in terms of $x_{ij}$ and $z_{ij}$ to $r_{i}.$ It implies if the $P$ gate is applied to qubit $j$ of the tableau, then $\forall i\in \{1,2,\dots,2n\}$, i.e., for all rows of the tableau, we set 
\begin{align*}
    z_{ij} &:= \bar{x}_{ij}.z_{ij}+x_{ij}.\bar{z}_{ij} = x_{ij} \oplus z_{ij}\\
    r_{i} &:= r_{i}\oplus x_{ij}.z_{ij}
\end{align*}
$x_{ij}$ remains unchanged. These Boolean operations translate to column operations on the tableau. Note that, all operations on the tableau are performed $\mod 2$.

\begin{figure}[h]
\centering
     \subfloat[]{
        \begin{karnaugh-map}*[2][2][1][$z_{ij}$][$x_{ij}$]
\maxterms{0,1}
\autoterms[1]
\implicant{2}{3}
\end{karnaugh-map}
        }
        
        \subfloat[]{
       \begin{karnaugh-map}*[2][2][1][$z_{ij}$][$x_{ij}$]
\maxterms{0,3}
\autoterms[1]
\implicant{1}{1}
\implicant{2}{2}
\end{karnaugh-map}
        }
        
        \subfloat[]{
       \begin{karnaugh-map}*[2][2][1][$z_{ij}$][$x_{ij}$]
\maxterms{0,2,1}
\autoterms[1]
\implicant{3}{3}
\end{karnaugh-map}
        }
        \caption{Karnaugh maps to derive column addition rules to perform on the tableau. (a), (b) and (c) show the K-maps for $x_{ij}'$, $z_{ij}'$ and $r_{ij}'$ bits respectively for the $P$ gate.}
        \label{fig:K-map}
    
\end{figure}

\begin{figure}
	\centering
	\includegraphics[scale=0.6]{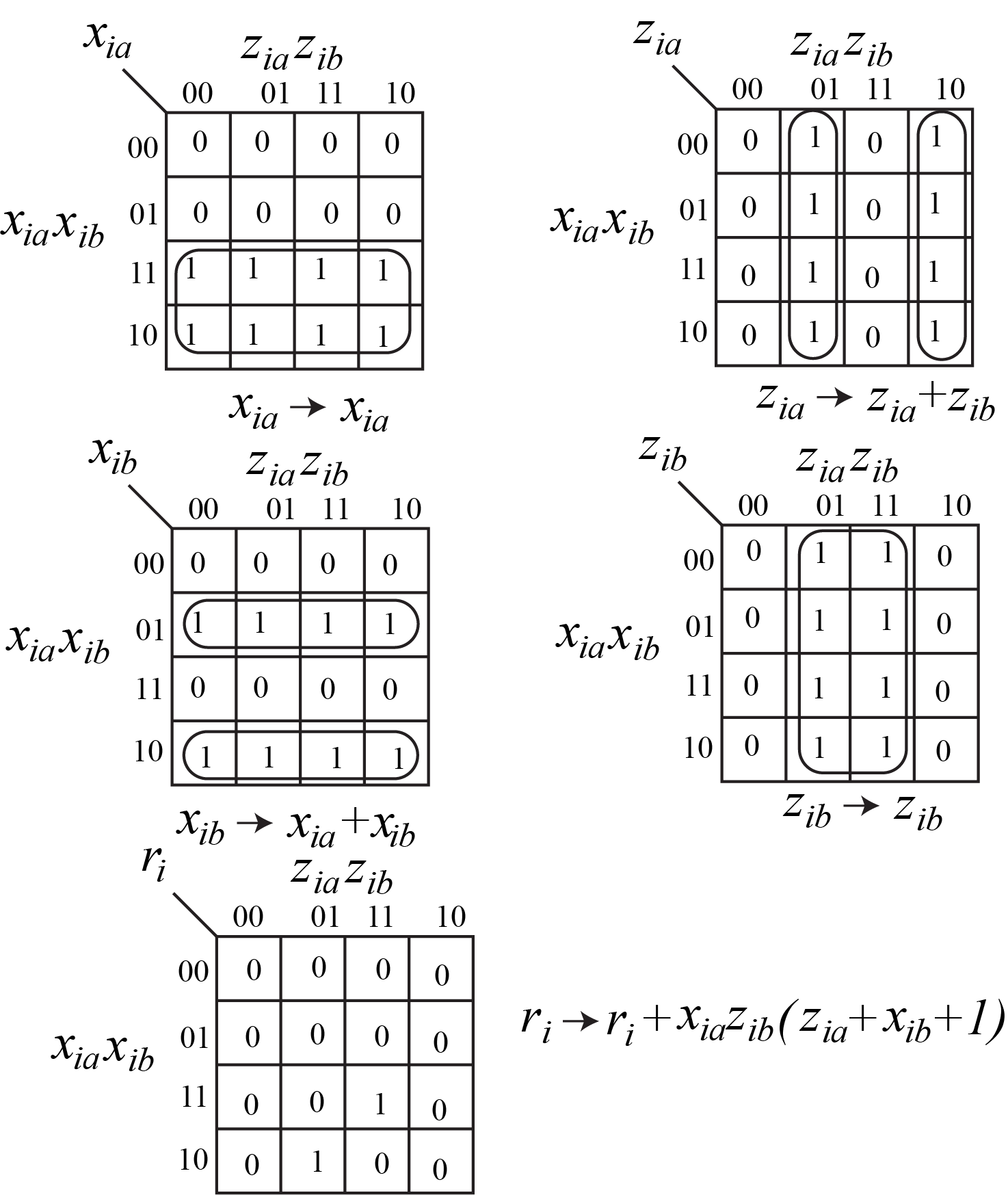}
	\caption{Karnaugh maps for CNOT gate such that qubit $a$ is the control and qubit $b$ is the target.}
	\label{fig:CNOTkmap}
\end{figure}

For 2-qubit gates, we get 4x4 K-maps and the Boolean expressions can be derived in a similar fashion. For example, using Section\ref{subsec:UnitaryOps}, FIG.~\ref{fig:CNOTkmap} demonstrates the transformations of $x, z,$ and $r$ when $CNOT_{a,b}$ is applied on qubits $a$ and $b$. 

\subsection{Measurements}
In this section, we will discuss tableau operation to implement measurements. We first define the symplectic inner product of any two rows $R_i$ and $R_j$ as follows
\begin{align*}
    R_i.R_j &= x_{i_1}z_{j_1}\oplus\dots\oplus x_{i_n}z_{j_n}\oplus x_{j_1}z_{i_1}\oplus\dots\oplus x_{j_n}z_{i_n}
\end{align*}
Rows $R_i$ and $R_j$ commute if $ R_i.R_j=0$ and anti-commute if $R_i.R_j=1$.

If the measurement operator anti-commutes with the stabilizer generator(s), then the rules to calculate the generators after measurement involve the multiplication of stabilizers (refer Section \ref{subsec:StabMeasure}). If stabilizer $S_h$ is multiplied with $S_k$, the corresponding modified row in the tableau $R_h'= R_h.R_k$  is calculated using the following subroutine - 
\newline \textbf{rowsum(h, k)}:  There are two parts to this subroutine. The first is calculating the transformed stabilizer. It is done by setting $x_{hj}'=x_{hj}\oplus x_{kj}$ and $z_{hj}'=z_{hj}\oplus  z_{kj} \forall j \in \{1,2,\dots,n\}$. This is equivalent to adding the first $2n$ elements of row $R_i$ to row $R_h$. Adding the first $2n$ columns of any two rows of the tableau is equivalent to taking the product of corresponding Pauli matrices ignoring the phase. The second part calculates the phase bit $r_{h}'$. The phase of $R_h'$ which is given by can take the values $\{\pm 1, \pm i\}$ i.e., $i^p$. We need to calculate the exponent $p$ which depends upon the individual phases of $R_h$ and $R_k$, as well as the phase that arises from the product of Pauli matrices encoded in $R_h$ and $R_k$. 

Now, let us define a function $g(x_1, z_1, x_2, z_2)$ that determines the power to which $i$ should be raised to when Pauli matrix $x_2z_2$ is multiplied by $x_1z_1$. For example, if $x_1z_1=00 \equiv I$, the phase of $x_2z_2$ remains unchanged. $\implies g=0$. If $x_1z_1=01 \equiv Z$, phase becomes $+i(-i)$ if $x_2z_2=10(11)$ $\implies g=x_2(1-2z_2)$. Similarly, if $x_1z_1=10 \equiv X$, $g=z_2(2x_2-1)$; if $x_1z_1=11 \equiv Y$, $g=z_2-x_2$. 

Then $p$ is calculated using the following equation - 
\begin{align}
\label{eq:rowsum}
	p&= 2r_h+2r_k+\sum_{j=1}^{n}g (x_{kj} , z_{kj} , x_{hj} , z_{hj})
\end{align}
As $p$ depends upon the initial phases of $R_h$ and $R_k$, we add $2r_h+2r_k$. The phase bit $r$ takes values from $\{0,1\}$, which means $i^{2r}$ takes the values $\{-1,1\}$. The last term in eq. (\ref{eq:rowsum}) signifies the product of phases arising due to the multiplication of Pauli matrices corresponding to all $n$ qubits in the stabilizers $S_h$ and $S_k$. If $p\equiv$ 0 mod 4 $\implies i^p=1$, we set $ r_h' \coloneqq 0$, and if $p\equiv$ 2 mod 4, $r_h'\coloneqq 1$. 

In the following sections, we describe the algorithm to perform all three Pauli basis measurements on the tableau. For any other basis, we can apply appropriate rotation on the qubit, followed by measurement in the Pauli-Z basis. For this algorithm, we must add an additional $(2n+1)$-th row to the tableau as a scratch place.
\subsubsection{Pauli-Z basis measurement}
From Section \ref{subsec:StabMeasure}, we know that the rules to modify stabilizers differ depending on whether the measured observable commutes with all stabilizer generators. If qubit $a$ is measured in Pauli-Z basis, 
\newline \textbf{Step 1 - }first check if there exists a $b\in \{n+1,\dots,2n\}$ such that $x_{ba}=1$ as it implies there exists a stabilizer generator with a $X_a$ or $Y_a$ Pauli operators. Hence, all stabilizers won't commute with the observable $Z_a$.
\newline \textbf{Step 2 - }
\newline \textit{(a) If $Z_a$ commutes with all generators -} i.e., no such $b$ exists. In this case, to identify the deterministic outcome, set the $(2n+1)$-th row identically to zero. Call rowsum($2n+1,i+n$) $\forall i \in \{1,\dots,n\}$ such that $x_{ia}=1$. This step adds the phases of all stabilizers that contain $Z_a$. Return $r_{2n+1}$ as the measurement outcome keeping the rest of the tableau unchanged.
\newline \textit{(b) If $Z_a$ anti-commutes with at least one generator - }i.e. such a $b$ exists. If more than one such $b$ exists, WLOG choose the smallest value. Call rowsum$(i,b)$ $\forall i\in\{i,\dots,2n\}$ such that $i\neq b$, $i \neq b-n$ and $x_{ia}=1$. This step is equivalent to multiplying one anti-commuting generator to all other anti-commuting generators and corresponding destabilizers to restore the commutation relations. Next, set the entire $(b-n)^{th}$ row as the $b^{th}$ row and set the first $2n$ columns of the $b^{th}$ row identically zero and set $z_{ba}=1$. It ensures that the $b^{th}$ row and its destabilizer anticommute and qubit $a$ is in the eigenstate of $Z_a$. Set $r_b$, the measurement outcome, to be 0 or 1 with equal probability.

The algorithm steps remain the same for Pauli-X and -Y basis measurements except for minor changes. We briefly discuss the algorithm in the following sections.

\subsubsection{Pauli-X basis measurement}
\noindent \textbf{Step 1 - }first check if there exists a $b\in \{n+1,\dots,2n\}$ such that $z_{ba}=1$.
\newline \textbf{Step 2 - }
\newline \textit{(a) If $X_a$ commutes with all generators -} i.e., no such $b$ exists. Set the $(2n+1)$-th row identically zero. Call rowsum($2n+1,i+n$) $\forall i \in \{1,\dots,n\}$ such that $z_{ia}=1$. Return $r_{2n+1}$ as the measurement outcome keeping the rest of the tableau unchanged.
\newline \textit{(b) If $X_a$ anti-commutes with at least one generator - }i.e. such a $b$ exists. Call rowsum$(i,b)$ $\forall i\in\{i,\dots,2n\}$ such that $i\neq b$, $i \neq b-n$ and $z_{ia}=1$. Next, set the entire $(b-n)^{th}$ row as the $b^{th}$ row and set the first $2n$ columns of the $b^{th}$ row identically zero and set $x_{ba}=1$. Set the measurement outcome $r_b$ to be 0 or 1 with equal probability.

\subsubsection{Pauli-Y basis measurement}
\noindent \textbf{Step 1 - }first check if there exists a $b\in \{n+1,\dots,2n\}$ such that $x_{ba}\oplus z_{ba}= 1$.
\newline \textbf{Step 2 - }
\newline \textit{(a) If $Y_a$ commutes with all generators -} i.e., no such $b$ exists. Set the $(2n+1)$-th row identically zero. Call rowsum($2n+1,i+n$) $\forall i \in \{1,\dots,n\}$ such that $x_{ia}\oplus z_{ia}=1$. Return $r_{2n+1}$ as the measurement outcome keeping the rest of the tableau unchanged.
\newline \textit{(b) If $Y_a$ anti-commutes with at least one generator - }i.e. such a $b$ exists. Call rowsum$(i,b)$ $\forall i\in\{i,\dots,2n\}$ such that $i\neq b$, $i \neq b-n$ and $x_{ia}\oplus z_{ia}=1$. Next, set the entire $(b-n)^{th}$ row as the $b^{th}$ row and set the first $2n$ columns of the $b^{th}$ row identically zero and set $x_{ba}=z_{ba}=1$. Set the measurement outcome $r_b$ to be 0 or 1 with equal probability.

\subsection{Cluster state simulator}
\label{sub:matlab_sim}
We have used the tableau method in conjunction with the theory of Clifford operations on cluster states discussed in  \cite{MercedesThesis, van2003graphical, van2004efficient} to develop a MATLAB-based simulator that can not only efficiently perform Clifford operations on all the stabilizer states but also converts a stabilizer state into cluster state in addition to visualizing the cluster states as graphs. The code for this simulator can be found here \cite{code}.

\begin{figure}
    \centering
    \includegraphics[scale=0.4]{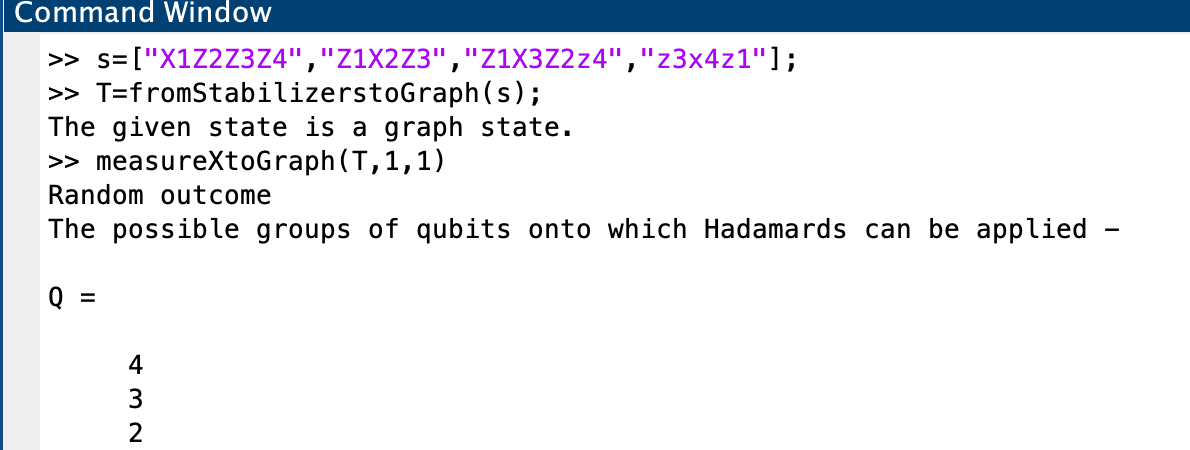}
    \caption{Screenshot of the MATLAB cluster state simulator}
    \label{fig:screenshot}
\end{figure}

We discussed in Section~\ref{sub:tableau} that the $x$-sub-matrix of the stabilizer part of the tableau ($S_x$) is a full rank matrix for cluster states and the $z$-sub-matrix of the stabilizer part of the tableau ($S_z$) is the adjacency matrix of the cluster state. Our code uses this fact to identify if the state encoded in the given tableau is a cluster state or not. We find the rank $m$ of this submatrix $S_x$ for an $n$ qubit stabilizer state. Remember that the tableau operations are binary or mod2 operations and hence, we need to calculate Galois field $\text{GF(2)}$ rank. If $m<n$, the simulator identifies that the given stabilizer state is not a cluster state and outputs a message - \texttt{Not a graph state.} If the user chooses to convert the stabilizer state to a cluster state, the simulator outputs all possible combinations of qubits that would undergo Hadamard gates to take the stabilizer state to a cluster state as shown in Fig.~\ref{fig:screenshot}. The user can choose one such combination of qubits or ask the simulator to display the outcomes for all combinations. The steps to identify the qubits that would undergo Hadamard operations are as follows - 
\begin{enumerate}
    \item Find all possible rank $m$ sub-matrices of $S_x$ using Gaussian elimination. If there are multiple rank $m$ sub-matrices, then there are multiple ways to convert the stabilizer state into a cluster state.
    \item The columns correspond to qubits in the tableau. As a result, for a given rank $m$ sub-matrix, Hadamard gates are applied to all the columns of $S_x$ (equivalently, qubits) that are not part of the sub-matrix. This works because the Hadamard operation converts $Z$ to $X$, increasing the rank of $S_x$. This step slows down the simulator as finding all possible combinations of $m\times m$ sub-matrices takes exponential time.
\end{enumerate}
If $m=n$, and if $S_z$ valid adjacency matrix such that all the diagonal entries of $S_z$ are zero, the simulator displays a message saying  - \texttt{The graph state is uniquely defined and no Hadamard gates needed.} If not all diagonal elements of $S_z$ are zero, it implies the qubits corresponding to non-zero entries have Pauli-Y operators in their stabilizers. The simulator prints all such qubits and then applies the phase gate to these qubits to convert $Y$ to $X$.

\begin{figure*}[htb]
\centering
\subfloat[\label{fig:matlab_clusterState}]{%
  \includegraphics[scale = 0.14]{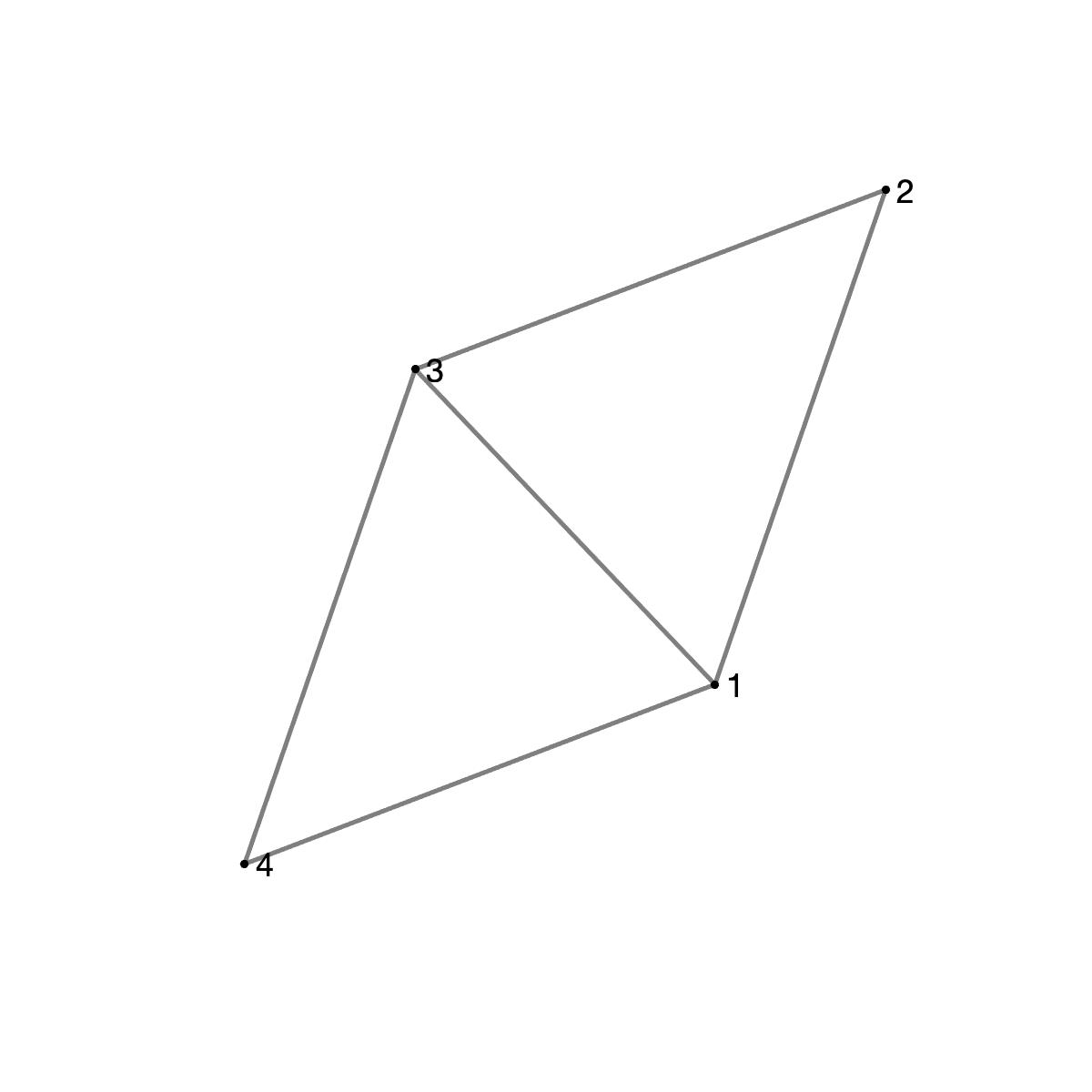}%
}
\subfloat[\label{fig:mx}]{%
  \includegraphics[scale = 0.14]{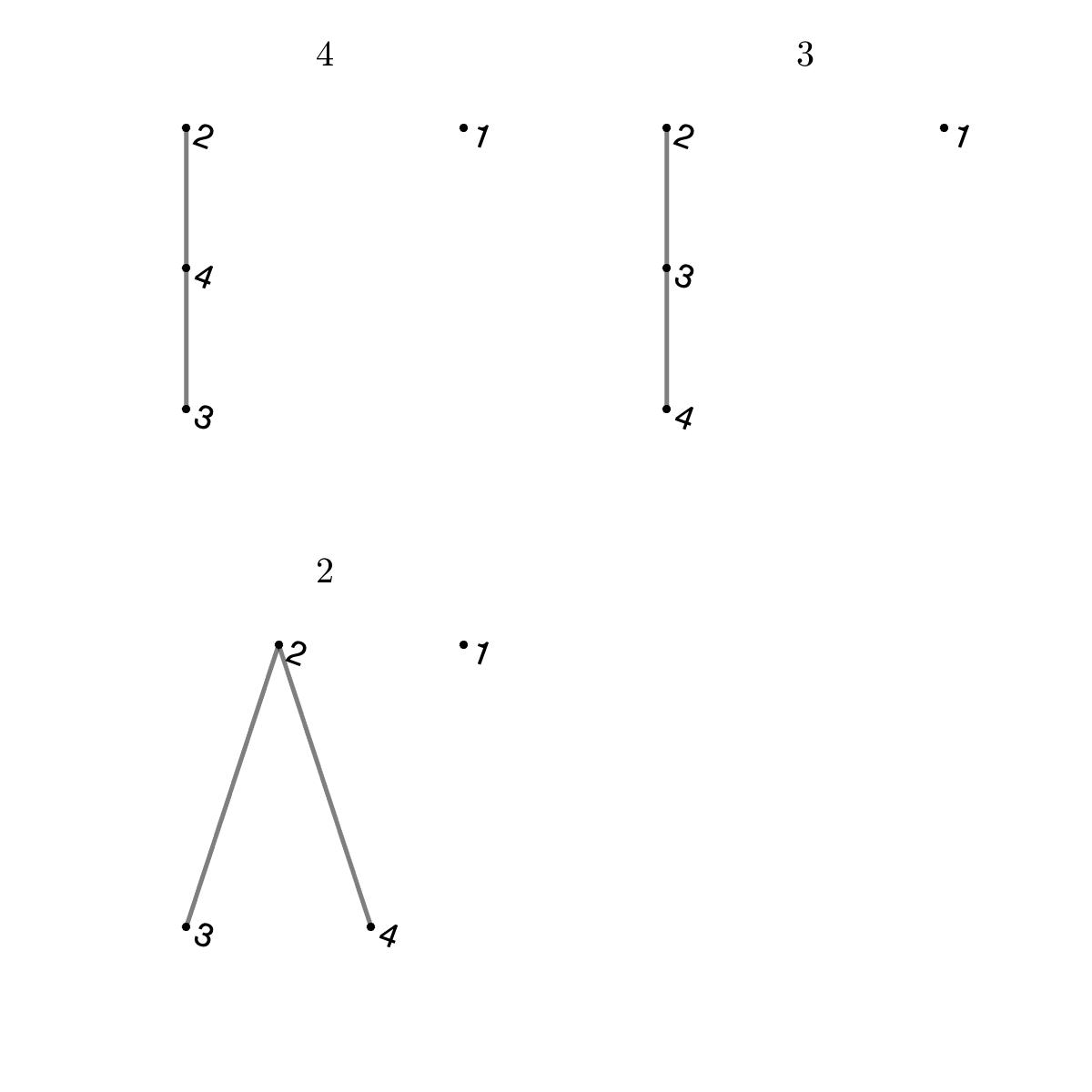}%
}
\subfloat[\label{fig:mz}]{%
  \includegraphics[scale = 0.14]{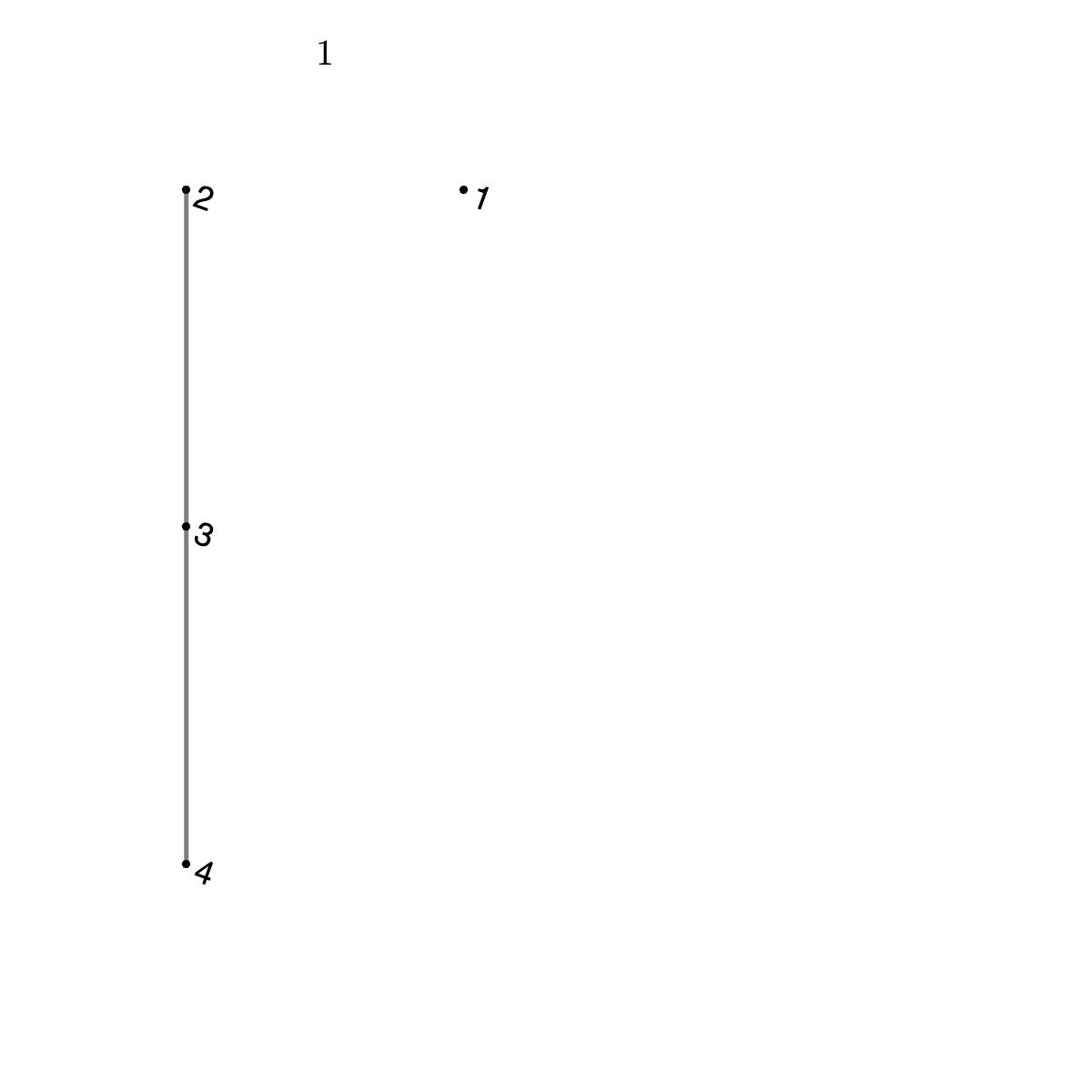}%
}
\caption{Outputs generated using the MATLAB Cluster state simulator (a) A cluster state drawn using the MATLAB program as shown in Fig.~\ref{fig:screenshot} (b) All possible Hadamard operations and the resultant cluster states when qubit 1 is measured in Pauli X basis. The number at the top of each graph is the qubit that undergoes the Hadamard operation. (c) All possible Hadamard operations and the resultant cluster states when qubit 1 is measured in Pauli Z basis}
\label{fig:CHP_pauli}
\end{figure*}

\begin{figure*}[htb]
\centering
\includegraphics[scale=0.6]{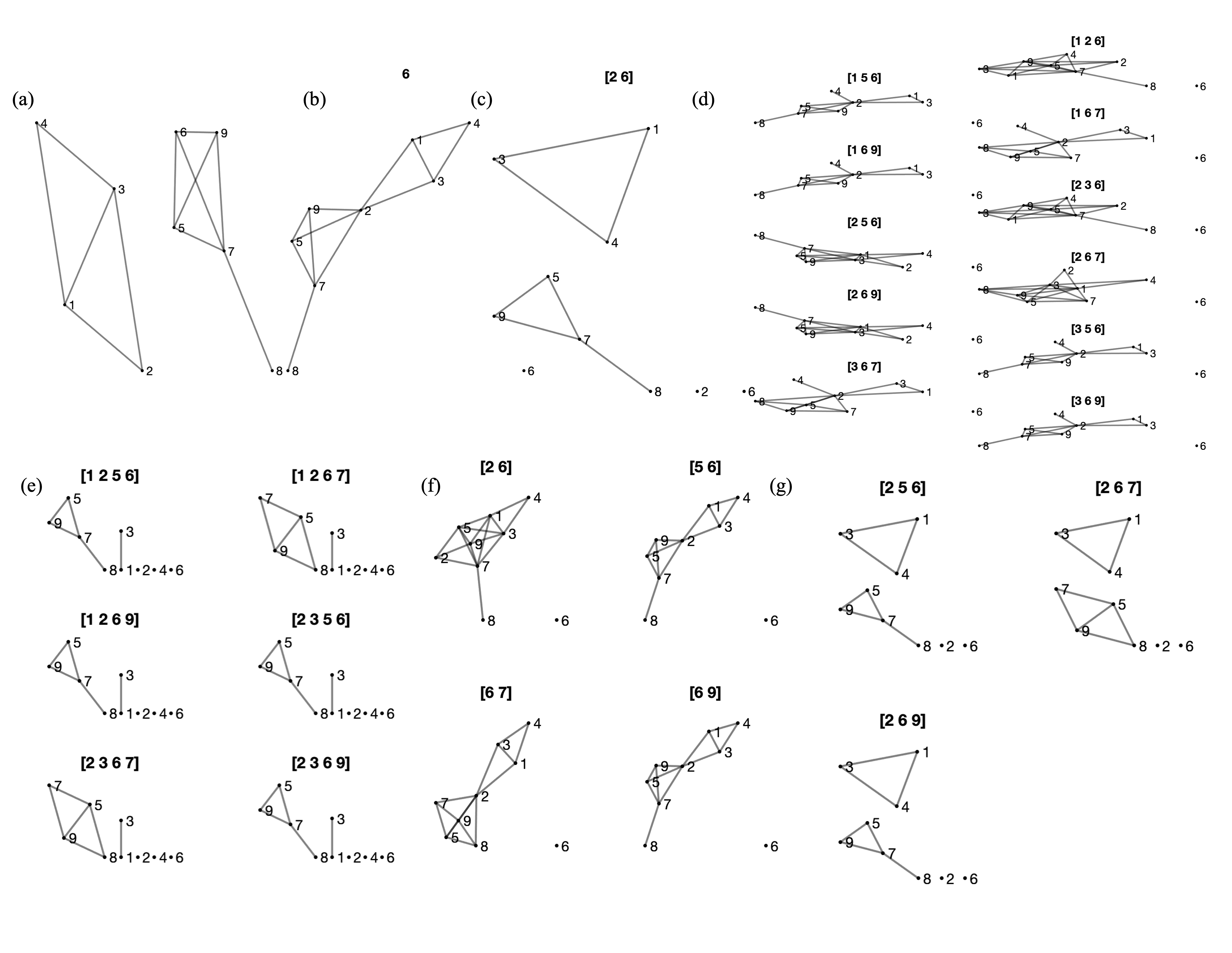}
\vspace{-4em}
\caption{Outputs calculated using the MATLAB cluster state simulator when type-I fusions (discussed in Section~\ref{subsubsec:typeI}) are performed (a) Qubits 2 and 6 from two cluster states undergo type-I fusion with qubit 2 being the control qubit and 6 the target qubit (b) Resulting cluster state when the first fusion from TABLE~\ref{tab:fusion_rulesT1} succeeds. The neighbors of qubit 6 get connected to qubit 2 followed by the removal of qubit 6 as described by the graph rules in TABLE~\ref{tab:fusion_rulesT1}. (c) When the first fusion from TABLE~\ref{tab:fusion_rulesT1} fails, it is equivalent to performing Pauli-Z measurements on qubits 2 and 6. (d)All possible output states, when the fourth fusion from TABLE~\ref{tab:fusion_rulesT1} succeeds, can be converted to a cluster state by performing Hadamards two qubits from the set $\{2\}\cup\mathcal{N}_2\cup\mathcal{N}_6$. (e) When the fourth fusion fails, it is the same as performing Pauli-X measurements on qubits 2 and 6 followed by a Hadamard on one neighbor of each. (f) When the third fusion from TABLE~\ref{tab:fusion_rulesT1} is successful, the resulting state is obtained by applying Hadamard on a qubit in $\{c\}\cup\mathcal{N}_t$. (g) Failure of fusion 3 is the same as measuring qubits 2 and 6 in Pauli-Z and X bases, respectively.}
\label{fig:chp_fusionT1}
\end{figure*}

\begin{figure*}[htb]
\centering
\includegraphics[scale = 0.75]{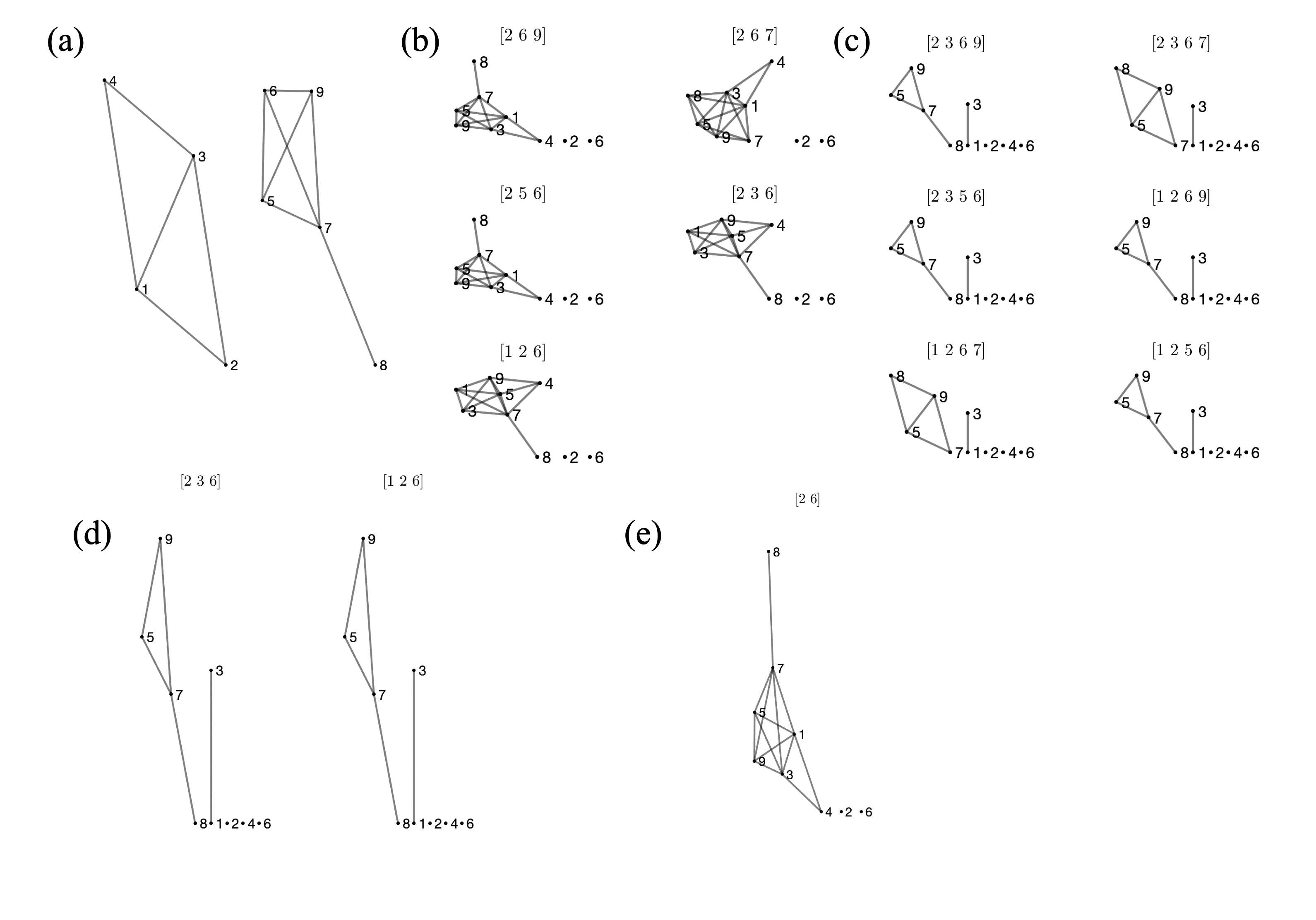}
\caption{Outputs calculated using the MATLAB cluster state simulator when two types of fusions are performed (a) Qubits 2 and 6 from two cluster states undergo fusion with qubit 2 being the control qubit and 6 being the target qubit (b) Resulting cluster state when the third fusion from TABLE~\ref{tab:fusion_rules} i.e.,  $R_c=H$, $R_t=I$, succeeds. All the neighbors of qubit 2 get connected to all the neighbors of qubit 6 followed by the removal of qubits 2 and 6 as described by the graph rules in TABLE~\ref{tab:fusion_rules}. (c) When the third fusion from TABLE~\ref{tab:fusion_rules} fails, it is equivalent to performing Pauli-X measurement on qubit 2, the control qubit and Pauli-Z basis measurement on qubit 6, the target qubit. The resultant state can be converted to a cluster state by performing Hadamard on either qubit 1 or 3 as shown in the square brackets at the top of each graph. The remaining two entries show the qubits that were measured. (d)The resulting state when the first fusion from TABLE~\ref{tab:fusion_rules} i.e.,  $R_c=R_t=H$, succeeds can be converted to a cluster state by performing Hadamard on one of the neighbors of qubits 2 or 6. (e) When the first fusion fails, it is the same as performing Pauli-X measurements on qubits 2 and 6 followed by a Hadamard on one neighbor of each.}
\label{fig:chp_fusion}
\end{figure*}

The snippets of the simulator performing various operations on cluster states are shown in Figs.~\ref{fig:screenshot},~\ref{fig:CHP_pauli},~\ref{fig:chp_fusionT1}, and ~\ref{fig:chp_fusion}.
Although it is always possible to calculate the results of stabilizer operations by hand, it can become tedious quickly as the number of qubits increases and the quantum operations become complex. For example, in Fig.~\ref{fig:chp_fusion}, there are many possible ways the state can undergo Hadamarad operation. Moreover, it is not always possible to translate the effect of quantum operations into nicely defined classical rules. This is when having a simulator such as ours can pay off. On the other hand, this simulator can also derive graph theoretic rules to describe quantum operations by simulating multiple test cases and deducing a pattern to formulate the rules. We believe this simulator will be useful in developing protocols for quantum computing and shared entanglement generation in quantum networks.

\section{Linear Optics}
\label{sec:LO}
Photonic qubits have a low probability of decoherence error and the single qubit gates have high fidelity \cite{o2007optical}. Thousands of linear optical (LO) elements can be embedded on integrated photonic chips~\cite{baehr2012myths,wang2018multidimensional,adcock2019programmable,harris2017quantum} to generate photonic entangled states and perform operations such as fusions. Integrated silicon chip has also generated error-protected qubits~\cite{vigliar2020error}. Photon loss errors and probabilistic entangling operations are the major source of hurdles for the linear optical hardware but they are being overcome by theoretical advances in cluster state generation~\cite{pant2019percolation,MercedesThesis,gimeno2015three} and computation methods~\cite{FBQC,morley2019loss,varnava2006loss}. This makes LO photonic qubits the leading candidate for quantum computation, quantum repeaters, etc. In this section, we discuss the linear optical (LO) realizations of the Clifford circuits. 

Two of many forms in which a qubit can be encoded in photons are the single-rail and dual-rail encodings. In the single-rail qubit, the degree of freedom of the photon used to encode the qubit is the photon number. The computational basis states $\ket{0}$ and $\ket{1}$ are defined as the absence or presence of a single photon in an optical mode, respectively. This encoding is not used with optical-frequency traveling-wave photons as single qubit gates require photon addition and/or subtraction, which are very difficult to realize. Single rail encoding is popular however in microwave-frequency superconducting qubits. Moreover, if the qubit in state $\ket{1}$ suffers photon loss, it becomes $\ket{0}$. As a result, photon loss becomes a Pauli-$X$ error on the qubit. For dual-rail encoding, as we will see below, photon loss causes an `erasure' error, where the qubit's quantum state, regardless of what it was before being subject to loss, goes into a state orthogonal to the Hilbert space span of the qubit. In the theory of quantum error correction, it is easier to correct erasure errors than for Pauli errors, giving another reason why single rail encoding is not preferred.

In the dual-rail encoding of photonic qubits, a qubit is encoded in two optical modes or degrees of freedom of a single photon, such as polarization, spectral, temporal
or spatial modes. Consider two orthogonal optical modes in the vacuum state, $\ket{0}_1$ and $\ket{0}_2$. In the dual rail basis, the computational basis states $\ket{0}, \ket{1}$ of a qubit corresponds to a single photon being in one of these two modes. These two modes are identical in the remaining degrees of freedom of the photon.  \begin{align*}
    \ket{0} &= a_1^{\dagger}\ket{0,0}_{1,2}= \ket{1,0}_{1,2}\\
    \ket{1} &= a_2^{\dagger}\ket{0,0}_{1,2}= \ket{0,1}_{1,2}
\end{align*}
For example, in the case of spatial encoding with two waveguides, the state
of the dual-rail qubit depends on which waveguide the photon is in. 

All linear optical unitary operations on dual-rail qubits can be performed using beamsplitters and phase-shifters. For $N$ optical modes, it has been shown that any unitary can be implemented using at most $N(N-1)/2$ beamsplitters and phase-shifters~\cite{reck1994experimental}. Note that, photon loss brings the qubit's state outside of the dual-rail Hilbert space and hence, can be treated as an erasure error. As discussed in Sections~\ref{subsubsec:typeI} and~\ref{subsec:LOtypeII}, Photon loss in the dual-rail qubits can be partially or fully heralded.

In this section, we will focus on dual-rail polarization encoding and discuss the LO circuits to realize single and two-qubit unitaries, measurements, and fusions. If
the qubit is encoded in polarization modes, we use the orthogonal horizontal and vertical polarization of
the photon to encode the computational basis states. 
\begin{equation}
    \begin{split}
         \ket{0} &= a_H^{\dagger}\ket{0,0}_{H,V}= \ket{1,0}_{H,V}=\ket{H}\\
    \ket{1} &= a_V^{\dagger}\ket{0,0}_{H,V}= \ket{0,1}_{H,V}=\ket{V}\\
    \end{split}
    \label{eq:dualRaildef}
\end{equation}

If the photon is in horizontal (vertical) polarization the qubit is in state $\ket{0(1)}$. Photons encoded in diagonal and anti-diagonal polarization correspond to the eigenstates of Pauli-X matrix ($\ket{D}=\frac{\ket{H}+\ket{V}}{\sqrt{2}}$, $\ket{A}=\frac{\ket{H}-\ket{V}}{\sqrt{2}}$) and those encoded in right- and left-circular polarization are the eigenstates of Pauli-Y matrix ($\ket{L}=\frac{\ket{H}+i\ket{V}}{\sqrt{2}}$, $\ket{R}=\frac{\ket{H}-i\ket{V}}{\sqrt{2}}$).   

In the following sections, we first describe the single qubit operations on polarization-encoded dual-rail qubit. We then discuss how heralded two-qubit entangling gates are realized and the probabilistic nature of these gates. We introduce the LO circuit for a heralded CZ gate and rotated variations for the CNOT and the CZ gates. These variations can build large dual-rail photonic entangled states using fewer single photons. We also give step-by-step instructions to convert a quantum circuit into a LO circuit. We then use these instructions to build the LO circuit for two- and multi-qubit fusions. Moreover, we analyze the projections all these entangling circuits make when they succeed or fail.  
\subsection{Unitary Operations and Measurements}

For a polarization-encoded qubit, all single qubit gates or any rotation on the Bloch sphere can be deterministically implemented using half-wave plates (HWP) and quarter-wave plates (QWP)~\cite{kokBook}. An HWP rotated by an angle $\theta$ performs the unitary $U_{hwp}(\theta)=\begin{bmatrix}\cos{2\theta} & \sin{2\theta}\\\sin{2\theta} & -\cos{2\theta}\end{bmatrix}$~\cite{collett2005field}. The Pauli-Z gate is implemented by the (nonrotated) HWP. The Hadamard and Pauli-X are then implemented by rotating the HWP by $22.5^{\circ}, 45^{\circ}$, respectively. The Hadamard gates can also be built using a $45^{\circ}$ linear polarizer~\cite{collett2005field, browne2005Fusion, MercedesThesis}. A QWP is used to convert the $\ket{D}, \ket{A}$ states into $\ket{L}, \ket{V}$ states. This corresponds to rotation around the $z$-axis in Bloch sphere~\cite{Nielsen2010}.
A QWP with fast horizontal axis~\cite{collett2005field} applies $U_{qwp_h}=e^{-i\pi/4}\begin{bmatrix}1 & 0\\0 & i\end{bmatrix} = e^{-i\pi Z/4}$. This unitary is the Phase gate. The QWP with fast vertical axis~\cite{collett2005field} applies $U_{qwp_v}=e^{-i\pi/4}\begin{bmatrix}1 & 0\\0 & -i\end{bmatrix} = e^{i\pi Z/4}= U_{qwp_h}^{\dagger}$.  Note that, the unitary matrices applied by the wave plates are their Jones matrices~\cite{collett2005field}.
 
The unitary matrix that rotates a qubit by an angle $\theta$ around the $y$-axis of the Bloch sphere is $R_{Y}(\theta)=  e^{-i\theta Y/2} =  \begin{bmatrix}\cos{\theta/2} & -\sin{\theta/2}\\\sin{\theta/2} & \cos{\theta/2}\end{bmatrix} $~\cite{Nielsen2010}. It can be implemented using a HWP rotated by angle $\theta/4$ and a non-rotated HWP - 
\begin{equation}
    R_{Y}(\theta)= U_{hwp}(\theta/4)U_{hwp}(0) 
    \label{eq:polRY}
\end{equation}. 

Rotation by an angle $\theta$ around the $x$-axis of the Bloch sphere is applied by the unitary $R_{X}(\theta)=  e^{-i\theta X/2} =  \begin{bmatrix}\cos{\theta/2} & -i\sin{\theta/2}\\-i\sin{\theta/2} & \cos{\theta/2}\end{bmatrix} $~\cite{Nielsen2010}. $R_{X}(\theta)$ can be constructed using HWPs and QWPs -
\begin{equation}
\begin{split}
       R_{X}(\theta) &= e^{i\pi Z/4}R_{Y}(\theta)e^{-i\pi Z/4} \\ &= U_{qwp_h}^{\dagger} U_{hwp}(\theta/4)U_{hwp}(0)  U_{qwp_h}\\
\end{split}
\label{eq:polRx}
\end{equation}

An arbitrary single-qubit gate $U$ (i.e. rotation in the Bloch sphere) can be decomposed into rotations around $x-$ and $y-$ axes of the Bloch sphere - $U = e^{i\alpha}R_X(\beta)R_Y(\gamma)R_X(\delta)$~\cite{Nielsen2010}. Using eqs.~(\ref{eq:polRY}) and ~(\ref{eq:polRx}), we can write 
\begin{equation}
\begin{split}
    U &= e^{i\alpha} U_{qwp_h}^{\dagger} U_{hwp}(\beta/4)U_{hwp}(0)  U_{qwp_h} U_{hwp}(\gamma/4)U_{hwp}(0) \\& \times U_{qwp_h}^{\dagger} U_{hwp}(\delta/4)U_{hwp}(0)  U_{qwp_h}
    \end{split}
\label{eq:polUniversal}
\end{equation}
i.e., any single-qubit gate on polarization-encoded qubits can be deterministically performed using HWP and QWPs.

A linear optical element called the polarization beam splitter (PBS) is used in two-qubit entangling gates. It is made up of a birefringent material such that the transmission angle for two orthogonal polarizations is different. A PBS has two inputs. The PBS we use is H-V oriented, i.e., it always transmits horizontally polarized photons and always reflects vertically polarized photons. It separates the H-V polarizations in the input signal and converts polarization encoding to spatial encoding and v.v.~\cite{kokBook,MercedesThesis}. If the polarization-encoded qubits $1$ and $2$ are passed through a PBS, their transformation can be written in terms of the creation operators as 
\begin{align*}
    a^{\dagger}_{H_1}\rightarrow a^{\dagger}_{H_1}&\quad\quad a^{\dagger}_{H_2}\rightarrow a^{\dagger}_{H_2}\\
    a^{\dagger}_{V_1}\rightarrow a^{\dagger}_{V_2}&\quad\quad a^{\dagger}_{V_2}\rightarrow a^{\dagger}_{V_1}
\end{align*}
For example, a PBS will transform states as follows -
\begin{align*}
    &\ket{H_1H_2} = a^{\dagger}_{H_1}a^{\dagger}_{H_2} \ket{0,0,0,0}\rightarrow a^{\dagger}_{H_1}a^{\dagger}_{H_2} \ket{0,0,0,0}= \ket{H_1H_2}\\
&\ket{H_1V_2}= a^{\dagger}_{H_1}a^{\dagger}_{V_2}\ket{0,0,0,0}\rightarrow a^{\dagger}_{H_1}a^{\dagger}_{V_1}\ket{0,0,0,0}=\ket{1,1,0,0}\\
&\ket{V_1H_2} = a^{\dagger}_{V_1}a^{\dagger}_{H_2}\ket{0,0,0,0}\rightarrow a^{\dagger}_{V_2}a^{\dagger}_{H_2}\ket{0,0,0,0} = \ket{0,0,1,1}\\
&\ket{V_1V_2} =a^{\dagger}_{V_1}a^{\dagger}_{V_2}\ket{0,0,0,0}\rightarrow a^{\dagger}_{V_1}a^{\dagger}_{V_2}\ket{0,0,0,0} = \ket{V_1V_2}
\end{align*}
Here, $\ket{0,0,0,0}$ is a short-hand notation for $\ket{0,0,0,0}_{H_1,V_1,H_2,V_2}$. 

Recall that a dual-rail qubit is defined as the presence of \textit{exactly one} photon in one of the two modes. When the input state is $\ket{V_1H_2}$ or $\ket{H_1V_2}$, the output state has two photons in the orthogonal modes of the same qubit. This output state is out of the Hilbert space of the dual-rail photonic qubit. This implies that the PBS ``works" for only half of the input states by keeping the output states within the Hilbert space. 

Now, if we measure at least one of the modes of the output state of the PBS using a photon number resolving detector (PNRD), we can determine whether the PSB worked or not based on the measurement result. A PNRD calculates the exact number of photons in the signal. Assuming the PNRD is ideal, if it measures exactly one photon, we know that the output state was within the Hilbert space and the PBS was ``successful". Otherwise, if the PNRD measures zero or two, the PBS has ``failed". This is an example of a heralded measurement, i.e., this measurement helps us deduce the success or failure of an operation and the resulting state. This circuit fails for half of the input state and hence, has a probability of success of 50\%. 

\begin{figure}
    \centering
    \includegraphics[scale=.65]{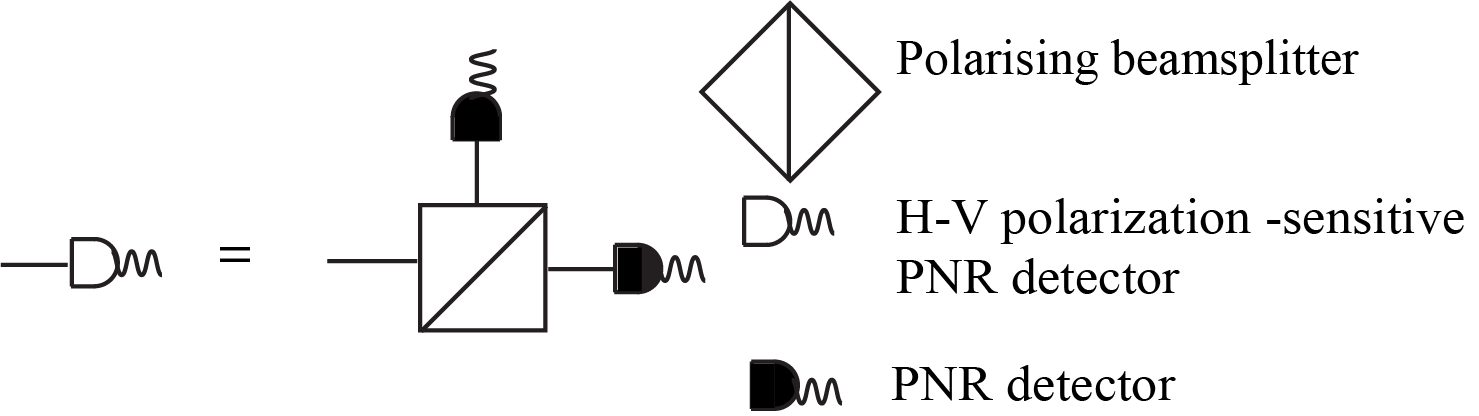}
    \caption{An $H-V$ polarization resolving detector can be built using a PBS and two-photon number resolving detectors}
    \label{fig:PNRD}
\end{figure}

If a PBS is placed at the inputs of two PNRDs, as shown in FIG.~\ref{fig:PNRD}, this setup measures the number of photons in both $H$ and $V$ polarizations of the signal. We refer to this modified detector as the ``$H-V$ polarization-resolving" PNRD. From now on, the term detector implies the $H-V$ polarization-resolving PNRD unless specified otherwise. A $D-A$ polarization-resolving PNRD can be built by placing a $22.5^{\circ}$ rotated HWP or a $45^{\circ}$ linear polarizer between the PBS and the PNRD. 

As stated earlier, the LO implementations of all the two-qubit entangling gates need PBS. We now add heralded measurements as another essential component of LO entangling gates, making these gates probabilistic in nature. 

\subsubsection{LO Entangling Gates}
\label{subsubsec:typeI}

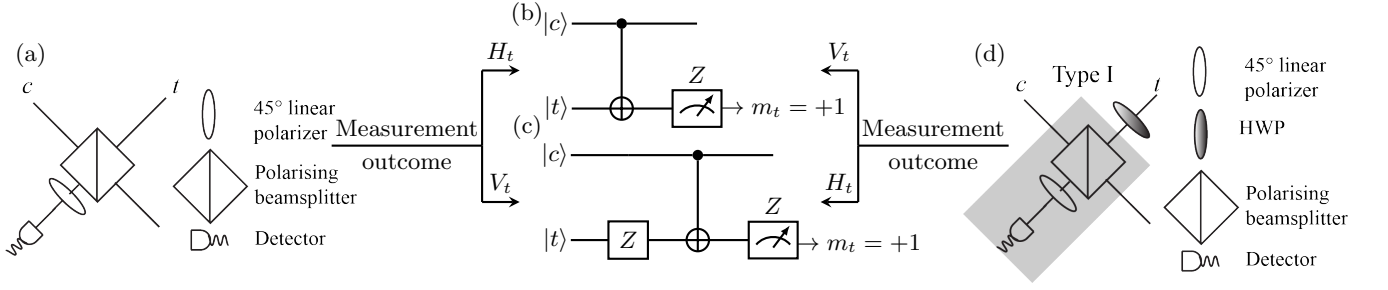
\begin{figure*}
		\centering
		
	\begin{tikzpicture}[darkstyle/.style={circle,inner sep=0pt,minimum size=0.4cm,draw=blue,fill=blue!20!,font=\sffamily\small\bfseries},
 edge_style/.style={draw=black,  thick}]
 
     \node[inner sep=0pt] (meter) at (10-.75,2.75)
    {\begin{quantikz}
		\lstick{$\ket{c}$}   & \ctrl{1} &\qw\\
		\lstick{$\ket{t}$} & \targ{} & \meter{$Z$}\\
    \end{quantikz}};
    \node at (12.25,0.7) {$\rightarrow m_t=+1$};
    \node at (11.25,2.5) {$\rightarrow m_t=+1$};
    \node[inner sep=0pt] (meter) at (10-.25,1)
    {\begin{quantikz}
		\lstick{$\ket{c}$} &\qw  & \ctrl{1} &\qw\\
		\lstick{$\ket{t}$} &\gate{Z} & \targ{} & \meter{$Z$}\\
    \end{quantikz}};
    \node at (1.25,3.2) {(a)};
    \node at (8.-.15,3.75) {(b)};
    \node at (8.-.15,2.25) {(c)};
    \node at (14,3.2) {(d)};
       \draw [-stealth,thick] (7.25,3) --  node[above]{$H_t$}(8-.25,3);
       \draw [thick] (5.25,2) --  node[above]{Measurement} node[below]{outcome}(7.25,2);
       \draw [thick] (7.25,2) -- (7.25,3);
       \draw [thick] (7.25,2) -- (7.25,1.25);
       \draw [-stealth,thick] (7.25,1.25) --  node[above]{$V_t$}(8-.25,1.25);
     \node[inner sep=0pt] (meter) at (3.25,1.75)
    {\includegraphics[scale = 0.65]{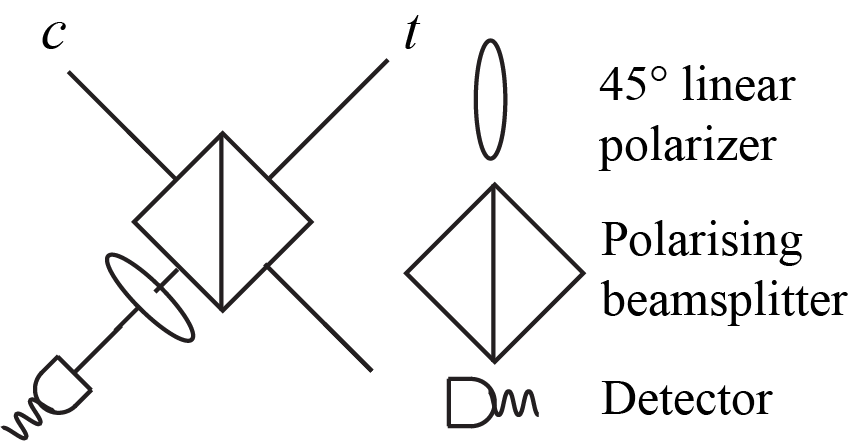}};
     
     \draw [-stealth,thick] (12.25,3) --  node[above]{$V_t$}(11.75,3);
       \draw [thick] (14.25,2) --  node[above]{Measurement} node[below]{outcome}(12.25,2);
       \draw [thick] (12.5-.25,2) -- (12.5-.25,3);
       \draw [thick] (12.5-.25,2) -- (12.5-.25,1.25);
     \draw [-stealth,thick] (12.5-.25,1.25) --  node[above]{$H_t$}(11.75,1.25);
     
     \node[inner sep=0pt] (meter) at (16.2,1.75)
    {\includegraphics[scale = 0.65]{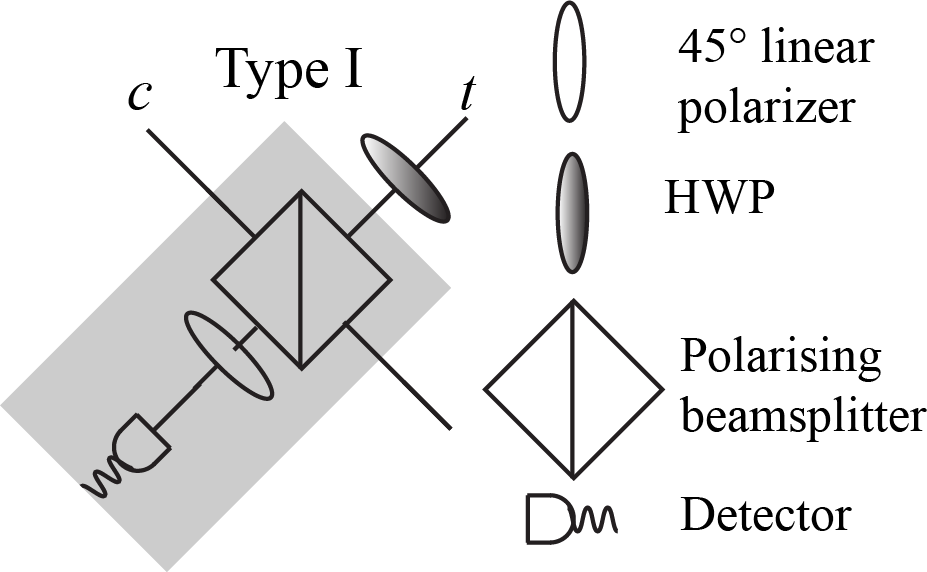}};

		\end{tikzpicture}
  
		\caption{Type-I fusion (a) The LO circuit. If a single photon is detected, it heralds the Kraus operators of one of the two quantum circuits in (b) and (c) depending upon the polarization of the measurement. If two photons are detected, it is equivalent to performing Pauli-Z measurement on both the input qubits to the type-I fusion. (d) An equivalent type-I fusion circuit, constructed using (c), with the same Kraus operators as (a) but the mapping of measurement outcomes to Kraus operators is different than (a). The grey box highlights the type-I fusion circuit from (a).}
		\label{fig:LOcnot}
	\end{figure*}

We first analyze the LO circuit shown in FIG.~\ref{fig:LOcnot} by calculating the evolution of the creation operators of the four two-qubit basis states as they traverse the circuit. For example, if the input state to the LO circuit is $\ket{H_cH_t} = a_{H_c}^{\dagger}a_{H_t}^{\dagger} \ket{0,0,0,0}$, the PBS doesn't affect this state. After the $45^{\circ}$ polarizer, the state becomes $\ket{H_c}\big(\frac{\ket{H_t}+\ket{V_t}}{\sqrt{2}}\big)= a_{H_c}^{\dagger}\big(\frac{a_{H_t}^{\dagger}+a_{V_t}^{\dagger}}{\sqrt{2}}\big)\ket{0,0,0,0}$. For brevity, we will drop $\ket{0,0,0,0}$ and write only the operators from here onward. 

Similarly, if the input state is $\ket{H_cV_t} = a_{H_c}^{\dagger}a_{V_t}^{\dagger}$, the state after PBS is $a_{H_c}^{\dagger}a_{V_c}^{\dagger}$. As the polarizer acts only on the $t$ mode, it doesn't affect $a_{H_c}^{\dagger}a_{V_c}^{\dagger}$. The output states corresponding to the four dual-rail two-qubit input states are as follows - 
\begin{align*}
    a^{\dagger}_{H_c}a^{\dagger}_{H_t} &\rightarrow a^{\dagger}_{H_c}\big(\frac{a_{H_t}^{\dagger}+a_{V_t}^{\dagger}}{\sqrt{2}}\big) \\
a^{\dagger}_{H_c}a^{\dagger}_{V_t}&\rightarrow a_{H_c}^{\dagger}a_{V_c}^{\dagger}\\
a^{\dagger}_{V_c}a^{\dagger}_{H_t}&\rightarrow \big(\frac{a_{H_t}^{\dagger}-a_{V_t}^{\dagger}}{\sqrt{2}}\big)\big(\frac{a_{H_t}^{\dagger}+a_{V_t}^{\dagger}}{\sqrt{2}}\big)\\
a^{\dagger}_{V_c}a^{\dagger}_{V_t}&\rightarrow a^{\dagger}_{V_c}\big(\frac{a_{H_t}^{\dagger}-a_{V_t}^{\dagger}}{\sqrt{2}}\big)\\
\end{align*}


The target ($t$) mode is measured such that the LO circuit succeeds if exactly one photon is detected by the PNRD. The circuit fails if zero or two photons are detected. Only two ($\ket{H_cH_t}$, $\ket{V_cV_t}$) of the four two-qubit basis states ($\ket{H_cH_t}$, $\ket{H_cV_t}$, $\ket{V_cH_t}$, and $\ket{V_cV_t}$) satisfy this condition, making the success probability of this circuit $50\%$. 

If a polarization-sensitive PNRD is used, there are five possible photon detection patterns - single photon detected in mode $H_t$ or $V_t$, two photons detected s.t. both are in either mode $H_t$ or $V_t$, and no photon detected. Each of these patterns corresponds to a Kraus operator given below~\cite{browne2005Fusion} - 
\begin{equation}\label{eq:typeIProjectors}
\begin{tabular}{|l|l|}
\hline
\multicolumn{1}{|c|}{\centering\textbf{Measurement outcome}}
& \multicolumn{1}{c|}{\textbf{Kraus Operator}}\\
\hline
 \centering $H_t$    & $a^{\dagger}_{H_c}\ket{0,0}\bra{0,0,0,0}a_{H_c}a_{H_t}$  \rule[1ex]{0pt}{3ex}\\ &$+a^{\dagger}_{V_c}\ket{0,0}\bra{0,0,0,0}a_{V_c}a_{V_t}$ \rule[0.0ex]{0pt}{3ex}\\ 
  \centering $V_t$    & $a^{\dagger}_{H_c}\ket{0,0}\bra{0,0,0,0}a_{H_c}a_{H_t}$  \rule[1ex]{0pt}{3ex}\\ &$-a^{\dagger}_{V_c}\ket{0,0}\bra{0,0,0,0}a_{V_c}a_{V_t}$ \rule[0.0ex]{0pt}{3ex}\\ 

  \centering $H_t^2$    & $\ket{0,0}\bra{0,0,0,0}a_{V_c}a_{H_t}$\rule[1.0ex]{0pt}{3ex}\\ 
  \centering $V_t^2$    & $-\ket{0,0}\bra{0,0,0,0}a_{V_c}a_{H_t}$ \rule[01.0ex]{0pt}{3ex}\\ 
  \centering zero photon    & $a^{\dagger}_{H_c}a^{\dagger}_{V_c}\ket{0,0}\bra{0,0,0,0}a_{H_c}a_{V_t}$ \rule[2.0ex]{0pt}{3ex}\\ 
\hline
\end{tabular}
\end{equation}
Here, measurement outcome $p_t^i$ s.t. $p\in\{H,V\}$ implies that $i$ with $p$ polarization of mode $t$ photons were detected.

We now compare this LO circuit with the quantum circuits shown in FIG.~\ref{fig:LOcnot}.  The operator for the CNOT gate is $\ket{00}\bra{00}+\ket{01}\bra{01}+\ket{10}\bra{11}+\ket{11}\bra{10}$. If the target qubit is measured in the computational basis after going through CNOT such that the measurement outcome is +1 as shown in FIG.~\ref{fig:LOcnot}(b), the corresponding Kraus operator is $\ket{0}\bra{00}+\ket{1}\bra{11}$. We refer to this Kraus operator as $K_{1}$. If the modes $c$ and $t$ from the LO circuit are mapped to the control and target of the CNOT, respectively, the LO applies the Kraus operator $K_{1}$ when the measurement outcome is $H_t$ (see Eq.~\ref{eq:typeIProjectors}). Similarly, by comparing the Kraus operators, we can see that when the measurement outcome of the LO circuit is $V_t$, it realizes the quantum circuit in FIG.~\ref{fig:LOcnot}(c). When the LO circuit fails, the resultant states of the input qubits, $c$ and $t$, have either zero or two photons and hence are outside of the dual-rail qubit Hilbert space. However, from the Kraus operators, we observe that failure also performs the projections $\pm\bra{0,0,0,0}a_{V_c}a_{H_t} = \bra{1,0,0,1}$ when two photons are detected and $\bra{0,0,0,0}a_{H_c}a_{V_t} = \bra{0,1,1,0}$ when no photon is detected. These projections can be mapped to $\pm\bra{10}$ and $\bra{01}$, respectively in the two-qubit dual-rail Hilbert space. As a result, type-I fusion failure can be mapped as an effective Pauli-Z measurement on the dual-rail qubits $c$ and $t$.

\textit{Graph rule: }This LO circuit is commonly known as the type-I fusion \cite{browne2005Fusion, MercedesThesis}. When qubits from two cluster states with $m$ and $n$ qubits each are input to the type-I circuit, if the operation succeeds, the resulting state is a single $(m+n-1)$ qubit cluster state, thus `fusing' the two cluster states. The geometry of the resulting state is obtained using the graph rule - connect all the neighbors of the target qubit to the control qubit and delete the target qubit. From Sections~\ref{sub:clusterCNOT} and~\ref{sec:PauliZ}, the graph rule for type-I fusion matches that for the quantum circuit shown in FIG.~\ref{fig:LOcnot}(b). The outputs of FIG.~\ref{fig:LOcnot}(b) and FIG.~\ref{fig:LOcnot}(c) differ only in phase and hence the two quantum circuits have the same graph rule. When the LO circuit fails, it is the same as performing Pauli-Z measurement on the control and target qubit. Thus the resulting state is obtained by deleting the control and target qubits from the cluster states. 


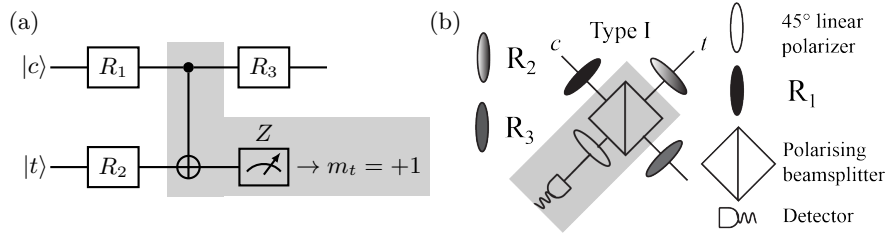
\begin{figure*}
		\centering
		
	\begin{tikzpicture}[darkstyle/.style={circle,inner sep=0pt,minimum size=0.4cm,draw=blue,fill=blue!20!,font=\sffamily\small\bfseries},
 edge_style/.style={draw=black,  thick}]

    \node at (1,3) {(a)};
    \node at (6.6,3) {(b)};
     \path [fill=black!18] (2.9,0.7) rectangle (3.65,2.75);
\path [fill=black!18] (3.6,0.7) rectangle (6.4,1.75);

     \node[inner sep=0pt] (meter) at (9.5,1.75)
    {\includegraphics[scale = 0.65]{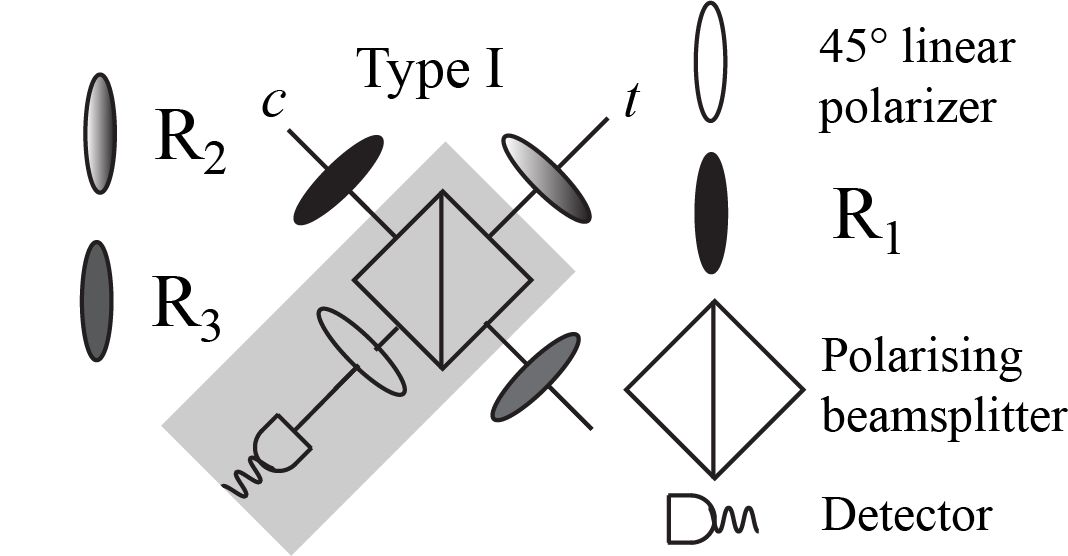}};

 \node[inner sep=0pt] (meter) at (3.,1.5)
    {\begin{quantikz}
		\lstick{$\ket{c}$}   &\gate{R_1}  & \ctrl{1} &\gate{R_3}&\qw \\
		\lstick{$\ket{t}$}  & \gate{R_2} & \targ{}  &  \meter[fill=black!18]{$Z$} \\
    \end{quantikz}};
    \node at (5.45,1.1) {$\rightarrow m_t=+1$};
   
		\end{tikzpicture}
  
		\caption{A general rotated type-I fusion (a) quantum circuit. The grey highlighted area is the non-rotated type-I fusion part of the circuit. (b) the LO circuit that implements (a) when the measurement outcome is $H_t$.}
		\label{fig:T1LOrecipe}
	\end{figure*}

We now use the type-I circuit as the foundation to build the LO circuit for a quantum circuit of the form shown in FIG. \ref{fig:T1LOrecipe}(a). The steps to build the corresponding LO circuit are as follows (refer Fig. \ref{fig:T1LOrecipe}(b)) - 
\begin{enumerate}
    \item Read the quantum circuit from left to right.  As there are two qubits in the quantum circuit, start with two modes. Assign one of the optical modes to be the control qubit and the other to be the target qubit. 
    \item  Decompose the single qubit unitaries $R_1$ and $R_2$ into HWPs and QWPs using Eq.~\ref{eq:polUniversal} and put them on the corresponding optical modes.
    \item Place the type-I fusion circuit such that the control and target qubits of the circuit match with the corresponding optical modes.
    \item Put the LO element (made of HWPs and QWPs) on the mode corresponding to the control qubit that applies the unitary $R_3$.
    \item Place an $H-V$ detector on the target mode.
\end{enumerate}

Using this recipe, we can build the LO circuit for FIG.~\ref{fig:LOcnot}(c) as shown in FIG~\ref{fig:LOcnot}(d). This LO circuit implements the Kraus operators as the one in FIG~\ref{fig:LOcnot}(a) but for opposite measurement results. In general, our recipe generates the LO circuit which maps to the given quantum circuit when the detector outcome is $H_t$.

\begin{figure*}
		\centering
		
	\begin{tikzpicture}[darkstyle/.style={circle,inner sep=0pt,minimum size=0.4cm,draw=blue,fill=blue!20!,font=\sffamily\small\bfseries},
 edge_style/.style={draw=black,  thick}]

    \node at (-5.4,3) {(a)};
    \node at (6.6,3) {(b)};

     \node[inner sep=0pt] (meter) at (9,1.75)
    {\includegraphics[scale = 0.65]{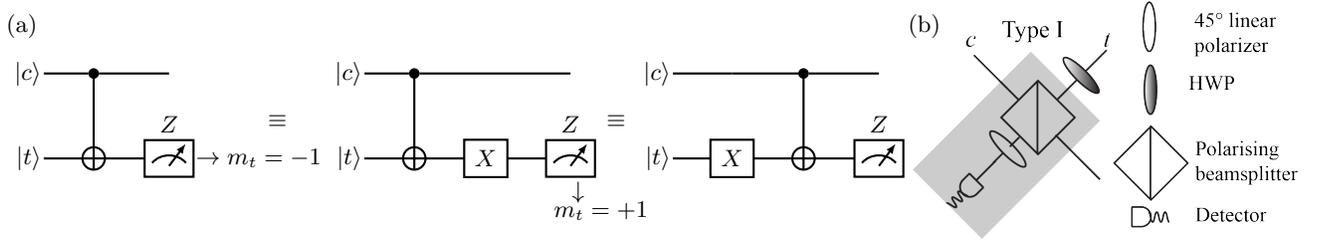}};

     \node[inner sep=0pt] (meter) at (0.5,1.5)
    {\begin{quantikz}
		\lstick{$\ket{c}$}  & \ctrl{1} &\qw &\qw\\
		\lstick{$\ket{t}$}  & \targ{} & \gate{X} & \meter{$Z$}\\
    \end{quantikz}};
 \node[inner sep=0pt] (meter) at (4.6,1.5)
    {\begin{quantikz}
		\lstick{$\ket{c}$}  &\qw  & \ctrl{1} &\qw\\
		\lstick{$\ket{t}$}  & \gate{X} & \targ{}  & \meter{$Z$}\\
    \end{quantikz}};
\node at (-2.,1.75) {$\equiv$};
\node at (2.5,1.75) {$\equiv$};
    \node[inner sep=0pt] (meter) at (-4.3,1.5)
    {\begin{quantikz}
		\lstick{$\ket{c}$}  & \ctrl{1} &\qw\\
		\lstick{$\ket{t}$} & \targ{} & \meter{$Z$}\\
   \end{quantikz}};
     \node at (2.,0.79) {$\downarrow$};
     \node at (2.3,0.55) {$m_t=+1$};
     \node at (-2.25,1.25) {$\rightarrow m_t=-1$};
		\end{tikzpicture}
  
		\caption{Rotated type-I fusion that applies the Kraus operators $\ket{0}\bra{01}\pm\ket{1}\bra{10}$ when successful. (a) The corresponding quantum circuit (b) The LO circuit is built by placing the $45^{\circ}$ rotated HWP before the type-I circuit.}
		\label{fig:LOcnot2}
	\end{figure*}

Note that, for the quantum circuit to LO circuit recipe to work, if the given quantum circuit is not in the form shown in FIG.\ref{fig:T1LOrecipe}(a), we need first to rewrite it s.t. it matches FIG.\ref{fig:T1LOrecipe}(a). For example, the circuit in Fig.\ref{fig:LOcnot2} (a) implements CNOT followed by measurement of the target qubit in Pauli-Z basis s.t. the measurement outcome is -1. It gives rise to the Kraus operator $\ket{0}\bra{01}+\ket{1}\bra{10}$. We first decompose this circuit into elements whose LO implementation is known, as shown in Fig.\ref{fig:LOcnot2} (a).  As this is a two-qubit operation, we start with two optical modes. We read the quantum circuit from left to right and add the corresponding components to the LO circuit. We add a $45^{\circ}$ rotated HWP on one of the optical modes for the Pauli X gate that is left of the CNOT. This mode is now set as the target for the CNOT. We then add the type-I fusion circuit for CNOT followed by measurement on target in the quantum circuit. This circuit succeeds when exactly one photon is detected and it has 50$\%$ success probability and has the same graph rule as the type-I fusion.  

We now study the linear optical circuit for the other two-qubit entangling gate - the CZ gate. The circuit in FIG.~\ref{fig:LOCZ}(a) when successful, realizes the Kraus operators $\ket{+}\bra{0+}\pm\ket{-}\bra{1-}$, where the polarization of the measurement result decides the sign in the Kraus operator and the corresponding quantum circuits are shown in FIG.~\ref{fig:LOCZ}(b)-(c). When this circuit fails, it has the effect of measuring the control and target qubits (for the LO circuit) in Pauli-Z and Pauli-X bases, respectively. Note that, in this LO circuit the detector measures the $D-A$ polarization instead of $H-V$ polarization, where $\ket{D} = \frac{\ket{H}+\ket{V}}{\sqrt{2}}$ and $\ket{A} = \frac{\ket{H}-\ket{V}}{\sqrt{2}}$. This circuit succeeds when exactly one photon is detected and has 50$\%$ success probability similar to the type-I fusion. The Kraus operators for this circuit are - 

\begin{equation}\label{eq:typeIrotatedProjectors}
\begin{tabular}{|l|l|}
\hline
\multicolumn{1}{|c|}{\centering Measurement outcome}
& \multicolumn{1}{c|}{Kraus Operator}\\
\hline
 \centering $D_t$    & $a^{\dagger}_{H_c}\ket{0,0}\bra{0,0,0,0}a_{H_c}a_{D_t}$  \rule[1ex]{0pt}{3ex}\\ &$+a^{\dagger}_{V_c}\ket{0,0}\bra{0,0,0,0}a_{V_c}a_{A_t}$ \rule[0.0ex]{0pt}{3ex}\\ 
  \centering $A_t$    & $a^{\dagger}_{H_c}\ket{0,0}\bra{0,0,0,0}a_{H_c}a_{D_t}$  \rule[1ex]{0pt}{3ex}\\ &$-a^{\dagger}_{V_c}\ket{0,0}\bra{0,0,0,0}a_{V_c}a_{A_t}$ \rule[0.0ex]{0pt}{3ex}\\ 

  \centering $D_t^2$    & $\ket{0,0}\bra{0,0,0,0}a_{V_c}a_{D_t}$\rule[1.0ex]{0pt}{3ex}\\ 
  \centering $A_t^2$    & $-\ket{0,0}\bra{0,0,0,0}a_{V_c}a_{D_t}$ \rule[01.0ex]{0pt}{3ex}\\ 
  \centering zero photon    & $a^{\dagger}_{H_c}a^{\dagger}_{V_c}\ket{0,0}\bra{0,0,0,0}a_{H_c}a_{A_t}$ \rule[2.0ex]{0pt}{3ex}\\ 
\hline
\end{tabular}
\end{equation}

As discussed earlier, when the type-I fusion fails, it performs Pauli Z measurements on the input qubits. This breaks down the cluster states into small fragments. Hence, the type-I fusion cannot always be recursively used on the same pair of cluster states~\cite{browne2005Fusion}. The LO circuit in FIG.~\ref{fig:LOCZ}(a) partially overcomes this problem as when it fails, it performs Pauli-X measurement on one of the input states, instead of Pauli-Z measurement. This measurement converts an $n$-qubit cluster state into an $n-1$ qubit state with redundantly encoded qubits (refer Section~\ref{sec:PauliX}), rather than diving the state into two halves as the Pauli-Z measurement does. 

\textit{Graph rule:} This fusion first performs a CZ between the two input qubits and then performs Paui-X measurement on one of them. It implies that if the input states are cluster states, the output state after this fusion will be a stabilizer state that is local Hadamard equivalent to a cluster state. Let $G$ be the graph of the cluster state obtained by adding an edge between qubit $c$ and $t$ from the two input cluster states. Then if $t$ is measured in X basis, and a Hadamard is applied on one of its neighbors $u$, in graph $G$, it gives us the the cluster state after the fusion. The graph theoretic rule for transformation under this fusion is then  (using TABLE.~\ref{tab:graph_rules}) $(((G.t).u).t)\setminus\{t\}$.

We can improve the type-I LO circuit further as shown in FIG.~\ref{fig:rotatedType1}(a). This circuit when successful gives rise to the Kraus operators - $\ket{+}\bra{++}\pm\ket{-}\bra{--}$ and applies Pauli-X measurement on both the input qubits on failure. If the input to this circuit is two cluster states collectively described using graph $G$, the output state is not a cluster state. It can be converted to a cluster state by applying Hadamards on a pair of qubits from the set $(\{c\}\cup \mathcal{N}_c\cup\mathcal{N}_t$). The graph rules for all possible pairs of qubits to undergo Hadamard are as follows:
\begin{itemize}
    \item $c$ and $u\in\mathcal{N}_t$: Perform $(((G.t).u).t)\setminus\{t\}$. Invert the edges from $\mathcal{N}_c$ to the set of qubits $(\{u\}\cup\mathcal{N}_u)$. 
    \item $c$ and $v\in\mathcal{N}_c$: Perform $(((G.v).c).v)\setminus\{c,t\}$. Invert the edges from $(\{v\}\cup\mathcal{N}_v\cup\{c\})$ to the qubits in $\mathcal{N}_t$.
    \item $v\in\mathcal{N}_c$ and $u\in\mathcal{N}_t$: Perform $((((((G.v).c).v).t).u).t)\setminus\{c,t\}$. Connect $(\{c\})$ to the qubits in $\{v\}\cup\mathcal{N}_v\mathcal{N}_v$.
\end{itemize}Let  or qubits $c$ and $v\in\mathcal{N}_c$.
This fusion effectively does Pauli-X measurements on $c,t$ on failure. 

The graph theoretical rules for transforming cluster states undergoing the type-I fusion and its rotated variations are listed in TABLE.~\ref{tab:fusion_rulesT1}.

\begin{figure*}
		\centering
		
	\begin{tikzpicture}[darkstyle/.style={circle,inner sep=0pt,minimum size=0.4cm,draw=blue,fill=blue!20!,font=\sffamily\small\bfseries},
 edge_style/.style={draw=black,  thick}]
 
\node[inner sep=0pt] (meter) at (10-.75,2.75)
    {\begin{quantikz}
		\lstick{$\ket{c}$}   & \ctrl{1} &\qw\\
		\lstick{$\ket{t}$}   & \ctrl{} & \meter{$X$}\\
    \end{quantikz}};
    \node[inner sep=0pt] (meter) at (10-.25,1)
    {\begin{quantikz}
		\lstick{$\ket{c}$}   &\qw  & \ctrl{1} &\qw\\
		\lstick{$\ket{t}$}   &\gate{X} & \ctrl{} & \meter{$X$}\\
    \end{quantikz}};
    \node at (1-.5,3.2) {(a)};
    \node at (8.,3.75) {(b)};
    \node at (8.,2.25) {(c)};
    \node at (14-.5,3.2) {(d)};
       \draw [-stealth,thick] (7,3) --  node[above]{$D_t$}(8-.25,3);
       \draw [thick] (5,2) --  node[above]{Measurement} node[below]{outcome}(7,2);
       \draw [thick] (7,2) -- (7,3);
       \draw [thick] (7,2) -- (7,1.25);
       \draw [-stealth,thick] (7,1.25) --  node[above]{$A_t$}(8-.25,1.25);
     \node[inner sep=0pt] (meter) at (3,1.75)
    {\includegraphics[scale = 0.65]{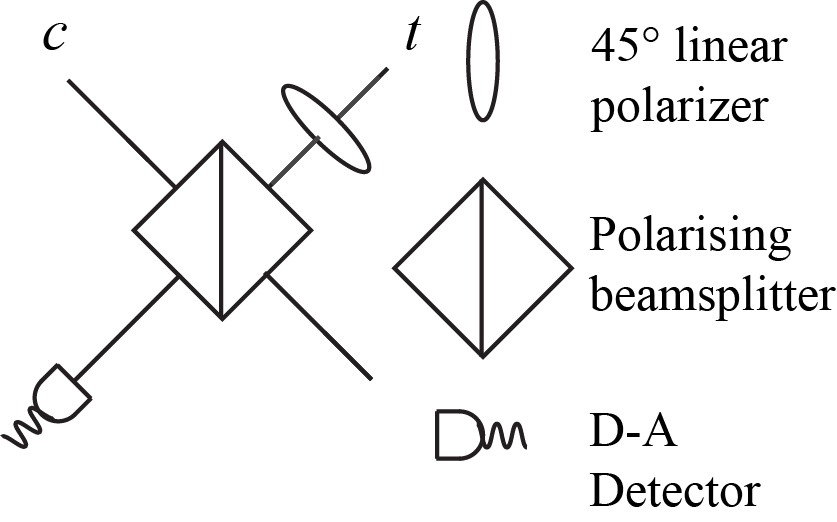}};
     
     \draw [-stealth,thick] (12+.25,3) --  node[above]{$A_t$}(11.75-.25+.25,3);
       \draw [thick] (14+.25,2) --  node[above]{Measurement} node[below]{outcome}(12+.25,2);
       \draw [thick] (12.5+.25-.5,2) -- (12.5+.25-.5,3);
       \draw [thick] (12.5+.25-.5,2) -- (12.5+.25-.5,1.25);
     \draw [-stealth,thick] (12.5-.5+.25,1.25) --  node[above]{$D_t$}(11.75-.25+.25,1.25);
     \node at (12.25,0.7) {$\rightarrow m_t=+1$};
    \node at (11.25,2.5) {$\rightarrow m_t=+1$};
     \node[inner sep=0pt] (meter) at (15.7+.25,1.75)
    {\includegraphics[scale = 0.65]{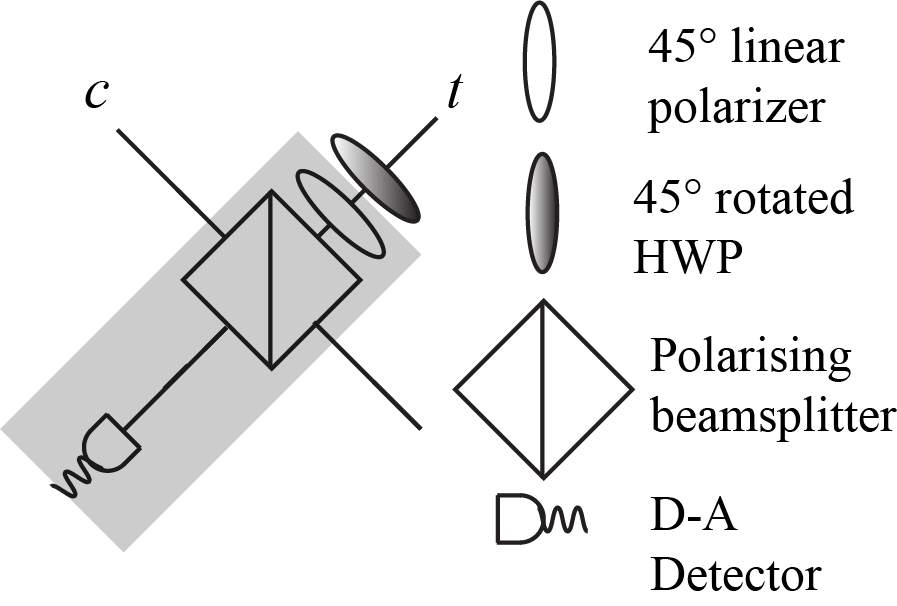}};

		\end{tikzpicture}
  
		\caption{Rotated type-I fusion (a) The LO circuit constructed using a $45^{\circ}$ linear polarizer, a PBS, and a D-A polarization resolving detector. If a single photon is detected, it heralds the Kraus operators of one of the two quantum circuits in (b) and (c) depending upon the polarization of the measurement. If two photons are detected, it is equivalent to performing a Pauli-X measurement on both the target qubit and a Pauli-Z measurement on the control qubit. (d) An equivalent type-I fusion circuit, constructed using (c), with the same Kraus operators as (a) but an opposite mapping of measurement outcomes to Kraus operators. The grey box highlights the circuit from (a).}
		\label{fig:LOCZ}
	\end{figure*}
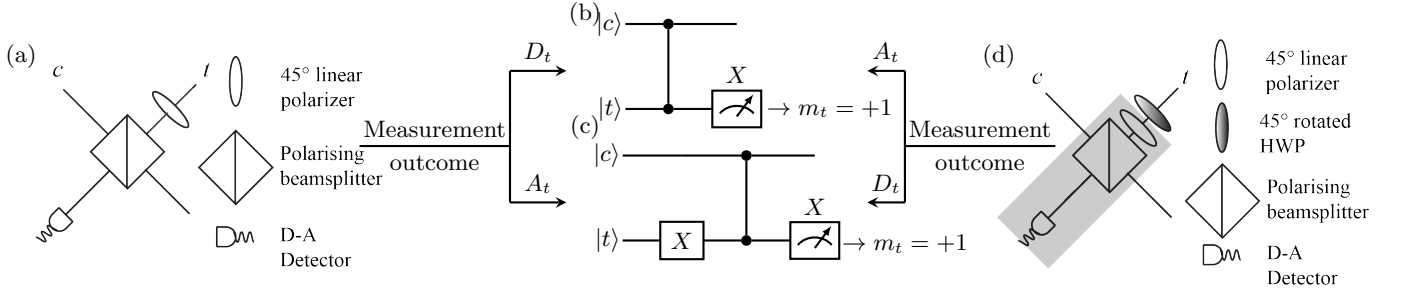
\begin{figure*}
		\centering
		
	\begin{tikzpicture}[darkstyle/.style={circle,inner sep=0pt,minimum size=0.4cm,draw=blue,fill=blue!20!,font=\sffamily\small\bfseries},
 edge_style/.style={draw=black,  thick}]
 
\node[inner sep=0pt] (meter) at (10.25,2.75)
    {\begin{quantikz}
		\lstick{}   & \gate{H}&\ctrl{1} &\qw\\
		\lstick{} & \qw & \ctrl{} & \meter{$X$}\\
    \end{quantikz}};
    \node[inner sep=0pt] (meter) at (10.25,0.75)
    {\begin{quantikz}
		\lstick{}   & \gate{H}  & \ctrl{1} &\qw\\
		\lstick{}   &  \gate{X} & \ctrl{} & \meter{$X$}\\
    \end{quantikz}};
    \node at (1-.7,3.2) {(a)};
    \node at (8.2,3.75) {(b)};
    \node at (8.2,2.) {(c)};
    \node at (12.7,0.3) {$\rightarrow m_t=+1$};
    \node at (12.7,2.3) {$\rightarrow m_t=+1$};
       \draw [-stealth,thick] (7.25,3) --  node[above]{$D_t$}(8,3);
       \draw [thick] (5.25,2) --  node[above]{Measurement} node[below]{outcome}(7.25,2);
       \draw [thick] (7.25,2) -- (7.25,3);
       \draw [thick] (7.25,2) -- (7.25,1.25);
       \draw [-stealth,thick] (7.25,1.25) --  node[above]{$A_t$}(8,1.25);
     \node[inner sep=0pt] (meter) at (3-.25,1.75)
    {\includegraphics[scale = 0.65]{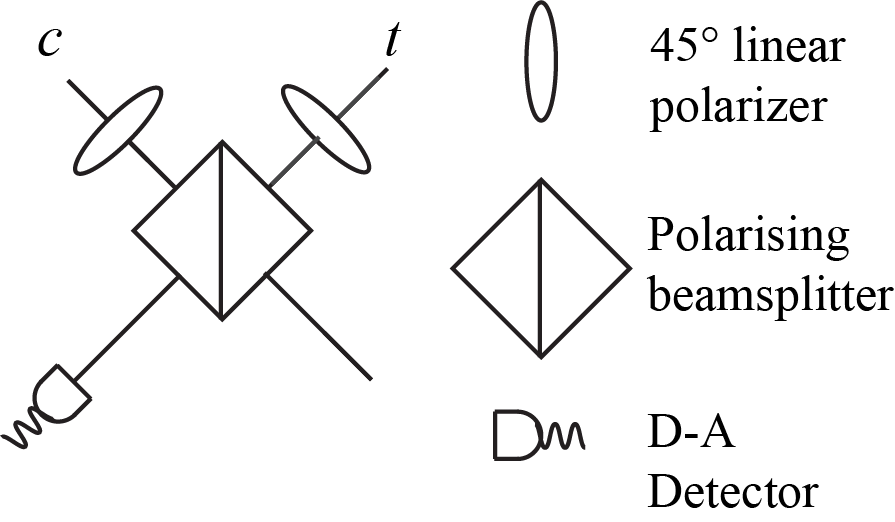}};

		\end{tikzpicture}
  
		\caption{(a) The LO circuit that when successful implements the quantum circuits in  (b) and (c) depending upon the polarization of the measurement and performs Pauli-X measurements on qubits $c$ and $t$ on failure. The Kraus operators for (b) and (c) are $\ket{+}\bra{++}+\ket{-}\bra{--}$ and  $\ket{+}\bra{++}-\ket{-}\bra{--}$, respectively. }
		\label{fig:rotatedType1}
	\end{figure*}
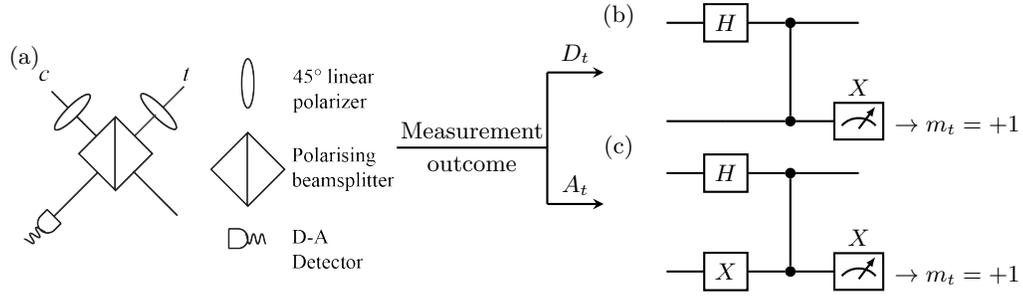

\begin{table*}[ht]
    \centering
  \begin{tabular}{?p{2cm}?p{1.8cm}|p{1.8cm}|p{4.2cm}?p{1.5cm}|p{1.8cm}| p{4.2cm}?}
    \noalign{\hrule height 1pt} 
     & 
      \multicolumn{3}{c?}{\textbf{Success}} &
      \multicolumn{3}{c?}{\textbf{Failure}} \\
      \hline\xrowht{10pt}
     \centering Circuit &  \centering Kraus Operator & \centering Unitary & \centering Graph rule & \centering  Project on & \centering Unitary & Graph rule \\
    \noalign{\hrule height 1pt}\xrowht{20pt}
   \centering \raisebox{-\totalheight/2}{ \includegraphics[scale=0.25]{figures/type1Fusion.png}}  & \centering $\ket{0}\bra{00}\pm\ket{1}\bra{11}$ & \centering - & Invert the edges from $\mathcal{N}_t$ to $c$ and $G\setminus \{t\}$.  & \centering $\pm\bra{10}$ or $\bra{01}$  & \centering - & Perform Pauli-Z measurements on $c$ and $t$. $G\setminus \{c,t\}$\\
    \hline\xrowht{20pt}
     \raisebox{-\totalheight/2}{ \includegraphics[scale=0.25]{figures/type1Fusion2.png}}  & \centering $\ket{0}\bra{01}\pm\ket{1}\bra{10}$ & \centering -  &  Invert the edges from $\mathcal{N}_t$ to $c$ and $G\setminus \{t\}$. & $\pm\bra{11}$ or $\bra{00}$ & \centering - & Perform Pauli-Z measurements on $c$ and $t$. $G\setminus \{c,t\}$\\
    \hline\xrowht{20pt}
     \raisebox{-\totalheight/2}{ \includegraphics[scale=0.25]{figures/LOCZ.png}}  & \centering $\ket{+}\bra{0+}\pm\ket{-}\bra{1-}$ & \centering $H_i$ s.t. $i\in\mathcal{N}_t\cup \{c\}$& Add an edge from $c$ to $t$. Let this new graph be $G'$. Then perform $(((G'.t).i).t)\setminus\{c,t\}$ & $\pm\bra{1+}$ or $\bra{0-}$  & \centering $H_i$ s.t. $i\in\mathcal{N}_t$ & Perform Pauli-Z and -X measurements on $c$ and $t$, respectively. $(((G.t).i).t)\setminus\{c,t\}$ \\
    \hline\xrowht{20pt}
     \vspace{-3em}\raisebox{-\totalheight/2}{ \includegraphics[scale=0.25]{figures/RotatedT1.png}}  & \centering $\ket{+}\bra{++}\pm\ket{-}\bra{--}$ & \centering $H$ on a pair of qubits in $\{c\}\cup\mathcal{N}_c\cup\mathcal{N}_t$ &  If $H$ is applied on \begin{itemize}
    \item $c$ and $u\in\mathcal{N}_t$: Perform $(((G.t).u).t)\setminus\{t\}$. Invert the edges from $\mathcal{N}_c$ to $(\{u\}\cup\mathcal{N}_u)$. 
    \item $c$ and $v\in\mathcal{N}_c$: Perform $(((G.v).c).v)\setminus\{c,t\}$. Invert the edges from $(\{v\}\cup\mathcal{N}_v\cup\{c\})$ to $\mathcal{N}_t$.
    \item $v\in\mathcal{N}_c$ and $u\in\mathcal{N}_t$: Perform $((((((G.v).c).v).t).u).t)\setminus\{c,t\}$. Connect $(\{c\})$ to $\{v\}\cup\mathcal{N}_v\mathcal{N}_v$.
    \end{itemize} & $\pm\bra{-+}$ or $\bra{+-}$  & \centering $H_v, H_u$ s.t. $u\in\mathcal{N}_t$ and $v\in\mathcal{N}_c$  &  Perform Pauli-X measurements on $c$ and $t$. $((((((G.c).v).c).t).u).t)\setminus\{c,t\}$\\
    \noalign{\hrule height 1pt}
  \end{tabular}
\caption{For the type-I fusion and its rotated variations, the corresponding Kraus operators and rules to evolve the original cluster state when the linear optical (LO) either succeeds or fails, are summarized in this Table. The first two fusions result in a cluster state when the input states are cluster states; all other fusions need local unitaries (as listed in the column ``Unitary") to convert the post-fusion state into a cluster state. Note that, $G$ is the collective graph of the two input cluster states and $\mathcal{N}_i$ refers to the neighborhood of qubit $i$ in $G$.}
\label{tab:fusion_rulesT1}
\end{table*}

\subsection{Two-Qubit Fusion} 
\label{subsec:LOtypeII}

We have now developed an understanding of single and two-qubit LO gates. In this section, we use Section~\ref{subsubsec:typeI} to build and analyze the LO implementation of two-qubit fusions. 

\begin{figure*}
		\centering
		
	\begin{tikzpicture}[darkstyle/.style={circle,inner sep=0pt,minimum size=0.4cm,draw=blue,fill=blue!20!,font=\sffamily\small\bfseries},
 edge_style/.style={draw=black,  thick}]
 
    
     \node[inner sep=0pt] (meter) at (2.6,1.75)
    {\includegraphics[scale = 0.65]{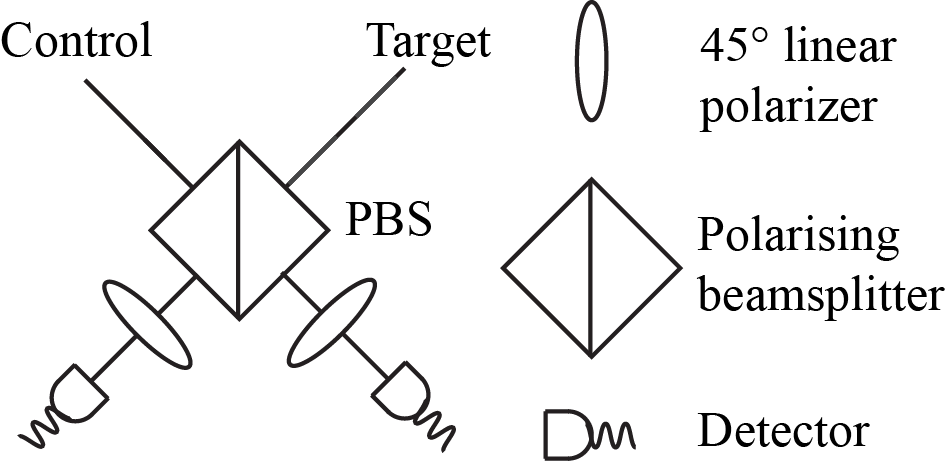}};
     \draw [-stealth,thick] (7.25,3) --  node[above]{$H_cH_t$ or} node[below]{$V_cV_t$}(8.25,3);
       \draw [thick] (5.25,2) --  node[above]{Measurement} node[below]{outcome}(7.25,2);
       \draw [thick] (7.25,2) -- (7.25,3);
       \draw [thick] (7.25,2) -- (7.25,1);
       \draw [-stealth,thick] (7.25,1) --  node[above]{$V_cH_t$} node[below]{or $H_cV_t$}(8.25,1);
     \node[inner sep=0pt] (meter) at (10,3)
    {\begin{quantikz}
		\lstick{$\ket{c}$} & \ctrl{1} & \gate{H} & \meter{$Z$} \\
	\lstick{$\ket{t}$}& \targ{} & \qw & \meter{$Z$}\\
    \end{quantikz}};
    \node at (12.65,2.5) {$\rightarrow m_t=+1$};
    \node at (12.65,3.75) {$\rightarrow m_c=+1$};
 \node at (12.65,1.4) {$\rightarrow m_c=-1$};
 \node at (12.65,.1) {$\rightarrow m_t=+1$};
 
     \node[inner sep=0pt] (meter) at (10,0.7)
    {\begin{quantikz}
		\lstick{$\ket{c}$} & \ctrl{1}& \gate{H}& \meter{$Z$} \\
	\lstick{$\ket{t}$} & \targ{}& \qw& \meter{$Z$}\\
    \end{quantikz}};
    \node at (0,3.2){(a)};
    \node at (8,4){(b)};
    \node at (8,1.75){(c)};
    
		\end{tikzpicture}
		\caption{Type-II fusion (a) The LO circuit succeeds when each detector detects one photon and fails otherwise. It projects on the Bell states $\frac{\ket{00}\pm\ket{11}}{\sqrt{2}}$ on success and performs Pauli-Z measurements when it fails. (b) When the LO circuit is successful, it implements quantum circuits (b) or (c) depending upon the photon measurement pattern.}
		\label{fig:LOtype2}
	\end{figure*}

Consider the LO circuit in FIG.~\ref{fig:LOtype2}(a) acting on two input modes - $c$ and $t$. If the input state is $a_{H_c}^{\dagger}a_{H_t}^{\dagger}$, after going through the PBS and the polarizers, it changes to $\big(\frac{a_{H_c}^{\dagger}+a_{V_c}^{\dagger}}{\sqrt{2}}\big)\big(\frac{a_{H_t}^{\dagger}+a_{V_t}^{\dagger}}{\sqrt{2}}\big)$. Similarly, we can write down the state just before detectors for the four two-qubit basis states -
\begin{align*}
    a_{H_c}^{\dagger}a_{H_t}^{\dagger}&\rightarrow \big(\frac{a_{H_c}^{\dagger}+a_{V_c}^{\dagger}}{\sqrt{2}}\big)\big(\frac{a_{H_t}^{\dagger}+a_{V_t}^{\dagger}}{\sqrt{2}}\big)\\
    a_{H_c}^{\dagger}a_{V_t}^{\dagger}&\rightarrow \big(\frac{a_{H_c}^{\dagger}+a_{V_c}^{\dagger}}{\sqrt{2}}\big)\big(\frac{a_{H_c}^{\dagger}-a_{V_c}^{\dagger}}{\sqrt{2}}\big)\\
     a_{V_c}^{\dagger}a_{H_t}^{\dagger}&\rightarrow \big(\frac{a_{H_t}^{\dagger}+a_{V_t}^{\dagger}}{\sqrt{2}}\big)\big(\frac{a_{H_t}^{\dagger}-a_{V_t}^{\dagger}}{\sqrt{2}}\big)\\
      a_{V_c}^{\dagger}a_{V_t}^{\dagger}&\rightarrow \big(\frac{a_{H_c}^{\dagger}-a_{V_c}^{\dagger}}{\sqrt{2}}\big)\big(\frac{a_{H_t}^{\dagger}-a_{V_t}^{\dagger}}{\sqrt{2}}\big)\\
\end{align*}
From the equations above, we can write down the projections applied by the LO circuit for all possible photon detection patterns - 
\begin{equation}\label{eq:type2Projectors}
\begin{split}
 \text{$H_cH_t$ or $V_cV_t$} &\rightarrow \frac{\bra{0,0,0,0}a_{H_c}a_{H_t} +\bra{0,0,0,0}a_{V_c}a_{V_t}}{\sqrt{2}}\\ 
 \\& = \frac{\bra{00} +\bra{11}}{\sqrt{2}}\\ 
 \text{$H_cV_t$ or $V_cH_t$}&\rightarrow \frac{\bra{0,0,0,0}a_{H_c}a_{H_t} -\bra{0,0,0,0}a_{V_c}a_{V_t}}{\sqrt{2}}\\\\& = \frac{\bra{00} -\bra{11}}{\sqrt{2}}\\ 
 \text{ $H_c^2$} &\rightarrow \bra{0,0,0,0}a_{H_c}a_{V_t} = \bra{01}\\
 \text{ $V_c^2$} &\rightarrow -\bra{0,0,0,0}a_{H_c}a_{V_t}= - \bra{01}\\
 \text{$H_t^2$} &\rightarrow \bra{0,0,0,0}a_{V_c}a_{H_t} = \bra{10}\\
\text{$V_t^2$} &\rightarrow -\bra{0,0,0,0}a_{V_c}a_{H_t} = -\bra{10}
\end{split}
\end{equation}

This circuit is said to be successful when exactly one photon is detected at each detector and has a 50$\%$ probability of success. When successful, it projects the input qubits $c$ and $t$ onto the Bell states $\frac{\ket{00}\pm\ket{11}}{\sqrt{2}}$. The fusion failure maps to Pauli-Z basis measurement on input qubits. This LO circuit is called the type-II fusion~\cite{browne2005Fusion, MercedesThesis}. The success probability of the fusion circuits can be improved beyond 50$\%$ using ancilla single photons or Bell pairs~\cite{grice2011arbitrarily, vanLoockBell, GrosshansBell,MercedesThesis}. These circuits are called the \textit{boosted} fusion circuits and are out of the scope of this paper. 

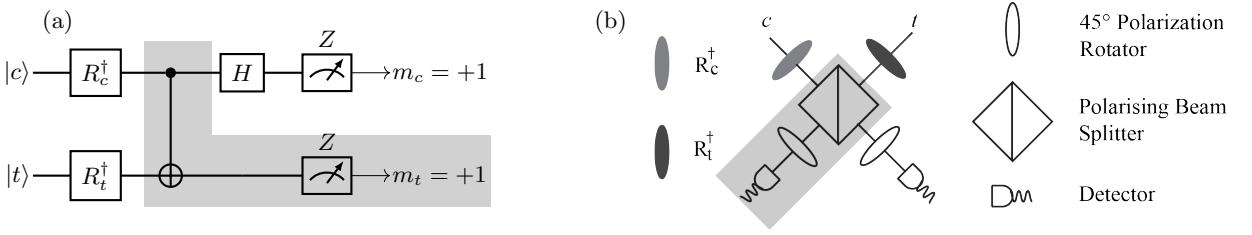
\begin{figure*}
		\centering
		
	\begin{tikzpicture}[darkstyle/.style={circle,inner sep=0pt,minimum size=0.4cm,draw=blue,fill=blue!20!,font=\sffamily\small\bfseries},
 edge_style/.style={draw=black,  thick}]
  
    \node at (-.75,3) {(a)};
    \node at (6.6,3) {(b)};
     \path [fill=black!18] (0.4,0.55) rectangle (1.3,2.75);
\path [fill=black!18] (1.3,0.55) rectangle (5,1.5);

     \node[inner sep=0pt] (meter) at (11,1.75)
    {\includegraphics[scale = 0.7]{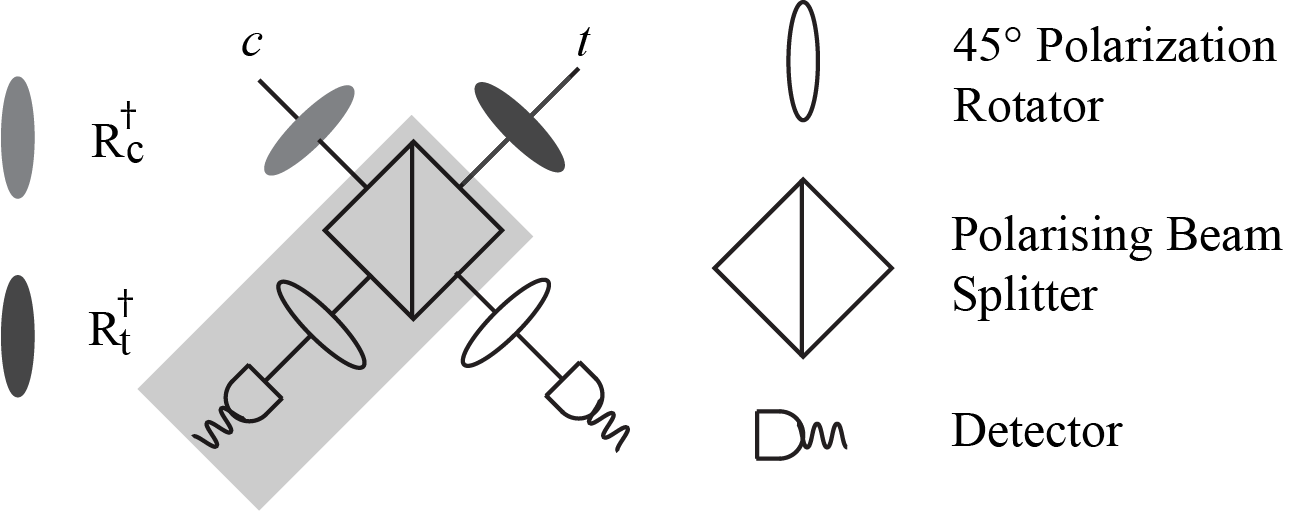}};

 \node[inner sep=0pt] (meter) at (1.75,1.5)
  {\begin{quantikz}
	\lstick{$\ket{c}$} & \gate{R_c^{\dagger}} & \ctrl{1} & \gate{H} & \meter{$Z$} \arrow[r] & \rstick{$m_c=+1$}\\
	\lstick{$\ket{t}$} & \gate{R_t^{\dagger}} & \targ{} & \qw & \meter[fill=black!18] {$Z$} \arrow[r] & \rstick{$m_t=+1$}\\
	\end{quantikz}};

		\end{tikzpicture}
		\caption{A generalized rotated type-II fusion - the grey boxes show the type-I fusions. (a) The quantum circuit that projects on $R_cR_t\frac{\ket{00}+\ket{11}}{\sqrt{2}}$. (b) The LO circuit built by first applying the rotations $R_c^{\dagger}$ and $R_t^{\dagger}$ to the input qubits, then a type-I fusion, followed by a $45^{\circ}$ linear polarizer and detector on the control qubit. }
		\label{fig:LOtype2general}
	\end{figure*}

We now analyze the rotated fusion circuit as shown in FIG.~\ref{fig:LOtype2general}(a). We can build the corresponding LO circuit by following the quantum circuit to LO circuit recipe from Section~\ref{subsubsec:typeI}. The only addition to this recipe is the detector on the $c$ mode to mimic the Pauli-Z measurement on qubit $c$ in the quantum circuit. The LO circuit obtained using this recipe implements the given quantum circuit when the photon detection pattern is $H_cH_t$ or $V_cV_t$. 

All possible photon detection patterns and the corresponding projections applied by the LO-rotated type-II fusion circuit are as follows - 
\begin{equation}\label{eq:type2Projectors}
\begin{split}
 \text{$H_cH_t$ or $V_cV_t$} &\rightarrow \frac{\bra{00} +\bra{11}}{\sqrt{2}}R_c^{\dagger}R_t^{\dagger}\\ 
 \text{$H_cV_t$ or $V_cH_t$}&\rightarrow  \frac{\bra{00} -\bra{11}}{\sqrt{2}}R_c^{\dagger}R_t^{\dagger}\\ 
 \text{ $H_c^2$} &\rightarrow  \bra{01}R_c^{\dagger}R_t^{\dagger}\\
 \text{ $V_c^2$} &\rightarrow - \bra{01}R_c^{\dagger}R_t^{\dagger}\\
 \text{$H_t^2$} &\rightarrow \bra{10}R_c^{\dagger}R_t^{\dagger}\\
\text{$V_t^2$} &\rightarrow -\bra{10}R_c^{\dagger}R_t^{\dagger}
\end{split}
\end{equation}
In other words, the rotated type-II fusion circuit rotates the projections of the non-rotated type-II fusion and the rotation is determined by $R_c^{\dagger}R_t^{\dagger}$.

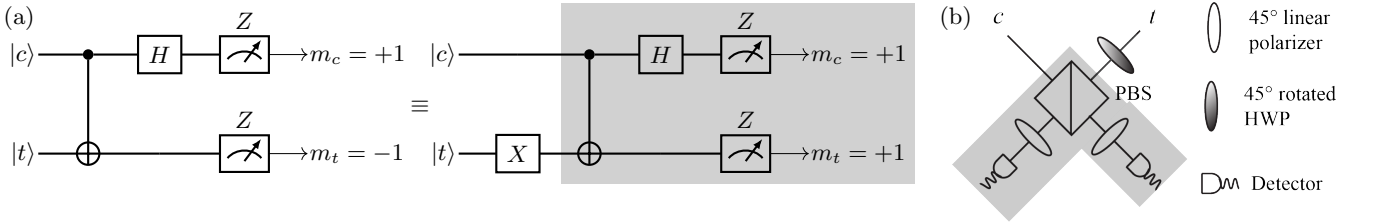
\begin{figure*}
		\centering
		
	\begin{tikzpicture}[darkstyle/.style={circle,inner sep=0pt,minimum size=0.4cm,draw=blue,fill=blue!20!,font=\sffamily\small\bfseries},
 edge_style/.style={draw=black,  thick}]

   \node at (-6.,1) {(a)};
   \node at (6.45,1) {(b)};
 \node[inner sep=0pt] (meter) at (-3.5,-.25)
    {\begin{quantikz}
		\lstick{$\ket{c}$} & \ctrl{1}& \gate{H}& \meter{$Z$} \arrow[r] & \rstick{$m_c=+1$} \\
	\lstick{$\ket{t}$} & \targ{}& \qw& \meter{$Z$} \arrow[r] & \rstick{$m_t=-1$}\\
    \end{quantikz}};
 
 \path [fill=black!18] (1.2,-1.2) rectangle (5.9,1.2);
 
    \node[inner sep=0pt] (meter) at (2.5,-.25)
    {$\equiv$\begin{quantikz}
		\lstick{$\ket{c}$} & \qw &\ctrl{1}& \gate[style={fill=black!18}]{H}& \meter[fill= black!18]{$Z$} \arrow[r] & \rstick{$m_c=+1$} \\
	\lstick{$\ket{t}$} & \gate{X} &\targ{}& \qw & \meter[fill= black!18]{$Z$} \arrow[r] & \rstick{$m_t=+1$}\\
    \end{quantikz}};

     \node[inner sep=0pt] (meter) at (8.95,-.25)
    {\includegraphics[scale = 0.65]{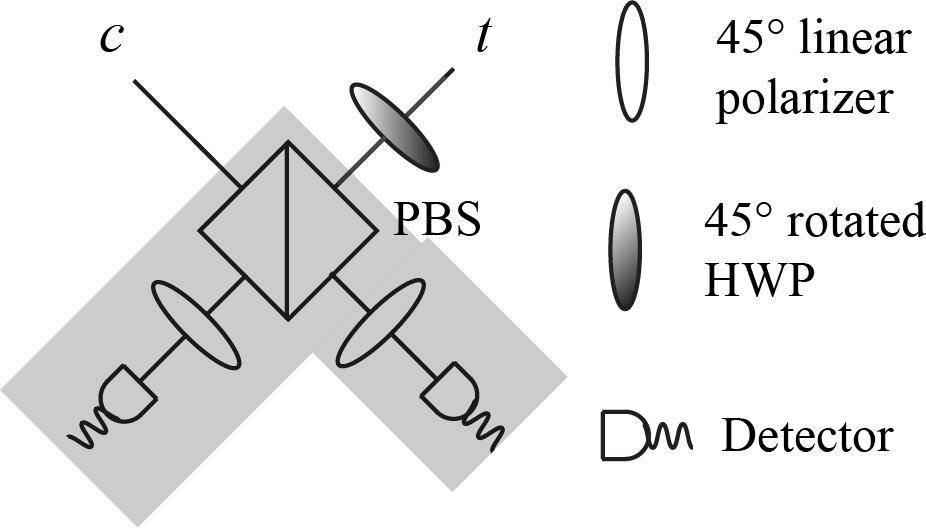}};

		\end{tikzpicture}
		\caption{Rotated type-II fusion (a) The quantum circuit projects onto the Bell state $\frac{\ket{01}+\ket{10}}{\sqrt{2}}$ (b) The LO circuit built from (a) that projects onto $\frac{\ket{01}\pm\ket{10}}{\sqrt{2}}$ on success. The grey boxes show the non-totated type-II fusion circuits.}
		\label{fig:LOType2Other}
	\end{figure*}

The type-II fusion circuit projects onto two of the four Bell states - $\frac{\ket{00}\pm\ket{11}}{\sqrt{2}}$. These two Bell states differ only in phase. Now, to figure out the LO circuit that projects onto the remaining two Bell states $\frac{\ket{01}\pm\ket{10}}{\sqrt{2}}$, which differ only in phase, we first write down the quantum circuit to project onto $\frac{\ket{01}+\ket{10}}{\sqrt{2}}$ as shown in FIG.~\ref{fig:LOType2Other}(a). Using the quantum circuit to LO circuit recipe, to build the corresponding LO circuit, we start with two modes. We then add a 90$^\circ$ polarizer on one of the modes for the Pauli-X gate in the quantum circuit. This mode now is set as the target $t$ qubit. We then add the non-rotated type-II fusion circuit to complete the construction as shown in ~\ref{fig:LOType2Other}(b). This LO circuit projects onto the Bell state $\frac{\ket{01}+\ket{10}}{\sqrt{2}}$, when the photon detection outcome is either $H_cH_t$ or $V_cV_t$. It projects onto the other Bell state $\frac{\ket{01}-\ket{10}}{\sqrt{2}}$, when the photon detection outcome is either $H_cV_t$ or $V_cH_t$. The fusion circuit fails when either of the detectors detects two photons. The failure can be mapped as the projections $\pm\bra{00}$ or $\pm\bra{11}$, depending upon the photon measurement pattern. In other words, if the fusion fails, it performs Pauli-Z basis measurement on the input qubits. 

Both type-I and type-II fusions can be used for growing large cluster states by fusing smaller states. Type-I fusion grows the size of the cluster state faster than type-II fusion because it measures one less qubit than type-II. However, type-II fusion is still favored for cluster state generation. The type-II fusion measures both the input qubits. If one or both of the input qubits are lost, i.e., if the photons are lost, the detectors will measure less than two photons. As a result, type-II fusion can \textit{herald} photon loss in the input qubits. In the type-I fusion circuit, only the target mode is detected. The photon loss in the control mode goes undetected or \textit{unheralded}. The cluster state generated using type-I fusion then has a higher probability of unheralded loss per qubit compared to the cluster state generated using type-II fusions. 

\subsection{Multi-qubit fusion}
\label{subsec:GHZproj3}
\begin{figure*}
		\centering
		
	\begin{tikzpicture}[darkstyle/.style={circle,inner sep=0pt,minimum size=0.4cm,draw=blue,fill=blue!20!,font=\sffamily\small\bfseries},
 edge_style/.style={draw=black,  thick}]

   \node at (-2.75,1.5) {(a)};
   \node at (5.75-1,1.5) {(b)};

   \node at (-2.75,1.5-3.75) {(c)};
   \node at (5.75-1,1.5-3.75) {(d)};

   \node at (-2.75,1.5-3.75-3.75) {(e)};
   \node at (5.75-1,1.5-3.75-3.75) {(f)};

   \node at (-2.75,1.5-3.75-3.75-3.75) {(g)};
   \node at (5.75-1,1.5-3.75-3.75-3.75) {(h)};

    \node at (12.5,0) {Projects onto };
     \node at (12.5,-.5) {$\frac{\ket{000}\pm\ket{111}}{\sqrt{2}}$};

      \node at (12.5,0-3.75) {Projects onto };
     \node at (12.5,-.5-3.75) {$\frac{\ket{001}\pm\ket{110}}{\sqrt{2}}$};
      \node at (12.5,0-3.75-3.75) {Projects onto };
     \node at (12.5,-.5-3.75-3.75) {$\frac{\ket{010}\pm\ket{101}}{\sqrt{2}}$};
      \node at (12.5,0-3.75-3.75-3.75) {Projects onto };
     \node at (12.5,-.5-3.75-3.75-3.75) {$\frac{\ket{011}\pm\ket{100}}{\sqrt{2}}$};

 \path [fill=black!18] (-1.8,-.7-3.75) rectangle (-1.1,1.4-3.75);
 \path [fill=black!18] (-1.8,-.7-3.75) rectangle (3.9,.45-3.75);

 \path [fill=black!18] (-1.8,-.7-3.75-3.75) rectangle (-1.1,1.4-3.75-3.75);
 \path [fill=black!18] (-1.8,-.7-3.75-3.75) rectangle (3.9,.45-3.75-3.75);

 \path [fill=black!18] (-1.8,-.7-3.75-3.75-3.75) rectangle (-1.1,1.4-3.75-3.75-3.75);
 \path [fill=black!18] (-1.8,-.7-3.75-3.75-3.75) rectangle (3.9,.45-3.75-3.75-3.75);

 \path [fill=black!18] (-1.8,-.7) rectangle (-1.1,1.4);
 \path [fill=black!18] (-1.8,-.7) rectangle (3.9,.45);
 
    \node[inner sep=0pt] (meter) at (.5,-.25)
    {\begin{quantikz}
		\lstick{$\ket{c}$}  & \ctrl{1} & \ctrl{2} & \gate{H} & \meter{$Z$} \arrow[r] & \rstick{$m_c=+1$}\\
		\lstick{$\ket{t_1}$} & \targ{} & \qw & \qw & \meter[fill = black!18]{$Z$} \arrow[r] & \rstick{$m_{t_1}=+1$}\\
		\lstick{$\ket{t_2}$} & \qw & \targ{} & \qw & \meter{$Z$} \arrow[r] & \rstick{$m_{t_2}=+1$}\\
    \end{quantikz}};

 \node[inner sep=0pt] (meter) at (.5,-4)
    {\begin{quantikz}
		\lstick{$\ket{c}$}  & \ctrl{1} & \ctrl{2} & \gate{H} & \meter{$Z$} \arrow[r] & \rstick{$m_c=+1$}\\
		\lstick{$\ket{t_1}$} & \targ{} & \qw & \qw & \meter[fill = black!18]{$Z$} \arrow[r] & \rstick{$m_{t_1}=+1$}\\
		\lstick{$\ket{t_2}$} & \qw & \targ{} & \qw & \meter{$Z$} \arrow[r] & \rstick{$m_{t_2}=-1$}\\
    \end{quantikz}};
    \node[inner sep=0pt] (meter) at (.5,-7.75)
    {\begin{quantikz}
		\lstick{$\ket{c}$}  & \ctrl{1} & \ctrl{2} & \gate{H} & \meter{$Z$} \arrow[r] & \rstick{$m_c=+1$}\\
		\lstick{$\ket{t_1}$} & \targ{} & \qw & \qw & \meter[fill = black!18]{$Z$} \arrow[r] & \rstick{$m_{t_1}=-1$}\\
		\lstick{$\ket{t_2}$} & \qw & \targ{} & \qw & \meter{$Z$} \arrow[r] & \rstick{$m_{t_2}=+1$}\\
    \end{quantikz}};
    \node[inner sep=0pt] (meter) at (.5,-11.5)
    {\begin{quantikz}
		\lstick{$\ket{c}$}  & \ctrl{1} & \ctrl{2} & \gate{H} & \meter{$Z$} \arrow[r] & \rstick{$m_c=+1$}\\
		\lstick{$\ket{t_1}$} & \targ{} & \qw & \qw & \meter[fill = black!18]{$Z$} \arrow[r] & \rstick{$m_{t_1}=-1$}\\
		\lstick{$\ket{t_2}$} & \qw & \targ{} & \qw & \meter{$Z$} \arrow[r] & \rstick{$m_{t_2}=-1$}\\
    \end{quantikz}};   
    
     \node[inner sep=0pt] (meter) at (8.95-1,-.25)
    {\includegraphics[scale = 0.65]{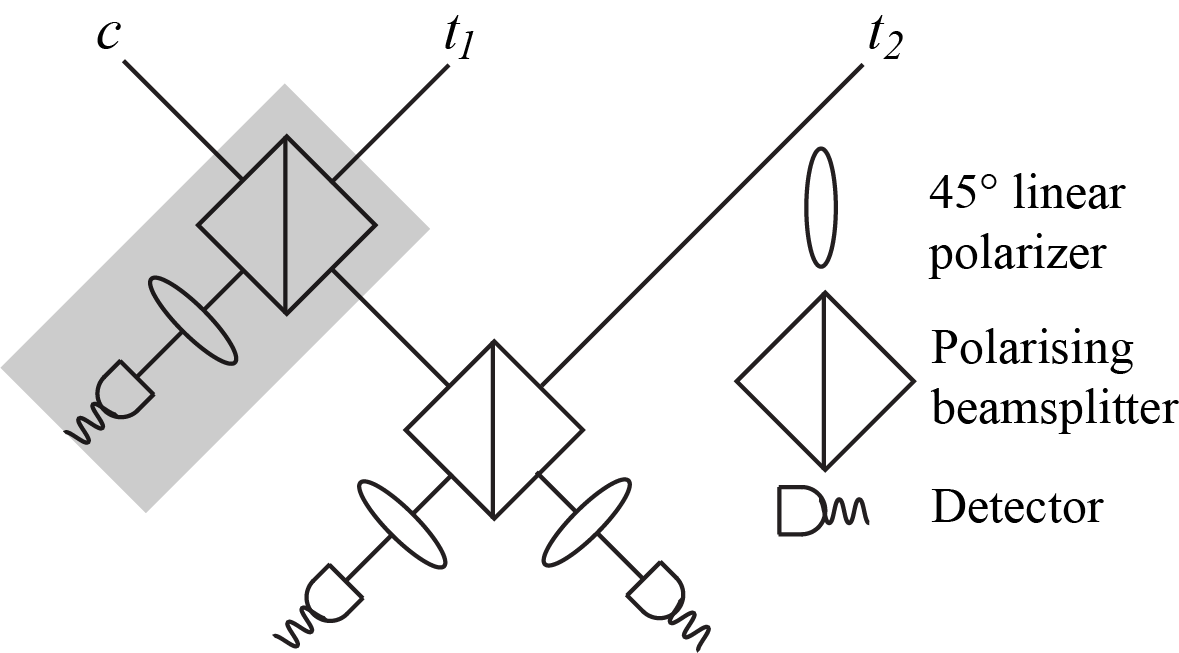}};
     \node[inner sep=0pt] (meter) at (8.95-1,-.25-3.75)
    {\includegraphics[scale = 0.65]{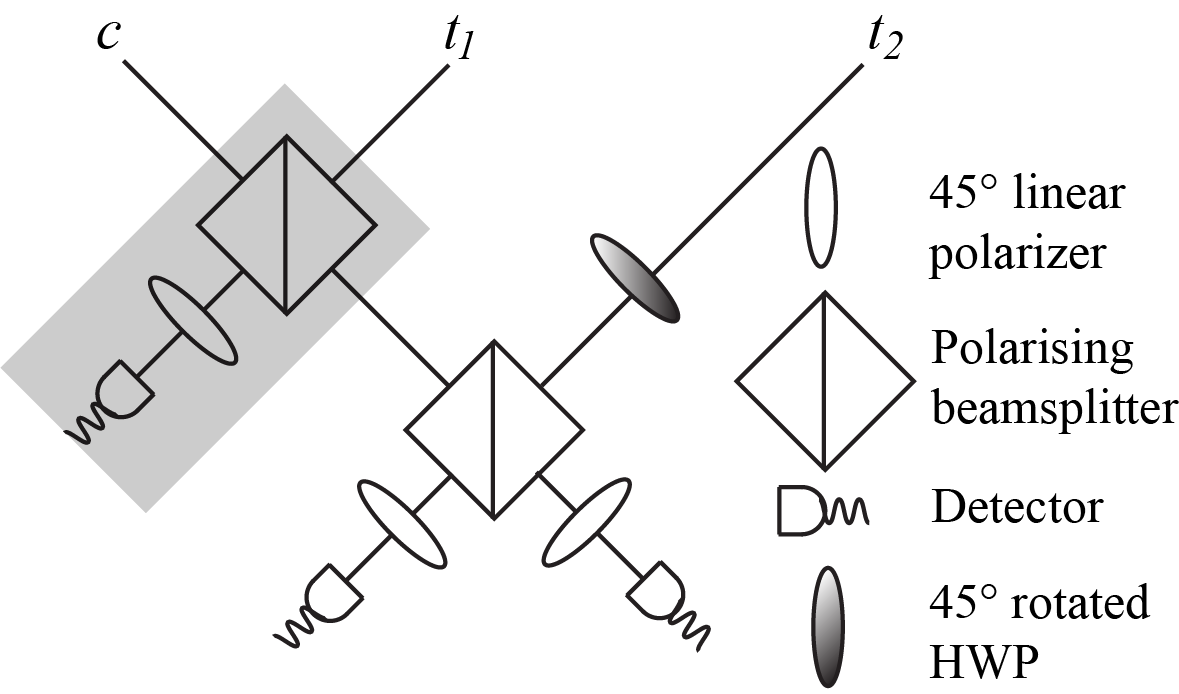}};
     \node[inner sep=0pt] (meter) at (8.95-1,-.25-3.75-3.75)
    {\includegraphics[scale = 0.65]{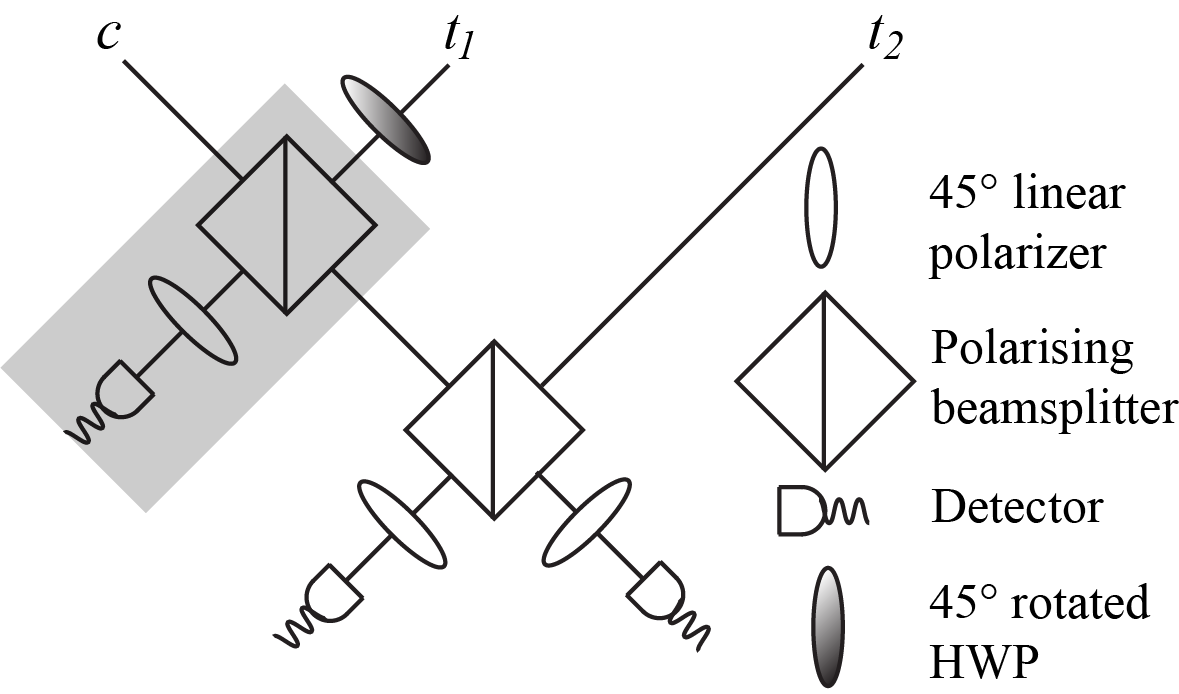}};
     \node[inner sep=0pt] (meter) at (8.95-1,-.25-3.75-3.75-3.75)
    {\includegraphics[scale = 0.65]{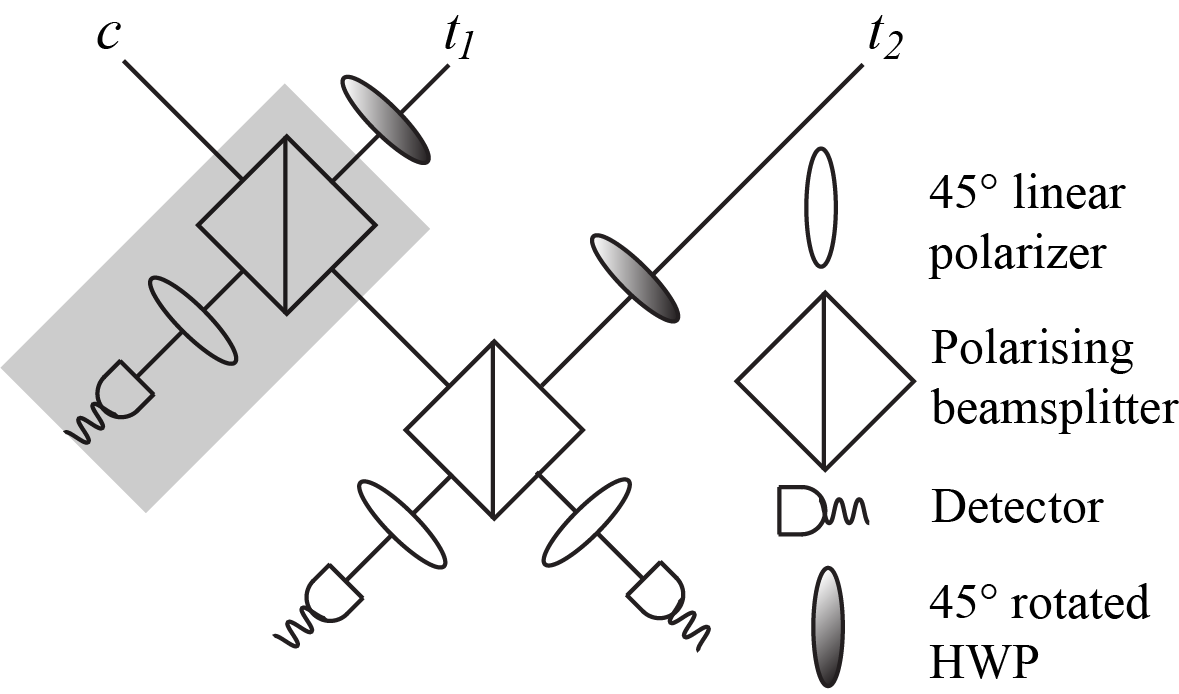}};
 
     
		\end{tikzpicture}
		\caption{3-qubit fusion - (a), (c), (e), (g) show the quantum circuits that project onto  $\frac{\ket{000}+\ket{111}}{\sqrt{2}}, \frac{\ket{001}+\ket{110}}{\sqrt{2}},\frac{\ket{010}+\ket{101}}{\sqrt{2}}$, and $\frac{\ket{100}+\ket{011}}{\sqrt{2}}$, respectively. The corresponding LO circuits are shown in (b), (d), (f), and (h). The LO circuits succeed when one photon is detected in each detector and fail otherwise. The success probability of this circuit is 1/4.}
		\label{fig:LOT3GHZ}
	\end{figure*}
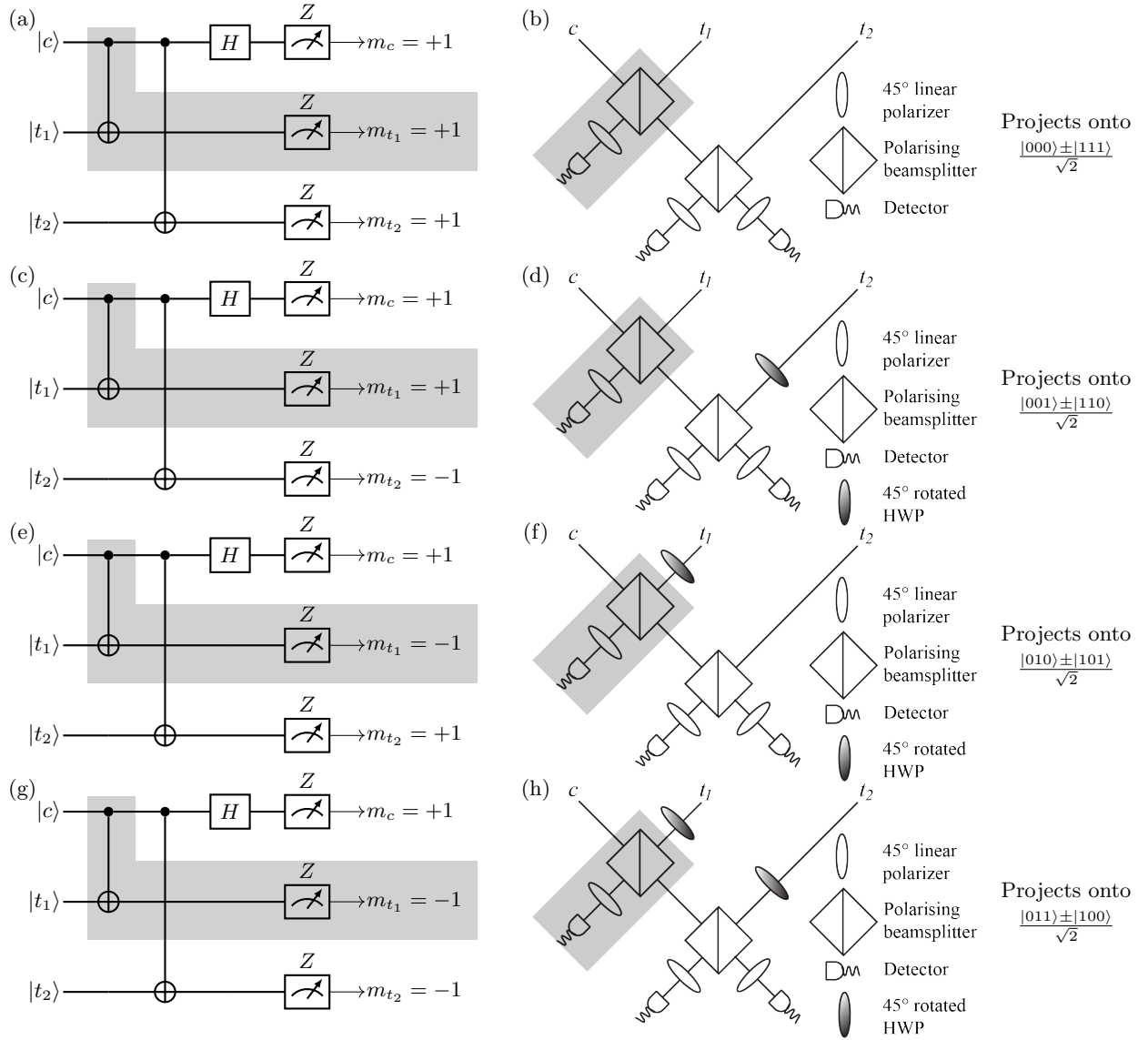

As discussed in Section~\ref{subsec:GHZproj1}, the concept of two-qubit fusion can be extended to perform joint projective measurement on the GHZ basis to stitch together more than 2 entangled states.  The quantum circuit for 3-qubit fusion (refer FIG.~\ref{fig:LOT3GHZ}(a)) can be decomposed into a type-I fusion between qubits $c$ and $t_1$, followed by a type two fusion between qubits $c$ and $t_2$. The LO 3-qubit fusion circuit is then built by concatenating the type-I and type-II fusion circuits as shown in FIG.~\ref{fig:LOT3GHZ}(b). This LO circuit succeeds when each detector detects exactly one photon and fails otherwise. When successful, it projects the input qubits onto the 3-qubit GHZ states - $\frac{\ket{000}_{c,t_1,t_2}\pm\ket{111}_{c,t_1,t_2}}{\sqrt{2}}$ (see TABLE~\ref{tab:LO3GHZprojections} for details). If an even(odd) number of vertically polarized photons are detected, the state projected onto is $\frac{\ket{000}_{c,t_1,t_2}+\ket{111}_{c,t_1,t_2}}{\sqrt{2}}(\frac{\ket{000}_{c,t_1,t_2}-\ket{111}_{c,t_1,t_2}}{\sqrt{2}})$.  The LO circuit performs Pauli-Z measurements on all three input qubits on failure. As this circuit is a concatenation of the type-I and type-II LO circuits, each of which succeeds with a probability of 1/2, the success probability of the 3-qubit fusion circuit is 1/4. 

 \begin{table*}
      \centering
  \begin{tabular}{?P{7cm}|P{10cm}?}
      \hline\xrowht{10pt}
     \textbf{Measurement outcome} & \textbf{Projection} \\
    \noalign{\hrule height 1pt}\xrowht{20pt}
     $H_cH_{t_1}H_{t_2}$ or $H_cV_{t_1}V_{t_2}$ or $V_cH_{t_1}V_{t_2}$ or $V_cV_{t_1}H_{t_2}$ & $\frac{\bra{0,0,0,0,0,0,0}a_{H_c}a_{H_{t_1}}a_{H_{t_2}}+\bra{0,0,0,0,0,0,0}a_{V_c}a_{V_{t_1}}a_{V_{t_2}}}{\sqrt{2}}$    = $\frac{\bra{000}+\bra{111}}{\sqrt{2}}$\\
    \hline\xrowht{30pt}
   $V_cV_{t_1}V_{t_2}$ or $V_cH_{t_1}H_{t_2}$ or $H_cV_{t_1}H_{t_2}$ or $H_cH_{t_1}V_{t_2}$ & $\frac{\bra{0,0,0,0,0,0,0}a_{H_c}a_{H_{t_1}}a_{H_{t_2}}-\bra{0,0,0,0,0,0,0}a_{V_c}a_{V_{t_1}}a_{V_{t_2}}}{\sqrt{2}}$    = $\frac{\bra{000}-\bra{111}}{\sqrt{2}}$\\
    \hline\xrowht{20pt}
   $H_c^2H_{t_1}$ or  $H_c^2V_{t_1}$  & $\bra{0,0,0,0,0,0,0}a_{H_c}a_{H_{t_1}}a_{V_{t_2}} = \bra{001}$\\
    \hline\xrowht{20pt}
   $V_c^2H_{t_1}$ or  $V_c^2V_{t_1}$  & $-\bra{0,0,0,0,0,0,0}a_{H_c}a_{H_{t_1}}a_{V_{t_2}} = -\bra{001}$\\
   \hline\xrowht{20pt}
   $H_c^2H_{t_2}$ or  $H_c^2V_{t_2}$  & $\bra{0,0,0,0,0,0,0}a_{H_c}a_{V_{t_1}}a_{H_{t_2}} = \bra{010}$\\
    \hline\xrowht{20pt}
   $V_c^2H_{t_2}$ or  $V_c^2V_{t_2}$  & $-\bra{0,0,0,0,0,0,0}a_{H_c}a_{V_{t_1}}a_{H_{t_2}} = -\bra{010}$\\
     \hline\xrowht{20pt}
   $H_{t_1}^2H_{t_2}$ or  $H_{t_1}^2V_{t_2}$  & $\bra{0,0,0,0,0,0,0}a_{V_c}a_{H_{t_1}}a_{H_{t_2}} = \bra{100}$\\
    \hline\xrowht{20pt}
   $V_{t_1}^2H_{t_2}$ or  $V_{t_1}^2V_{t_2}$  & $-\bra{0,0,0,0,0,0,0}a_{V_c}a_{H_{t_1}}a_{H_{t_2}} = - \bra{100}$\\
   \hline\xrowht{20pt}
   $H_{t_1}H_{t_2}^2$ or  $V_{t_1}V_{t_2}^2$  & $\bra{0,0,0,0,0,0,0}a_{V_c}a_{V_{t_1}}a_{H_{t_2}} =  \bra{110}$\\
   \hline\xrowht{20pt}
   $H_{t_1}V_{t_2}^2$ or  $V_{t_1}H_{t_2}^2$  & $-\bra{0,0,0,0,0,0,0}a_{V_c}a_{V_{t_1}}a_{H_{t_2}} = - \bra{110}$\\
   \hline\xrowht{20pt}
   $H_c^3$ or  $H_cV_c^2$  & $\bra{0,0,0,0,0,0,0}a_{H_c}a_{V_{t_1}}a_{V_{t_2}} =  \bra{011}$\\
   \hline\xrowht{20pt}
   $V_c^3$ or  $H_c^2V_c$  & $-\bra{0,0,0,0,0,0,0}a_{H_c}a_{V_{t_1}}a_{V_{t_2}} =  -\bra{011}$\\
    \noalign{\hrule height 1pt}
    \end{tabular}
     \caption{Measurement outcomes and the corresponding projections applied by the 3-GHZ projection circuit in FIG.~\ref{fig:LOT3GHZ}(b)}
     \label{tab:LO3GHZprojections}
 \end{table*}

\begin{figure}
    \centering
    \includegraphics[scale=0.5]{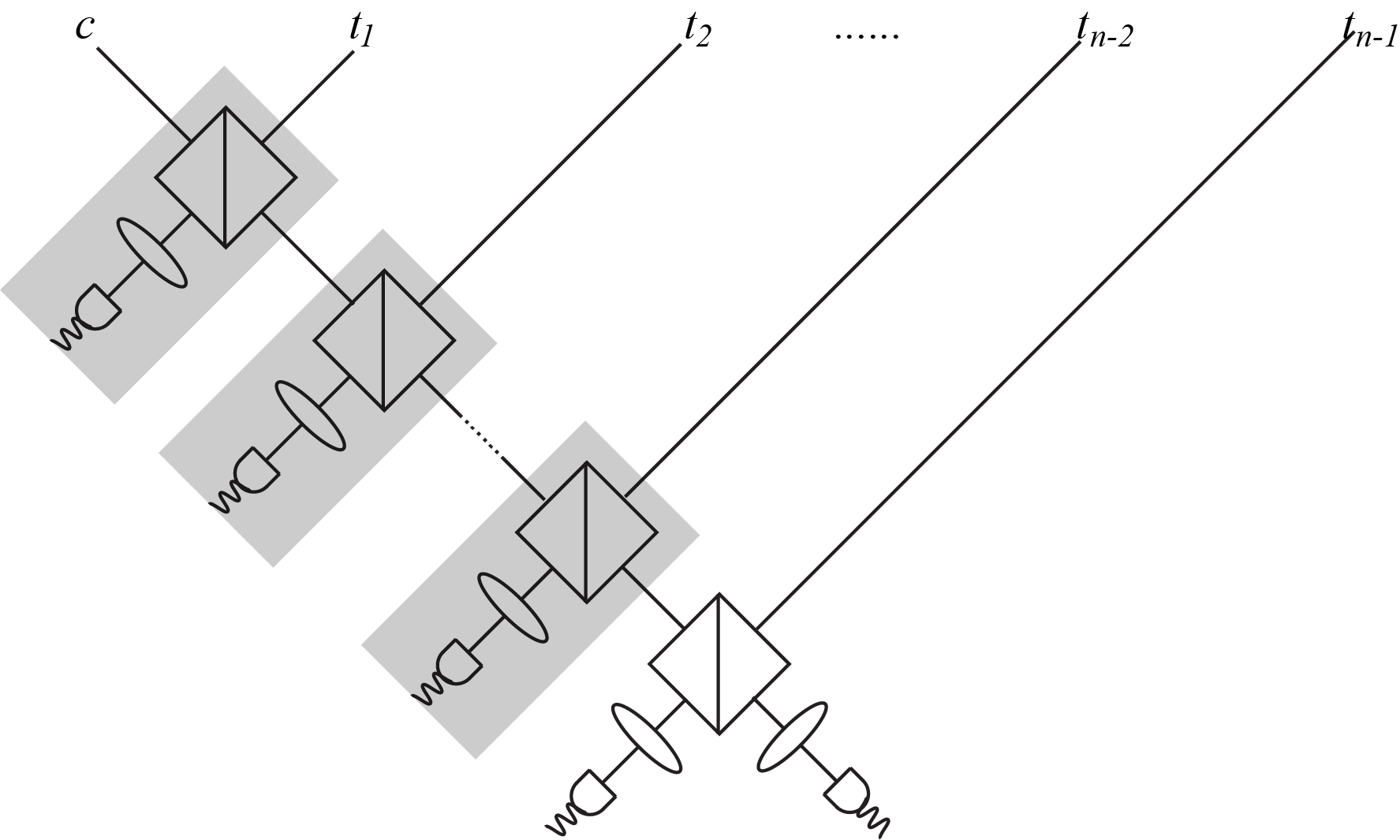}
    \caption{The LO circuit that performs $n-$qubit fusion, i.e., projects on $\frac{\ket{0}^n\pm\ket{1}^n}{\sqrt{2}}$ when successful and performs Pauli-Z measurements on failure. It is created by putting $(n-1)$ type-I fusion circuits (grey boxes) in succession, followed by a type-II circuit at the end.}
    \label{fig:LOnGHZProj}
\end{figure}

The LO circuit for $n$-qubit fusion can be created by concatenating $n-1$ type-I fusions and adding a type-II fusion in the end as shown in FIG.~\ref{fig:LOnGHZProj}. This circuit succeeds with probability $1/2^{n-1}$~\cite{MercedesThesis}.

 By doing an analysis similar to FIG.~\ref{fig:LOType2Other}, we can work out the LO circuits for projections on the remaining six 3-qubit GHZ states as shown in FIG.~\ref{fig:LOT3GHZ}. We now describe steps to calculate the LO circuits for projection onto all $2^n$ n-qubit GHZ states.
 \begin{enumerate}
     \item Divide the $2^n$ GHZ states into groups of two, such that the GHZ states in the same group differ only in phase. There will be $2^{n-1}$ such groups.
     \item  Pick one GHZ state from every group and write down the quantum circuit that projects onto it. For a group $i$, let this chosen GHZ state be $\ket{\psi_i}$
     \item Build the LO circuits corresponding to the quantum circuit for the $2^{n-1}$ groups.
 \end{enumerate}
 For a group $i$, the LO circuit obtained by following these steps projects onto $\ket{\psi_i}$ when each detector detects one photon and an even number of the detected photons have vertical polarization. If there are an odd number of vertically polarized detected photons, the LO circuit projects onto the other GHZ state in the same group. The failure of any of the $2^{n-1}$ LO circuits can be mapped as Pauli-Z measurement on the input qubits. 
 
Like the type-I and II fusions, the multi-qubit fusion can also be rotated and the corresponding graph rules for projection onto cluster states can be worked out. We don't discuss those circuits and the graph rules here due to the complexity of that analysis. We however believe that we have given the reader all the necessary tools to do that analysis when needed. 

\section{Summary}\label{sec:conclusions}
The stabilizer formalism is a very powerful tool to understand measurement-based and fusion-based quantum computation, design quantum error correcting codes and efficiently simulate Clifford operations on a classical computer. In this paper, we have given a complete guide to manipulating stabilizer states using Clifford unitary operations, and single and multi-qubit joint projective measurements (fusions). We have then reviewed cluster states using the stabilizer formalism. We have given graph theoretic rules for all three Pauli measurements on cluster states along with some widely used operations such as rotated BSMs (fusions) and local complementation. We have proved these graph theoretic rules in the Appendix.

We have reviewed the tableau method of encoding stabilizer generators developed and efficiently simulating Clifford operation on a classical computer using this method \cite{aaronson2004improved}. We discussed the new method that calculates tableau transformation rules for any single or multi-qubit Clifford operations using Kaurnaugh maps. We have discussed how the tableau transformation rules can be derived from the steps to transform stabilizers under Pauli measurements and extended the tableau formalism to include tableau transformation rules for Pauli-Y measurements. Based on the tableau method and refs.~\cite{MercedesThesis,van2003graphical,van2004efficient}, we have developed a MATLAB simulator, which performs Clifford operations and Pauli measurements on stabilizer states. 

Finally, we have reviewed linear optical (LO) measurements and unitary operations. We have in-depth discussed the origin of the probabilistic nature of LO entangling gates. We have discussed the type-I, type-II, and multi-qubit fusions used to grow the size of cluster states. We have also given a variation of the type-I fusion that performs Pauli-X measurements on the input qubits on failure and hence overcomes one of the major drawbacks in the type-I circuit. We have also developed a recipe to build a LO circuit corresponding to any quantum circuit with single and two-qubit gates.

\begin{acknowledgments}
This work was co-funded by the Army Research Office (ARO) MURI on Quantum Network Science, funded under grant number W911NF2110325, and the National Science Foundation (NSF) grant numbers CIF-1855879 and CNS-1955834. The authors also acknowledge the Center for Quantum Networks (CQN), cooperative agreement number EEC-1941583, for associated support, in useful interactions with faculty and students. The authors acknowledge Don Towsley for valuable discussions. AP thanks Yu Shi, Filip Rozpedek and Kenneth Goodenough for their feedback on the manuscript. SG thanks Babak Saif for being a patient bouncing board to discuss graphical rules for Pauli measurements.
\end{acknowledgments}


\bibliography{bibFile}

\appendix
\section{Derivation of the graph rule for local complementation on cluster states}
\label{apx:LC}

The unitary for local complementation on qubit $k$ is - 
\begin{align*}
U_{\text{LC}} &= e^{-i\frac{\pi}{4}X_k}\prod_{j \in \mathcal{N}_k}e^{i\frac{\pi}{4}Z_j}\\
&=\frac{I-iX_k}{\sqrt{2}}\prod_{j \in \mathcal{N}_k}\frac{I+iZ_j}{\sqrt{2}}
\end{align*}
Let 
\begin{align*}
    U_1 = \frac{I-iX}{\sqrt{2}} \qquad U_2 = \frac{I+iZ}{\sqrt{2}} 
\end{align*}
The Pauli operators transform under $U_1$ and $U_2$ as follows -

\begin{equation}
  \begin{split}
    X &\xrightarrow{{U_1}} X\\
    Y &\xrightarrow{{U_1}} Z \\
     Z &\xrightarrow{{U_1}} -Y\\
  \end{split}
\quad\quad
  \begin{split}
    X &\xrightarrow{{U_2}} -Y\\
    Y &\xrightarrow{{U_2}} X \\
    Z &\xrightarrow{{U_2}} Z\\
  \end{split}
\end{equation}

When local complementation is done on qubit $k$ of a cluster state, the only generators that are affected are those of the neighbors of qubit $k$. 
\begin{align}
\label{eq:A2}
    X_k\prod_{i\in \mathcal{N}_k}Z_i &\xrightarrow{{U_{LC}}} X_k\prod_{i\in \mathcal{N}_k}Z_i \\
    \label{eq:A3}
   X_j\prod_{i\in \mathcal{N}_j}Z_i &\xrightarrow{{U_{LC}}} Y_kY_j\prod_{i\in \mathcal{N}_j\setminus\{k\}}Z_i\quad \forall j\in \mathcal{N}_k
   \\  \label{eq:A4} & \equiv Z_kX_j\prod_{i\in \mathcal{N}_j\setminus\{k\}}Z_i\prod_{i\in \mathcal{N}_k\setminus\{j\}}Z_i\quad \forall j\in \mathcal{N}_k
\end{align}
We can transform~\ref{eq:A3} by multiplying it with~\ref{eq:A2} to get~\ref{eq:A4}. We can equivalently rewrite~\ref{eq:A4} as - 
\begin{align}
    X_j\prod_{i\in D(\mathcal{N}_j,\mathcal{N}_k)\setminus\{j\}}Z_i\quad \forall j\in \mathcal{N}_k \quad \forall j\in \mathcal{N}_k
    \label{eq:A5}
\end{align}
such that $D(\mathcal{N}_j,\mathcal{N}_k) = (\mathcal{N}_j\cup \mathcal{N}_k)\setminus(\mathcal{N}_j\cap \mathcal{N}_k)$. $D(\mathcal{N}_j,\mathcal{N}_k)$ is the symmetric difference of the two sets $\mathcal{N}_j$ and $\mathcal{N}_k$. Note that, $D(\mathcal{N}_j,\mathcal{N}_k)=D(\mathcal{N}_k,\mathcal{N}_j)$. $D(\mathcal{N}_j,\mathcal{N}_k)$ has only those qubits that were neighbors of either $j$ or $k$ but not both before local complementation i.e., if $j$ had an edge with another neighbor of $k$, it is deleted and if it didn't, a new edge is now added, inverting the local neighborhood of $k$. 
\section{Derivation of the graph rule for Pauli-X measurement on cluster states}
\label{apx:MeasureX}
If a qubit, say qubit 1 of a cluster state with stabilizer generators
\begin{align}
     \begin{split}
    &X_1 \prod_{i\in \mathcal{N}_1}Z_i\\
   &X_a \prod_{i\in \mathcal{N}_a}Z_i \quad\forall a\in \mathcal{N}_1\\
  &X_b \prod_{i\in \mathcal{N}_b}Z_i \quad\forall b\in \mathcal{N}\setminus\mathcal{N}_1\\
  \end{split}
  \label{eq:B1}
\end{align}
 where $\mathcal{N}$ is the set of all qubits in the cluster state, is measured in Pauli-X basis, the resulting generators calculated using the rules given in Section~\ref{subsec:StabMeasure} are -
\begin{align}
\begin{split}
    &X_1\\
    &\prod_{i\in \mathcal{N}_1}Z_i\\
     &X_aX_c\prod_{i\in \mathcal{N}_a}Z_i\prod_{i\in \mathcal{N}_c}Z_i\quad \forall a\in \mathcal{N}_1\setminus\{c\}\\
     &X_b \prod_{i\in \mathcal{N}_b}Z_i \quad\forall b\in \mathcal{N}\setminus\mathcal{N}_1\\
\end{split}
\label{eq:B2}
\end{align}
Note that, the generators that anti-commute with $X_1$ are the ones of the neighbors of qubit 1. Hence, we have replaced the generator, $S_c = X_c\prod_{i\in \mathcal{N}_c}Z_i$ for a $c\in \mathcal{N}_1$, which anti-commutes with $X_1$ with $X_1$ and multiplied every other generator anti-commuting generator with $S_c$.

We can rewrite the generators in ~\ref{eq:B2} as follows using the definition $D(\mathcal{N}_a,\mathcal{N}_c) = (\mathcal{N}_a\cup \mathcal{N}_c)\setminus(\mathcal{N}_a\cap \mathcal{N}_c)$ from Appendix~\ref{apx:LC} - 
\begin{align}
\begin{split}
    &X_1\\
    &\prod_{i\in \mathcal{N}_1}Z_i\\
    &X_aX_c\prod_{i\in D(\mathcal{N}_a,\mathcal{N}_c)} Z_i\forall a\in \mathcal{N}_1\setminus\{c\}\\
    &X_b \prod_{i\in \mathcal{N}_b}Z_i \quad\forall b\in \mathcal{N}\setminus\mathcal{N}_1\\
\end{split}
\label{eq:B3}
\end{align}
Now, if $a\in \mathcal{N}_1\setminus\{c\}$ also satisfies $a\in \mathcal{N}_c$, $D(\mathcal{N}_a,\mathcal{N}_c)$ contains both $a$ and $c$. As a result, $X_aX_c\prod_{i\in D(\mathcal{N}_a,\mathcal{N}_c)} = X_aX_cZ_aZ_c\prod_{i\in D(\mathcal{N}_a,\mathcal{N}_c)\setminus\{a,c\}}Z_i$. ~\ref{eq:B3} can be written as - 
\begin{align}
\begin{split}
    &X_1\\
    &\prod_{i\in \mathcal{N}_1}Z_i\\
     &X_aX_c\prod_{i\in D(\mathcal{N}_a,\mathcal{N}_c)}Z_i\quad \forall a\in \mathcal{N}_1\setminus(\mathcal{N}_c\cup\{c\})\\
     &-Y_eY_c\prod_{i\in D(\mathcal{N}_e,\mathcal{N}_c)\setminus\{e,c\}}Z_i\quad \forall e\in (\mathcal{N}_1\cap\mathcal{N}_c)\\
     &X_b \prod_{i\in \mathcal{N}_b}Z_i \quad\forall b\in \mathcal{N}\setminus\mathcal{N}_1\\
\end{split}
\label{eq:B3a}
\end{align}

In order to convert the stabilizer state in~\ref{eq:B3a} to a cluster state, we apply Hadamard on $c$,
\begin{align}
\begin{split}
    &X_1\\
    &X_c\prod_{i\in \mathcal{N}_1\setminus\{c\}}Z_i\\
     &X_aZ_c\prod_{i\in D(\mathcal{N}_a,\mathcal{N}_c)}Z_i\quad \forall a\in \mathcal{N}_1\setminus(\mathcal{N}_c\cup\{c\})\\
     &Y_eY_c\prod_{i\in D(\mathcal{N}_e,\mathcal{N}_c)\setminus\{e,c\}}Z_i\quad \forall e\in (\mathcal{N}_1\cap\mathcal{N}_c)\\
     &X_d X_c\prod_{i\in \mathcal{N}_d\setminus\{c\}}Z_i \quad\forall d\in \mathcal{N}_c\setminus\mathcal{N}_1\\
      &X_b \prod_{i\in \mathcal{N}_b}Z_i \quad\forall b\in \mathcal{N}\setminus(\mathcal{N}_c\cup\mathcal{N}_1)\\
\end{split}
\label{eq:B4}
\end{align}
Multiply the fourth and the fifth generators with the second generator in eq.~\ref{eq:B4} - 
\begin{align}
\begin{split}
    &X_1\\
    &X_c\prod_{i\in \mathcal{N}_1\setminus\{c\}}Z_i\\
     &X_aZ_c\prod_{i\in D(\mathcal{N}_a,\mathcal{N}_c)\setminus\{c\}}Z_i\quad \forall a\in \mathcal{N}_1\setminus(\mathcal{N}_c\cup\{c\})\\
     &X_eZ_c\prod_{i\in D(D(\mathcal{N}_e,\mathcal{N}_c),\mathcal{N}_1)\setminus\{e,c\}}Z_i\quad \forall e\in (\mathcal{N}_1\cap\mathcal{N}_c)\\
     &X_d\prod_{i\in D(\mathcal{N}_d,\mathcal{N}_1)}Z_i \quad\forall d\in \mathcal{N}_c\setminus\mathcal{N}_1\\
      &X_b \prod_{i\in \mathcal{N}_b}Z_i \quad\forall b\in \mathcal{N}\setminus(\mathcal{N}_c\cup\mathcal{N}_1)\\
\end{split}
\label{eq:B5}
\end{align}
This is a cluster state modulo single qubit Pauli-Z operators on all $a\in \mathcal{N}_1\setminus\{c\}, a \in\mathcal{N}_c$.

Now we show that ~\ref{eq:B5} is indeed same as the set of generators after applying the graph theoretic rule for Pauli-X measurements discussed in Section~\ref{sec:PauliX} i.e., ~\ref{eq:B5} has the same geometry as $((G.1).c).1)\setminus 1$, where $G$ is the graph topology of the cluster state we started with. We first write the generators of the cluster state - 
\begin{align}
     \begin{split}
    &X_1 Z_c\prod_{i\in \mathcal{N}_1\setminus\{c\}}Z_i\\
    &X_c Z_1\prod_{i\in \mathcal{N}_c\setminus\{1\}}Z_i\\
   &X_a Z_1\prod_{i\in\mathcal{N}_a\setminus\{1\}}Z_i \quad\forall a\in \mathcal{N}_1\setminus(\mathcal{N}_c\cup\{c\})\\
    &X_e Z_1Z_c\prod_{i\in\mathcal{N}_e\setminus\{1,c\}}Z_i \quad\forall e\in (\mathcal{N}_1\cap\mathcal{N}_c)\\
   &X_d Z_c\prod_{i\in\mathcal{N}_d\setminus\{c\}}Z_i \quad\forall d\in \mathcal{N}_c\setminus\mathcal{N}_1\\
  &X_b \prod_{i\in \mathcal{N}_b}Z_i \quad\forall b\in \mathcal{N}\setminus(\mathcal{N}_1\cup\mathcal{N}_c)\\
  \end{split}
\end{align}
Here, we have divided the qubits into different categories based on if they are neighbors with qubits $1$, $c$, both, or none. Using Appendix~\ref{apx:LC}, the generators after local complementation on qubit 1 are - 
\begin{align}
     \begin{split}
   &X_1 Z_c\prod_{i\in \mathcal{N}_1\setminus\{c\}}Z_i\\
    &X_c Z_1\prod_{i\in D(\mathcal{N}_c,\mathcal{N}_1)\setminus\{1,c\}}Z_i\\
   &X_a Z_cZ_1\prod_{i\in D(\mathcal{N}_a,\mathcal{N}_1)\setminus\{1,c,a\}}Z_i \quad\forall a\in \mathcal{N}_1\setminus(\mathcal{N}_c\cup\{c\})\\
    &X_e  Z_1\prod_{i\in D(\mathcal{N}_e,\mathcal{N}_1)\setminus\{1,e\}}Z_i \quad\forall e\in (\mathcal{N}_1\cap\mathcal{N}_c)\\
   &X_d Z_c\prod_{i\in\mathcal{N}_d\setminus\{c\}}Z_i \quad\forall d\in \mathcal{N}_c\setminus\mathcal{N}_1\\
  &X_b \prod_{i\in \mathcal{N}_b}Z_i \quad\forall b\in \mathcal{N}\setminus(\mathcal{N}_1\cup\mathcal{N}_c)\\
  \end{split}
\end{align}
After this step, the new neighborhood of $c$, $\mathcal{N'}_c = D(\mathcal{N}_c,\mathcal{N}_1)\setminus\{c\}$. Let $D(\mathcal{N}_c',\mathcal{N}_x) = (\mathcal{N'}_c\cup \mathcal{N}_x)\setminus(\mathcal{N'}_c\cap \mathcal{N}_x)$. As a result, local complementation on $c\in \mathcal{N}_1$ yields 
\begin{align}
     \begin{split}
  &X_1 Z_c\prod_{i\in D(\mathcal{N}_1,\mathcal{N}_c')\setminus\{1\}}Z_i\\
    &X_c Z_1\prod_{i\in D(\mathcal{N}_c,\mathcal{N}_1)\setminus\{1,c\}}Z_i\\
   &X_a Z_c\prod_{i\in D(D(\mathcal{N}_a,\mathcal{N}_1),\mathcal{N}_c')\setminus\{c,a\}}Z_i \quad\forall a\in \mathcal{N}_1\setminus(\mathcal{N}_c\cup\{c\})\\
    &X_e  Z_1\prod_{i\in D(\mathcal{N}_e,\mathcal{N}_1)\setminus\{e\}}Z_i \quad\forall e\in (\mathcal{N}_1\cap\mathcal{N}_c)\\
   &X_d Z_1Z_c\prod_{i\in D(\mathcal{N}_d,D(\mathcal{N}_c,\mathcal{N}_1))\setminus\{1,c,d\}}Z_i \quad\forall d\in \mathcal{N}_c\setminus\mathcal{N}_1\\
  &X_b \prod_{i\in \mathcal{N}_b}Z_i \quad\forall b\in \mathcal{N}\setminus(\mathcal{N}_1\cup\mathcal{N}_c)\\
  \end{split}
\end{align}
$D(\mathcal{N}_1,\mathcal{N}_c') = \mathcal{N}_c$ and $D(D(\mathcal{N}_a,\mathcal{N}_1),\mathcal{N}_c')= D(\mathcal{N}_a,\mathcal{N}_c)$. The simplified expressions for the generators then are -
\begin{align}
     \begin{split}
  &X_1 Z_c\prod_{i\in \mathcal{N}_c}Z_i\\
    &X_c Z_1\prod_{i\in D(\mathcal{N}_c,\mathcal{N}_1)\setminus\{1,c\}}Z_i\\
   &X_a Z_c\prod_{i\in D(\mathcal{N}_a,\mathcal{N}_c)\setminus\{c\}}Z_i \quad\forall a\in \mathcal{N}_1\setminus(\mathcal{N}_c\cup\{c\})\\
    &X_e  Z_1\prod_{i\in D(\mathcal{N}_e,\mathcal{N}_1)\setminus\{e\}}Z_i \quad\forall e\in (\mathcal{N}_1\cap\mathcal{N}_c)\\
   &X_d Z_1Z_c\prod_{i\in D(\mathcal{N}_d,D(\mathcal{N}_c,\mathcal{N}_1))\setminus\{1,c,d\}}Z_i \quad\forall d\in \mathcal{N}_c\setminus\mathcal{N}_1\\
  &X_b \prod_{i\in \mathcal{N}_b}Z_i \quad\forall b\in \mathcal{N}\setminus(\mathcal{N}_1\cup\mathcal{N}_c)\\
  \end{split}
  \end{align}
After another local complementation on 1 with neighborhood $\mathcal{N}_c$, we get - 
\begin{align}
     \begin{split}
  &X_1 Z_c\prod_{i\in \mathcal{N}_c}Z_i\\
    &X_c Z_1\prod_{i\in D(D(\mathcal{N}_c,\mathcal{N}_1),\mathcal{N}_c)\setminus\{1,c\}}Z_i\\
   &X_a Z_c\prod_{i\in D(\mathcal{N}_a,\mathcal{N}_c)\setminus\{c\}}Z_i \quad\forall a\in \mathcal{N}_1\setminus(\mathcal{N}_c\cup\{c\})\\
    &X_e  Z_1\prod_{i\in D(D(\mathcal{N}_e,\mathcal{N}_1),\mathcal{N}_c)}Z_i \quad\forall e\in (\mathcal{N}_1\cap\mathcal{N}_c)\\
   &X_d Z_1\prod_{i\in D(D(\mathcal{N}_d,D(\mathcal{N}_c,\mathcal{N}_1)),\mathcal{N}_c)\setminus\{1,c,d\}}Z_i \quad\forall d\in \mathcal{N}_c\setminus\mathcal{N}_1\\
  &X_b \prod_{i\in \mathcal{N}_b}Z_i \quad\forall b\in \mathcal{N}\setminus(\mathcal{N}_1\cup\mathcal{N}_c)\\
  \end{split}
\end{align}
Using $D(D(c,1),c)= \mathcal{N}_1$ and $ D(D(\mathcal{N}_d,D(\mathcal{N}_c,\mathcal{N}_1)),\mathcal{N}_c)\setminus\{1,c,d\}=D(\mathcal{N}_d,\mathcal{N}_1)$ - 
\begin{align}
     \begin{split}
  &X_1 Z_c\prod_{i\in \mathcal{N}_c}Z_i\\
    &X_c Z_1\prod_{i\in \mathcal{N}_1\setminus\{c\}}Z_i\\
   &X_a Z_c\prod_{i\in D(\mathcal{N}_a,\mathcal{N}_c)\setminus\{c\}}Z_i \quad\forall a\in \mathcal{N}_1\setminus(\mathcal{N}_c\cup\{c\})\\
    &X_e  Z_1\prod_{i\in D(D(\mathcal{N}_e,\mathcal{N}_1),\mathcal{N}_c)}Z_i \quad\forall e\in (\mathcal{N}_1\cap\mathcal{N}_c)\\
   &X_d Z_1\prod_{i\in D(\mathcal{N}_d,\mathcal{N}_1)}Z_i \quad\forall d\in \mathcal{N}_c\setminus\mathcal{N}_1\\
  &X_b \prod_{i\in \mathcal{N}_b}Z_i \quad\forall b\in \mathcal{N}\setminus(\mathcal{N}_1\cup\mathcal{N}_c)\\
  \end{split}
  \label{eq:B12}
\end{align}
After performing Pauli-Z measurement on qubit 1, we get - 
\begin{align}
     \begin{split}
  &Z_1 \\
    &X_c \prod_{i\in \mathcal{N}_1\setminus\{c\}}Z_i\\
   &X_a Z_c\prod_{i\in D(\mathcal{N}_a,\mathcal{N}_c)\setminus\{c\}}Z_i \quad\forall a\in \mathcal{N}_1\setminus(\mathcal{N}_c\cup\{c\})\\
    &X_e \prod_{i\in D(D(\mathcal{N}_e,\mathcal{N}_1),\mathcal{N}_c)}Z_i \quad\forall e\in (\mathcal{N}_1\cap\mathcal{N}_c)\\
   &X_d \prod_{i\in D(\mathcal{N}_d,\mathcal{N}_1)}Z_i \quad\forall d\in \mathcal{N}_c\setminus\mathcal{N}_1\\
  &X_b \prod_{i\in \mathcal{N}_b}Z_i \quad\forall b\in \mathcal{N}\setminus(\mathcal{N}_1\cup\mathcal{N}_c)\\
  \end{split}
\end{align}
This is Hadamard equivalent ($H_1$) to the generators after Pauli-X measurement on qubit 1 followed by Hadamard on $c$ as shown in~\ref{eq:B5}. 
\section{Derivation of the graph rule for Pauli-Y measurement on cluster states}
\label{apx:MeasureY}
If say, qubit 1 of a cluster state with generators given in eq.~\ref{eq:B1} is measured in Pauli-Y basis, the output generators are -
\begin{align}
\begin{split}
    &Y_1\\
     &X_cZ_1\prod_{i\in \mathcal{N}_c\setminus\{1\}}Z_iX_1Z_c\prod_{i\in \mathcal{N}_1\setminus\{c\}}Z_i\quad \forall c\in \mathcal{N}_1\\
     &=Y_cY_1\prod_{i\in \mathcal{N}_c\setminus\{1\}}Z_i\prod_{i\in \mathcal{N}_1\setminus\{c\}}Z_i\quad \forall c\in \mathcal{N}_1\\
     &X_b \prod_{i\in \mathcal{N}_b}Z_i \quad\forall b\in \mathcal{N}\setminus\mathcal{N}_1\\
\end{split}
\label{eq:C2}
\end{align}
Here, we have replaced the generator $S_1 = X_1\prod_{i\in \mathcal{N}_1}Z_i$ with $Y_1$ and multiplied every other generator anti-commuting generator with $S_1$ as per the rules from Section~\ref{subsec:StabMeasure}. After eliminating $Y_1$ and applying a $P^{\dagger}$ gate to all the neighboring qubits of 1, we get 
\begin{align}
\begin{split}
    &Y_1\\
     &X_c\prod_{i\in D(\mathcal{N}_1,\mathcal{N}_c)\setminus\{1,c\}}Z_i\quad \forall c\in \mathcal{N}_1\\
     &X_b \prod_{i\in \mathcal{N}_b}Z_i \quad\forall b\in \mathcal{N}\setminus\mathcal{N}_1\\
\end{split}
\label{eq:C3}
\end{align}
From the equation above it is clear that the local neighborhood of qubit 1 has been inverted due to Pauli-Y measurement.
\section{Stabilizer transformation rules for stabilizer measurements}
\label{apx:StabMeasure}
The density matrix of an $n$-qubit stabilizer pure state with generators $\mathcal{G} = \langle S_1,S_2,\dots,S_n\rangle$ is given by~\cite{aaronson2004improved} -
\begin{equation}\label{eq:E1}
\begin{split}
    \rho &= \frac{1}{2^n}\prod_{i=1}^n(I+S_i)\\
    &=\frac{1}{2^n}\bigg(I+\sum_{i=1}^nS_i+\prod_{i,j}S_iS_j+\dots+\prod_{i=1}^nS_i\bigg)
   \end{split}
\end{equation}
Here, we wish to calculate the density matrix of the state in~\ref{eq:E1} after the measurement of  a single qubit Pauli measurement operator $M$. The projectors for the measurement operator $M$ are $\Pi_{m} = \frac{I+ mM}{2}$, s.t., $m\in\{\pm 1\}$ corresponding to the +1 and -1 eigenstates, respectively~\cite{Nielsen2010,gottesman1998heisenberg}. The density matrix transforms as follows due to the measurement -
\begin{align}
    \rho' = \frac{\Pi_m\rho\Pi_m^{\dagger}}{2^n\Tr(\Pi_m^{\dagger}\Pi_m\rho)}
    \label{eq:E2}
\end{align}
Here, we derive the stabilizer transformation rules assuming $M$ anti-commutes with two of $n$ generators - $S_1$ and $S_2$. But this derivation can be easily generalized for an arbitrary number of generators. We now expand Eq.~\ref{eq:E2} as follows -
\begin{align}
     \rho' = \frac{1}{2^n\Tr(\Pi_m^{\dagger}\Pi_m\rho)}\bigg(\frac{I+ mM}{2}\bigg)\bigg(\prod_{i=1}^n(I+S_i)\bigg)\bigg(\frac{I+ mM^{\dagger}}{2}\bigg)
     \label{eq:E3}
\end{align}
For a generator $S_i$
\begin{equation}\label{eq:E4}
\begin{split}
    &\bigg(\frac{I+ mM}{2}\bigg)S_i\bigg(\frac{I+ mM^{\dagger}}{2}\bigg) \\&= \begin{dcases*}
        \frac{S_i(I+ mM)}{2}, & if $ [M,S_i]=0$,\\
        0, & if $\{M,S_i\}=0$. 
        \end{dcases*}
\end{split}
\end{equation}
and $\Tr(\Pi_m^{\dagger}\Pi_m\rho)=1/2$.
Now, we rewrite Eq.~\ref{eq:E3} using Eqs.~\ref{eq:E1} and ~\ref{eq:E4}
\begin{equation}\label{eq:E5}
\begin{split}
     \rho' &= \frac{1}{2^{n-1}}\bigg(\frac{I+ mM}{2}\bigg)\bigg(I+\sum_{i=1}^nS_i+\prod_{i,j}S_iS_j+\dots
     \\&+\prod_{i=1}^nS_i\bigg)\bigg(\frac{I+ mM}{2}\bigg)\\
    &= \frac{1}{2^{n-1}}\bigg(\frac{I+mM}{2}\bigg)\bigg(I+\sum_{i=3}^nS_i+S_1S_2+\prod_{\mathclap{\substack{i,j\\ i,j\notin \{1,2\}}}}S_iS_j+
     \\&S_1S_2\sum_{i=3}^{n}S_i+\prod_{\mathclap{\substack{i,j,k\\ i,j,k\notin \{1,2\}}}}S_iS_jS_k+\dots+\prod_{i=1}^nS_i\bigg)
\end{split}
\end{equation}
i.e., the only surviving terms are the ones with even number of the anti-commuting generators $(S_1,S_2)$.
We can simply $\rho'$ as
\begin{align}
    \rho' &= \bigg(\frac{I+ mM}{2}\bigg)\bigg(\frac{I+ S_1S_2}{2}\bigg)\prod_{i=3}^{n}\bigg(\frac{I+ S_i}{2}\bigg)
\end{align}

As a result, the generators after measurement are - $\langle mM, S_1S_2, S_3,S_4,\dots, S_n\rangle$.

We now generalize the stabilizer transformation rules to include an $l$-qubit stabilizer joint-measurement. For this $l$-qubit measurement operator $M=\langle M_1,M_2,\dots, M_l\rangle$. This measurement operator has $2^l$ eigenstates. The corresponding $2^l$ projectors are 
\begin{align*}
    \Pi_{\mathbf{m}} = \prod_{k=1}^{l}\bigg(\frac{I+ m_kM_k}{2}\bigg)
\end{align*}
Here, $\mathbf{m} = (m_1,m_2,\dots, m_l)$ is a vector of length $l$ made of the measurement outcomes $m_k = \{\pm 1\}$.

Let the set of generators $\mathcal{G}_{i}\subseteq \mathcal{G}$ be the set of generators that anti-commute with the operator $M_i$. Let $\mathcal{G}' = \mathcal{G}\setminus(\mathcal{G}_{1}\cup\mathcal{G}_{2}\cup\dots\cup\mathcal{G}_{l})$ be the set of generators that commute with all operators in $M$. The density matrix post-measurement - 
\begin{equation} 
\begin{split}
     \rho' &= \frac{1}{2^n}\bigg[\prod_{k=1}^{l}\bigg(\frac{I+  m_kM_k}{2}\bigg)\bigg] \bigg(I+\sum_{i=1}^nS_i+\prod_{i,j}S_iS_j+
     \\& \dots+\prod_{i=1}^nS_i)\bigg)\bigg[\prod_{k=1}^{l}\bigg(\frac{I+  m_kM_k}{2}\bigg)\bigg]\\
    &= \frac{1}{2^n}\bigg[\prod_{k=1}^{l}\bigg(\frac{I+  m_kM_k}{2}\bigg)\bigg] \bigg(I+\sum_{i\in\mathcal{G}'}S_i+\sum_{k=1}^{l}\sum_{S_i,S_j\in \mathcal{G}_{k}} S_iS_j
    \\&+\prod_{\mathclap{\substack{S_i,S_j\\ S_i,S_j\in\mathcal{G}'}}}S_iS_j+ \sum_{k=1}^{l}\sum_{S_i,S_j\in \mathcal{G}_{k}} S_iS_j\big(\sum_{S_{i'}\in\mathcal{G}'}S_{i'}\big)+\prod_{\mathclap{\substack{S_i,S_j,S_p\\ S_i,S_j,S_p\in\mathcal{G}'}}}S_iS_jS_p+
    \\&+\dots+\prod_{k=1}^l\quad\prod_{\mathclap{\substack{S_{ik}\in\mathcal{G}_k\\\forall S_{jk}\in\mathcal{G}_k\setminus\{S_{ik}\}}}}S_{ik}S_{jk}\big(\prod_{S_i\in \mathcal{G}'}S_i\big)\bigg)
\end{split}
\label{eq:E7}
\end{equation}
The density matrix can be simplified as -
\begin{equation} 
\begin{split}
    \rho'&= \bigg[\prod_{k=1}^{l}\bigg(\frac{I+  m_kM_k}{2}\bigg)\bigg] \bigg[\prod_{k=1}^l\quad\prod_{\mathclap{\substack{S_{ik}\in\mathcal{G}_k\\ \forall S_{jk}\in\mathcal{G}_k\setminus\{S_{ik}\}}}}\bigg(\frac{1+S_{ik}S_{jk}}{2}\bigg)\bigg]
    \\&\times\prod_{S_i\in\mathcal{G}'}\bigg(\frac{1+S_i}{2}\bigg)
\end{split}
\label{eq:E8}
\end{equation}
i.e., the corresponding generator are - $\mathcal{G}_M\cup\mathcal{G}'_{1}\cup\mathcal{G}'_{2}\cup\dots\cup\mathcal{G}'_{l}\cup\mathcal{G}'$ s.t $\mathcal{G}_{M}= \{m_1M_1,m_2M_2,\dots,m_lM_l\}$. We derive $\mathcal{G}'_{k}, \forall k\in\{1,2,\dots,l\}$ from $\mathcal{G}_k = \{S_{1k},S_{1k},\dots,\}$, the set of generators that anti-commute with $M_k$, by multiplying a (randomly chosen) generator in $\mathcal{G}_k$ and multiplying remaining elements of $\mathcal{G}_k$ with it. -  $\mathcal{G}_{k'} = \{S_{1k}S_{2k},S_{1k}S_{3k},\dots\}$.

\section{Derivation of the graph rule for two-qubit fusions on cluster states}
\label{apx:fusion}
If we have two cluster states with stabilizer generators 
\begin{align}
     \begin{split}
    &X_1 \prod_{i\in \mathcal{N}_1}Z_i\\
   &X_a \prod_{i\in \mathcal{N}_a}Z_i \quad\forall a\in \mathcal{N}_1\\
  &X_b \prod_{i\in \mathcal{N}_b}Z_i \quad\forall b\in \mathcal{N}\setminus\mathcal{N}_1\\
  \end{split}
  \label{eq:D1}
\end{align}
and 
\begin{align}
     \begin{split}
    &X_{1'} \prod_{i\in \mathcal{N}_{1'}}Z_i\\
   &X_{a'} \prod_{i\in \mathcal{N}_{a'}}Z_i \quad\forall a'\in \mathcal{N}_{1'}\\
  &X_{b'} \prod_{i\in \mathcal{N}_{b'}}Z_i \quad\forall b'\in \mathcal{N}\setminus\mathcal{N}_{1'}\\
  \end{split}
  \label{eq:D1'}
\end{align}
and WLOG fuse qubits 1 and 1' by projecting them on $\frac{\ket{00}\pm\ket{11}}{\sqrt{2}}$, we can write down the stabilizer generator of the resultant state, given the fusion is successful, using the multi-qubit joint stabilizer measurement rules given in Section~\ref{subsec:GHZproj1}. This projection jointly measures the operators $X_1X_{1'}$ and $Z_1Z_{1'}$.
Rewriting the generators such that exactly one generator anti-commutes with $X_1X_{1'}$ and $Z_1Z_{1'}$, each.
\begin{align}
     \begin{split}
    &X_1X_{1'} \prod_{i\in \mathcal{N}_1}Z_i\prod_{i\in \mathcal{N}_{1'}}Z_i\\
   &X_c \prod_{i\in \mathcal{N}_c}Z_i \quad c\in \mathcal{N}_1\\
    &X_cX_a \prod_{i\in D(\mathcal{N}_a,\mathcal{N}_c)}Z_i \quad\forall a\in \mathcal{N}_1\setminus\{c\}\\
  &X_b \prod_{i\in \mathcal{N}_b}Z_i \quad\forall b\in \mathcal{N}\setminus\mathcal{N}_1\\
  \end{split}
  \label{eq:D2}
\end{align}
and 
\begin{align}
     \begin{split}
    &X_{1'} \prod_{i\in \mathcal{N}_{1'}}Z_i\\
   &X_cX_{a'} Z_1Z_{1'}\prod_{i\in D(\mathcal{N}_{a'},\mathcal{N}_c)\setminus\{1,{1'}\}}Z_i \quad\forall a'\in \mathcal{N}_{1'}\\
  &X_{b'} \prod_{i\in \mathcal{N}_{b'}}Z_i \quad\forall b'\in \mathcal{N}\setminus\mathcal{N}_{1'}\\
  \end{split}
  \label{eq:D2'}
\end{align}

After the measurement - 
\begin{align}
     \begin{split}
    &X_1X_{1'} \prod_{i\in \mathcal{N}_1}Z_i\prod_{i'\in \mathcal{N}_{1'}}Z_{i'}\\
   &X_1X_{1'}\\
    &X_cX_a \prod_{i\in D(\mathcal{N}_a,\mathcal{N}_c)}Z_i \quad\forall a\in \mathcal{N}_1\setminus\{c\}\\
  &X_b \prod_{i\in \mathcal{N}_b}Z_i \quad\forall b\in \mathcal{N}\setminus\mathcal{N}_1\\
    &Z_1Z_{1'}\\
   &X_cX_{a'}Z_1Z_{1'} \prod_{i\in D(\mathcal{N}_{a'},\mathcal{N}_c)\setminus\{1,{1'}\}}Z_i \quad\forall a'\in \mathcal{N}_{1'}\\
  &X_{b'} \prod_{i\in \mathcal{N}_{b'}}Z_i \quad\forall b'\in \mathcal{N}\setminus\mathcal{N}_{1'}\\
  \end{split}
  \label{eq:D3}
\end{align}

Eliminating $X_1X_{1'}$, and $Z_1Z_{1'}$ from the generators - 
\begin{align}
     \begin{split}
   &X_1X_{1'}\\
    &Z_1Z_{1'}\\
   &\prod_{i\in \mathcal{N}_1}Z_i\prod_{i'\in \mathcal{N}_{1'}}Z_{i'}\\
    &X_cX_a \prod_{i\in D(\mathcal{N}_a,\mathcal{N}_c)}Z_i \quad\forall a\in \mathcal{N}_1\setminus\{c\}\\
  &X_b \prod_{i\in \mathcal{N}_b}Z_i \quad\forall b\in \mathcal{N}\setminus\mathcal{N}_1\\
   &X_cX_{a'} \prod_{i\in D(\mathcal{N}_{a'},\mathcal{N}_c)\setminus\{1,{1'}\}}Z_i \quad\forall a'\in \mathcal{N}_{1'}\\
  &X_{b'} \prod_{i\in \mathcal{N}_{b'}}Z_i \quad\forall b'\in \mathcal{N}\setminus\mathcal{N}_{1'}\\
  \end{split}
  \label{eq:D4}
\end{align}
This state is not a cluster state. We need to apply a Hadamard on one of the qubits in the neighborhood of qubit 1 or 1'. WLOG, we apply Hadamard on qubit $c\in\mathcal{N}_c$. We rewrite Eq.~\ref{eq:D4} by splitting the qubits into into three categories - neighbors of both 1 and $c$, neighbors of 1 but not that of $c$, and neighbors of $c$ but not that of $1$.
\begin{align}
     \begin{split}
   &X_1X_{1'}\\
    &Z_1Z_{1'}\\
   &Z_c\prod_{i\in \mathcal{N}_1\setminus\{c\}}Z_i\prod_{i'\in \mathcal{N}_{1'}}Z_{i'}\\
    &X_cX_eZ_c \prod_{i\in D(\mathcal{N}_e,\mathcal{N}_c)\setminus\{c\}}Z_i \quad\forall e\in (\mathcal{N}_1\cap\mathcal{N}_c)\\
    &X_cX_a \prod_{i\in D(\mathcal{N}_a,\mathcal{N}_c)}Z_i \quad\forall a\in \mathcal{N}_1\setminus(\mathcal{N}_c\cup\{c\})\\
  &X_d Z_c\prod_{i\in \mathcal{N}_d\setminus\{c\}}Z_i \quad\forall d\in \mathcal{N}_{c}\setminus\mathcal{N}_1\\
  &X_b \prod_{i\in \mathcal{N}_b}Z_i \quad\forall b\in \mathcal{N}\setminus\mathcal{N}_1\\
   &X_cX_{a'} \prod_{i\in D(\mathcal{N}_{a'},\mathcal{N}_c)\setminus\{1,{1'}\}}Z_i \quad\forall a'\in \mathcal{N}_{1'}\\
  &X_{b'} \prod_{i\in \mathcal{N}_{b'}}Z_i \quad\forall b'\in \mathcal{N}\setminus\mathcal{N}_{1'}\\
  \end{split}
  \label{eq:D5}
\end{align}

After $H_c$
\begin{align}
     \begin{split}
   &X_1X_{1'}\\
    &Z_1Z_{1'}\\
   &X_c\prod_{i\in \mathcal{N}_1\setminus\{c\}}Z_i\prod_{i'\in \mathcal{N}_{1'}}Z_{i'}\\
    &Z_cX_eX_c \prod_{i\in D(\mathcal{N}_e,\mathcal{N}_c)\setminus\{c\}}Z_i \quad\forall e\in (\mathcal{N}_1\cap\mathcal{N}_c)\\
    &Z_cX_a \prod_{i\in D(\mathcal{N}_a,\mathcal{N}_c)}Z_i \quad\forall a\in \mathcal{N}_1\setminus(\mathcal{N}_c\cup\{c\})\\
  &X_d X_c\prod_{i\in \mathcal{N}_d\setminus\{c\}}Z_i \quad\forall d\in \mathcal{N}_{c}\setminus\mathcal{N}_1\\
  &X_b \prod_{i\in \mathcal{N}_b}Z_i \quad\forall b\in \mathcal{N}\setminus\mathcal{N}_1\\
   &Z_cX_{a'} \prod_{i\in D(\mathcal{N}_{a'},\mathcal{N}_c)\setminus\{1,{1'}\}}Z_i \quad\forall a'\in \mathcal{N}_{1'}\\
  &X_{b'} \prod_{i\in \mathcal{N}_{b'}}Z_i \quad\forall b'\in \mathcal{N}\setminus\mathcal{N}_{1'}\\
  \end{split}
  \label{eq:D6}
\end{align}
Multiplying the fourth and sixth generators with the third generator to eliminate $X_c$ -
\begin{align}
     \begin{split}
   &X_1X_{1'}\\
    &Z_1Z_{1'}\\
   &X_c\prod_{i\in \mathcal{N}_1\setminus\{c\}}Z_i\prod_{i'\in \mathcal{N}_{1'}}Z_{i'}\\
    &X_eZ_c \prod_{i\in D(\mathcal{N}_1,D(\mathcal{N}_e,\mathcal{N}_c))\setminus\{c\}}Z_i \prod_{i'\in \mathcal{N}_{1'}}Z_{i'}\quad\forall e\in (\mathcal{N}_1\cap\mathcal{N}_c)\\
    &Z_cX_a \prod_{i\in D(\mathcal{N}_a,\mathcal{N}_c)}Z_i \quad\forall a\in \mathcal{N}_1\setminus(\mathcal{N}_c\cup\{c\})\\
  &X_d \prod_{i\in D(\mathcal{N}_1,\mathcal{N}_d)}Z_i\prod_{i'\in \mathcal{N}_{1'}}Z_{i'} \quad\forall d\in \mathcal{N}_{c}\setminus\mathcal{N}_1\\
  &X_b \prod_{i\in \mathcal{N}_b}Z_i \quad\forall b\in \mathcal{N}\setminus\mathcal{N}_1\\
   &Z_cX_{a'} \prod_{i\in D(\mathcal{N}_{a'},\mathcal{N}_c)\setminus\{1,{1'}\}}Z_i \quad\forall a'\in \mathcal{N}_{1'}\\
  &X_{b'} \prod_{i\in \mathcal{N}_{b'}}Z_i \quad\forall b'\in \mathcal{N}\setminus\mathcal{N}_{1'}\\
  \end{split}
  \label{eq:D7}
\end{align}
From the equation above, it is easy to see that all the qubits in $\mathcal{N}_{1'}$ are now neighbors of $\mathcal{N}_c\cup\{c\}$ and from Eq.~\ref{eq:B5}, the neighborhood of 1 has changed to $((G.1).c).1$. Consequently, the cluster state described by the generators above is same as the one obtained using rules in Table.~\ref{tab:fusion_rules}.

We now write the generators for when the projection on $\frac{\ket{0i}\pm\ket{1(-i)}}{\sqrt{2}}$. This projection jointly measures the operators $X_1Z_{1'}$ and $Z_1Y_{1'}$. Rewriting the generators of the two cluster state from Eq.~\ref{eq:D1}-\ref{eq:D1'} such that exactly one generator anti-commutes with each of the measurement operators.
\begin{align}
     \begin{split}
    &X_1 \prod_{i\in \mathcal{N}_1}Z_iX_{c'} \prod_{i\in \mathcal{N}_{c'}}Z_i \quad {c'}\in \mathcal{N}_1\\
   &X_c \prod_{i\in \mathcal{N}_c}Z_i \quad c\in \mathcal{N}_1\\
   &X_a X_c\prod_{i\in D(\mathcal{N}_a,\mathcal{N}_c)}Z_i \quad\forall a\in \mathcal{N}_1\setminus\{ c\}\\
  &X_b \prod_{i\in \mathcal{N}_b}Z_i \quad\forall b\in \mathcal{N}\setminus\mathcal{N}_1\\
    &X_{1'}X_cX_{c'} \prod_{i\in \mathcal{N}_{1'}}Z_i\prod_{i\in \mathcal{N}_c}Z_i\prod_{i\in \mathcal{N}_{c'}}Z_i  \quad c' \in \mathcal{N}_{1'}\\
   &X_{c'} \prod_{i\in \mathcal{N}_{c'}}Z_i \quad {c'}\in \mathcal{N}_1\\
   &X_{a'} X_{c'}\prod_{i\in D(\mathcal{N}_{a'},\mathcal{N}_{c'})}Z_i \quad\forall {c'}\in \mathcal{N}_1'\setminus\{ c'\}\\
  &X_{b'} \prod_{i\in \mathcal{N}_{b'}}Z_i \quad\forall b'\in \mathcal{N}\setminus\mathcal{N}_{1'}\\
  \end{split}
  \label{eq:D41}
\end{align}
The second and the sixth generators anti-commute with $X_1Z_{1'}$ and $Z_1Y_{1'}$, respectively. After the measurement the generators are, 
\begin{align}
     \begin{split}
     &Z_1Y_{1'}\\
    &X_1 Z_{1'}\\
    &X_{c'}\prod_{i\in \mathcal{N}_1}Z_i \prod_{j\in \mathcal{N}_{c'}\setminus\{1'\}}Z_j \quad {c'}\in \mathcal{N}_1\\
   &X_a X_c\prod_{i\in D(\mathcal{N}_a,\mathcal{N}_c)}Z_i \quad\forall a\in \mathcal{N}_1\setminus\{ c\}\\
  &X_b \prod_{i\in \mathcal{N}_b}Z_i \quad\forall b\in \mathcal{N}\setminus\mathcal{N}_1\\
    &X_cY_{c'} \prod_{i\in D(\mathcal{N}_{1'},\mathcal{N}_{c'})\setminus\{1',c'\}}Z_i\prod_{j\in \mathcal{N}_c\setminus\{1\}}Z_j \quad c' \in \mathcal{N}_{1'},c\in \mathcal{N}_{1}\\
  &X_{a'} X_{c'}\prod_{i\in D(\mathcal{N}_{a'},\mathcal{N}_{c'})}Z_i \quad\forall {c'}\in \mathcal{N}_1'\setminus\{ c'\}\\
  &X_{b'} \prod_{i\in \mathcal{N}_{b'}}Z_i \quad\forall b'\in \mathcal{N}\setminus\mathcal{N}_{1'}\\
  \end{split}
  \label{eq:D42}
\end{align}
Multiplying the sixth generator with the second generator - 
\begin{align}
     \begin{split}
    &Z_1Y_{1'}\\
    &X_1 Z_{1'}\\
    &X_{c'}\prod_{i\in \mathcal{N}_1}Z_i \prod_{j\in \mathcal{N}_{c'}\setminus\{1'\}}Z_j \quad {c'}\in \mathcal{N}_1\\
   &X_a X_c\prod_{i\in D(\mathcal{N}_a,\mathcal{N}_c)}Z_i \quad\forall a\in \mathcal{N}_1\setminus\{ c\}\\
  &X_b \prod_{i\in \mathcal{N}_b}Z_i \quad\forall b\in \mathcal{N}\setminus\mathcal{N}_1\\
    &Y_cZ_{c'} \prod_{i\in D(\mathcal{N}_{1},\mathcal{N}_{c})\setminus\{1,c\}}Z_i\prod_{j\in \mathcal{N}_{1'}\setminus\{c'\}}Z_j \quad c' \in \mathcal{N}_{1'},c\in \mathcal{N}_{1}\\
  &X_{a'} X_{c'}\prod_{i\in D(\mathcal{N}_{a'},\mathcal{N}_{c'})}Z_i \quad\forall {c'}\in \mathcal{N}_1'\setminus\{ c'\}\\
  &X_{b'} \prod_{i\in \mathcal{N}_{b'}}Z_i \quad\forall b'\in \mathcal{N}\setminus\mathcal{N}_{1'}\\
  \end{split}
  \label{eq:D43}
\end{align}
Now we split the qubits in the set $\mathcal{N}_1\setminus\{ c\}$ into two groups - those that are neighbors of both 1 and $c$ and those that are neighbors of 1 but not $c$. 
\begin{align}
     \begin{split}
    &Z_1Y_{1'}\\
    &X_1 Z_{1'}\\
    &X_{c'}\prod_{i\in \mathcal{N}_1}Z_i \prod_{j\in \mathcal{N}_{c'}\setminus\{1'\}}Z_j \quad {c'}\in \mathcal{N}_1\\
   &X_a X_c\prod_{i\in D(\mathcal{N}_a,\mathcal{N}_c)}Z_i \quad\forall a\in (\mathcal{N}_1\setminus(\mathcal{N}_c\cup\{ c\})\\
   &X_e X_cZ_eZ_c\prod_{i\in D(\mathcal{N}_e,\mathcal{N}_c)}Z_i \quad\forall e\in \mathcal{N}_1\cap\mathcal{N}_c\\
  &X_b \prod_{i\in \mathcal{N}_b}Z_i \quad\forall b\in \mathcal{N}\setminus\mathcal{N}_1\\
    &Y_cZ_{c'} \prod_{i\in D(\mathcal{N}_{1},\mathcal{N}_{c})\setminus\{1,c\}}Z_i\prod_{j\in \mathcal{N}_{1'}\setminus\{c'\}}Z_j \quad c' \in \mathcal{N}_{1'},c\in \mathcal{N}_{1}\\
  &X_{a'} X_{c'}\prod_{i\in D(\mathcal{N}_{a'},\mathcal{N}_{c'})}Z_i \quad\forall {c'}\in \mathcal{N}_1'\setminus\{ c'\}\\
  &X_{b'} \prod_{i\in \mathcal{N}_{b'}}Z_i \quad\forall b'\in \mathcal{N}\setminus\mathcal{N}_{1'}\\
  \end{split}
  \label{eq:D43}
\end{align}
We now multiply the fourth and fifth generators with the seventh generator and the eighth generator with the third one.
\begin{align}
     \begin{split}
    &Z_1Y_{1'}\\
    &X_1 Z_{1'}\\
    &X_{c'}\prod_{i\in \mathcal{N}_1}Z_i \prod_{j\in \mathcal{N}_{c'}\setminus\{1'\}}Z_j \quad {c'}\in \mathcal{N}_1\\
   &Y_a Z_c\prod_{i\in D(\mathcal{N}_a,\mathcal{N}_1)\setminus\{a\}}Z_i \prod_{j\in \mathcal{N}_{1'}\setminus\{c'\}}Z_j\quad\forall a\in (\mathcal{N}_1\setminus(\mathcal{N}_c\cup\{ c\})\\
   &Y_e \prod_{i\in D(\mathcal{N}_e,\mathcal{N}_1)}Z_i\prod_{j\in \mathcal{N}_{1'}\setminus\{c'\}}Z_j \quad\forall e\in \mathcal{N}_1\cap\mathcal{N}_c\\
  &X_b \prod_{i\in \mathcal{N}_b}Z_i \quad\forall b\in \mathcal{N}\setminus\mathcal{N}_1\\
    &Y_cZ_{c'} \prod_{i\in D(\mathcal{N}_{1},\mathcal{N}_{c})\setminus\{1,c\}}Z_i\prod_{j\in \mathcal{N}_{1'}\setminus\{c'\}}Z_j \quad c' \in \mathcal{N}_{1'},c\in \mathcal{N}_{1}\\
  &X_{a'}\prod_{i\in \mathcal{N}_{a'}\cup\mathcal{N}_{1})}Z_i \quad\forall {c'}\in \mathcal{N}_1'\setminus\{ c'\}\\
  &X_{b'} \prod_{i\in \mathcal{N}_{b'}}Z_i \quad\forall b'\in \mathcal{N}\setminus\mathcal{N}_{1'}\\
  \end{split}
  \label{eq:D44}
\end{align}
After applying $P^{\dagger}$ to all qubits in $\mathcal{N}_1$, we get
\begin{align}
     \begin{split}
    &Z_1Y_{1'}\\
    &X_1 Z_{1'}\\
    &X_{c'}\prod_{i\in \mathcal{N}_1}Z_i \prod_{j\in \mathcal{N}_{c'}\setminus\{1'\}}Z_j \quad {c'}\in \mathcal{N}_1\\
   &X_a Z_c\prod_{i\in D(\mathcal{N}_a,\mathcal{N}_1)\setminus\{a\}}Z_i \prod_{j\in \mathcal{N}_{1'}\setminus\{c'\}}Z_j\quad\forall a\in (\mathcal{N}_1\setminus(\mathcal{N}_c\cup\{ c\})\\
   &X_e \prod_{i\in D(\mathcal{N}_e,\mathcal{N}_1)}Z_i\prod_{j\in \mathcal{N}_{1'}\setminus\{c'\}}Z_j \quad\forall e\in \mathcal{N}_1\cap\mathcal{N}_c\\
  &X_b \prod_{i\in \mathcal{N}_b}Z_i \quad\forall b\in \mathcal{N}\setminus\mathcal{N}_1\\
    &X_cZ_{c'} \prod_{i\in D(\mathcal{N}_{1},\mathcal{N}_{c})\setminus\{1,c\}}Z_i\prod_{j\in \mathcal{N}_{1'}\setminus\{c'\}}Z_j \quad c' \in \mathcal{N}_{1'},c\in \mathcal{N}_{1}\\
  &X_{a'}\prod_{i\in \mathcal{N}_{a'}\cup\mathcal{N}_{1})}Z_i \quad\forall {c'}\in \mathcal{N}_1'\setminus\{ c'\}\\
  &X_{b'} \prod_{i\in \mathcal{N}_{b'}}Z_i \quad\forall b'\in \mathcal{N}\setminus\mathcal{N}_{1'}\\
  \end{split}
  \label{eq:D45}
\end{align}
Note that, the neighbors of 1 have undergone neighborhood inversion and all the neighbors of 1 are now connected to all the neighbors of 1'. These stabilizers are consistent with the rules given in Table~\ref{tab:fusion_rules}.

The rules for the third fusion with projection $\frac{\ket{0+}\pm\ket{1-}}{\sqrt{2}}$ can be derived in a similar manner.
\section{Derivation of the graph rule for three-qubit fusions on cluster states}
\label{apx:3fusion}
If we have two cluster states with stabilizer generators 
\begin{align}
     \begin{split}
    &X_1 \prod_{i\in \mathcal{N}_1}Z_i\\
   &X_a \prod_{i\in \mathcal{N}_a}Z_i \quad\forall a\in \mathcal{N}_1\\
  &X_b \prod_{i\in \mathcal{N}_b}Z_i \quad\forall b\in \mathcal{N}\setminus\mathcal{N}_1\\
  \end{split}
  \label{eq:D1}
\end{align}, 
\begin{align}
     \begin{split}
    &X_{1'} \prod_{i\in \mathcal{N}_{1'}}Z_i\\
   &X_{a'} \prod_{i\in \mathcal{N}_{a'}}Z_i \quad\forall a'\in \mathcal{N}_{1'}\\
  &X_{b'} \prod_{i\in \mathcal{N}_{b'}}Z_i \quad\forall b'\in \mathcal{N}\setminus\mathcal{N}_{1'}\\
  \end{split}
  \label{eq:D1'}
\end{align}
and
\begin{align}
     \begin{split}
    &X_{1''} \prod_{i\in \mathcal{N}_{1''}}Z_i\\
   &X_{a''} \prod_{i\in \mathcal{N}_{a''}}Z_i \quad\forall a''\in \mathcal{N}_{1''}\\
  &X_{b''} \prod_{i\in \mathcal{N}_{b''}}Z_i \quad\forall b''\in \mathcal{N}\setminus\mathcal{N}_{1''}\\
  \end{split}
  \label{eq:D1''}
\end{align}
and WLOG fuse qubits 1, 1' and 1'' by projecting them on $\frac{\ket{000}\pm\ket{111}}{\sqrt{2}}$, we can write down the stabilizer generator of the resultant state, given the fusion is successful, using the multi-qubit joint stabilizer measurement rules given in Section~\ref{subsec:GHZproj1}. This projection jointly measures the operators $X_1X_{1'}X_{1''}$, $Z_1Z_{1'}$ and $Z_1Z_{1''}$.
Rewriting the generators such that exactly one generator anti-commutes with $X_1X_{1'}$ and $Z_1Z_{1'}$, each.
\begin{align}
     \begin{split}
    &X_1X_{1'}X_{1''} \prod_{i\in \mathcal{N}_1}Z_i\prod_{i\in \mathcal{N}_{1'}}Z_i\prod_{k\in \mathcal{N}_{1''}}Z_k\\
   &X_c \prod_{i\in \mathcal{N}_c}Z_i \quad c\in \mathcal{N}_1\\
    &X_cX_a \prod_{i\in D(\mathcal{N}_a,\mathcal{N}_c)}Z_i \quad\forall a\in \mathcal{N}_1\setminus\{c\}\\
  &X_b \prod_{i\in \mathcal{N}_b}Z_i \quad\forall b\in \mathcal{N}\setminus\mathcal{N}_1\\
    &X_{1'} \prod_{i\in \mathcal{N}_{1'}}Z_i\\
   &X_cX_{a'} Z_1Z_{1'}\prod_{i\in D(\mathcal{N}_{a'},\mathcal{N}_c)\setminus\{1,{1'}\}}Z_i \quad\forall a'\in \mathcal{N}_{1'}\\
  &X_{b'} \prod_{i\in \mathcal{N}_{b'}}Z_i \quad\forall b'\in \mathcal{N}\setminus\mathcal{N}_{1'}\\
  &X_{1''} \prod_{i\in \mathcal{N}_{1''}}Z_i\\
   &X_cX_{a''} Z_1Z_{1''}\prod_{i\in D(\mathcal{N}_{a''},\mathcal{N}_c)\setminus\{1,{1'}\}}Z_i \quad\forall a''\in \mathcal{N}_{1''}\\
  &X_{b''} \prod_{i\in \mathcal{N}_{b''}}Z_i \quad\forall b''\in \mathcal{N}\setminus\mathcal{N}_{1''}\\
  \end{split}
  \label{eq:D2}
\end{align}

After the measurement - 
\begin{align}
     \begin{split}
     & \prod_{i\in \mathcal{N}_1}Z_i\prod_{i\in \mathcal{N}_{1'}}Z_i\prod_{k\in \mathcal{N}_{1''}}Z_k\\
   &X_1X_{1'}X_{1''}\\
    &X_cX_a \prod_{i\in D(\mathcal{N}_a,\mathcal{N}_c)}Z_i \quad\forall a\in \mathcal{N}_1\setminus\{c\}\\
  &X_b \prod_{i\in \mathcal{N}_b}Z_i \quad\forall b\in \mathcal{N}\setminus\mathcal{N}_1\\
    &Z_1Z_{1'}\\
   &X_cX_{a'} \prod_{i\in D(\mathcal{N}_{a'},\mathcal{N}_c)\setminus\{1,{1'}\}}Z_i \quad\forall a'\in \mathcal{N}_{1'}\\
  &X_{b'} \prod_{i\in \mathcal{N}_{b'}}Z_i \quad\forall b'\in \mathcal{N}\setminus\mathcal{N}_{1'}\\
  &Z_1Z_{1''}\\
   &X_cX_{a''} \prod_{i\in D(\mathcal{N}_{a''},\mathcal{N}_c)\setminus\{1,{1'}\}}Z_i \quad\forall a''\in \mathcal{N}_{1''}\\
  &X_{b''} \prod_{i\in \mathcal{N}_{b''}}Z_i \quad\forall b''\in \mathcal{N}\setminus\mathcal{N}_{1''}\\
  \end{split}
  \label{eq:D3}
\end{align}

This state is not a cluster state. We need to apply a Hadamard on one of the qubits in the neighborhood of qubit 1 or 1'. WLOG, we apply Hadamard on qubit $c\in\mathcal{N}_c$. We rewrite Eq.~\ref{eq:D4} by splitting the qubits into into three categories - neighbors of both 1 and $c$, neighbors of 1 but not that of $c$, and neighbors of $c$ but not that of $1$.
\begin{align}
     \begin{split}
  &X_1X_{1'}X_{1''}\\
    &Z_1Z_{1'}\\
    &Z_1Z_{1''}\\
   &Z_c\prod_{i\in \mathcal{N}_1\setminus\{c\}}Z_i\prod_{i'\in \mathcal{N}_{1'}}Z_{i'}\prod_{k\in \mathcal{N}_{1''}}Z_k\\
    &X_cX_eZ_c \prod_{i\in D(\mathcal{N}_e,\mathcal{N}_c)\setminus\{c\}}Z_i \quad\forall e\in (\mathcal{N}_1\cap\mathcal{N}_c)\\
    &X_cX_a \prod_{i\in D(\mathcal{N}_a,\mathcal{N}_c)}Z_i \quad\forall a\in \mathcal{N}_1\setminus(\mathcal{N}_c\cup\{c\})\\
  &X_d Z_c\prod_{i\in \mathcal{N}_d\setminus\{c\}}Z_i \quad\forall d\in \mathcal{N}_{c}\setminus\mathcal{N}_1\\
  &X_b \prod_{i\in \mathcal{N}_b}Z_i \quad\forall b\in \mathcal{N}\setminus\mathcal{N}_1\\
   &X_cX_{a'} \prod_{i\in D(\mathcal{N}_{a'},\mathcal{N}_c)\setminus\{1,{1'}\}}Z_i \quad\forall a'\in \mathcal{N}_{1'}\\
  &X_{b'} \prod_{i\in \mathcal{N}_{b'}}Z_i \quad\forall b'\in \mathcal{N}\setminus\mathcal{N}_{1'}\\
   &X_cX_{a''} \prod_{i\in D(\mathcal{N}_{a''},\mathcal{N}_c)\setminus\{1,{1'}\}}Z_i \quad\forall a''\in \mathcal{N}_{1''}\\
  &X_{b''} \prod_{i\in \mathcal{N}_{b''}}Z_i \quad\forall b''\in \mathcal{N}\setminus\mathcal{N}_{1''}\\
  \end{split}
  \label{eq:D5}
\end{align}

After $H_c$
\begin{align}
     \begin{split}
   &X_1X_{1'}X_{1''}\\
    &Z_1Z_{1'}\\
    &Z_1Z_{1''}\\
   &X_c\prod_{i\in \mathcal{N}_1\setminus\{c\}}Z_i\prod_{i'\in \mathcal{N}_{1'}}Z_{i'}\prod_{k\in \mathcal{N}_{1''}}Z_k\\
    &Z_cX_eX_c \prod_{i\in D(\mathcal{N}_e,\mathcal{N}_c)\setminus\{c\}}Z_i \quad\forall e\in (\mathcal{N}_1\cap\mathcal{N}_c)\\
    &Z_cX_a \prod_{i\in D(\mathcal{N}_a,\mathcal{N}_c)}Z_i \quad\forall a\in \mathcal{N}_1\setminus(\mathcal{N}_c\cup\{c\})\\
  &X_d X_c\prod_{i\in \mathcal{N}_d\setminus\{c\}}Z_i \quad\forall d\in \mathcal{N}_{c}\setminus\mathcal{N}_1\\
  &X_b \prod_{i\in \mathcal{N}_b}Z_i \quad\forall b\in \mathcal{N}\setminus\mathcal{N}_1\\
   &Z_cX_{a'} \prod_{i\in D(\mathcal{N}_{a'},\mathcal{N}_c)\setminus\{1,{1'}\}}Z_i \quad\forall a'\in \mathcal{N}_{1'}\\
  &X_{b'} \prod_{i\in \mathcal{N}_{b'}}Z_i \quad\forall b'\in \mathcal{N}\setminus\mathcal{N}_{1'}\\
  &Z_cX_{a''} \prod_{i\in D(\mathcal{N}_{a''},\mathcal{N}_c)\setminus\{1,{1''}\}}Z_i \quad\forall a''\in \mathcal{N}_{1''}\\
  &X_{b''} \prod_{i\in \mathcal{N}_{b''}}Z_i \quad\forall b''\in \mathcal{N}\setminus\mathcal{N}_{1''}\\
  \end{split}
  \label{eq:D6}
\end{align}
Multiplying the fifth and seventh generators with the fourth generator to eliminate $X_c$ -
\begin{align}
     \begin{split}
    &X_1X_{1'}X_{1''}\\
    &Z_1Z_{1'}\\
    &Z_1Z_{1''}\\
   &X_c\prod_{i\in \mathcal{N}_1\setminus\{c\}}Z_i\prod_{i'\in \mathcal{N}_{1'}}Z_{i'}\prod_{k\in \mathcal{N}_{1''}}Z_k\\
    &X_eZ_c \prod_{i\in D(\mathcal{N}_1,D(\mathcal{N}_e,\mathcal{N}_c))\setminus\{c\}}Z_i \prod_{i'\in \mathcal{N}_{1'}}Z_{i'}\quad\forall e\in (\mathcal{N}_1\cap\mathcal{N}_c)\\
    &Z_cX_a \prod_{i\in D(\mathcal{N}_a,\mathcal{N}_c)}Z_i \quad\forall a\in \mathcal{N}_1\setminus(\mathcal{N}_c\cup\{c\})\\
  &X_d \prod_{i\in D(\mathcal{N}_1,\mathcal{N}_d)}Z_i\prod_{i'\in \mathcal{N}_{1'}}Z_{i'} \quad\forall d\in \mathcal{N}_{c}\setminus\mathcal{N}_1\\
  &X_b \prod_{i\in \mathcal{N}_b}Z_i \quad\forall b\in \mathcal{N}\setminus\mathcal{N}_1\\
   &Z_cX_{a'} \prod_{i\in D(\mathcal{N}_{a'},\mathcal{N}_c)\setminus\{1,{1'}\}}Z_i \quad\forall a'\in \mathcal{N}_{1'}\\
  &X_{b'} \prod_{i\in \mathcal{N}_{b'}}Z_i \quad\forall b'\in \mathcal{N}\setminus\mathcal{N}_{1'}\\
   &Z_cX_{a''} \prod_{i\in D(\mathcal{N}_{a''},\mathcal{N}_c)\setminus\{1,{1''}\}}Z_i \quad\forall a''\in \mathcal{N}_{1''}\\
  &X_{b''} \prod_{i\in \mathcal{N}_{b''}}Z_i \quad\forall b''\in \mathcal{N}\setminus\mathcal{N}_{1''}\\
  \end{split}
  \label{eq:D7}
\end{align}
From the equation above, it is easy to see that all the qubits in $\mathcal{N}_{1'}$ and $\mathcal{N}_{1''}$ are now neighbors of $\mathcal{N}_c\cup\{c\}$ and from Eq.~\ref{eq:B5}, the neighborhood of 1 has changed to $((G.1).c).1$.

\end{document}